# FAREWELL TO WESTPHALIA

# FAREWELL TO WESTPHALIA

*Crypto Sovereignty and*
*Post-Nation-State Governance*

**JARRAD HOPE**      **PETER LUDLOW**

LOGOS PRESS ENGINE

*... dedicated to Julian Assange and to the memory of Hal Finney.*

This work is the product of our labour, but it is also the product of the labour of many others, including those who have helped us, argued with us, taught us, inspired us and encouraged us.

Among those we wish to thank are Carl Bennetts, Vitalik Buterin, Mike Lorrey, Troy Cross, Andrew Bailey, Bradley Rettler, Craig Warmke, Brady Dale and the team at the Institute of Free Technology. Special thanks to Mike Lorrey (aka Intlibber Braughtigan) for encouragement and for suggesting the title. Thanks to Marvin Jones for comments on an earlier draft and to Rick Delaney for next-level editing and guidance.

An earlier and very different version of Section 15.2 previously appeared under Ludlow's *nom de plume*, EJ Spode, in *Aeon* magazine as 'The great cryptocurrency heist'.

# CONTENTS

*Arise, you have nothing to lose but your barbed wire fences!*

**– Timothy May,**
*The Crypto Anarchist Manifesto*

# INTRODUCTION

## Pursuing Decentralised-Yet-Cooperative Governance

In 1648, the Peace of Westphalia marked the end of the Thirty Years' War – an event in which Catholics and Protestants had clashed across Northern Europe. The war had certainly been grim business. Precise numbers are difficult to come by, but apart from the two million soldiers that perished, it is estimated that rural areas of what is now Germany lost over 60% of their population due to war, starvation and disease. Cities lost around a third of their populations. An entry written in the margins of a family bible in Swabia recorded, 'We live like animals, eating bark and grass. No one could have imagined that anything like this would happen to us. Many people say that there is no God.'[1] By some accounts, the situation in the Rhineland had grown so desperate that people were resorting to cannibalism.

The treaties enshrined in the Peace of Westphalia not only ended the conflict but made sovereign nation states a permanent fixture in our world.[2] It was a solution in which all parties agreed that what nation

---

[1] Simon Adams, *The Thirty Years' War*, 2nd ed. (London, 1997), 160.

[2] Contrary to common wisdom, the Peace of Westphalia does not refer to a single treaty but is shorthand for two separate treaties signed in two separate locations on 24 October 1648: the Instrumentum Pacis Monasteriense or the Treaty of Münster, signed between the Holy

states do internally is their own business. The agreements successfully ended the Thirty Years' War; this cannot be denied. However, in view of the wars and genocide that swept across Europe since the Peace of Westphalia – including the Napoleonic Wars, the Franco–Prussian War and two World Wars in the twentieth century – one has to question how much was accomplished by the invention of the modern nation state. The question gains even more bite when we query the wisdom of European post-Westphalian colonial powers imposing their artificial nation-state boundaries onto territories around the world with complete indifference to previous tribal boundaries and conflicts.

Sovereign nation states are human technologies designed to facilitate the peaceful organisation of human beings, solving for their ideological, political and religious differences. However, they are technologies that are nearly 380 years old and, like all technology of that era, perhaps not the optimal solutions available today. To put things in perspective, the Peace of Westphalia was reached six years after the invention of the mechanical adding machine by Blaise Pascal and eight years before the invention of the pendulum clock by Christiaan Huygens. These are great inventions, to be sure, but we do not consider them to be the end of history. They have been followed by newer and superior versions of those technologies. Why should our technologies for political organisation be set in stone? Can we not do better? We believe that we can. Our solution will involve a thorough investigation into developments like blockchain technologies and smart contracts and showing how they can be productively applied to the project of cooperative human governance.

---

Roman Empire and the King of France, and the Instrumentum Pacis Osnabrugense or the Treaty of Osnabrück, signed between the Holy Roman Empire and Queen of Sweden. Arguably, there is a third treaty in this mix since Dutch autonomy from Spanish Habsburg Rule was not part of the Münster–Osnabrück settlements but was established in January of 1648 in a separate Treaty of Münster. For further discussion, see Andreas Osiander, 'Sovereignty, International Relations, and the Westphalian Myth', *International Organization*, 55/2 (2001), 251–87 <https://www.jstor.org/stable/3078632> [accessed 28 April 2023].

Since the Bitcoin white paper was published by Satoshi Nakamoto in 2008,[3] there has been plenty of talk about what blockchain technology can (and cannot) accomplish, and most of that talk has assumed that the principal application of blockchain technology will be in the financial sector (for example, as payment systems or as reserve assets or as asset exchanges). However, the central idea of this book is that applying blockchain technology to human governance will be, by far, its most important application.

Let us be more explicit about this central idea. First, let us define what we mean by 'human governance'.

'Human governance' refers to the systems and processes by which people manage and make decisions about their communities and implement those decisions to achieve some political, economic or cultural goals.

Next, there is the question of what blockchain technology is for, and on its deepest and most general level, its function can be described as a tool that allows humans to organise their activities in a way that is decentralised yet cooperative. This leads us back to the primary thesis of this book: the principal application of blockchain technology will be in the facilitation of human governance by providing platforms for decentralised-yet-cooperative human activities.

The key benefit of this new technology, so harnessed, will be that it will allow humans to resist systems of centralised power and develop alternatives that enable them to cooperate with each other in governing their affairs. In other words, it will facilitate decentralised cooperation in the context of human governance.

We know the problems with centralised systems of governance. They give rise to tyranny, they are susceptible to corruption and they present one single point of failure. Decentralised systems, on the other hand, are resilient in the sense that they have no single point of failure and are resistant to corruption.

---

[3] Satoshi Nakamoto, 'Bitcoin: A Peer-to-Peer Electronic Cash System' (2008) <https://bitcoin.org/bitcoin.pdf>.

While they may be resilient in this sense, the standard view is that decentralised systems lack efficiency. Many people think that when we avoid centralised authority, we are looking at the collapse of human organisation. How else can we get everyone on the same page? There needs to be a boss to coordinate us, no? The thought has been that centralised systems, despite their obvious flaws, are at least efficient systems for governance. This is the alleged resilience-versus-efficiency tradeoff with respect to human governance. As we will see, blockchains allow us to find a better solution to the tradeoff between resilience and efficiency – i.e. blockchains can offer a similar level of efficiency to centralised systems while still being resilient and corruption resistant.

As stated, the promise of blockchain technologies is that they allow authority to be decentralised, yet they enable people to organise themselves cooperatively. There are many reasons why this would be a positive outcome. To begin, let us consider the very simple example of government archives. Governmental archives are more secure when decentralised. They become anti-fragile. They become difficult to censor, difficult to destroy, difficult to tamper with and, importantly, difficult to hide from community members, thus making governance more transparent.

Transparency is a key issue. The French philosopher Jacques Derrida received a lot of criticism for being an obscurantist, but he was very clear on this point, at least: archives may preserve documents, but they can also be a place where documents go to disappear from view.[4] That is not a happy outcome. We want documents to be safe, but we do not want them hidden in a safe in the wall of a building somewhere. Important documents should be visible to all.

Transparent, immutable records, as we will see, are critical to good governance. However, so too are secure communications. Furthermore,

---

[4] Jacques Derrida, *Archive Fever: A Freudian Impression* (Chicago, IL, 1998).

it is important that the intentions of the government are transparent and that promises are kept. Decentralised governmental systems can facilitate these features via 'smart contracts' – contracts encoded as computer programs and deployed on the blockchain.

Bitcoin is just one illustration of how this sort of decentralising technology works, for at its most abstract level, Bitcoin is a protocol which is very decentralised and in which users are all on the same page (or more precisely, they are on pages that say the same thing). Specifically, there is no centralised ledger that keeps track of who owns what or who sent what to whom, but there is a distributed ledger, meaning that the information is held by many individuals on the network. Therefore, everyone can be confident that they are reading the same thing regarding who owns what and so on. Our point in this book is that decentralised cooperation can be applied to much more than digital currencies. It can be applied to all forms of human governance.

This groundbreaking idea does not actually begin with Satoshi's white paper but rather with several decades of important earlier work on distributed systems in computer science. One example we will discuss is the 'Paxos protocol', which was inspired by ideas about distributed organisation in a community of ancient Greek citizens that were constantly on the move.[5] In that and related research, the question was this: How do we organise a system that has multiple computer processors (multiple electronic brains, as it were) as opposed to a single processor (like the CPU in your laptop computer)? How does such a system stay organised for a single, unified purpose? What happens when some of the many processors in the system fail or begin operating at cross purposes? Will this not lead to a system with many points of failure and, thus, organisational collapse? As it turns out, the answer is no. But the system must be designed correctly.

---

[5] Leslie Lamport, 'The Part-Time Parliament', *ACM Transactions on Computer Systems*, 16/2 (1998), 133–69 <https://dl.acm.org/doi/10.1145/279227.279229> [accessed 5 November 2024].

You are already acquainted with some distributed systems that are successfully organised in such a way. Your brain is one such system. Unlike your laptop, there is no central processing unit running operations in your brain. Rather, there are many smaller nodes that are coordinated towards a common goal, and the entire system carries on even when multiple nodes fail. Research over the last several decades has helped illuminate the mathematics of such 'fault-tolerant' distributed systems, and blockchain technology is the product of that research. Applied to human governance, the promised outcome is a way for people to peacefully and effectively organise themselves without resorting to centralised authorities.

When we hear the word 'governance', we are apt to think of great governmental institutions like nation states and their centres of power; we may form a mental image of the capitol building in Washington, D.C. or the Kremlin in Moscow. Of course, human governance is much more extensive than what happens in the centres of power of nation states. Indeed, obviously, governance also applies to state legislatures and to city councils and, less obviously, to homeowner associations and even to condo boards. At the end of the day, nation states and their centres of political power are just the tip of the iceberg regarding human governance.

Some type of governance seems to be involved in the organisation of human flourishing (and failure thereof) at every level of granularity, from the United Nations, the Organization of American States, and the African Union to labour union meetings, meetings of church deacons, faculty meetings, and even scout troop meetings. And, of course, private organisations like corporations have systems of governance as well. Governance is everywhere. However, even these examples and cases like them do not begin to illuminate how vast and all-encompassing human governance is and how much activity it facilitates, obstructs and, ultimately, controls.

Let us start by looking at this in economic terms. The vast majority of human wealth is within the control of these human governance structures. Even if we focus solely on governance in a coarse-grained way – i.e. the governance that is done by what we recognise as official nation states – the amount of wealth made possible by governance is staggering. Tom W. Bell, in his book *Your Next Government?: From the Nation State to Stateless Nations*, cites statistics from a 2000 study by the World Bank that put the percentage of global wealth attributable to traditional government activity at 44%. By contrast, according to the study, only 5% of the world's wealth is attributable to natural assets like oil, gold and timber, and only 18% of the world's wealth is generated through manufacturing things like gasoline,  jewellery and lumber.[6] How is this even possible?

Rather than focus on the exact numbers, for the moment, let us consider the more abstract question of why governance is extremely important in the creation of wealth. Let us begin with the case of natural resources. Consider a gold mine, for example. If there is no governance, there is no one to control or even keep track of who owns the land, which means it belongs to whoever can grab it, and people will grab it just in case the cost of grabbing it is less than the cost of what can be extracted from it.

We seldom think about the role of government in tracking property ownership, but even in countries like Mexico, which currently has the twelfth largest economy in the world,[7] the system of property ownership can break down. Until a few years ago, there was a system in which a single *notaría* kept a record of property ownership, but notarías were notoriously susceptible to being bribed or intimidated into changing ownership records. You could lose your property with a change to a single document.

---

[6] Tom W. Bell, *Your Next Government? From the Nation State to Stateless Nations* (Cambridge, 2017)

[7] Caleb Silver, 'The Top 25 Economies in the World', *Investopedia*, 10 April 2024 <https://www.investopedia.com/insights/worlds-top-economies/> [accessed 23 October 2024].

In countries like the United States, there is an extra level of safety in the form of title insurance, but of course, insurance markets do not operate in a vacuum. Governments are involved in regulating insurance markets and requiring that insurance companies have the resources to pay what they have promised to pay and that they pay legitimate claims. Sometimes, disputes arise over insurance claims and land claims and every other sort of claim, and these claims require a court system that is expeditious and fair. This may seem like a small ask, but it is a huge one in many parts of the world.

Even if you can maintain control of the territory where your gold mine operates, you need to rely on governments (or something like them) for additional help. If you do not have your own electrical generation plant, you need to rely on a system that delivers power, and you need lines of communication. You need to bring workers and equipment to your mine, and you need to safely transport your extracted gold to a well-functioning (non-corrupt) market.

Of course, you do not need traditional governments to do all of this. You could rely on private militias and security firms, and you might work out a financial arrangement with whoever controls the roads and power grid. Maybe you prefer such a system. Perhaps if security, record keeping and so on were private matters, it might be more obvious to us just how much value is added by these activities (or subtracted if they are not handled well).

Our point is that even if these activities are taken away from traditional governments and placed in the hands of private enterprises free from traditional government oversight, there is still plenty of governance that has to take place. Whether that governance occurs in traditional governmental institutions or the private sector, immutable records must be preserved yet, at the same time, made available; disputes must be fairly adjudicated; and decisions must be enforced. The amount of wealth that relies on the proper execution of this form of governance is staggering.

Since we began with the example of a gold mine, let us use the example of gold to put things in perspective. As of 2020, around $11 trillion of our global wealth was tied up in above-ground gold. That is a lot. It is, as of this writing, more than ten times the value of all the Bitcoin in the world. However, it is also a drop in the bucket. The Fortune 500 companies have a combined value of $90 trillion.[8] Global real estate clocks in at $280 trillion. The sum total of all global wealth is around $380 trillion.[9]

When we ask how much of global wealth is created by government, we are really asking how much of that value would exist if there was no government (or something performing its governance function). Take the example of real estate; its global value is $280 trillion, but all of those deeds would not be worth the paper they are written on if there was no mechanism in place to ensure that you actually owned the property in question and that no warlord could simply snatch it from you.

It may seem farfetched to think real estate could simply be taken by the stroke of a pen or the brandishing of a sword, but this reaction stems from the fact that many readers of this book will live in relatively stable economies with stable governance structures. As we will see, there are many countries where ownership of property is tenuous, and history has countless examples of property being seized and records of ownership destroyed.

Another way to illustrate the situation is with Tom W. Bell's 'law bomb' thought experiment. Bell asks that we imagine something like a neutron bomb that did not harm people or physical objects but which could eliminate the rule of law. If such a bomb were to be detonated, much of the wealth we have acquired would be wiped out.[10] Property has little value when there can be no legal guarantee of ownership, and similarly, basic commerce, transportation and communication would have to be reinvented for a world without law.

---

[8] NGRAVE, 'Too Big to Fail? Crypto Market Size vs Traditional Assets', 2022 <https://medium.com/ngrave/too-big-to-fail-crypto-market-size-vs-traditional-assets-eff4b-b2ec529> [accessed 23 October 2024].

[9] Even here, we are ignoring the wealth generated by derivative markets, which dwarf everything we have discussed thus far. As of this writing, the amount of wealth in the derivative markets is over a quadrillion dollars.

Our point here is that the practice of governance, whether private or public or some hybrid, is critical to the maintenance and growth of wealth. As we will see, economic problems are not the only consequences of governance failures.

In the next chapter, we will discuss consequences like wars and genocides, and it is important to keep in mind that these are also examples of failures of governance, even if we do not always think of them that way. If human society governed itself effectively, these things would be far less frequent. Beyond this, there are many other consequences of failed governance that cut to the very heart of human flourishing. Human governance, if it is working well, not only enables the generation and preservation of wealth but also enables the flourishing of human cultures and assists individuals in securing their personal freedom and in their pursuit of happiness and wellbeing.

Governance, whether it comes in the form of public governments or other forms of human governance, is absolutely critical to every aspect of our lives. The trouble is that it often seems to be broken. The question is, what can we do about this? As we noted earlier, our aim in this book is to introduce new technologies and corresponding values that can help us govern better. Those technologies (in particular, blockchain technologies) will facilitate decentralised governance and human cooperation.

As stated, good governance can generate wealth for its people, and bad governance can debase the wealth of its people. Indeed, especially corrupt governance can be worse than no governance at all. And here is the problem: because there is so much money created by governing institutions and because so much money passes through governing institutions and (although we have not talked about it yet) because so much private sector wealth is regulated by such institutions, governing institutions of every stripe are an absolute magnet for corrupt individuals. Willie Sutton once famously said that he robbed banks 'because that is where the money is', but the problem is that Sutton was wrong about that. If you want to find the money, you have to find the central authority in control of organising economic activities. That is where the money really is. Corrupt individuals – our modern-day bank robbers – know this well.

---

[10] Bell, *Your Next Government?*

All of this leaves us with the question: Is it actually necessary for governments and other governing institutions and organisations to be susceptible to corruption and incompetence? Is corruption the inevitable consequence of human governance? We think not. And here we return to the issue of governance technology, for as we said, the tools that are used for governance today are obsolete.

Since Satoshi published the Bitcoin white paper in 2008, the world has seen an explosion of tools that leverage blockchain technology to provide universally transparent and immutable records of financial transactions and human activity. At the same time, we have developed other tools that allow us to carry out our private affairs in private. In effect, we now have tools that make government activity transparent and immutable and our personal business personal and private.

However, as we also noted earlier, the really profound idea grounding Satoshi's white paper was not just applicable to cryptocurrencies but also to how we can bring about decentralised governance. Most governance today has been aided by technologies that have helped centralised governance structures to prosper. These include technologies for surveillance and technologies to facilitate the application of force against a restive population. In fact, technology has been facilitating centralisation since the Bronze Age, as evidenced by bronze swords and armour helping to unify the Mesopotamian empire.

The revolutionary aspect of fault-tolerant distributed systems is the idea that we can decentralise authority, and while that generates more points of attack, we can utilise the mathematics of Byzantine fault tolerance to engineer systems that can absorb those attacks and survive despite them. We will go into some detail about Byzantine fault tolerance in Chapter 5, but for now, it is enough to understand that we can engineer our governing technologies in a way that avoids central points of attack and can absorb local failures. Indeed, we can emerge from those failures stronger.

Finally, it will be important to keep in mind that no technology can be successful on its own. Technologies succeed or fail (sometimes fail spectacularly) depending on whether they are designed to have human beings as part of the technological system and whether the attitudes of these humans align with the goals of the technology.

In this case, we think that the fundamental values that we should prize, and that are inherent in the new technologies we offer here, are those of decentralisation, cooperation, corruption resistance and transparency.

Whatever your view about governance, the real dividing lines are between systems of governance that are centralised and those that are not. Thus, the issue is not whether you are a socialist or a free-market capitalist but whether you are a centraliser or a decentraliser. Just as there can be decentralised socialism,[11] there can also be centralised capitalism. As we write this, Silicon Valley is full of centralised powers (from Google and Facebook to Apple and Microsoft) that hold great control over our lives, and they are nothing if not centralised centres of governance in the information age.

---

[11] And yes, there actually are versions of socialism that reject centralisation, perhaps most famously Marx's contemporary, Mikhail Bakunin, who is often classified as an anarchist socialist. In Bakunin's view, Marx wanted to seize the reins of centralised authority for his own ends. As he put the fundamental problem: 'Marx's program is a complete fabric of political and economic institutions, strongly centralized, and very authoritarian'. Prophetically, Bakunin wrote of Marx's project: 'There will therefore be no longer any privileged class, but there will be government and, note this well, an extremely complex government, which will not content itself with governing and administering the masses politically, as all governments do today, but which will also administer them economically, concentrating in its own hands the production and just division of wealth, the cultivation of land, the establishment and development of factories, the organization and direction of commerce, finally the application of capital to production by the only banker, the State.' See Mikhail Bakunin, 'Marxism, Freedom and the State', in *Selected Writings from Mikhail Bakunin* (St. Petersburg, FL, 2010).

We believe – and will argue – that decentralised communities and blockchain governance (at every level) are not only feasible but are on the immediate horizon. Furthermore, the seeds have already been planted. Our aim is to nurture those embryonic forms of blockchain governance and make their future adoption as frictionless as possible. Thus, we also aim to facilitate the development of new forms of blockchain governance. In Chapter 14, we provide a toolbox of resources for building blockchain governance at every level. This toolbox includes tools for communication, commerce and security.

Of course, you do not need to use our tools; you can make your own. In fact, we have made all of our tools open source, so you can copy them or modify them and mix and match them as you see fit. There is no right way to build a system of blockchain governance. Or rather, there are lots of promising ways to build blockchain communities and engineer new forms of blockchain governance at every level – even in your local homeowner association meetings. In the fullness of time, blockchain governance will take many forms, encoding diverse values and principles, and adopting different goals. Our aim here is to help that happen – to facilitate the weaving of a beautiful tapestry of diverse human projects, all with an eye to nurturing diverse human cultures, values, plans and public goods.

# NATION STATES ARE OBSOLETE GOVERNANCE TECHNOLOGIES

## 2.1  Preliminaries

In 1994, during the Rwandan genocide, Immaculée Ilibagiza, a member of the Tutsi tribe, hid in a secret space in her pastor's house, listening to Hutu tribesmen who were armed with machetes as they searched for her. She heard them say that they needed to terminate the Tutsi 'cockroaches', and then, ' "She's here . . . we know she's here somewhere. Find her – find Immaculée." '

Immaculée heard one of the Hutus brag that he had killed 399 cockroaches and wanted her to bring his record to an even 400. In her book *Left to Tell*, Immaculée described the experience:

> I tried to swallow, but my throat closed up. I had no saliva, and my mouth was drier than sand. I closed my eyes and tried to make myself disappear, but their voices grew louder. I knew that they would show no mercy, and my mind echoed with one thought: *If they catch me, they will kill me. If they catch me, they will kill me. If they catch me, they will kill me.*[1]

---

[1] Immaculée Ilibagiza and Steve Erwin, *Left to Tell: Discovering God Amidst the Rwandan Holocaust* (Carlsbad, CA, 2006).

When Immaculée finally left her hiding spot three months later, she learned about the damage that had been done to her family and tribe. Her father, her mother and two of her three brothers had been butchered. Uncounted neighbours perished as well. The estimates of her Tutsi tribespeople that had been killed ranged from half a million to a million.[2] The number of Tutsi women raped ranged from 200,000 to a quarter of a million.[3] The psychological damage done to survivors was incalculable.

In the wake of this atrocity, it was natural to talk about the evil of the Hutu murderers, and to be sure, the murderers cannot and should not be absolved. However, it also needs to be observed that these events may not have happened at all had two different tribes with diverse cultures and histories not been kettled together within the boundaries of a single nation state, with those boundaries drawn by colonial European powers.

Indeed, once kettled, the question became which of the two tribes would rule the other. When Germany controlled Rwanda, the minority Tutsi were placed in power, with consequences that set in motion a deep resentment that festered up to and through the horrific genocide of 1994. To put it another way, whatever the benefits to Europe that accrued from the Peace of Westphalia, the attempt to impose this political technology on Africa, ignoring tribal boundaries and creating new and artificial nation-state boundaries, has been disastrous.

Sadly, the Rwandan story is not unique. Even today, our political landscape is full of examples of ethnic groups forced to live together within nation states, with one group lording power over the other and, all too often, resulting in attempted genocide. Modern states are, as we write, the settings for numerous examples. The victims include the Rohingya in Myanmar, the Nuer in South Sudan, Christians and

---

[2] Luc Reydams, '"More than a Million": The Politics of Accounting for the Dead of the Rwandan Genocide', *Review of African Political Economy*, 48/168 (2021), 168–256.

[3] Gérard Prunier, *The Rwanda Crisis, 1959–1994: History of a Genocide*, 1st ed. (London, 1998).

Yazidis in Iraq and Syria, Christians and Muslims in the Central African Republic, Darfuris in Sudan, and the list goes on.

Even when people are not dying on a genocidal scale, rights are often trampled upon in nation states. If inhabitants have diverse values, and someone is in charge of enforcing one set of values, then someone's interests are being protected and someone else's are not. Thus, the result for nation states like the United States is what the media calls 'culture wars' – conflicts that generate lots of existential anguish and a few hate crimes and murders, but no mass killings. At least, not yet.

You might be thinking that all these examples of genocide and repression are indeed terrible, but are not nation states the only viable option? Is there not an alternative system of government that can avoid such outcomes or that can at least allow people to safely escape such situations when they implode? What is the alternative to nation states?

## 2.2   Nation states are not the only option

Sometimes, it seems that nation states have been with us forever. Before the United Nations, after all, there was the League of Nations, and before that, there were surely nations, no? But while nations and, in particular, nation states seem to be set in stone in our world today, it was not always so.

Indeed, it was not so long ago that people used alternative systems to organise themselves for political and economic purposes. There have been kingdoms and city states, empires and duchies, federations of city states, caliphates, palatinates, papal states, clans, tribes, and 'nations' of people sharing common heritage but not yet organised into states. Indeed, until the arrival of European powers, the entire continent of North America was principally organised into tribes, with occasional empires rising up and then fading away.[4]

---

[4] Felipe Fernandez-Armesto, *Before Columbus: Exploration and Colonization from the Mediterranean to the Atlantic, 1229-1492* (Philadelphia, PA, 1987).

The fact is that nation states are relatively recent inventions. They are a form of government that has its roots in the Protestant–Catholic wars that swept through Northern Europe after the Reformation and a form of government designed to solve a very particular problem: How do we contain conflicts over religion and ideology that might otherwise sweep across continents consuming everything in their path?

As we noted in Chapter 1, nation states as we know them were born in 1648 with the aforementioned Peace of Westphalia – the treaties that marked the end of the Thirty Years' War. Leo Gross, writing in the *American Journal of International Law* on the 300th anniversary of the Peace (shortly after the founding of the United Nations), described the Peace of Westphalia as 'the majestic portal which leads from the old into the new world.'[5] According to the standard view, the Peace of Westphalia secured peace by establishing the convention that we should not look too closely at what goes on within national boundaries. National boundaries could thus serve as bulkheads against waves of violence that might otherwise sweep across continents.

Fifty-six years later, Henry Kissinger echoed this take in his book, *World Order*:

> The Peace of Westphalia became a turning point in the history of nations because the elements it set in place were as uncomplicated as they were sweeping. The state, not the empire, dynasty, or religious confession was affirmed as the building block of European order. The concept of state sovereignty was established. The right of each signatory to choose its own domestic structure and religious orientation free from intervention was affirmed, while novel clauses ensured that minority sects could practice their faith in peace and be free. Beyond the immediate demands of the moment, the principles of a system of 'international relations' were taking shape, motivated by the common desire to avoid a recurrence of total war on the Continent.[6]

---

[5] Leo Gross, 'The Peace of Westphalia, 1648–1948', *American Journal of International Law*, 42/1 (1948), 20–41 <https://www.cambridge.org/core/journals/american-journal-of-international-law/article/abs/peace-of-westphalia-16481948/80489D3C080D4CDD97C7EDC0354DC37F> [accessed 29 April 2023].

Not everyone buys this interpretation of the Peace of Westphalia, and some have dubbed Leo Gross 'the Homer of the Westphalia myth',[7] but the critics are not so much troubled by the thought that nation states are inventions as they are with the concern that Gross was attempting to canonise the legitimacy of nation states by giving them a date of birth and connecting them with important international accords. Cormac Shine, writing in *History Today*, complained that the Westphalia myth is designed to 'make the formation of the existing settlement seem inevitable. Any alternatives outside the realm of sovereign states are discounted.'[8]

Similarly, in this book, our idea is that we should not think of nation states as inevitable or necessary. They are human inventions and, perhaps, not particularly good ones, despite their canonisation in international law. Two things should be kept in mind here. First, if the idea of nation states was to have a system of human political organisation that avoided wars of religion and ideology sweeping across continents (or the globe, for that matter), they have not been all that successful. The second thing is that sometimes nation states make the problem worse by kettling persons of diverse interests and backgrounds within artificial borders, where one group typically dominates the other (leading to extreme examples like the Rwandan genocide).

As we stated, it was not always like this. Tribes of people came into conflict with each other, to be sure, but they were not typically locked together within artificial territorial boundaries and told to choose which

---

set of interests would dominate that shared territory. A tribe might be forced out of a territory as, for example, when the Lakota forced the Cheyenne out of the Black Hills before Europeans arrived there, but the Cheyenne were still self-governing and still controlled their values, even if they were displaced from their previous territory.[9]

Here lies the problem: when persons of conflicting religions and ideologies are locked into the same space and forced to work things out, compromise is possible, but more likely, someone is going to be the loser. Maybe a strongman like Saddam Hussein will resolve the conflict between Sunnis and Shiites by dictatorial decree and military suppression of one group. Or maybe a government, like that of the Soviet Union, will outlaw all religious beliefs or at least create obstacles for believers. Or maybe an election will be held in which the larger group will prevail, leading to the minority group's values being disregarded. But whether the system of governance is democratic or dictatorial, there always seem to be winners and losers, and because there are losers, there is typically also a group of people that are left resentful and angry, even if not yet physically harmed.

One way to take this point is that we are trying to solve new problems today – not just the problems that drove the solution in the Peace of Westphalia. To be sure, we are still dealing with the ideological differences of the time, but their culture wars are not the same as ours today. And while the problems faced in Northern Europe in 1648 were arguably worse than those we face today in the West, we still face problems that call for solutions. It is not enough to kettle communities together and tell them to deal with it. What we are looking for are solutions that minimise distrust, allow communities to be self-determining and provide safe exit if necessary.

Often housing groups of people with radically different interests and goals and values, nation states are odd by their very nature. However, they are also odd in how they operate externally. Nation states do go to war. They do it all the time. And while tribes also went to war, when a nation state engages in kinetic conflict, it is really something to behold. Because some nation states are drawn with borders that have enormous territories and numbers of people, they typically also have vast resources to wage wars – wars in which weaker nations are turned into economic vassals. When large nation states find themselves in conflict with their peers, the result (as seen in those wars of the twentieth century) is typically the deaths of tens of millions of people.

---

[9] James R. Walker, *Lakota Society* (Lincoln, NE, 1992)

Indeed, some thinkers, like the conservative French philosopher and political theorist Bertrand de Jouvenel, have argued that nation states are, by their very nature, inevitable engines of violence and repression. Having fought in the resistance during World War II, he had seen more than enough warfare and observed that the modern nation state had become a meat grinder in its execution of warfare. When nation states went to war, Jouvenel argued, 'national resources' became targets. 'In this war everyone – workmen, peasants, and women alike – is in the fight, and in consequence everything, the factory, the harvest, even the dwelling-house, has turned target. As a result, the enemy to be fought has been all flesh that is and all soil, and the bombing plane has striven to consummate the utter destruction of them all.' As he put it, 'the whole nation becomes a weapon of war wielded by the state.'[10]

In Jouvenel's view, this sort of violence could not have happened in the pre-Westphalian age. Back then, kings might go to war, but they would need to tax and enlist the support of nobles, who often refused their support. However, in the new age, there were no such checks on a leader. In the age of nation states, wars simply became too easy to execute, and economic resources were rarely, if ever, denied.

The curious thing is that while political discourse can go on and on about the best system of government for nation states – the best way to organise democracy or theocracy or whichever alternative – and we can argue about which nation states are behaving well and which are part of the 'axis of evil', we seldom ask ourselves why we need nation states at all. Are they even a good idea?

---

[10] Bertrand de Jouvenel, *On Power: The Natural History of Its Growth* (Carmel, IN, 1993).

Let us set aside the issue of whether nation states prevent or actually contribute to genocides and other atrocities. Are they even that good of a system for trade, taxation, economic growth or really any aspect of human flourishing? Can it really be that an almost 380-year-old technology is the best tool we have for approaching these problems?

## 2.3   Are nation states bad at everything?

Our point here is not that nation states have been bad at stopping genocide and atrocities and religious oppression, which, let us remember, is the task for which they were originally created. Our point is that they seem to be bad at *everything*.

Consider the issue of national currencies. If nation states have failed to protect human rights and minimise acts of genocide around the world, are they at least good at setting up national currencies with which people can carry out economic activities? Sadly, no. Currencies, under the control of nation states, have been notorious failures. There has been a litany of famous cases in which currencies have collapsed utterly.

The most famous case, of course, is the collapse of the German currency in the Weimar Republic after World War I. As the pseudonymous writer Peruvian Bull notes in his book *The Dollar Endgame: Hyperinflation is Coming*, inflation reached rates of more than 30,000%, meaning that prices doubled every few days. People burned paper currency to stay warm, as it had less value than the wood from which it was made.

That might be the most famous case, but it was hardly an isolated example. Again, Bull observes that economists estimate that the annual inflation rate in Hungary reached 41.9 quadrillion percent after World War II. This means that prices in Hungary doubled approximately every fifteen hours. There is a seemingly endless supply of additional cases. In 2008, Zimbabwe famously had to start printing bills with denominations greater than a trillion. Other recent hyperinflation casualties included the Greek drachma (1941–1944), the Chinese yuan (1947–1949), the Chilean escudo (1971–1974), the Argentine peso (1975–1992), the Peruvian sol (1985–1991 and 1992–2003), the Yugoslavian dinar (1992–1995), the Belarusian ruble (1992–2003) and the Angolan kwanza (1999). Even when the result is not hyperinflation, a modest inflation rate of 2% can wipe out family wealth within 100 years if it is kept in the national currency.

Bull notes that 'an eventual collapse of fiat currency is the norm, not the exception' and adds:

> In a study of 775 fiat currencies created over the last 500 years, researchers found that approximately 599 have failed, leaving only 176 remaining in circulation. Approximately 20% of the 775 fiat currencies examined failed due to hyperinflation, 21% were destroyed in war, and 24% percent [sic] were reformed through centralized monetary policy. The remainder were either phased out, converted into another currency, or are still around today.

More poignantly, Peruvian Bull adds that 'the average lifespan for a pure fiat currency is only 27 years'[11] – much shorter than a typical human life.[12]

This does not mean that earlier forms of human organisation fared much better with monetary integrity. The Roman Emperor Diocletian diluted the amount of silver in the Roman coin, the denarius, leading to one of the first recorded examples of hyperinflation. However, our point is not that hyperinflation is unique to nation states; it is rather that nation states have not been effective in forestalling such phenomena. In fact, they have made hyperinflation quite common in today's world. All of this is to ask why currencies should be under the control of nation states at all.

---

[11] Peruvian Bull, *The Dollar Endgame: Hyperinflation Is Coming* (2023).
[12] For a list of specific cases, see <https://www.cato.org/sites/cato/files/pubs/pdf/hanke-krus-hyperinflation-table-may-2013.pdf>.

In her recent book *Money is Broken*, Lyn Alden makes an interesting observation about the inevitability of nation states debasing their currencies. Thanks to centralised national currencies, nation states can undertake actions that would otherwise call for the levying of new taxes. However, rather than face that political challenge, it is easier to print more money, thereby diluting the value of their currency with the effect of syphoning away the wealth of their citizens. Ultimately, this path continues until the wealth of the nation is exhausted.[13]

The other interesting observation that Alden makes is that the debasing of national currencies is central to the capacity of nation states to wage war. In many (perhaps most) cases, the loser of the war is the power that exhausts the wealth of its people first. In special cases, like the United States, which currently holds the world's reserve currency, the wealth of the whole world – at least every nation holding dollars – is tapped to execute its military operations. In this case, there is no threat of being exhausted first, but there is a definite danger that a state can lose its reserve currency status by constantly destabilising its currency at the expense of its holders and users.

One might think that the problem is not with nation states per se but rather with central bankers printing money or with the failure to link units of currency to gold or silver, but these are just symptoms of the problem. Of course, we know how to avoid hyperinflation; the problem is that nation states, when they have control of their currencies, feel that they are free to print money to avoid the economic problems that they face. These phenomena are symptoms of the deeper problem, which is that we are operating with obsolete methods of human organisation.

---

[13] Lyn Alden, *Broken Money: Why Our Financial System is Failing Us and How We Can Make it Better* (New York, NY, 2023)

## 2.4   Questioning sovereignty

This observation is not unique to us. For these reasons and others, the idea of the sovereign nation state has been unravelling for some time. Consider the issue of sovereignty itself to begin with. Not surprisingly, in response to events like the Rwandan genocide, people have begun to question the wisdom of state sovereignty. The growing recognition of universal human rights has challenged the idea that a nation's internal affairs are solely its own business. Quite sensibly, many thinkers now argue that human rights abuses should be addressed by the international community, even if they occur within the borders of a sovereign state.[14]

The concern about sovereignty has not only been driven by cases like Rwanda; there is a growing list of issues that are undermining the idea of national sovereignty. In the discussion that follows, we touch on a few of these issues, and our point is that these issues, whether or not you believe in them or consider them important, are driving wedges into the material integrity of nation states across the globe. They are breaking the Westphalian order.

For example, environmental issues, such as pollution, are transnational in scope and are leading to actions that undermine nation-state sovereignty. It is clear why this would happen. Poisoned air and water are not contained by borders. No one simply pollutes their own air or their own water because air and water do not recognise the boundaries of nation states. If poisonous chemicals are dumped into a nation's river systems, they soon find their way into the world's oceans. It is thus a problem that transcends the boundaries of Westphalian nation states. And new centralised powers in our world are undoing aspects of national sovereignty in order to address these issues.

---

[14] See Chapter 10 for further discussion on the new varieties of sovereignty that will emerge in a post-state future, including those that will reshape our understanding of human rights and how they should be defended.

Similarly, on the economic front, regional blocs and trade agreements can be used (and are being used) to limit a nation's ability to fully exercise its sovereignty. If you enter into a trade agreement with another country, there will be disputes, and it is folly to think that you alone get to call the shots when resolving those disputes. International agreements, as structured today, require the relinquishing of at least some sovereignty to international adjudicators.

Finally, the rise of global security threats has led some to argue that traditional sovereignty is out of touch with respect to the reality of conflict today. The rise of non-state actors and transnational threats, such as terrorism and narcotrafficking, has led many to conclude that no country can fully protect itself without ceding some of its security operations to other states or to new forms of centralised authority.

All of these concerns are, of course, reflected in recent international law. The growth of international institutions such as the International Criminal Court and the International Court of Justice has, by their very existence, carved out exceptions to absolute national sovereignty. These institutions are not toothless and they can and do limit a nation's ability to act unilaterally in certain areas, such as human rights and war crimes.

These are all cases in which nation states have had to cede some of their sovereignty to larger groups of nations or, at a minimum, to other nations. However, things cut in the other direction, too. Nation states are also losing sovereignty to smaller groups within their borders as well as to transnational groups that do not recognise national borders at all.

Nation states and familiar international organisations like the UN are no longer the only available option as far as global governance is concerned. There has been a Cambrian explosion in new forms of governance on the global stage, and there is an interesting question as to whether traditional nation states (and organisations of nation states) will even have a place in the new global environment. They are certainly no longer the only players at the table.

We can start with some dramatic cases. We often overlook the global scope and power of international drug cartels like the Cártel de Sinaloa (Sinaloa Cartel) in Mexico and the now-defunct Cártel de Cali (Cali Cartel) in Colombia. They operate across multiple borders and have significant economic and political power. It is fair to say that in many parts of Mexico, the cartels are the de facto governing authority. This is true even for the less powerful cartels.

Until his death in 2014, Nazario Moreno González – aka El Chayo,[15] aka El Dulce (The Candy), aka El Más Loco (The Craziest) and sometimes known as Emiliano Morelos Guevara[16] – ran an evangelical Christian mission and narcotics trafficking operation, La Familia Michoacana, which ultimately evolved into a narcotrafficking organisation known as Los Caballeros Templarios (The Knights Templar). Under his stewardship, the two successive cartels flourished economically, but they also served as a nongovernmental aid programme to the indigent and as the de facto government for many of the citizens of the Mexican state of Michoacán. As we shall see, they also created problems for people living within their area of control.

On the business end, González's cartels were involved in the manufacture and export (to the United States) of methamphetamines but also in the mining of iron ore for sale to China. The Mexican government said that the iron ore mining was illegal. That did not seem to phase González, who sold the ore anyway and distributed some of the proceeds to the poor in Michoacán. He also gave loans to farmers, funded schools and churches and, in general, provided the kinds of services that governments were supposed to provide but which the official Mexican government was not willing or able to do. It is no surprise that, on his death, he became an unofficial saint and to this day is venerated as San Nazario.[17]

---

[15] Chayo is the diminutive of Nazario.

[16] Emiliano as in Emiliano Zapata, Morelos as in José María Morelos and Guevara as in Che.

[17] On the spiritual end, Nazario required his followers to carry a 'spiritual manual' that provided helpful advice about self-improvement. For example: 'Don't view your obstacles as problems, but accept them and discover in them the opportunity to improve yourself.' His writings also called for his followers to abstain from the use of alcoholic beverages and

Cases like this call us to rethink where the real power resides in today's world. For example, in many parts of the world, terrorist organisations have more actual control over terrestrial regions than do the nation states in which they operate. Groups such as Al-Qaeda, ISIS and Boko Haram operate in multiple countries and have their own political ideologies and agendas, which are not aligned with any of the nation states in which they operate.

In a *Guardian* essay aptly titled 'The Demise of the Nation State', Rana Dasgupta notes that the adherents of such organisations 'have lost the enchantment for the old slogans of nation-building.' Instead, 'their political technology is charismatic religion, and the future they seek is inspired by the ancient golden empires that existed before the invention of nations.' The most telling part of the story is how terrorist organisations rework the plumbing of global governance, completely indifferent to the borders of nation states. They are not interested in seizing the state apparatus, but 'instead, they cut holes and tunnels in state authority, and so assemble transnational networks of tax collection, trade routes and military supply lines.' The networks they build are impressive, tunnelling under the purview of nation states 'from Mauritania in the west to Yemen in the east, and from Kenya and Somalia in the south to Algeria and Syria in the north.'[18]

Like the Mexican cartels, terrorist organisations also take on many roles of governing institutions, or at least try to. Thus, they take on infrastructure projects, project taxing authority and, like the Los Caballeros Templarios, provide social services to local populations. Whether it is out of the goodness of their hearts or to capture the

---

drugs, to not sell drugs to anyone except gringos, to treat women with respect, and to beat thieves. The manual, which became known in Michoacán as 'The Sayings of the Craziest One', called for humility, service, wisdom, brotherhood, courage, and obedience to God and his agents on Earth – namely, La Familia Michoacana and, subsequently, Los Caballeros Templarios.

[18] Rana Dasgupta, 'The Demise of the Nation State', *The Guardian*, 5 April 2018 <https://www.theguardian.com/news/2018/apr/05/demise-of-the-nation-state-rana-dasgupta> [accessed 30 April 2023].

hearts and minds of their subjects is beside the point. What is critical for our discussion is how nation states have become almost invisible to these organisations.[19]

Meanwhile, transnational corporations (TNCs) are doing a 'legitimate' version of the same thing – tunnelling under and around nation states and taking over the roles that nation states once performed. TNCs have become increasingly powerful and influential in global affairs, operating across multiple countries and regions, often with more economic and political clout than many states. Sometimes, they have come under criticism for flexing their power by involving themselves in what is traditionally the province of governments.

We can start with a relatively benign example. Shell, the multinational oil and gas company, has played a quasi-governmental role in the Niger Delta region of Nigeria, where it has operated for decades. While it did not exactly replace the Nigerian government, it did play a major role as a kind of police force working with the national government. It was accused of sharing intelligence with the Nigerian security forces, allowing them to carry out operations against local communities and activists. Shell was also accused of providing financial support and logistical assistance to the security forces, including transporting soldiers and equipment to areas where they were carrying out operations.[20]

Cases like this are just a small part of the story, given that corporations today are bypassing nation states in their traditional domain of controlling the flow of money within their borders. Those borders are nearly invisible to TNCs, and TNCs do their best to cut nation states out of the loop entirely. Dasgupta notes that TNCs are specifically designed to avoid nation-state taxation systems. In 2018, 94% of Apple's cash reserves were held offshore, which means that the amount of money Apple offshored was $250 billion – more than the combined

---

[19] Michael Weiss and Hassan Hassan, *ISIS: Inside the Army of Terror* (New York, NY, 2015).
[20] Amnesty International, *Amnesty International Report 2017/2018: The State of the World's Human Rights* (London, 2018) <https://www.amnesty.org/en/wp-content/uploads/2021/05/POL1067002018ENGLISH.pdf>.

foreign reserves of the British government and the Bank of England. However, it is not just about money. As Dasgupta notes, big data companies like Google and Facebook have 'already assumed many functions previously associated with the state, from cartography to surveillance.'[21]

Even some nongovernmental organisations (NGOs) have become quite powerful on the global stage. NGOs, of course, are independent organisations that advance specific causes, such as human rights, environmental protection or economic development. They often operate on a global scale and can have a significant impact on policymaking. Famous examples include Médecins Sans Frontières (better known in the Anglophone world as Doctors Without Borders), Oxfam International, CARE International, Save the Children, Amnesty International, Human Rights Watch and Transparency International. These are all perhaps worthy organisations, but they have been criticised for usurping roles that traditionally belonged to national governments.

For example, Dani Rodrik, an economist at Harvard University's John F. Kennedy School of Government, has argued that NGOs may undermine the development of strong, accountable government institutions. He claims that 'by substituting for government in the provision of basic public goods and services, NGOs risk diminishing the incentives and capacity of governments to deliver these services themselves.'[22]

Dean Karlan and Christopher Udry, codirectors of Kellogg's Global Poverty Research Lab, subsequently investigated whether it was true that NGOs could 'crowd out' governments. They summarised studies of aid groups in Ghana and Uganda and concluded that in those nations, 'government funding decreased by 6.8% in the sectors where the NGO was active, even as it increased by 7.4% in areas where the NGO was not focused. This indicated that money was flowing away from the government institutions that villagers had previously relied upon and into the new programmes and services sponsored by the NGO – yet these new programmes and services were less effective at improving people's well-being.'[23]

We are not here to judge whether NGOs, on balance, do good or bad work. Our point is merely that they end up doing the work that was traditionally done by governments, for better or for worse. They are just another example of how the function of governance on the global stage is being taken over by new actors.

We are also not the first to notice this. In fact, some authors have pointed to the rise of transnational policy networks (TPNs) or, as they are sometimes called, 'informal governance'. As Oliver Westerwinter, Kenneth W. Abbott and Thomas Biersteker note, 'There is a growing and increasingly broad-based consensus that it is no longer possible to focus exclusively, or even predominantly, on states and their interactions in intergovernmental organizations to comprehend, understand, and analyze contemporary global governance.'[24] Their view is that TPNs of informal governance mechanisms are replacing traditional organisations like the United Nations and the African Union because those organisations view the world at the wrong level of granularity – they believe that states still hold the power in today's world. However, as we look at case after case, we see this is not so.

For another challenge to nation-state sovereignty, we can examine the phenomenon of city networks. Some cities have formed networks to collaborate on issues such as urban planning, climate change and economic development. Examples include the C40 Cities Climate Leadership Group and the Global Parliament of Mayors (GPM) – an

---

organisation that has over 1,000 mayors and city leaders from more than 130 countries as members.

The GPM is a particularly interesting case. The organisation was inspired by a 2013 book by Benjamin Barber, entitled *If Mayors Ruled the World: Dysfunctional Nations, Rising Cities*. And one can see why mayors like the ideas in the book.

> Nation-states have made little progress toward global governance. Too inclined by their nature to rivalry and mutual exclusion, they seem quintessentially indisposed to cooperation and incapable of establishing global common goods. Moreover, democracy is locked in their tight embrace, and there seems little chance either for democratizing globalization or for globalizing democracy as long as its flourishing depends on rival sovereign nations. What then is to be done? The solution stands before us, obvious but largely uncharted: let cities, the most networked and interconnected of our political associations, defined above all by collaboration and pragmatism, by creativity and multiculture, do what states cannot. Let mayors rule the world. Since, as Edward Glaeser writes, 'the strength that comes from human collaboration is the central truth behind civilization's success and the primary reason why cities exist,' then surely cities can and should govern globally.[25]

And to be sure, this attitude is not just talk. Mayors have organised their own climate initiatives and their own approaches to problems of illegal immigration and drug trafficking.

Climate change is the most salient example. Former Chicago Mayor Rahm Emanuel issued a news release after a 2017 summit of mayors on climate change, stating, 'Even as Washington fails to act, cities have the power and will to take decisive action'. Former New York Mayor Michael Bloomberg, also a leading figure in the movement, put a sharper spin on the idea. 'All the U.S. cities signing the Chicago Climate Charter [...] sends a strong signal to the world that we will keep moving forward toward our Paris goal, with or without Washington.'[26]

---

[25] Benjamin R. Barber, *If Mayors Ruled the World: Dysfunctional Nations, Rising Cities* (New Haven, CT, 2013).

Of course, this did not pass without criticism. Mark Anderson, a climate change sceptic writing in the *UK Column*, complained that 'these "climate mayors" have little or no qualm about usurping the role of national leaders. [...] this "grassroots globalism", if you will, appears to be a revolutionary means of undercutting national authority from the bottom up.'[27]

As you have probably guessed, our point is not about climate change or the rightness of mayors and networks of mayors seizing sovereignty from nation states. It is, of course, that this is yet another example of how the sovereignty of nation states is unravelling.

As nation states unravel, or rather, as the very idea of the nation state unravels, new forms of governance are emerging in unlikely places. For instance, there is the important example of so-called 'stateless nations'. These are ethnic or cultural groups that have distinct identities and aspirations for self-rule but do not have a recognised nation state. Examples include the Kurds, Palestinians and Basques. Do they have their own state-independent form of governance? Absolutely yes.

Dasgupta draws attention to the fact that these stateless nations are constructing governance mechanisms and their own forms of sovereignty from the detritus of the collapse of nation states. He notes that 'several ethnic groups, meanwhile – such as the Kurds and the Tuareg – which were left without a homeland after decolonisation, and stranded as persecuted minorities ever since, have also exploited the rifts in state authority to assemble the beginnings of transnational territories.' He adds that 'it is in the world's most dangerous regions that today's new political possibilities are being imagined.'[28]

---

[26] Office of the Mayor, 'Mayor Emanuel and Global Mayors Sign the Chicago Climate Charter at the North American Climate Summit', *Chicago.org*, 2017 <https://www.chicago.gov/city/en/depts/mayor/press_room/press_releases/2017/december/Chicago-ClimateSummitCharter.html> [accessed 28 October 2024].

[27] Mark Anderson, 'Mayors Usurping Nation States in Quest to Fight "Climate Change"', *UK Column*, 20 December 2017 <https://www.ukcolumn.org/article/mayors-usurping-nation-states-quest-fight-climate-change> [accessed 30 April 2023].

We have not even arrived at what is probably the most impactful example – special economic zones or special enterprise zones (SEZs). These are designated geographic areas within a country that are given preferential treatment to attract foreign investment and promote economic growth. Examples include Shenzhen in China, Dubai in the UAE and the Suez Canal Economic Zone in Egypt. Quite voluntarily, nation states have ceded authority over law enforcement, taxation and the provisioning of infrastructure to these SEZs.

Earlier, we mentioned the book *Your Next Government* by Tom W. Bell. It goes into great detail about the recent rise of SEZs and their place in today's world. However, one of the critical remaining questions is why nation states would cede so much authority to SEZs, and the answer seems to be that nation states, as evidence of their collapsing authority, are all too happy to cede the actual business of governing – providing services, fixing potholes and so on – as long as they retain the critical component of governance – the ability to control how favours are distributed and thus how favours are returned. This is a point we will return to later when we dive deeper into the forms of corruption that have often accompanied the presence of SEZs around the world.

We should also mention the rise of micronations. These are selfproclaimed independent nations that are not recognised by the international community. They often operate on a small scale and have unique political systems, such as the Principality of Sealand in the UK or the Republic of Molossia in the US. More recently, people have proposed and are currently involved in trial efforts to make new micronations in the form of seasteading operations in international waters.[29]

---

[28] Dasgupta, 'The Demise of the Nation State'.

We could cite additional examples, but you get the point. Nation states are losing their sovereignty (sometimes voluntarily), and many of their fundamental functions – security, taxation, monetary policy, climate initiatives, welfare, industrial policy and so on – are being taken over by other governmental institutions. And the source of this pressure is the aforementioned Cambrian explosion of organisations and institutions that are taking on the role of traditional nation states – TNCs, NGOs, terrorist organisations, networked cities, stateless nations, drug cartels, SEZs, micronations and more.

The aggregate effect of these new organisations on the future of the nation state should not be dismissed. As Sean McFate, writing in *The New Rules for War*, put it:

> The Westphalian Order is dying. Today states are receding everywhere, a sure sign of disorder. From the weakening European Union to the raging Middle East, states are breaking down into regimes or are manifestly failing. They are being replaced by other things, such as networks, caliphates, narco-states, warlord kingdoms, corporatocracies, and wastelands. The Fragile States Index, an annual ranking of 178 countries that measures state weakness using social science methods, warned in 2017 that 70 percent of the world's countries were 'fragile.' This trend continues to worsen. The ability of the United Nations or the West to police the situation fades each year, while nonstate actors grow more powerful. International relations are returning to the chaos of pre-Westphalian days.[30]

---

[29] 'The Seasteading Institute' <https://www.seasteading.org/> [accessed 28 October 2024].
[30] Sean McFate, *The New Rules of War: Victory in the Age of Durable Disorder* (New York, NY, 2019), 42.

Whether we consider these developments to be good or bad, the important point is this: it is now myopic to only consider traditional governance structures like nation states when we think about the principal actors in human governance on the global stage. The point is that a lot of individuals and a lot of institutions have seen the writing on the wall for national sovereignty, but the question is, have they also seen the writing on the wall for the idea of the nation state itself? Not exactly.

## 2.5   Searching for alternatives

There is a prevailing view that the nation state can still be saved – that it can coexist with all these competitors and that what we need is a global level of governance to tie things together, keeping nation states in the game but with diminished roles.

According to this view, global governance would complement and enhance the existing system of nation states.[31] Sometimes, this takes the form of what Anne-Marie Slaughter has called 'networked governance'.[32] She proposes a model of governance that involves multiple actors, including states, non-state actors and international organisations, collaborating to solve problems through networks of connections.

Similarly, Robert Keohane has developed the concept of 'complex interdependence'. His idea is that as the world becomes more interconnected, the traditional system of states interacting through military force and economic coercion is no longer sufficient to address global challenges. As a solution, he proposes a model of governance based on cooperation and mutual dependence among states – this is the complex interdependence idea. Key to this is the need for global governance mechanisms to manage this interdependence.[33]

---

[31] David Held and Anthony McGrew, eds., *Governing Globalization: Power, Authority and Global Governance* (Cambridge, 2002).
[32] Anne-Marie Slaughter, *The Chessboard and the Web: Strategies of Connection in a Networked World* (New Haven, CT, 2017).

For the most part, such theorists, while critical of traditional notions of sovereignty, have argued that more layers of centralised governance, such as global or regional governance, are necessary to address the challenges posed by globalisation and promote cooperation among states. They see these forms of governance as complementary to, rather than a replacement for, the existing system of nation states. However, from a centraliser-versus-decentraliser perspective, one could say that they are more of the same.

In other words, even if we adopt a system of global governance structures designed to ameliorate the problems with national sovereignty, those structures can create precisely the same problems at a second-order level. They are, if you will, meta states and, as such, inherit the same problems that traditional nation states have had. They represent the same points of failure, invitations to corruption and opportunities for centralised hegemony.

For example, the United Nations, for all the good it does in the world, is very much a centralised institution. It is no stranger to bureaucratic failures and incidents of corruption and seems unable to prevent or halt episodes of violence and genocide. The same can be said of all international organisations and courts. They are just another layer of centralisation, and adding a meta layer does not solve the problems inherent to centralisation if the meta layer is centralised as well. A cocktail of such organisations, mixed with traditional nation states, is not the answer.

Even if we like the cocktail, we need to consider its individual ingredients. Just because they are not nation states does not mean that they cannot have the same weaknesses as nation states. They can all be either centralised or decentralised in their internal structure.

---

[33] Robert O. Keohane, *After Hegemony: Cooperation and Discord in the World Political Economy* (Princeton, NJ, 2005).

In view of the chequered history of the nation state and the prolif-eration of alternative forms of governance, we think it is time to rethink traditional conceptions of government at the global (and local) level. Let us face it: nation states were not a particularly good idea at their inception, and after 377 years of tinkering with the formula, they are still a bad idea. The good news is that there is no universal law insisting that we organise ourselves within nation states or even that traditional nation states need to be part of the new equation. Maybe they were a noble experiment; we can argue about that. But the point is that today, we have alternatives – better alternatives. And we should avail ourselves of those better alternatives.

Those better alternatives are not necessarily the new players in the international order. Some of those new players are simply offering new attempts at centralised solutions. Some of them are actually very old attempts at centralised solutions (as when terrorist organisations call for caliphates, for example). They are symptoms of the underlying prob-lem rather than the solution to that problem.

So, what is the alternative? Are we going to advocate for a new version of kingdoms? Tribes? City states? Empires? Those alternatives were not particularly successful either. Do not worry; we are not going to propose rolling back the clock to an imaginary age of better forms of government. What we are proposing is the utilisation of blockchain technologies as the scaffolding of new, more effective forms of gov-ernment. Our goal is to show how blockchain governance can lay the foundations for a broad range of governmental alternatives, including a reimagining of the nation state.

In the subsequent chapters, we will sketch an alternative to tra-ditional governance – one in which people are organised in a decen-tralised-yet-cooperative manner, one in which governments operate transparently and in which people retain their privacy, and one in which individuals maintain their sovereignty and in which they are free to exit from communities that do not align with their values. If all this sounds too good to be true, it means you have been paying attention. It does sound too good to be true, and we have a lot of work to do to explain how all of this is possible.

To see where we are headed, recall the issue that started this chapter – the way in which nation states tend to kettle together people of conflicting interests and values. The good news is that we do not actually have to lock people of conflicting values together within a single, unified governance structure. Sunnis and Shiites, Catholics and Protestants, Tutsis and Hutus do not need to be forced together and told they must fight over who controls the levers of power.

Our idea is that any start over should utilise communities organised around blockchain technology, taking communities in the very broad sense of any group of individuals that need to cooperate towards some common interest. The central idea driving our thesis is that the use of blockchain technology can enable distributed-yet-cooperative organisation. This will open up a whole new basket of questions, of course. How will all these blockchain communities be organised with each other? What happens when they come into conflict? Do they overlap in function? How are we to understand sovereignty going forward?

We will get to these questions soon enough, but first, we need to examine the nature of the problem with nation states and why many of the new alternative forms of governance fail to provide a solution.

# CHAPTER 3

# POST-STATE GOVERNANCE

## 3.1 Preliminaries

In 1695, the British established an outpost on the Caribbean island of New Providence in Bermuda. Originally called Charles Town, its name was eventually changed to Nassau, which it still goes by today. When the War of the Spanish Succession broke out in 1701, multiple conflicts ensued across Europe, spilling over into the Western Hemisphere and eventually the Caribbean. When subsequent Franco–Spanish raids on Nassau forced the British to abandon the outpost in 1706, Caribbean pirates formed a pirate republic, took possession of the town and made it their base until the British eventually got around to retaking control in 1718. In its twelve-year existence, it provided plenty of material for future Hollywood movies. Citizens included Edward 'Blackbeard' Teach, 'Calico Jack' Rackham, Charles Vane, Stede Bonnet, Benjamin Hornigold and 'Black Sam' Bellamy.

Even before taking possession of Nassau, the pirates had established something of a virtual governmental system – quite decentralised and not yet physically located. After claiming the area, they formed what would become known as the Republic of Pirates. It had its own government and its own pirates' code (like the one reproduced below) governing affairs. Like the pirate code established onboard ships, it was *relatively* democratic. Booty from a conquest was shared, and sailors voted for their captains. Blackbeard was elected magistrate of Nassau and seems to have been in charge of the Republic in some way or other for a while.[1]

Different pirate enclaves had different codes, but all of them were quite interesting. The code of the Nassau Republic of Pirates seems not to have survived, but we can provide a sense of how such codes worked with an example. This one from Bartholomew 'Black Bart' Roberts was written in 1722, just four years after the Republic folded:

> ARTICLE I. Every man shall have an equal vote in affairs of moment. He shall have an equal title to the fresh provisions or strong liquors at any time seized, and shall use them at pleasure unless a scarcity may make it necessary for the common good that a retrenchment may be voted.

> ARTICLE II. Every man shall be called fairly in turn by the list on board of prizes, because over and above their proper share, they are allowed a shift of clothes. But if they defraud the company to the value of even one dollar in plate, jewels or money, they shall be marooned. If any man rob another he shall have his nose and ears slit, and be put ashore where he shall be sure to encounter hardships.

> ARTICLE III. None shall game for money either with dice or cards.

> ARTICLE IV. The lights and candles should be put out at eight at night, and if any of the crew desire to drink after that hour they shall sit upon the open deck without lights.

> ARTICLE V. Each man shall keep his piece, cutlass and pistols at all times clean and ready for action.

> ARTICLE VI. No boy or woman to be allowed amongst them. If any man shall be found seducing any of the latter sex and carrying her to sea in disguise he shall suffer death.

---

[1] Colin Woodard, *The Republic of Pirates: Being the True and Surprising Story of the Caribbean Pirates and the Man Who Brought Them Down* (New York, NY, 2008).

ARTICLE VII. He that shall desert the ship or his quarters in time of battle shall be punished by death or marooning.

ARTICLE VIII. None shall strike another on board the ship, but every man's quarrel shall be ended on shore by sword or pistol in this manner. At the word of command from the quartermaster, each man being previously placed back-to-back, shall turn and fire immediately. If any man do not, the quartermaster shall knock the piece out of his hand. If both miss their aim they shall take to their cutlasses, and he that draweth first blood shall be declared the victor.

ARTICLE IX. No man shall talk of breaking up their way of living till each has a share of l,000. Every man who shall become a cripple or lose a limb in the service shall have 800 pieces of eight from the common stock and for lesser hurts proportionately.

ARTICLE X. The captain and the quartermaster shall each receive two shares of a prize, the master gunner and boat-swain, one- and one-half shares, all other officers one and one quarter, and private gentlemen of fortune one share each.[2]

The Republic of Pirates was one of the many examples that Hakim Bey used in *T.AZ.: The Temporary Autonomous Zone* to illustrate his idea that temporary autonomous zones (TAZs) would be temporary 'islands in the net', coming into and out of existence.[3] However, the idea that TAZs would have short lives presupposed the ability of nation states and their security forces to push the TAZ out of existence. Today, as fissures in the Westphalian order continue to grow, it becomes less obvious that nation states are permanent features in our political order. It may well be that future pirate enclaves will not be as impermanent as

---

[2] 'Pirate Code of Conduct', *Elizabethan Era*, 2023 <https://www.elizabethan-era.org.uk/pirate-code-conduct.htm> [accessed 28 October 2024].
[3] Hakim Bey, *T.A.Z.: The Temporary Autonomous Zone, Ontological Anarchy, Poetic Terrorism* (New York, NY, 1991).

Hakim Bey imagined. In any case, nation states are now abandoning elements of their terrestrial sovereignty.

Whether that abandonment is temporary or permanent, the question for us is this: What does it look like today when alternative forms of governance achieve a degree of sovereignty over physical territory? It will not look like the Republic of Pirates, although there will be some similarities: it will involve a network of individuals, coordinated and bound by shared principles, who have seized an opportunity to gain sovereign control over their lives at a time when Westphalian nation states are in crisis. These crises have only intensified since the days of the Caribbean pirates, and our resources for coordination across great distances are much more developed than those available to Black Beard and Calico Jack.

This leads us to the topic of territorial control in the post-Westphalian order. We already have some idea of what it will look like, and we have discussed some examples of this phenomenon, namely special economic zones (SEZs). In the next section of this chapter, we want to dive deeper into this topic, looking at various forms of alternative governance that are currently emerging as nation states are dissolving. Then, in the following section, we paint a vision in which communities organised around blockchains emerge as the most promising alternative form of governance in the post-Westphalian era.

## 3.2   Some examples from the Cambrian explosion

In the previous chapter, we noted many ways in which the Westphalian political order is fragmenting, giving rise to new forms of governance. In some cases, the new forms are officially sanctioned and the product of intentional effort by legislatures and rulers. However, there are plenty of examples where new forms of government are unsanctioned (drug cartels, for example) and other, less criminal, examples where the governing alternative is not really the product of planned actions and certainly not plans of the state but in which the state, more or less, throws up its hands and lets the new form of governance continue to operate.

A striking example of the latter occurred in the village of Cherán in the Michoacán region of Mexico. In this case, the locals revolted against their official government *and* La Familia Michoacana (which, alongside some legacy Mexican governmental power, had become the de facto government). Like other Michoacán communities, Cherán suffered from crime, corruption and violence (thanks in no small part to La Familia Michoacana drug cartel). Kidnappings, extortion and murders were quite common, but in this particular case, the illegal logging of the forest surrounding the village set off the 2011 revolt. Here, it is important to understand that the forest was sacred to the 12,000 locals, who were largely members of an indigenous population known as the Purépecha and spoke the Purépecha language.

One of the community leaders, speaking to the *Los Angeles Times*, explained the motivation for the villagers' revolt as follows: 'To defend ourselves, we had to change the whole system – out with the political parties, out with City Hall, out with the police and everything. We had to organize our own way of living to survive.'

In April 2011, armed with rocks and fireworks, Purépecha women and men attacked a busload of illegal loggers linked to La Familia Michoacana. The cartel members were armed with automatic weapons, but the local vigilantes, by virtue of their sheer numbers and passion, somehow managed to prevail. They took control of the town, expelled the corrupt legacy officials and police, and barricaded the roads leading to the oak forest on the neighbouring mountain – a mountain that had fallen victim to the chainsaws of armed gangs supported by corrupt authorities and cartel gang members. This micro-revolution subsequently evolved into a programme of independent community policing, which now encompasses 20,000 inhabitants and over 27,000 hectares of communal land.

The Mexican government was forced to let this new administration continue, in part because indigenous communities possess the right to self-government and self-policing under the Mexican constitution. This having been said, it was only after protracted legal battles that the national government finally recognised autonomous Cherán as a legally self-governing indigenous community.

Lourdes Cárdenas, writing for *Truthout* in 2016, summed up the new form of governance as follows: 'In Cherán's unique form of government, the real power lies wholly with the people. There is not a single decision taken without consensus, from who will get a local job in construction, to the allocation of public services and overseeing the spending of the budget. The authority of the community's assembly is above any other local governmental body.'[4]

Perhaps more controversially, political parties and campaigns have been outlawed in the town. Cherán's model of direct democracy, as described by an article in *The Guardian*, offers 'a simple solution to the vote-buying and patronage which plague Mexican democracy.' This system may well have saved their sacred forest and brought peace to Cherán. In any case, by 2017, the town boasted the lowest homicide rate in the entire state of Michoacán, perhaps even in all of Mexico.[5]

This case might seem like an outlier since it involves indigenous peoples with special rights to independence carved out by the Mexican constitution. However, similar movements are forming in less dramatic forms.

Neighbourhood watch groups, after all, are also ways for residents to take on at least some of the role of policing in their communities. There are also tools to assist in this, including communication tools that make policing events more accessible and transparent. For example, a mobile application called Citizen (formerly Vigilante) tracks crime reports in real time and sends push notifications to local subscribers. Do tools like this replace police departments? Of course not – not yet.

---

[4] Lourdes Cárdenas, 'Life Without Politicians: A Mexican Indigenous Community Finds Its Own Way', *Truthout*, 2016 <https://truthout.org/articles/life-without-politicians-a-mexican-indigenous-community-finds-its-own-way/> [accessed 7 September 2023].
[5] David Agren, 'The Mexican Indigenous Community That Ran Politicians Out of Town', *The Guardian*, 4 March 2018 <https://www.theguardian.com/world/2018/apr/03/mexico-indigenous-town-banned-politicians-cheran> [accessed 28 October 2024].

However, our point is that this is a gradual process by which, bit by bit, the functions of state governance are being taken on by other non-official forms of governance.

Indeed, there are more dramatic examples of this phenomenon in the United States involving communities in which the actual kinetic form of policing is not in the hands of traditional governments but in the hands of other groups. For example, Sidney Torres, who made a fortune in waste management, started a policing operation in New Orleans in 2015 called the French Quarter Task Force. Torres invested in a fleet of GPS-equipped Polaris Rangers (which the *New York Times Magazine* describes as 'militarized golf carts') and an application that allows residents to summon the Task Force and view where they are on a map in real time. It is worth noting that he formed the Task Force in part thanks to encouragement (or perhaps baiting: the mayor told him to 'put your money where your mouth is').[6] Our point is that we have yet another example in which a legacy government – in this case, a city government – was willing to cede its sovereign control over policing, and the replacement was up and running in short order. It is easy to see that efforts like this, combined with projects like the Citizen application, could change policing and, ultimately, governance as we know it.

Aside from these new, smaller forms of governance that are emerging and becoming more important at the dawn of the post-state era, there are also larger organisations that are taking on the role of states. One prominent example is the European Union, which has assumed numerous roles – economic, cultural, social services and so on – that used to be the province of individual states. As they will also be a part of any post-state cocktail, we want to devote a few paragraphs to discussing them and how they will interact with state and, more importantly, non-state actors as nation states continue to dissolve.

---

[6] David Amsden, 'Who Runs the Streets of New Orleans?', *New York Times Magazine*, 30 July 2015 <https://www.nytimes.com/2015/08/02/magazine/who-runs-the-streets-of-new-orleans.html> [accessed 28 October 2024].

The phenomenon of centralised umbrella governance that sits over national governments is quite common (we have already mentioned the European Union and, earlier, the Organization of American States). Where things get interesting is when these umbrella organisations codify their authority into international law, creating another layer of constitutions and subsuming those of nation states. This codification of covering laws or constitutions goes by several names, including 'transnational constitutionalism', 'multi-layered constitutionalism', 'international constitutionalism' and, with respect to the European Union, 'European constitutionalism' and, with respect to the British Commonwealth, 'Commonwealth constitutionalism'. The phenomenon has drawn substantial attention in international law.[7]

Given the complexities of international law and the inherent difficulty of attempting to get transnational constitutions of whatever stripe to dovetail with nation-state constitutions and other transnational constitutions, the topic is challenging. For example, Commonwealth constitutionalism has to dovetail not only with British law but also with international law and (until recently) with European Union law.

At the same time nation states are ceding some of their sovereignty to larger transnational governmental units with constitutions of their own, they are also ceding legal authority to smaller governmental units like special taxing districts, special enterprise zones and a host of other local governmental units. There is plenty to be said about special taxing districts and special enterprise zones, and these smaller governmental units are becoming an important part of the post-Westphalian puzzle.

As we write this chapter, there is an ongoing struggle between the Walt Disney Company and Florida Governor Ron DeSantis. The dispute originates in what was initially called the Reedy Creek Improvement District (RCID) and concerns the alleged 'wokeness' of

---

[7] For a discussion of all of these approaches, see Roger Masterman and Robert Schütze, *The Cambridge Companion to Comparative Constitutional Law* (Cambridge, 2019); Part V

Disney. The RCID was formed by an agreement made between Disney and the State of Florida in 1967, in which Disney Company would take on many of the functions of a local government. These functions included municipal services like power, water, roads, fire protection and so on. When Disney bought the property, none of these services were available.

The scope of governance within the Reedy Creek framework was quite impressive. The Reedy Creek website listed the following facts concerning its role in governance as of 2023. Their responsibilities included:

- 134 miles of roadways and 67 miles of waterways built and maintained
- 250,000 daily guests
- 6 to 8-minute response time for fire and EMS
- 60,000 tons of waste managed
- 30 tons of aluminum, paper, steel cans, cardboard and plastic containers recycled every year
- 22,800 water samples collected by RCID scientists from 1,500 locations on the property for testing every year
- 90,000 analyses conducted to make sure that water quality meets or exceeds state and national standards. Water draining from the south end of the District is generally cleaner than when it entered Reedy Creek at its north end.
- 2,000 vendors, suppliers and contractors used to provide a high level of public services for visitors[8]

Since the dispute with Governor DeSantis, the governing body is now known as the Central Florida Tourism Oversight District, and

---

[8] RCID, 'About', *Reedy Creek Improvement District*  www.rcid.org/about/ [accessed 29 June 2023].

Disney no longer appoints the organisation's five-member Board of Supervisors. Those supervisors are now appointed by the Office of the Governor of Florida. Certain powers were also removed by the Florida Legislature, including the power to construct a nuclear power plant, airport and stadium.

This case study is interesting for several reasons, not least of which is the tension that holds between the Disney property in Orlando and the Florida governor's office. It shows that, as with our Republic of Pirates in the first section of this chapter, the traditional state may come to reclaim its sovereignty – or at least some of it.

Of course, the Disney Company is not without weapons of its own, some of them legal, some of them stemming from the popularity of the Disney brand and some of them being that the State of Florida would rather not be involving itself with 60,000 tons of waste management and 250,000 daily guests. Most importantly, Florida politicians would very much prefer not having to put their names on the tax initiatives necessary to pay for it all.

Thus, in the *Disney v. DeSantis* case, we quickly run into the limits of political will. Governments, paradoxically, would rather avoid the messy business of day-to-day governance, and if a corporation or a community steps forward and offers to do the dirty work, legacy governments are often happy to cede some of their authority. In other words, the Empire may strike back, but it does not win in the end; in this case, it may not even want to win.

The Disney–DeSantis dispute is perhaps best understood as an aberration, owing to the career interests of the current governor and his unsuccessful attempt to win the Republican nomination by fighting against the 'wokeness' of the Disney Corporation. Otherwise, this dispute might not have happened at all, and the evidence for this is that there are hundreds, perhaps thousands, of other such districts in Florida that continue to operate without controversy.

As far back as 1982, there were approximately 1,000 special taxing districts in Florida, which was more than the state's total number of cities and counties combined. They were 1,000 black boxes full of governmental mystery, by some accounts. Noting the rise of such districts in the middle of the twentieth century, John C. Bollens referred to special district governments as the 'dark continent' of American politics – a reference to the expression used to talk about precolonial Africa, in which many facets of Africa were a mystery, not least its various forms of tribal governance.[9]

A working definition of a special taxing district, from David M. Hudson, is that it is a 'local unit of special government, except district school boards and community college districts, created pursuant to general or special law for the purpose of performing prescribed special functions, including urban service functions, within limited boundaries.' Of course, as the name implies, these districts can levy their own taxes to pay for these services.[10]

The Disney special taxing district may not even be the most well-known example in Florida, with the most famous, or at least most notorious, being a district known as The Villages. The Villages is a retirement community of over 130,000 senior citizens, famous for allegedly having one of the highest STD rates in the United States (although this is disputed).[11] What is not disputed is that the governance of The Villages maintains most of the services generally associated with government, including police, fire, utilities and so on, but also matters like animal shelters and, of course, lots of golf courses. We will return to The Villages later to discuss a recent political conflict, but for now, we simply want to note that this is but one of many cases in which the State of Florida ceded governmental authority.

Of course, as we mentioned earlier, nation states do this as well in the form of the aforementioned SEZs. It would take a monumental effort to come up with a taxonomy of all the different kinds of these 'special' zones dotting the governmental landscape around the world, and it may well be that no one has a firm handle on the scope and variety that exist today. Our job here is not to nail down existing flavours of such zones but rather to note that their composition is quite fluid, as is their source of origin. They promise to make a significant contribution to the tapestry of post-state governance. However, the problem is that, on their own, they are not the answer. It is not enough to break governance into smaller pieces – that just gives us more (if smaller) centres of authority, with all of the baggage that comes with centralisation. In the next section, we will explore this problem.

## 3.3   Sharding is not enough

'Governance' has been our term for strategies that humans use to organise their collective activities and decision making. Sometimes, that organisation is top-down and sometimes, it is bottom-up, but however it is structured, it permeates nearly every aspect of our lives. Governance is ubiquitous.

We have explored some of the many kinds of institutions that are replacing the governance role of nation states, and we suggested that the very idea of the state may be dissolving before our eyes. In its place, we are beginning to see a complex web of different levels of governance and different forms of sovereignty. We have speculated that even the notion of territorial sovereignty is dissolving and soon will not be the province of any single level or form of government.

You might think that this is a great state of affairs, as only good can come from moving from large centralised powers to smaller centres of power. However, this is not exactly correct. Decentralisation without coordination is not really a solution to anything. In fact, it is not really decentralisation so much as a phenomenon that we call 'sharding'.

'Sharding' is a term that is used in online video games to refer to cases in which gameplay proceeds independently on different servers. Like a piece of glass that has broken into smaller pieces (shards of glass), each of the smaller pieces has the same material composition as the original. Sharded games have the same mechanics and graphics, although the gameplay evolves in different ways. Similarly, we can take a centralised power and break it into pieces, but if those pieces are little centralised fiefdoms themselves, we have not really solved any problems. We have, if anything, just created more of the same.

Of course, we can argue about whether sharded centralised authority is at least better than the original or not but, in our view, this is an unambitious approach to the problem. Sharding does not eliminate corruption. We need technologies that help shards become cooperative, transparent forms of governance. In this section, we aim to highlight some of the many forms of sharded governance that exist today – SEZs, homeowner associations (HOAs) and so on – and some of the problems to which they have given rise. We will then make the case that many of these problems can be ameliorated by bringing blockchain governance to these shards – to SEZs and HOAs and so on.

Some review may be helpful here. As we noted earlier, there are currently many levels of official government, ranging from collections of traditional nation states (like the EU and the UN) to their member states (as in the United States of America and the United States of Mexico) to any number of smaller political units, including counties, municipalities, congressional districts and, in cities like New York, boroughs. Each level of government has its own concerns, its own methods and its own way of organising itself.

While these levels of government are the most visible forms of human government, it is important to keep in mind that humans are engaged in many more forms of governance. As we have noted, there are scout troops, churches, condo boards, HOAs, school boards, event planning committees, social clubs and many forms of corporate governance. As Westphalian governance structures dissolve, these smaller units may become more important. This is not to say that they are going to be much better than large centralised authorities. They may even be worse.

The problem is that the sharding of traditional forms of government into smaller pieces accomplishes little if those smaller pieces are still centralised in organisational structure. At best, we get smaller pieces of centralisation. The problem with governance at any level is that if it is centralised, and if it has control over financial resources or over our behaviour and our rights, then it is going to be a magnet for corruption and abusive leaders. This is true whether we are talking about nation states or cities or HOAs or church administrations. Centralisation is an attack vector at every level of governance, whether it concerns large or small governance institutions. Maybe with larger institutions, the failures are more spectacular, but the cumulative effect of sharded centralised governance may be worse in the aggregate.

Here is another way to think about it. There is certainly plenty of corruption at the federal level of government in the United States, but there is precious little reason to think that there is less corruption in the aggregate at the state level. Illinois alone has sent four of its governors to prison for corruption since 1969, and it is far from being viewed as the most corrupt state. However, there is no reason to pick on states like Illinois. There are plenty of cases of corrupt county commissioners, mayors, sheriffs, assembly persons and so on.

Our point in talking about all these different levels of government and different kinds of governance in a post-state world is not simply to impress you with their ubiquity. It is rather to observe that all of these levels of government and kinds of governance rely on tools to get their jobs done. Some of these tools involve the keeping of records – for example, minutes of meetings or spreadsheets of accounts. Other tools might include strategies for voting (raising hands versus secret ballots). Still, other tools might facilitate the distribution and control of financial resources (cookie jars versus checking accounts or investment portfolios). However, we believe that just as nation states (themselves obsolete technologies) are lumbering along using obsolete technology, so too all these different levels of human organisation are using obsolete technologies.

Of course, for certain purposes, vintage technologies work just fine. Maybe the proceeds from a bake sale can be recorded using a pencil on the back of an envelope. And even that might be too much technology in some cases – as Avon Barksdale said in the HBO series *The Wire*, you do not want to be 'takin' notes on a criminal fuckin' conspiracy'.[12] However, sound, decentralised record keeping would be a huge step forward for many of the governance applications we have discussed, whether that be church business, condo board business or HOA business.

We recognise that talking about these lower levels of human governance may not be that exciting. It takes a lot of mental stamina to start grinding through governance failures in things like SEZs and HOAs, but when we take a closer look, what we see can be pretty horrifying and thus worthy of our attention. Let us consider the example of HOAs.

As of this writing, it is difficult to quantify the number of HOAs worldwide, but estimates suggest that there are approximately 50,000 in both Florida and California alone.[13] That number is apt to have increased dramatically by the time you read this book. As the *Miami Herald* reported in a story titled 'HOAs from hell: homes associations that once protected residents now torment them', 80% of new housing starts today are in HOAs.[14]

While we know little about what goes on in most of these individual HOAs, the combination of centralised organisation, a lack of transparency and tempting pots of money has predictably led to some spectacular scandals. For example, the board of the Hammocks Community

Association in the West Kendall suburbs of Miami, Florida, recently faced criminal charges for stealing $2 million worth of their HOA's maintenance fees. Another article in the *Miami Herald* (this one titled 'Wild allegations at Miami homeowners association show why Florida needs HOA crackdown') suggested that the situation was even worse than it sounds: 'It involves charges of racketeering, money laundering, fabricating evidence and using shell companies.' During a recall election, the association board threw out two-thirds of the ballots. According to the Miami-Dade State Attorney prosecuting the case, the board was 'a criminal enterprise'.[15]

Our point in discussing cases of scandal at the lower levels of governance is not to take some moral high ground and judge the offenders (although there are some quite reprehensible people included among them). Our central point is that if we wish to minimise cases of corruption and other vectors of attack, these smaller forms of governance also need to avail themselves of blockchain technology. But what would this look like?

To illustrate, let us continue to consider HOAs. The problem encountered in the case of the Hammocks Community Association is that the day-to-day financial operations of the association board were not transparent. No one could see what the board was doing until it was far too late. But second, when it became necessary to vote, the voting mechanisms were also opaque. It became possible to discard two-thirds of the ballots. This might have been mitigated if the voting had taken place using an onchain smart contract with a publicly accessible audit trail immutably recorded on the blockchain. But how might this work?

Recall that in the Introduction we introduced the idea of smart contracts – contracts that are encoded as computer programs and deployed on the blockchain. If the mechanisms of voting (and vote counting) are encoded in onchain computer programs, then the integrity of the system will be visible to all. Our basic idea is that the core activities (including financial activities) of an HOA would take place transparently onchain. Proposals would be put forward to the community; votes would be made via a smart contract or some trusted online tabulator. How do we know the records will be reliable? Because they will be recorded on a distributed ledger grounded on a trustworthy global blockchain like Ethereum. Clearly, an onchain HOA does not solve every problem, but if the core target of criminality (the money) can be constantly audited by the community, then the other aspects of criminality (money laundering, racketeering and so on) become more difficult to engage in.

---

[15] Miami Herald Editorial Board, 'Wild Allegations at Miami Homeowners Association Show Why Florida Needs HOA Crackdown', *Miami Herald*, 17 November 2022 <https://www.aol.com/wild-allegations-miami-homeowners-association-214951325.html> [accessed 28 October 2024].

To illustrate this idea, let us return to the case of The Villages, the special taxing district with 130,000 residents that we discussed earlier in this chapter. While The Villages has many good things to offer its residents (it is sometimes characterised as Disneyland for senior citizens), not surprisingly, it has become a target for corruption. In this particular case, the issue stems from the relationship between the developers of The Villages and legacy politicians, and this is going to be an issue that any new independent governance structure will face – the desire of legacy institutions to get their hands on it.

The Villages began its history in the 1970s as a trailer park. Its founder, Harold Schwartz, was soon joined by family members – surname Morse – and over the subsequent years, they began expanding the district from a handful of mobile homes to the booming suburban community that today does around $2 billion in annual revenue. However, with those years of success also came entrenched power. The Morse family today owns the local newspaper (*The Villages Daily Sun*), owns the local radio station and owns *The Villages Magazine*, which covers events in the community.

Outsiders rarely peek into the affairs of The Villages, and when they do, it is usually to make fun of 'boomers' and their golf cart parades and their allegedly off-the-charts STD rates. However, that changed when *The Intercept* scrutinised the politics of The Villages more closely in February of 2023. The title of the article by Ryan Grim was definitely an eye-opener: 'The Villages vendetta: How a grassroots revolt in the iconic retirement community ended with a 72-year-old political prisoner'.

The reported events began in 2019 when residents of The Villages were surprised by a 25% hike in their property taxes. They were not merely surprised but deeply concerned, as most senior citizens live on fixed incomes and have their expenses calculated for the long term. What was unusual about this particular tax increase was that it was not designed to pay for amenities or upgrades to the existing housing units. Rather, it was designed to help subsidise the expansion of The Villages developments to the south. Ordinarily, a development of that sort would trigger impact fees for the developer – fees designed to pay for things like schools, fire departments, emergency medical services, police, parks and local government buildings. However, in this case, the developers paid for none of that, and the impact fees were being subsidised by current residents of The Villages.

This led to an election in which three individuals, Craig Estep, Oren Miller and Gary Search (running on the 'EMS' ticket), campaigned for county commission – the only real oversight board for The Villages. EMS won the election but lost the war, as developers working for The Villages enlisted the support of the Florida Legislature and none other than Governor Ron DeSantis to bring the hammer down on EMS. The EMS-led commission voted to raise the impact fees for the developers by 75%, with the plan to roll back the property tax increases. However, friends of the developers in the state capital immediately passed legislation making it illegal for a county commission to raise impact fees on developers. The community's newspaper, *The Villages Daily Sun*, happily celebrated the victory with multiple snarky subheaders: 'Law stymies freshmen commissioners', 'Conservatives lead charge', 'Understanding economic theory' and 'Locals applaud new law'.[16]

---

[16] Ryan Grim, 'The Villages Vendetta: How a Grassroots Revolt in the Iconic Retirement Community Ended With a 72-Year-Old Political Prisoner', *The Intercept*, 2 May 2023 <https://theintercept.com/2023/02/05/ron-desantis-florida-villages-oren-miller/> [accessed 28 October 2024].

The story in *The Intercept* is worth the read as it goes into subsequent actions that were taken against the EMS coalition, including the jailing of Oren Miller for allegedly using his wife to communicate with another commission member (a violation of Florida's 'Sunshine' Law that requires said communications to take place exclusively in public meetings). Those details do not really matter to us, but two things do matter. The first is the observation that the kinds of government structures that are replacing traditional governance structures are no more democratic simply because the unit of governance is smaller. Whether it is a nation state or an HOA or a special taxing district, governance is centralised. Thus, it is resistant to democratic initiative and serves a small group. However, the second point to observe is that when corruption comes to these smaller forms of government, it typically comes from, or at least with the generous help of, legacy governments. The Florida Legislature and the Florida state government could not resist dabbling in the affairs of The Villages to ensure that the developers (significant campaign donors) were protected against the rabble.

The moral of this section is that when we talk about centralised governance being a target of corruption, that corruption need not be some local warlord or cartel leader like Pablo Escobar. In many cases, the source of the corruption is the legacy political system. As in the case of the Disney special taxing district, legacy governments are happy to have smaller forms of government do the dirty work of fixing potholes and hauling garbage, but there is an element of power and sovereignty that they are not willing to give up, whether that element is control over a corporation's support of the LGBTQ community or just ensuring that big donors are protected. This means that getting from point A to point B (in this case, from a world of centralised governance to a world of decentralised governance) is not a trivial matter. External forces prefer that you be centralised because, if you are centralised, you are easier to manipulate. In other words, governmental sharding is permitted as long as the shard is itself centralised and, thus, controllable.

Obviously, we think that another way – a better way – is possible, and we have indicated that newer and more successful forms of governance will deploy new technologies that enable decentralised-yet-cooperative governance. There is a lot at stake, and we owe a more complete account of our alternative vision. We begin laying out this vision in the following three chapters, explaining what blockchain technologies are, how they work, and how they can be leveraged to enable better governance and greater human flourishing.

# CHAPTER 4

# NEW CONCEPTUAL FOUNDATIONS

## 4.1  Preliminaries

In the previous two chapters, we discussed the many new (and old) forms of post-state governance that have emerged in recent decades. Many of these forms of governance are not good solutions to the failures of Westphalian governance. Rather, they are *symptoms* of these failures. It goes without saying that we do not think that drug cartels, paramilitary organisations and terrorist organisations are offering positive developments in human governance. All of which raises the question of what we should be looking for. What are the positive possibilities? To what should we aspire?

To see what our aspirations should be, perhaps it is best to start by thinking in terms of the failures we do not want to repeat. We have discussed these failures at some length already. We certainly do not want a system in which people of diverse values are kettled together within a geographical territory and forced to comply with a set of values that they do not share. People should not be imprisoned by their geography nor by whichever tyrant controls events within arbitrarily drawn borders. People should be free to exit hostile governmental systems. People should also be able to live free of corrupt systems in which centralised authorities take advantage of their position of power for their personal benefit.

Beyond these desiderata, we believe people should be free to politically organise themselves with whom they wish, independent of terrestrial borders. They should be self-sovereign in that they should be free to join the political system they choose and live under the values that they endorse. Another way to put this is that if they are self-sovereign, they should also be free to choose their own governing principles, which is to say that they should be free to be part of governing systems that align with their principles.

Thinking in this way is possible partly because of the revolution in communication technologies and our ability to connect (almost instantly) with people anywhere in the world. The thought is that rather than dividing governments according to boundaries established by rivers and oceans, perhaps we should divide them up in other ways. Perhaps governments can be divided according to the ways in which people network and organise with each other, whether that be with the aid of digital communications or face-to-face relations.

Having said all this, it is important to ask where these ideas are coming from, what grounds them conceptually and, perhaps just as importantly, what drives them in terms of their praxis – what makes these ideas agents of change rather than just inert academic talking points?

It is interesting that in many past revolutions in governance, people could turn to philosophers and other thinkers for new ideas about how to proceed. The philosophers John Locke and Jean-Jacques Rousseau played important roles in the transition from monarchies to republics in the eighteenth century. Similarly, in the past, these ideas were instantiated by groups of people taking action – by protest and, in some cases, by taking up arms. But who can we turn to now for conceptual foundations, and who are the agents of change that can bring about a new order in our digitally interconnected world?

In our view, the thinkers that laid the foundations for the coming forms of governance were not individual philosophers as was historically the case, but rather groups of people working on the cutting edge of the digital revolution at the close of the twentieth century. There were several groups that saw the coming challenges of online life and called out the dangers but that also saw a path forward. More importantly, these groups also developed tools to clear the path and engaged in various forms of activism to raise awareness of the coming problems.

Groups like the Cypherpunks argued for the importance of individual privacy and then developed tools to make it possible. Meanwhile, hacktivist groups emphasised the importance of governmental transparency and developed tools to pierce veils of secrecy maintained by centralised powers. In this chapter, we want to explore the history of these groups, what they accomplished and how their work can help us build alternative forms of governance for a post-Westphalian order.

## 4.2   The Cypherpunks and privacy

At the dawn of the early Internet, there emerged a group of activists known as the Cypherpunks.[1] Their primary concern was that the Internet could very easily become a tool of universal surveillance and oppression, and they developed tools to ensure that privacy could be maintained online. One such tool that grew out of the Cypherpunk movement was Pretty Good Privacy, a military-grade encryption protocol developed by Phillip Zimmerman that allowed individuals to communicate freely without fear of surveillance by the numerous tyrants and dictators and bad actors around the world. However, the Cypherpunks also articulated a philosophy that has helped to steer the Internet in subsequent years. This contribution was important because, although the Internet is hardly a stellar protector of individual privacy, matters could have been much worse.

One of the movement's key documents was 'The Cypherpunk Manifesto', written by Eric Hughes in 1993 and anthologised in *Crypto Anarchy, Cyberstates, and Pirate Utopias*. Hughes' general thesis was that for an open society to function, certain things had to be kept private – for example, conversations or exchanges between individuals. Indeed, 'Privacy is necessary for an open society in the electronic age.'[2]

---

[1] The term Cypherpunk was coined by Jude Milhon, aka St. Jude, best known as the 'Editrix' of the fringe culture magazine *Mondo 2000*.

As Hughes viewed it, an open society requires people to be able to transact and communicate with whom they want without the whole world knowing the contents of the exchange or even whom the exchange was between. Why is this critical to an open society? Because we might fear the consequences of communicating with someone who is unpopular or we might be afraid to exchange new ideas in public as those new ideas might lead to backlash and discrimination if they are unfamiliar or challenge widely accepted viewpoints. There is a reason, after all, that Leonardo da Vinci wrote a number of his manuscripts in code (mirror writing and his own shorthand). Similarly, Charles Darwin closely guarded his draft of *The Origin of the Species* for a decade. It was just too provocative in the moment. As Hughes points out in his manifesto, 'Privacy is the power to *selectively* [emphasis added] reveal oneself to the world.'

Such historical examples were not lost on Hughes. As he noted in the manifesto:

> People have been defending their own privacy for centuries with whispers, darkness, envelopes, closed doors, secret handshakes, and couriers. The technologies of the past did not allow for strong privacy, but electronic technologies do.

This led him to the actions that the Cypherpunks would perform to make privacy on the Internet possible. The role of the Cypherpunks would be simple:

> Cypherpunks write code. We know that someone has to write software to defend privacy, and since we can't get privacy unless we all do, we're going to write it.[3]

---

[2] Eric Hughes, 'The Cypherpunk Manifesto', in *Crypto Anarchy, Cyberstates, and Pirate Utopias* (Cambridge, MA, 2001), 485.
[3] Ibid.

The Cypherpunks did not merely write code; they also established one of the earliest online communities – indeed, one of the earliest online distributed communities. In 1992, the Cypherpunks set up a mailing list to discuss topics of interest, mostly involving cryptography and its related political concerns. In 1997, they established a distributed mailing list so as to avoid reliance on a single point of potential failure.

The Cypherpunks' mailing list was a who's who of early Internet and web3 pioneers. List members included the core Cypherpunks like Eric Hughes, Tim May and John Gilmore but also Julian Assange (of WikiLeaks fame), multiple founders of the Electronic Frontier Foundation, Richard Stallman of the Free Software Foundation, Nick Szabo (who invented the smart contract), Satoshi Nakamoto (who created Bitcoin) and Hal Finney (who most probably – we believe – was Satoshi Nakamoto).

Of course, privacy is just one part of the equation for Cypherpunks. An open society requires people to have secure communications, but it also requires them to have access to the actions and deliberations of the government; it requires governmental transparency, and this leads us to the role of hacktivists.

## 4.3   The hacktivists and transparency

At the same time that the Cypherpunks were concerned about the privacy of individuals, other groups were concerned about the transparency of governments and other institutions that hold power over us (for example, private intelligence agencies). They spoke out against such cloaked government activity and, more importantly, developed tools to cast light on the methods that traditional centralised powers use to oppress others. Just as importantly, they developed tools to puncture the illusion of invincibility of these centralised powers. And finally, they showed that political action can be just as real and visceral when mediated by digital communications as it is when it takes place face to face.

First, let us explain what we mean by 'hacktivist'. 'Hacktivist' is a portmanteau of the words 'hacker' and (political) 'activist', which raises the question of what a hacker is. To fully grasp what 'hacker' means, we should first be clear about what it does not mean. Sometimes people take hackers to be folks that engage exclusively in illicit computer activities for some sort of personal gain or perhaps just to be chaotic evil agents in the online world. However, our view of what 'hacker' means is more aligned with the first definition given in the New Hacker's Dictionary:

> hacker n. [. . .] 1. A person who enjoys exploring the details of programmable systems and how to stretch their capabilities, as opposed to most users, who prefer to learn only the minimum necessary.[4]

Packed into this definition is the thought that computer technologies should not be sealed inside black boxes that no one can access, hidden from view and unchangeable, but rather should be accessible and modifiable. A hacker celebrates the idea of modifying technologies to 'stretch their capabilities'. This hacktivist idea, of course, lies at the foundation of the free software movement as articulated by Richard Stallman (not 'free' as in 'free beer' but 'free' in the sense that one should be at liberty to modify the software – i.e. to hack it). Hackers, in this sense, are individuals who like to extend and repurpose existing technologies, often in ways not intended by the creators of the technology.

Coming back to our portmanteau, 'hacktivism' thus refers to the idea that we can repurpose and modify existing technologies to achieve a new sociopolitical end. There are many ways that hacktivism could thus unfold, and the best way to illustrate the possibilities is to offer some historical examples from the early days of hacktivism.

---

[4] Various, 'New Hacker's Dictionary', *New Hacker's Dictionary*, 2002 <https://archive.org/stream/jarg422/jarg422.txt>.

### 4.3.1   WANK Worm

According to Julian Assange, the 'WANK Worm' is the first instance of hacktivism. On 16 October 1989, during the Cold War, when nuclear war was an immediate possibility, hackers hit NASA computers with the 'WANK Worm'. Two days prior to the launch of the plutonium-fueled Galileo space probe from the Kennedy Space Station, NASA employees logged on to see a humorous but, at the same time, frightening welcome screen: 'Your computer has been officially WANKed. You talk of times of peace for all, and then prepare for war' and 'Remember, even if you win the rat race, you're still a rat.' The machines of the US Department of Energy and NASA worldwide had been penetrated by the anti-nuclear WANK (WORMS AGAINST NUCLEAR KILLERS) worm.[5]

The WANK Worm event was not just a simple protest. It was also an effort to pierce the illusion of invincibility of certain institutions. Partly, it was to remind those powers that they should avoid hubris – their technologies are not as secure and reliable as they may think. Of course, it also demonstrated that perhaps the public is not as helpless against nation states and the technologies they employ as one might assume.

This example involved deploying hacktivism to embarrass the powers that be, but other hacktivist efforts had more direct objectives. One such case was the Hong Kong Blondes.

### 4.3.2   The Hong Kong Blondes

The Hong Kong Blondes was an underground network of Chinese students spread across at least three continents. It was started by Blondie Wong, who had reportedly witnessed his father being stoned to death during the 1966–1976 Cultural Revolution. The group initially

---

[5] Julian Assange, 'The Curious Origins of Political Hacktivism', *CounterPunch*, 25 November 2006 <https://www.counterpunch.org/2006/11/25/the-curious-origins-of-political-hacktivism/> [accessed 28 October 2024].

protested censorship and the violations of human rights that occurred in China.

In their most famous action, the Hong Kong Blondes launched cyberattacks against the 'Great Wall' – a series of firewalls put in place to block access to Western Internet sites. With members operating inside and outside of China, the group claimed to have found significant security holes within Chinese government computer networks and claimed to have defaced government websites, torn down firewalls and even disabled Chinese communication satellites. They also worked to forewarn political dissidents of imminent arrests.[6]

### 4.3.3 WikiLeaks

Probably the most famous hacktivist organisation was WikiLeaks, run by Julian Assange (initially in concert with Daniel Domscheit-Berg).

The basic idea behind WikiLeaks was simply to publish information concerning centralised powers sourced via internal leaks or the hacking of communications, often by finding security vulnerabilities or by 'social engineering' (for example, tricking someone into giving up a password).[7]

Probably the most famous group of documents published by WikiLeaks involved classified documents that had been provided by Chelsea Manning, then a private in the US Army in Afghanistan. Among those documents was a video shot from a helicopter gunship opening fire on a photographer followed by a van full of children in Afghanistan. It was an example of hacktivists providing a look behind the curtains and into the actions of the state in which they reside.[8]

---

[6] Oxblood Ruffin, 'Blondie Wong And The Hong Kong Blondes', 2015 <https://medium.com/emerging-networks/blondie-wong-and-the-hong-kong-blondes-9886609dd34b> [accessed 28 October 2024].

[7] The original plan was to run WikiLeaks like a wiki, which is to say that the editing of the content would be carried out by users rather than a central core staff, which is what happened.

[8] WikiLeaks, 'Collateral Murder', *WikiLeaks*, 4 May 2010 <https://collateralmurder.wikileaks.org/> [accessed 28 October 2024].

The 'Collateral Murder' video was not even the most influential document from Manning's leak. Even more significant were State Department documents that included a secret cable written in 2008 by Ambassador Robert F. Godec, which seemed to make it vivid that the external world saw the corruption of the Tunisian president as clearly as the Tunisians did.

Godec's description of the Tunisian situation in the leaked cable seemed to validate all the complaints that Tunisians had been making about their government:

> Beyond the stories of the First Family's shady dealings, Tunisians report encountering low-level corruption as well in interactions with the police, customs, and a variety of government ministries. [. . .] With those at the top believed to be the worst offenders, and likely to remain in power, there are no checks in the system.

Perhaps the most powerful statement in the cable was when Godec said that the Tunisian government seemed to believe that 'what's yours is mine'.[9]

Just days later, after a shopkeeper set himself on fire in protest, more protests broke out, leading to the collapse of the Tunisian government, followed by revolutions across the Arab world. The series of events became known as the Arab Spring and then-new technologies were deployed to assist protesting groups to organise demonstrations and circumvent attempts by state security systems to block communications with the global community.[10]

---

[9] WikiLeaks, 'Cable: 08TUNIS679_a', *WikiLeaks*, 23 June 2008 <https://wikileaks.org/plusd/cables/08TUNIS679_a.html> [accessed 28 October 2024].
[10] Burcu Bakioglu and Peter Ludlow, 'Can WikiLeaks and Social Media Help Fuel Revolutions? The Case of Tunisia', *The South African Civil Society Information Service* <https://sacsis.org.za/site/article/607.1> [accessed 28 October 2024].

It must be noted that not everyone was convinced that digitally mediated communications played a significant role in the Arab Spring. Famously, Malcolm Gladwell criticised it as 'Facebook activism' and argued that it lacked the 'strong ties' necessary to foment revolutionary fervour.[11] However, Gladwell was showing his disconnect from computer-mediated communication. Certainly, to the average Facebook user with little at stake in the uprisings, a few words posted in support of the cause are relatively meaningless. However, for people with skin in the game – people who were under the control of the Tunisian government, for example – it made little difference whether the communications were electronically mediated or not. The stakes were just as high, and the dangers were just as great.

For example, Slim Amamou, one of the more visible Tunisian bloggers online, was arrested and held during the Arab Spring because of his alleged connection with the international activist movement Anonymous. Azyz Amamy, who had covered the Tunisian protests from the beginning, was arrested. His Blogger and Facebook accounts were both deactivated. Hamadi Kaloutcha, a blogger and activist, was arrested at his home. During his arrest, police confiscated his computer equipment, suggesting that while Gladwell may not have been impressed by the online activism, the Tunisian police were concerned about it just the same. Around the same time as the others, the arrest of Hamada Ben Aoun, a rapper who had recently released two songs on his Facebook account criticising the Tunisian regime and its social policies, further demonstrates the authorities' concern over online activism.[12]

Our point is that online political activism is just as visceral, just as real and just as dangerous as face-to-face activism. When there is skin in the game, it does not matter if the communication is taking place through analogue sound waves or through digital electronic communications. Just as much is at stake, and the interpersonal connections are just as real – certainly real enough for people to be risking jail time and even death.

---

[11] Malcolm Gladwell, 'Small Change', *The New Yorker*, 27 September 2010 <https://www.newyorker.com/magazine/2010/10/04/small-change-malcolm-gladwell> [accessed 28 October 2024].
[12] Bakioglu and Ludlow, 'Can WikiLeaks and Social Media Help Fuel Revolutions?'

These Cypherpunks and hacktivists and the online activists of the Arab Spring are not merely interesting because of what they accomplished; they are of interest to us here because they showed how digital technologies could be deployed by non-state groups to express and defend important values, even if the defence of those values runs counter to the interests of nation states and their agents. In particular, they showed that technologies can be employed to defend the privacy of individuals, to pierce the veils of secrecy that centralised powers deploy, and to organise fellow citizens for change and a new political order.

Most importantly, they showed us ways to organise ourselves that are indifferent to the traditional boundaries of nation states. They showed that these communications and collaborations could carry on even when nation states did not look favourably on them and even when nation states used violence to block them.

## 4.4   Digitally mediated, borderless governance

A final element to add is the evolution of the idea that individuals can not only organise themselves online but that these online groups can take on the roles traditionally assumed by nation states. This was the unifying idea of the collection of essays anthologised by Peter Ludlow (one of the coauthors of this book) in the book *Crypto Anarchy, Cyberstates, and Pirate Utopias*. As early as the year 2000, new technological tools showed us that online governance systems were a very real alternative to traditional forms of government.

At the time *Crypto Anarchy, Cyberstates, and Pirate Utopias* was published, blockchain technology was still some years away. However, there were already efforts to bring about a viable digital currency, particularly by David Chaum and his DigiCash project.[13] Online communities were already forming, encryption technologies were being deployed in those communities and a new form of government – the cyberstate – seemed possible.

Another important idea that emerged at the time was the thought that these possibilities of online worlds might provide the opportunity for people to experiment with new forms of governance, for people to freely sample different forms of governance and for there to be a kind of competition for different forms of governance, each being accessible to a broad range of individuals. John Stuart Mill talked about a marketplace of ideas;[14] you could call this a marketplace of governmental systems.

First, an important essay by the legal scholars David Johnson and David Post, reprinted in the *Crypto Anarchy* volume, noted that the new lines of communication were orthogonal to the boundaries of traditional states:

> Global computer-based communications cut across territorial borders, creating a new realm of human activity and undermining the feasibility – and legitimacy – of applying laws based on geographic boundaries. While these electronic communications play havoc with geographic boundaries, a new boundary, made up of the screens and passwords that separate the virtual world from the 'real world' of atoms, emerges. This new boundary defines a distinct Cyberspace that needs and can create new law and legal institutions of its own.[15]

---

[13] Kirsten R. Schmitt, 'DigiCash: Meaning, History, Implications', *Investopedia*, 27 August 2023 <https://www.investopedia.com/terms/d/digicash.asp> [accessed 28 October 2024].
[14] John Stuart Mill, *On Liberty* (London, 1859).
[15] David R. Johnson and David G. Post, 'Law and Borders: The Rise of Law in Cyberspace', in *Crypto Anarchy, Cyberstates, and Pirate Utopias* (Cambridge, MA, 2001), 145–96.

In a separate essay, David Post developed the idea that there could and should be competition between these governance systems:

> Although each individual network can be constrained from 'above' in regard to the rule-sets it can, or cannot, adopt, the aggregate range of such rule-sets in cyberspace will be far less susceptible to such control. A kind of competition between individual networks to design and implement rule-sets compatible with the preferences of individual internetwork users will thus materialize in a new and largely unregulated, because largely unregulatable, market for rules. The outcome of the individual decisions within this market – the aggregated choices of individual users seeking particular network rule-sets most to their liking – will therefore, to a significant extent, determine the contours of the 'law of cyberspace'.[16]

Since these essays were written in the 1990s, newer technologies – in particular, blockchain technologies – can now be leveraged to create new forms of governance that enable the governed to retain their self-sovereignty. They can enable us to maintain our privacy and demand that governments be transparent. Most importantly, they facilitate exit from governmental systems that do not conform to their citizens' values. The possibility of exit thus allows governing systems to compete for citizens. Rather than having citizenship determined by terrestrial boundaries, it can be determined by our sense of mutual alignment in terms of politics and values.

One version of this alternative picture is what Ludlow called a 'cyberstate', and US entrepreneur and former CTO of the crypto exchange Coinbase Balaji Srinivasan has called a 'network state' – a state that is formed by the online digital interconnections of people around the world.[17] In this book, we will be looking at a more fine-grained set of solutions since we do not see that the concept of a state is all that helpful. Whether it is a network state or a nation state, it is still a state and thus carries all the baggage that states do. Our approach is to consider alternative forms of organisation as the model; we can organise around the concept of communities rather than states.

---

[16] David G. Post, 'Anarchy, State, and the Internet: An Essay on Lawmaking in Cyberspace', in *Crypto Anarchy, Cyberstates, and Pirate Utopias* (Cambridge, MA, 2001), 197–212.

Successful communities might scale up into groups numbering in the billions or be as small as a dozen friends. Our idea is that these communities can be decentralised yet cooperative thanks to their organisation around blockchain technologies. We call them sovereign blockchain communities because they will be politically sovereign and because the blockchain will become the central nervous system of these communities and their governance. The question, of course, is how these new technologies work – a topic we take up in the next two chapters.

---

# CHAPTER 5

# TECHNICAL FOUNDATIONS FOR DECENTRALISED COOPERATION

## 5.1  Preliminaries

In 1678, thirty years after the Peace of Westphalia, the German philosopher Gottfried Wilhelm Leibniz wrote to the Duke of Hanover proposing a permanent archive of documents for governance purposes. When that fell on deaf ears, Leibniz made the proposal again in 1680 to the Duke's successor. Again, it fell on deaf ears. But who is Leibniz? What was his proposal? And what does it have to do with the future of governance?

In addition to his writing on metaphysics (for example, on monads), Leibniz was known for his debates with Isaac Newton's associate Samuel Clarke about the nature of space and motion, for his work on the philosophy of religion, and for his work on the theory of knowledge. Leibniz is perhaps best known for independently inventing the calculus (indeed, inventing a version of the notation we use today). Although not his most well-known work, he wrote a tremendous amount on political philosophy, law and government. And these are the writings that concern us here – especially, as they relate to government archives.

When Leibniz wrote to the successive Dukes of Hanover, he described his archive project as 'a place where writings useful for government are kept in such a way that they rest intact and unaltered for future information, and that in that occasion they can be used as certified proofs in justice.' Contemporary Brazilian philosopher Ulysses Pinheiro provided a helpful summary of the project, noting that the idea was to create 'an instrument with a fundamental concern over the future actions of a Prince. It has the epistemic value of a virtual mechanism guiding the Prince's actions.'

Leibniz was apparently obsessed with his idea for an archive. In his proposal to the Dukes of Hanover, he went into great detail on how they should secure records. He proposed that the structure housing the records be protected so as to adequately preserve the documents from mould, mice and worms, as well as from fire and attack of enemies. He added that its internal vaults should be reinforced, and its doors should be made of iron. The most important original documents should be kept in a safe and hidden in a wall.[1]

One thing that Leibniz, for all his genius (and despite himself being very much a decentraliser), apparently did not envision was that these records could be decentralised and thus made even more secure. But how are they more secure?

As we will see, centralised record repositories create points of failure in governance. They provide vectors of attack for internal and external enemies, they are magnets for corruption, and they are susceptible to destruction from natural causes (for example, from mould, mice, worms and fire). Decentralisation, we will argue, is necessary to make archives secure. It is not surprising that Leibniz failed to see this despite him being a decentraliser regarding political sovereignty and even in his work on metaphysics (his use of monads, in particular, was a kind of decentralising project). His proposal was more than three centuries before Satoshi's white paper, after all, and no one has perfect foresight.

---

[1] Ulysses Pinheiro, 'Leibniz on the Concepts of Archive, Memory, and Sovereignty', *Für Unser Glück Oder Das Glück Anderer*, 3 (2016), 309–21 <https://www.academia.edu/39198781/Leibniz_on_the_Concepts_of_Archive_Memory_and_Sovereignty> [accessed 16 April 2023].

In this chapter and the next, we will develop some of Leibniz's ideas in ways we think he would like were he alive today. In this chapter, we will discuss immutable records and how they are only made possible by decentralised protocols, and we will lay out the theory of secure decentralised protocols. In the next chapter, we will explore how certain immutable records, in the form of onchain smart contracts, can be leveraged into 'virtual mechanisms', not merely 'guiding the Prince's actions' but executing those actions automatically.

## 5.2  Immutable records

In May 1747, Don Giuseppe Rapaccioli, a parish priest in Macinesso, Italy, was supervising the renovation of a field near his church when he came across a large bronze plaque containing Latin inscriptions. Today called the Tabula Alimentaria Traiana, it measures 1.38 metres high and 2.86 metres long. It is, to this day, the largest bronze inscription from antiquity that has been found. Further study of the Tabula dated it to 112 CE, during the reign of the Roman Emperor Trajan (Traiana signifying its relation to Trajan).[2]

In the early second century CE, Trajan instituted an alimentary programme in towns throughout the Italian peninsula. This programme encouraged landowners to take out loans from the government by mortgaging their property. The interest on these loans was paid in the form of a food stipend for children in the respective areas. In other words, landowners could borrow money from the government using their land as collateral, provided that the interest was paid as assistance to local poor children. Emperor Trajan thus used short-term bribes to incentivise landowners to look after local children in the long term.[3]

---

[2] Gianluca Bottazzi, 'Varsi E La Tabula Alimentaria Di Veleia', *ValcenoStoria*, 2020 <https://www.valcenostoria.it/2020/10/06/varsi-ela-tabula-alimentaria-di-veleia-by-gianluca-bottazzi-valcenostoria/> [accessed 24 January 2024].

In this particular case, the bronze Tabula codified the details of Trajan's alimentary project in Veleia, a rural northern Italian town near the Roman colony of Placentia (now known as Piacenza). The recorded information on the plaque included the debtors' names, the location and value of their property, and the loans they received for that mortgaged land (generally, a little over 8% of the land's appraised value).

This bronze plaque had certain important features that we are interested in. It was public; everyone in the town could see it, and everyone knew what was owed and what the landowners were expected to pay for the children. Second, it was, for all practical purposes, immutable. It was cast in bronze, after all. Not one detail had changed between the time it was lost and its discovery by Father Rapaccioli in 1747.[4]

Not only was the information in the Tabula public and functionally immutable, but it also had a 'cybernetic' effect (in the sense of the ancient Greek origins of the term). By making responsibilities public, it helped to ensure that those responsibilities were carried out. Everyone knew what was expected. Thus, it was not merely information but information that had a governing function. Of course, the encoded information could not guarantee that Trajan's policy would be carried out. Yet, whoever made the Tabula presumably assumed it would encourage people to fulfil their obligations or possibly feel shame if they failed to live up to them.

We can also imagine a scenario in which recorded information has a direct causal role by being encoded in a computer program that might reliably act on the information. We will develop this in the next chapter, but for now, let us focus on the importance of secure information and think about what happens when reliable records are not available.

---

[3] Here, we set aside questions about the nobility of Trajan's programme, which was probably motivated by a concern that rural regions of the empire would become depopulated.

[4] Strictly speaking, it was not absolutely immutable. Obviously, changes can be made to bronze plaques; in this case, changes were indeed made. The Tabula was updated through the years before it was lost. In fact, there has been a dispute about whether variation in orthography on the Tabula is a result of error or changes in Latin over the years the Tabula was in use.

Detailed records have been vital to human governance since the earliest civilisations on record. The ancient Sumerians left behind a detailed record in the form of cuneiform writing on clay tablets. Subsequent civilisations in the Near East (the Babylonians, Assyrians, Elamites, Hurrians, Kassites and Hittites) also made extensive use of cuneiform records to extend the scope of their governmental control, so much so that we now know their legal record keeping as cuneiform law. The most famous case of this legal record keeping is the Code of Hammurabi, which remains one of the most well-preserved legal texts of the era. Cuneiform law became not only a way of regularising the administration of justice but also a means of tying together the administration of perhaps the world's earliest empire and projecting its force throughout Mesopotamia.[5]

It is worth emphasising the role that proper record keeping plays in the administration of justice, for we seldom think about just how important reliable records are. A more contemporary illustration may bring this point home.

In 2008, a group of students from the Pepperdine Caruso School of Law went to Uganda, participating in a project to modernise the Ugandan legal system. One of the most salient facts that stood out to the students was the poor state of the legal records. A campus newspaper reported on their findings: 'Archives are designed to serve as an area where law precedent is carefully stored, indexed and readily available to the judges. This archive, however, was located deep in the basement of the High Court's building and consisted of hundreds of unorganized case files dumped into a dark and dirty closet.'[6]

The first point of note here is that if there is no reliable record of legal decisions, there is no meaningful notion of judicial precedent. Every case must be argued anew, based not on precedent but on whatever reasoning the local justice chooses to apply. Apart from the inconsistency that this generates in legal decisions, it slows the process considerably if every case needs to be reasoned from scratch.

This leads to a related point that might not occur to First-World citizens. There is a financial cost to record keeping. In the Western world, judges and courts have nearly unlimited access to clerks and stenographers and other ways of recording deliberations and decisions. However, in places like Uganda, this cannot be taken as a given. This too was noted by the Pepperdine article: 'This East African country doesn't have sufficient funding to provide judges with support staff or clerks; they must research, develop and write the decisions on their own. Additionally, High Courts have no court reporters to record trial proceedings, so the judges must hand-write notes.' As Pepperdine law student Greer Illingworth put it, ' "the proceedings can only move as fast as the judge takes notes." '[7]

Situations like this do not merely slow down the proceedings; they leave thousands of people in perilous legal limbo. As the students discovered, there were more than a few remanded minors living in squalid conditions with no reliable record of their case histories, including the charges against them, nor any prospect of timely resolutions of their cases. Failed record keeping put them into dangerous legal limbo.

Failed record keeping can also lead us to lose track of governmental atrocities, as the students learned while exploring the dingy basement that served as a repository for Uganda's legal system: 'Students stumbled across a turquoise book marked "Confidential" amid the piles of legal papers. Curious, they soon realised that they had found hundreds of execution orders given by Idi Amin, military dictator and president of Uganda from 1971–1979.'[8]

---

[7] Ibid.
[8] Ibid.

It may be surprising just how fragile record keeping has been through the centuries. For example, during revolutionary periods, one of the first and most important objectives for the revolutionaries is to find and destroy the records of the existing regime. Among other things, records contain files on dissidents and details of their arrests and activities. Destroying such records can protect dissidents from government retaliation.

Another target has been records of property ownership. In the earlier Russian revolution of 1905 (what Lenin called 'the dress rehearsal' for the revolution of 1917), peasants attacked the estates of landowners and destroyed physical property and, notably, debt records.[9] Similarly, in the Mexican Revolution, Emiliano Zapata's revolutionaries destroyed the previous regime's records of land ownership.

The case of the Mexican Revolution is particularly apt because records (and the lack of records) played a role on both sides of the conflict. Perhaps initiating the revolution were the actions of the *haciendas* – large estates that seized ancestral land claims from poor farmers, a crime made possible by the lack of permanent records of ownership. When Zapata's men attacked the haciendas, among the principal targets were the newly minted records of ownership. A fascinating collection of papers assembled by Carlos Aguirre and Javier Villa-Flores documents numerous examples of this sort of archival conflict in Latin America. It is a tradition that begins with the destruction of Mayan records by the conquistadors in 1530, an event that was followed by the destruction of twenty-seven Mayan codices by Diego de Landa, the Bishop of Yucatán, in 1566.[10] Reporting on the atrocity, de Landa himself noted, 'We found a large number of these books, and, as they contained nothing in which there was not to be seen superstition and the lies of the devil, we burned them all, which they regretted to an amazing degree and which caused them great affliction.'[11]

---

[9] Mark D. Steinberg, *The Russian Revolution, 1905–1921*, 1st ed. (Oxford, 2017).
[10] Carlos Aguirre and Javier Villa-Flores, *From the Ashes of History: Loss and Recovery of Archives and Libraries in Modern Latin America* (Chapel Hill, NC, 2015).

The history of archival conflict proceeded through the destruction of records of the Spanish Inquisition to the destruction of municipal archives during an uprising in Mexico City in 1630. It continued all the way through the twentieth century, during which the destruction of archives remained a robust tradition during times of political conflict. We have already mentioned Zapata's destruction of records of property ownership; this is a strategy that was repeated in El Salvador in 1932 when Red Commanders destroyed the property records in every town they occupied.[12]

Of course, economic records such as ownership and debt are not the only targets of archival destruction; cultural records are routinely destroyed as well. This was certainly the goal in the destruction of the Mayan codices in the sixteenth century. As Aguirre and Villa-Flores point out, the problem is not just with intentional efforts to destroy records but also the fact that records can be subject to lack of attention, natural disasters and so on.

Earlier, we mentioned the case of the poorly maintained legal records in Uganda. However, there are many examples of this in Latin America as well, including the Lima–Callao earthquake of 1746 and, more recently, the tragic 2015 fire in the National Museum of Brazil that destroyed 92.5% of its repository of 20 million items, including its archive of indigenous languages. An employee of the Federal University of Rio de Janeiro announced the tragedy to online linguists:

> Folks, there's nothing left from the Linguistics division. We lost all the indigenous languages collection: the recordings since 1958, the chants in all the languages for which there are no

---

[11] Inga Clendinnen, *Ambivalent Conquests: Maya and Spaniard in Yucatan, 1517–1570* (Cambridge, 1987), 70.
[12] Aguirre and Villa-Flores, *From the Ashes of History*.

native speakers alive anymore, the Curt Nimuendajú archives:
papers, photos, negatives, the original ethnic-historic-linguistic
map localizing all the ethnic group in Brazil, the only record
that we had from 1945. The ethnological and archeological
references of all ethnic groups in Brazil since the 16th cen-
tury… An irreparable loss of our historic memory. It just hurts
so much to see all in ashes.[13]

What does the loss of cultural information have to do with gov-
ernance? Well, culture plays a role in the unification of people for a
common cause. Good governments often take on the responsibility of
preserving cultural records, not just because it is a good thing to do, but
because the very identity and existence of the government often depend
on these cultural archives (think of the role the French government plays
in maintaining the Louvre or the role the Italian government plays in
maintaining the Uffizi). Even if we think that governments should have
nothing to do with preserving national culture, there are still institu-
tions that will be charged with protecting cultural heritage, and those
institutions also need proper governance, even if it is self-governance.

It is important to understand that the loss of archives is not the
only pain point. A similar effect can be achieved by restricting access to
archives. To put it another way, you do not need to destroy the archives
if you can instead make them inaccessible. And for sure, there are plenty
of information archives throughout the world that remain inaccessible,
either because they are classified or private or behind a paywall – a
point that was made clear by Aaron Swartz in his *Guerilla Open Access
Manifesto*.[14]

Why does all this matter? Because archives and access to the infor-
mation contained in those archives are critical to freedom, democracy and, certainly, our ability to know whether our government is functioning as it promised (whether that government is democratically chosen or not). Jacques Derrida summarised the importance of archives to governance as follows: 'There is no political power without control of the archive, if not of memory. Effective democratization can always be measured by this essential criterion: the participation in and access to the archive, its constitution, and its interpretation.'[15]

---

[13] <https://m.facebook.com/story.php?story_fbid=10107422574817418&id=5740086> [accessed 28 October 2024].
[14] Peter Ludlow, 'Aaron Swartz Was Right', *The Chronicle of Higher Education*, 2013 <https://www.chronicle.com/article/aaron-swartz-was-right/> [accessed 2 January 2024].

In his discussion of Leibniz's proposal for an archive in conjunction with the philosopher's views of national sovereignty, Pinheiro argued that, to some extent, the very existence of the state is a function of the archives it keeps.[16] If this sounds implausible, it should not. If the archives contain evidence of treaties and land ownership and citizenship and economic contracts, then at the very minimum, one can reconstruct the state, its borders and its interests from a well-maintained archive. Were the archive to disappear due to an 'archive bomb', it is not at all clear that the state would be able to persist in its pre-bombing form. Indeed, on reflection, an archive bomb would be at least as devastating as Tom W. Bell's law bomb because the law is dependent on the existence of functioning archives.

If we think of a government as a body, then information archives are its DNA; they are the information within the cells of the organisation that guides it in what organs to make and where and how they should be regulated. Similarly, archives are not just things that states keep. Archives are critical to the very *identity* of the state. We believe this also holds true at every level of governance.

Now, we get to the real challenge. Archives are not only under threat from conquistadors and revolutionaries and fire and earthquakes. Because they are under centralised control, they are also vulnerable to the corruption and ineptitude of record keepers. Is there a way to

---

decentralise record keeping and thus make records safer? This is one of the great promissory notes of blockchain technology.

## 5.3   Decentralised cooperation

One of the more infamous examples of record destruction happened on 7 December 1985, when Pablo Escobar paid the left-wing guerrilla group M-19 (the 19 April Movement) to invade Colombia's Supreme Court building in Bogotá and destroy records related to criminal cases being built against Escobar and other *narcotraficantes*. If Pablo's son is to be believed, this was accomplished for the low price of $1 million – peanuts to an international drug trafficker.[17] The damage went far beyond the documents in the case against Escobar. According to Mark Bowden in his book *Killing Pablo*, the attack destroyed documents for 6,000 pending criminal cases and 'crippled the Colombian legal system'.[18]

We can, of course, attempt to armour our central points of failure (as Leibniz proposed to the Dukes of Hanover), but sadly, these attempts are inherently vulnerable to all manner of attack. Pablo Escobar chose the crude method of a guerrilla military operation, but softer methods are often just as successful.

There is a reason why the Defense Advanced Research Projects Agency (DARPA) funded the early stages of the Internet. A decentralised (or at least more decentralised) network is less vulnerable to attack. In theory, it should even survive a nuclear war. Similarly, if we want our documents and records of value to survive attack, then we should consider decentralised networks. However, slogans about decentralisation aside, this turns out to be easier said than done.

---

[17] Whether Escobar's son is a reliable source is another question. In the same 2014 interview, he said that Pablo was not killed but committed suicide. Whatever the actual price, the important point is that a centralised store of records creates a point of vulnerability. The Spanish language interview can be found online here: <https://www.wradio.com.co/noticias/actualidad/pablo-escobar-se-suicido-no-lomataron-su-hijo/20141106/nota/2495321.aspx> [accessed 30 October 2024].

[18] Bowden is a journalist that is perhaps better known for his book *Blackhawk Down.*

If a centralised repository of records represents a point of failure, then it would be natural to assume that a decentralised network represents multiple points of failure. However, the difference is that a decentralised network can be fault tolerant. Nodes in the network can fail, but the network keeps humming. You see this in the case of the Internet. If an Internet node goes down, the entire Internet does not fail. As we will see in the next section, fault tolerance takes on a whole new dimension with regards to matters of money and control.

Most of the institutions of trust that we engage with today are centralised. For example, our national governments are typically central governing bodies that sit in a national capital. Most nations issue their own currency, which is under the control of a central bank. So too, our various institutions of financial trust are centralised. We trust large banks to accurately record how much money is in our accounts and our financial transactions. If we wire $100 to your account from our account, our bank debits our account $100. If we use the same bank, our bank simply credits your account $100 – no coins or dollar bills or bars of gold move. There is just a simple change on a ledger that the bank keeps. If we use different banks then our bank will credit the account of your receiving bank, and this will be recorded by a centralised interbank protocol like SWIFT.

Similarly, if we buy land, its ownership is recorded by some single official record keeper. In the United States, that would be a title company. In Mexico (prior to 2016), it would have been a local notaría. Perhaps you worry that your title company in the United States might misfile something or overlook a dispute about who owns your property. That is why we have title insurance. However, as with everything else, title insurance is recorded with a centralised institution.

The problem with centralisation is that it represents a massive point of failure for any system, whether it is governmental or financial. Pablo Escobar showed us how centralised legal records represent a point of failure, but this is true for all kinds of record keeping. If 'official' financial records are stored in a single centralised location, then those records can be destroyed by targeting that single location. Centralised records are also targets for corruption. If a record of property ownership is controlled by a single notaría, the notaría becomes a target for bribery and extortion and thus becomes the point of failure for the property ownership system. If the official copy of the records is moved to a central government office, then that government office becomes a target for corruption.

The same can be said for national currencies, government policies and other societal institutions. If there is a central authority, then that central authority becomes an attack vector for enemies and an opportunity for individuals to game the system for their own benefit. And it may be a very banal form of corruption at the end of the day. After all, central bankers have friends, and they are apt to be part of a wealthy class of individuals. Thus, central bankers are likely to take on the attitudes of their small class of friends. Does this make them corrupt or evil? Well, that is not the issue; more pressing is whether such decisions should rest in the hands of a few central authorities – authorities that are vulnerable to falling under the influence of a small number of individuals. It does not matter whether that influence is driven by force or corruption or simply the interests of close friends. The end result is the same.

We all know that corruption and more benign forms of influence exist, and we certainly know that there are countries and institutions where corruption and the exercise of political influence are particularly rampant. However, rather than criticise the tools that we use for governance, we are often critical of the morals of corrupt and inappropriately influenced individuals. For sure, there are some morally reprehensible people out there engaging in all sorts of corruption. However, corruption has been a thing for the entire history of human governance. While it might give us a sense of satisfaction to be critical of the individuals caught in the corruption and influence game, it would be much more effective to avoid governance architectures that invite such corruption and influence peddling.

It is worth noting that many individuals have played the corruption game because they had no choice. As Colombian officials found in the days of Pablo Escobar, there was often an ultimatum: *plata* or *plombo* (silver or lead) – 'take the bribe and do what we want, or you and your family may suffer physical harm.' We can blame people who cave to this pressure if we want, but it makes far more sense to attack the root of the problem, which is centralised governance structures.

You might think that human governance cannot be any other way than centralised because, after all, someone has to keep these records. If everyone kept their own record, surely there would be disputes. How would we even resolve those disputes if not through some central authority that had access to the correct or official record of ownership?

It is one thing to say that you do not like centralised records, but it is quite another matter to figure out what the decentralised alternative might be. The first thought is that we could simply make copies of the record and distribute it to everyone in the network. However, this raises the question of what happens when disputes arise. What if our records are inconsistent? The temptation is to think that there should be some centralised official record that we can check against, but now we are back in the business of centralised record keeping, and that is the very thing we are trying to get away from.

Centralised record keeping presents numerous points of weakness, but as we are about to see, decentralised networks have their own problems that we will need to address. One such problem is the 'double-spending problem'. Suppose that we offer to buy a widget from you for $100. You check the shared distributed ledger and confirm that we have $101 in our account, so we should be able to pay. You send us the widget, and you deduct the $100 from our account. However, what you did not realise is that just as we were buying the widget from you, we were buying a $100 cog from Smith. Smith also consulted their copy of the ledger, saw that we had $101 in our account and sent us the cog. Clearly, it is not possible for both you and Smith to be paid. Who is right?

This is the problem: if there are many nodes in the network, there can be bad actors at any of those nodes. Does decentralisation not just increase our headaches when it comes to failure and corruption?

## 5.4   Byzantine generals, decentralisation and Satoshi

The double-spending problem is just one instance of a class of problems that every decentralised network has to confront. In 1982, Leslie Lamport, Robert Shostak and Marshall Pease released a research paper titled 'The Byzantine Generals Problem',[19] which gave the class of problems the name by which we know it today.

Here is one version of the problem: several Byzantine generals must participate in a coordinated attack on a city, each at the head of an army. It is a critical point that the attack must be coordinated. If too few armies attack, they will be crushed. However, now we encounter several problems. One problem is that the generals must communicate with each other in what is effectively a hostile environment. What if the messenger gets intercepted? Well, you could wait for a responding message that your initial message was received. But how does the other general know you received the response? Do you have to respond that you got the confirmation? For a coordinated attack, it is not enough for everyone to receive the attack message; everyone has to know that everyone received it, and they have to know that everyone knows that everyone received it and so on.

However, there are more problems. A coordinated attack may also fail due to traitorous generals. Or perhaps, it will fail merely due to incompetent generals sleeping off their hangovers. Therefore, we must plan our attack in a way that is 'fault tolerant' – the attack must be successful even if there are weak links among our generals. This is 'Byzantine fault tolerance'.

---

[19] Leslie Lamport, Robert Shostak and Marshall Pease, 'The Byzantine Generals Problem', *ACM Transactions on Programming Languages and Systems*, 4/3 (1982).

This problem has been with us for as long as we have had to engineer distributed systems. Research on Byzantine fault tolerance (although not under that name) began in the 1950s and mainly revolved around the aviation industry. To see why, consider an aeroplane with multiple computers that might fail. You do not want one failure to bring down the whole system, but how do we engineer around these inevitable failures?

In the 1970s, researchers at Draper Laboratory published a technical report on the fault-tolerant multiprocessor (FTMP) – a multiprocessor computer that eliminates single-fault vulnerability for aircraft modules. During the same decade, Honeywell developed the multi-microprocessor flight control system (MMFCS), which focused on the detection of Byzantine failures. Then, in 1981, SRI International published a technical report for aircraft control computers called 'The software-implemented fault tolerance /SIFT/ approach to fault tolerant computing'.

As noted, these papers were connected with the aerospace industry, where there are plenty of distributed systems and where some failure is inevitable but the wrong kind of failure could be catastrophic. It was not for nothing that the research behind the 'Byzantine Generals Problem' paper by Lamport, Shostak and Pease was funded by NASA, the Ballistic Missile Defense Systems Command and the Army Research Office.

In 1998, Lamport wrote another important paper in which he offered a solution to the Byzantine Generals Problem in the form of the Paxos protocol. The point of departure for Lamport's paper is the Aegean island of Paxos, which in ancient times was (we are told) run by a 'part-time' parliament. It was part-time because trade was a bigger priority than governance for the governing citizens; no one in Paxos could afford to be a full-time member of the island's parliament – they all needed to be travelling and trading. This meant that the government had to find a way to function even while its voting leaders and members came and went, and each would need their own record of parliamentary decisions while they travelled.

The analogy to distributed systems and the problem of Byzantine fault tolerance should be clear. Members of the parliament would be voting on important measures at different times. Many people would not have access to a centralised ledger, and in point of fact, if people were allowed to vote while travelling, even a central ledger at the capital would not be completely up to date.

No one knows how Paxos carried out this feat, so Lamport spun a story of an imaginary protocol in which records of votes would be copied and distributed. Each member of parliament would have their own version of a complete record of all the votes submitted thus far – i.e. state descriptions of the protocol. People could then update their copy of the record with their votes using indelible ink. When disputes arose over conflicting ledgers, there would be votes to determine which modified state description would be deemed official – the official description would be the one that ultimately received the majority vote (this vote could take some time for obvious reasons). In effect, you would not only be voting on policies but also on the official record of additions to the record of votes and policies.[20]

Notice that this is quite a good solution to some varieties of the Byzantine Generals Problem. If you have a way of tracking which decisions have a majority vote and everyone has a record of that, then if a majority of the armies is sufficient to carry out the attack, you are in a good position. The solution only breaks down if most of the generals are corrupt.

The Paxos protocol resolved some issues with distributed networks, but it did not resolve all of them, and we mentioned one such problem earlier – the double-spend problem. Recall that this is a case where we have $101 and purchase a $100 widget from you and a $100 cog from Smith, but neither of you is aware of the scam because you are both working from records (state descriptions) that show that I have enough money to pay for the product.

---

[20] Lamport, 'The Part-Time Parliament'.

This leads us to the 2008 publication of Satoshi Nakamoto's Bitcoin white paper and the subsequent implementation of Bitcoin in 2009. Many people understandably think of Bitcoin as being a virtual coin of sorts ('coin' is in its name, after all), but that is a very weak metaphor. If you are a so-called 'whole coiner', who owns one bitcoin (BTC), there is no coin sitting in a repository somewhere. It is much better to think of the Bitcoin protocol as a decentralised record or ledger that keeps track of who holds what – a ledger that is immutable (like the Tabula Alimentaria Traiana) and which is visible to all (as was the Tabula) but which, unlike the Tabula, is a decentralised record. It is not located in one place, but control of the record is distributed across the network. This has the advantage that it does not have a single point of attack, but it still must encounter the Byzantine Generals Problem, and in particular, the double-spend problem, which heretofore had not been solved. Here lies the monumental genius of Satoshi's white paper.

Rather than having a single centralised authority that determines the official version of the blockchain, that determination can be a rotating responsibility. Furthermore, we can prove whether a proffered version of the blockchain is the official version. There are many ways to execute this general strategy successfully. However, it is worth beginning with Satoshi's approach, which Bitcoin still uses to this day: 'proof of work'.

There are three fundamental questions that a distributed ledger has to answer. First, how do we select which version of the ledger is the official version? Second, how do we prove that a proffered version of the ledger is the officially selected version? Third, how does the protocol protect itself against bad actors? Satoshi's proposal solves all three problems with one idea: the aforementioned proof-of-work protocol.

The person who gets to declare the official version of the ledger is the first person who solves a certain cryptographic puzzle. The complexity of the puzzle is weighted in such a way that a new block (a new official version of the ledger) is produced roughly every ten minutes. The output of that effort answers question number two because that output is a 'cryptographic hash' that can tell us in short order whether a proffered version of the ledger is the new official version.[21] If one digit in the ledger has changed (one digit in the entire history of the ledger), then the hash will be completely different. Think of the hash as perfectly reliable evidence that the blockchain you are looking at is the official version at a specific date and time. The people who deploy computer hardware attempting to solve the puzzle and create the blocks (and earn rewards for doing so) are called 'miners'.[22]

We now have a way of selecting the 'official' block, and we have a way of proving whether we are looking at that official block. But what about the bad actors part? The protocol incentivises those participating in the network's operation to maintain the ledger's integrity in the form of block rewards, distributed in BTC. Bad actors can be penalised – they can be denied their block rewards. Given their massive energy expenses and investment in equipment, a rational actor with finite resources would consider such a strategy too costly, given their massive energy expenses and investment in equipment, when they could simply deploy the same resources and earn profit by following the network rules.

Similarly, a problem could arise if bad actors were able to create the official blocks most of the time. However, the only way to achieve that is to acquire enough hashing power to solve the cryptographic puzzles more often than all other participants combined. To do that, they would need to amass computer hashing power greater than half of the sum total of all the other computers in the network. Strictly speaking, this is not impossible, but it would be astoundingly expensive to acquire that much hashing power and pay for the energy necessary to solve the hashing puzzles. Estimates made in 2021 put the energy consumption of Bitcoin as comparable to that of Finland,[23] so anyone wanting to corrupt the blockchain would have to be prepared to expend more than half that much in energy consumption alone. Plus, there is the cost of acquiring control of all the necessary mining equipment.

---

[21] We can think of the cryptographic hash as being like an electronic signature that can verify the integrity of some file (in this case the ledger). If one element in the file is changed, then the correct hash (the signature) will not be produced. The signature thus serves as a proof that the ledger has not been corrupted.

[22] The computers used to do the hashing are also called 'miners'.

This was perhaps the most profound insight of Satoshi's proposal. You do not have to make attacks impossible; you just have to make them cost-ineffective. It is not impossible to break Bitcoin; it is just not worth the expense. As mentioned, proof of work is just one way to engineer Byzantine fault tolerance and resolve the double-spending problem. There are a number of alternative protocols available now, but one important class of strategies that we should remark on here is 'proof-of-stake' protocols.

With proof of work, the idea was to secure the network by making the costs of being a bad actor prohibitive; one would have to acquire more than half the hashing power of the network. To date, this has been a successful strategy, but it has also drawn criticism over its energy usage. Whether, in the big picture, it actually poses an environmental problem is up for debate (a good case can be made that it helps make renewable energy sources profitable).[24] However, it does raise the question of whether other options are available. Proof of stake (PoS) is one such option.

---

[23] Jon Huang, Claire O'Neill and Hiroko Tabuchi, 'Bitcoin Uses More Electricity Than Many Countries: How Is That Possible?', *New York Times*, 9 March 2021 <https://www.nytimes.com/interactive/2021/09/03/climate/bitcoin-carbon-footprint-electricity.html> [accessed 28 October 2024].

[24] The idea is that for many renewable energy sources like wind, much of the energy production goes to waste, but the energy produced in off-peak hours can be used to mine Bitcoin, radically improving the profitability of the renewable energy method used.

In the case of PoS, the network is secured by block-creating nodes staking cryptocurrency (for example, ETH) as a kind of guarantee that they will maintain the system's integrity honestly. Instead of getting to create a block by solving a cryptographic puzzle, your entitlement to create a block is a function of how much of the asset you have staked. The more you stake, the better your chances of being the node that gets to create the block. There is some computing involved, but only as much as is necessary to assemble the block and generate a cryptographic hash for it. We will get into the finer details of this later, including questions about the degree to which PoS really is decentralised and how it can be gamed, but for now, we just want to focus on the general idea and how it qualifies as a version of decentralised governance.

There are, of course, other protocols, but for the moment, we set those aside because there is a more pressing issue. We have seen how immutable public archives are important and also how it is crucial to their security that they be decentralised. We have also seen that in the last few decades, we have developed general strategies for decentralised computation, culminating in the proof-of-work and proof-of-stake protocols. But this is only a glimpse of what these new technologies can offer us. Not only can we look forward to decentralised ledgers, but as we will see in the next chapter, decentralised smart contracts will also open up a whole new world of possibilities for human governance.

# NEW TOOLS FOR HUMAN GOVERNANCE

## 6.1 Preliminaries

In the previous chapter, we looked at a proposal that philosopher Gott-fried Wilhelm Leibniz made for a permanent, immutable archive, and we argued that it could be strengthened if the archive was decentralised. In this chapter, we want to explore a second idea of Leibniz's.

When he was twenty years old, in 1666, Leibniz published his dissertation, which was on the idea for a formal universal language in which inferences could be carried out by algorithmic proofs (much as they are in computer languages today).[1] What is most interesting for our purposes is that Leibniz not only had his idea of an archive of government information, but he also seemed to believe that some of this information could be written in his universal language and that this information, so encoded, could be used to guide government actions. That may not seem like a big idea on the surface. However, we believe it was an idea so far ahead of its time that it would take a further 330 years to see its full expression in Satoshi Nakamoto's invention of the blockchain (in the form of Bitcoin) and Nick Szabo's invention of smart contracts (self-executing contracts that are instantiated as computer programs).

---

[1] Massimo Mugnai and Han van Ruler, *Leibniz: Dissertation on Combinatorial Art* (Oxford, 2020).

Leibniz believed that the correct action might be computed and left for the prince to carry out. What might these actions entail? Leibniz only provided clues in his dissertation, but the relevant expressions over which he thought actions could be computed included war, peace, wagons, money, vassals, truce, allies, booty, bridges, gunpowder, attack, parley, clients, routes, cannonballs, security, treaty, neutrals, agreement, enemies, ships, medicine and counsels. In short, anything that might be relevant to seventeenth-century statecraft.

Did Leibniz imagine that actual actions could be automated as well, or was the output to be merely written guidance? Probably the latter, but were he alive today, he no doubt would be attracted to the idea that his system could automate government actions. Indeed, it is worth noting that he proposed one of the very first physical computing machines – his 'stepped reckoner',[2] which could do multiplication and division.[3]

Leibniz's project seems to have been well known at the time, and it came in for criticism from no less than Jonathan Swift in his famous novel *Gulliver's Travels*. You may recall the passage in which Gulliver visits the Grand Academy of Lagado, where he encounters a strange mechanism called 'The Engine'. The Engine had a large wooden frame with a network of wires. On the wires were small papercovered wooden cubes with symbols written on each side. Students turned a crank, and scribes wrote down the output. The professor claimed that they thereby could 'write books in philosophy, poetry, politics, laws, mathematics and theology, without the least assistance from genius or study.' In other words, it was seventeenth-century ChatGPT. Gulliver left, unimpressed.[4] However, Leibniz apparently did not care what Gulliver (i.e. Swift) thought. In any case, Swift's satire did not persuade Leibniz to abandon his project.

---

[2] Paul E. Donne, 'History of Computation: 16-19th Century Work', *Paul E. Dunne*, 2000 <https://intranet.csc.liv.ac.uk/~ped/teachadmin/histsci/htmlform/lect3.html> [accessed 5 May 2023].

[3] Did it work? Leibniz conceived the mechanism in 1672, but it was not built until 1694. Its intricate gearwork seems to have been ambitious for the fabrication technology of the time; it did not work reliably. According to Paul Donne's History of Computation course notes, the problem was not diagnosed until 1893, when an investigation showed that a design flaw in the carrier mechanisms created problems when carrying tens.

Of course, Swift's passage now seems amazingly prophetic, even if his goal was to poke fun at Leibniz. Much of what we aim to do in this book is to execute a contemporary version of the project as envisioned by Leibniz and skewered by Swift. To do that, we are going to go deep into the idea of smart contracts deployed on the blockchain and how those smart contracts can take on the role of governance.

In Section 6.2, we will explore the idea that states and other communities might be organised around 'smart contracts' – by which we mean decentralised, immutable computer programs that are visible to community members and encode the intentions and thus future actions of the state or community. In the following section, we take up the very important question of how the blockchain imports information from the extra-blockchain world – information that will be critical to many smart contracts. This will involve discussing so-called 'oracles' and their role in providing this information to the blockchain. In 6.4, we will develop the idea of communities built around smart contracts further by introducing the concept of a 'decentralised autonomous organisation', or DAO. In Section 6.5, we will take up the topic of 'impact DAOs' – DAOs that are designed not merely to assist in community management but to create positive externalities as well. Finally, in Section 6.6, we tie everything together and give examples of how we might apply these new technologies to human governance.

## 6.2   Smart contracts and human governance

To refresh our memories, when we refer to 'human governance', we mean the systems and processes by which people manage and make decisions about their communities and implement those decisions to achieve some political, economic or cultural goal. Here, we take political in the very broad sense suggested by its origin in the Latin term *politicus*, meaning 'of citizens or the state'. In other words, we take political goals to be those goals that involve some plan for the future conduct of the state or some group of organised citizens. Maybe that goal is peace or war or a robust economy or, at lower levels, fixing potholes, reliable trash removal or weekly yoga classes on the roof of the condo building.

---

[4] Jonathan Swift, *Gulliver's Travels* (London, 2003).

Different levels of human governance, of course, need different tools, but some tools seem to be required at nearly every level. These include tools for recording reliable records, transparent and reliable channels of communication between governing officials and the community, safe and secure handling of funds with a reliable audit trail of those funds, ways of articulating community goals, and ways of incentivising community members to align with and pursue those goals. Based on remarks made earlier in this book, we hope it is clear how blockchain technology is optimally designed to facilitate most of these goals. Secure communications, as we will see, require other measures. However, a few additional observations might be helpful here.

Let us begin by revisiting the idea of an immutable public ledger. Such a ledger can do many things for us. Firstly, it can record what belongs to whom, and this is, in effect, what Bitcoin and Ethereum do. They are (among other things) immutable ledgers recording ownership (by wallet address) of crypto assets and the history of ownership transfers.

This basic technology can do a lot more than keep track of asset ownership, however. It can also be a record for use in inventory control or in tracking the movement of produce from farm to factory to store. It can provide a secure record of the movement of products and parts all around the globe and, more importantly, be accessible to all parts of the globe. On top of this, it can record the history of every state change of that record should we need to audit the history of these records. Critically, everyone can know that everyone else has access to the very same record. It could even record the amount of labour that goes into product parts and then the amount of labour that goes into a product assembled from those parts. Blockchains are value agnostic. They do not care what you value, but whatever you value, they can record it in an immutable, decentralised-yet-reliable shared record. This brings us to the matter of smart contracts.

In late 2013, five years after the Bitcoin white paper was published, nineteen-year-old Vitalik Buterin wrote the Ethereum white paper.[5] Building on ideas put forward by Satoshi in forum posts and elsewhere, Buterin observed that just as you can record a ledger and other static documents on the blockchain, hypothetically, you can use a blockchain to store and execute computer programs (few people realise that Satoshi included a Bitcoin script – specifically, the OP_PUSHDATA4 opcode allowed up to 4.3 GB of data to be pushed onto the stack in the original Bitcoin protocol).[6] Buterin's profound idea was that you can record *any* computer program on the blockchain. A moment's reflection should show why this is so. You can write out any computer program if you have the patience. However, some programming languages are 'Turing complete', meaning they can encode any computable function. Thus, if you have a way of executing those programs, you can do anything a digital computer can – and you can do it on the blockchain.

This is effectively what Ethereum is – a blockchain-based platform that can perform any computable function. It is, as Camila Russo put it in the title of her book about the development of Ethereum, an *Infinite Machine*.[7] But because it is on the blockchain, it

---

is visible to all. It is transparent. What could one do with such a universal computer on the blockchain? One thing it can support is self-executing contracts. This is actually an idea that predates Bitcoin and Ethereum. As we mentioned earlier, in the 1990s, Nick Szabo proposed the concept of a smart contract, which he described as 'a set of promises, specified in digital form, including protocols within which the parties perform on these promises.'[8] Of course, the similarities to Leibniz's idea of proving (computing) courses of action for a prince should be apparent – the only difference is that Szabo made explicit the idea that the execution of governance policies could be automated.

Let us take a hypothetical example of a smart contract. Sometimes, when we make a large purchase, such as real estate, we utilise escrow companies to hold the money until the transaction is resolved. For example, you want to buy a house for a million dollars, but you do not wish to send the money until you get clear title to the house, and the homeowner does not want to relinquish title until the money lands in their account. This is where escrow companies come in – yet another layer of centralised trust. A mutually trusted third party holds the payment until the title is transferred to the buyer before releasing the money to the seller.

An Ethereum smart contract could supplant escrow services. A smart contract on the Ethereum blockchain can be designed to hold a document – for example, the title to the house – until payment is sent to the contract. You do not need trusted third parties because everyone can study the smart contract online and see that it will

---

[8] Nick Szabo, 'Smart Contracts: Building Blocks for Digital Markets', *Phonetic Sciences, Amsterdam*, 2006 <https://www.fon.hum.uva.nl/rob/Courses/InformationInSpeech/CDROM/Literature/LOTwinterschool2006/szabo.best.vwh.net/smart_con tracts_2.html> [accessed 27 May 2023].

automatically release the title the instant the money is received by the smart contract.[9]

This brings us back to Leibniz's idea of a virtual mechanism guiding a prince's actions. Did Leibniz design the mechanism to provide helpful advice only, or did he consider that the actual execution of policies could also be automated? Whatever Leibniz had in mind, it is certainly true that smart contracts can now automate aspects of governance. Government policies could be automated on the blockchain, visible to all, auditable by all and guaranteed to be executed.

One way in which this might be realised is that elements of a government's constitution could be programmed directly into the governmental smart contracts. As a hypothetical example, one might program a smart contract constitution such that the state budget has to be balanced. Alternatively, one might program the smart contract constitution so that a minimum percentage of GDP goes to universal health care. One could also program the smart contract so that the military budget could not exceed (or perhaps always exceed) a certain percentage of GDP. Or one could program the smart contract so that the government had to maintain minimum reserves in bitcoin. The options are endless.

Of course, constitutions are silent on most aspects of governance, but the day-to-day business of lawmaking could give way to smart contract design and execution. Thus, the entire process, from voting to policy execution, can be built into smart contracts. For example, let us imagine a hypothetical onchain community consisting of various digital nomads scattered around the world. If you wish, we can suppose that our community also has secured pieces of physical territory worldwide. Our hypothetical community provides services, such as health care and health insurance, for its members. In exchange for this, it taxes its members some amount, let us say 0.1% of all transactions that the members engage in on the blockchain. But now, suppose that some members think that this is too much, and a proposal is made to reduce the tax to 0.08%, with a correlative cut in benefits.[10]

---

[9] As we will see, it is a bit more complicated than this. There is the issue of assessing the safety of the smart contract, for example, and there is also the question of how things like titles make their way onto the blockchain – that is, how assets are tokenised.

In traditional nation states, people campaign for tax cuts all the time. Maybe the government will deliver on its promises or perhaps it will not. No one knows for sure. However, as we noted, with blockchain governance, we can vote directly for a policy, which is itself encoded in a blockchain smart contract. In this case, there would be a smart contract governing the taxation policy for the community (let us say it is a function that takes some percentage of transactions and then transfers that to the contract for health care and health insurance payments). The policy that receives the most votes is the policy that will go into effect. Will the contract really deliver on that policy? Well, we can inspect the code and see. Once the smart contract is activated, that is the policy.

How do we know that our votes will be correctly recorded? We live, after all, in the age of competing claims of voter fraud, double voting, lost ballots, hanging chads and other obstacles to reliable elections. Certainly, this is the state of conventional democratic mechanisms. Even when they do not fail, there often remains suspicion that they have failed. And why should there not be such a suspicion, given that the mechanisms of voting are centralised and are not visible to all? They are, at best, visible to election observers, who may or may not be reliable referees and who may or may not be in a position to actually assess what is happening.

However, voting too can be built into the smart contract constitution of our community. Voting can take place on a smart contract that is immutable and accessible to all. Therefore, we can inspect the contract and see if it is programmed to count every vote and if the vote outcome indeed leads to the execution of the desired smart contract. There is no need to wait for politicians to act on their promises after the election. The contract executes once the voting is closed or on a timeframe built into the smart contract.

---

[10] You might think that this rate of taxation would be draconian for someone engaged in potentially large numbers of high-speed transactions, but there are ways around this. Transactions could be processed in batches or conducted on a layer-two protocol – an option we will discuss later.

For example, let us suppose that our voting power depends on how many state-issued cryptocurrency tokens we hold in an anonymous wallet. After we vote, we can confirm on the blockchain that our vote was recorded and that the totals were added correctly. If we wish that our system be 'one-person, one-vote', then we need mechanisms to ensure each person has exactly one vote (or the same voting power), and this leads to challenges that involve tying a crypto identity (for example, a crypto wallet) to a 'real-world' identity.

The key point is that in our thought experiment, all the critical democratic mechanisms have been encoded directly into our immutable public blockchain. This allows us to vote directly and securely for the tax mechanism we desire, with the assurance (by the laws of mathematics and the theory of computation) that if our vote prevails, the policy will be enacted.

Of course, this is just one possible way to execute blockchain governance. Maybe you prefer a community where taxation is not subject to votes. Maybe you want the taxation to be hardcoded directly into the definition of the community itself. Or maybe you prefer that some elected leader set tax rates. Or maybe you do not want leaders to be chosen by election. All of these are possibilities with blockchain-based communities.

## 6.3  Oracles

One critical element to the success of blockchain technologies is the ability of the blockchain to 'know' things about the external world. Or if you prefer, it must represent states of the non-blockchain world. For example, a number of so-called 'stablecoins' exist that aim to emulate the value in terms of purchasing power of the US dollar. For that to happen, they must have offchain information about the dollar's value. Similarly, a sports betting application must have information about the results of sports contests.

Of course, it is easy enough to set up a traditional API that provides a direct feed of these values from a centralised source (for example, a news organisation like the BBC). However, that is not particularly aligned with the goal of decentralisation. If the decentralised blockchain is drawing its information from centralised sources, what have we really accomplished?

Data feeds that provide outside information to the blockchain are known as 'oracles', and we need oracles that, like the blockchain, are decentralised. In later chapters, we will go into detail about the challenges facing decentralised oracles, but for now, we can provide a gloss of what the principal strategy will be.

The real breakthrough of blockchain technology is that any human institution can be decentralised if we wish. For example, an oracle need not be a single source of information; it could be a group of individuals participating in a decentralised-yet-cooperative project to provide and verify offchain information. Instead of trusting a single news source, multiple people could survey multiple sources of information. Instead of trusting a newsfeed for a score, someone could go to the game, someone could contact the league office and so on. Here, there is not a single source but multiple sources, and everyone participating in the effort would have something at stake in the oracle's reliability and success. As noted, conceptual issues arise concerning oracles, but for now, we simply want to introduce them and their role within the basic governance technology stack. We will examine them more thoroughly later in the book.

## 6.4   DAOs: What are they and how do they work?

In Section 6.2, we explored the concept of smart contracts and showed some of their interesting potential applications – indeed, applications that already exist. Smart contracts have many extraordinary use cases, but for our purposes, their most interesting move beyond currencies and finance; we can also establish entire organisations that are encoded on the blockchain in the form of DAOs.

DAOs are the decision-making components of online communities. They are not organised top-down like traditional organisations. They are, as you might have guessed, decentralised. There are different ways of building DAOs. Some might be built around a community of individuals who make key decisions driving the organisation. Others might forgo the role of human intervention altogether. Let us start with the latter, very interesting idea.

In its simplest form, the idea behind a fully automatic DAO is that several smart contracts can be combined into a larger contract that can function as a corporation or organisation on the blockchain. As we will see, this idea is quite radical.

In principle, none of the usual trusted business partners are required in a DAO: you do not need employees, managers, human resource officers, CFOs or CEOs. In some cases, all of those jobs, and in other cases, most of those jobs, can be replaced by smart contracts. For example, if you designed a smart contract to do what a hedge fund like Renaissance Technologies does, it would no longer be necessary for shareholders to pay massive bonuses to hedge-fund executives who are trusted to make decisions about investors' money. In theory, at least, those executives could be replaced by a bundle of transparent, preset instructions stored on the blockchain. Perhaps this would include an AI program on the blockchain, constantly seeking arbitrage opportunities with advanced mathematics. Of course, for many applications, no AI is required, and some DAOs can be built around simple and straightforward algorithms.

On the other end of the spectrum from this completely automated DAO would be the idea of a DAO that is under the control of humans. All of its deliberations would take place offchain (for example, on a video conference call), and all of its actions take place offchain as well. What would make it a DAO, in this case, would be that the decisions made are still recorded and executed onchain and are informed by information that is stored onchain. We might think of this and the previous example as being the extreme limiting cases of a DAO. Between these two extremes, there are endless combinations of DAOs organised around different compositions of automation and human intervention. A few examples can illustrate some of the possibilities.

Consider a decentralised exchange (DEX), which is a blockchain-based platform on which you can exchange one type of cryptocurrency (or any other tokenised asset) for another. In protocols like Uniswap, this would involve using automated market makers, in which the asset price for the exchange is set algorithmically, depending on the incoming bids for a token in the exchange and the number of tokens held by the contract (the tokens held in the contract constitute the liquidity pool for the currency pair being traded). Suppose we want to exchange crypto asset B for crypto asset A. As the supply of A decreases in the smart contract, the algorithm will increase the price of A relative to B. In this way, people are subsequently incentivised to provide B tokens in exchange for A tokens. The algorithm takes the place of the order book in traditional centralised exchanges.[11]

A DEX would not have to be organised around a DAO. Maybe a centralised corporation can maintain it. However, it can also take the form of a DAO in which holders of the protocol's governance token make decisions concerning the protocol. In the case of Uniswap, that token is UNI. Token holders can debate and vote on upgrades to the staking contracts. The Uniswap DAO, for example, has subsequently introduced version two and version three staking contracts.

---

[11] The protocol relies on people seeking arbitrage opportunities to bring the exchange rate on the DEX back in line with the broader market. The liquidity is provided by liquidity providers that stake liquidity pairs in the smart contract in exchange for financial remuneration. For more information on AMMs and liquidity providers, see 'What Is an Automated Market Maker (AMM)?', *Coinbase Learn* <https://www.coinbase.com/en-es/learn/advanced-trading/what-is-an-automated-market-maker-amm> [accessed 29 October 2024].

Another application of DAOs has involved the issuance of over-collateralised loans. For example, the protocol MakerDAO (renamed to Sky while this book was in production) takes in collateral in the form of cryptocurrencies like ETH and then issues loans in the form of the stablecoin DAI, the value of which is pegged to the US dollar. Another way to describe this is that when you deposit ETH as security, a certain quantity of DAI is created, or 'minted'. When you later choose to return your DAI in exchange for ETH, the DAI is destroyed (or 'burned', as per the parlance of those involved in the blockchain sector). However, the key thing is that DAI is able to remain pegged to the dollar by ensuring that it always has sufficient collateral to support DAI's value of $1.

We said that the MakerDAO protocol is overcollateralised, meaning that you must maintain more collateral in the smart contract than the borrowed amount. As of this writing, the minimum collateralisation in MakerDAO is 170%. If you wanted to borrow 10,000 DAI, you must first deposit $17,000 worth of ETH as collateral. Thus, while your $17,000 of ETH is locked in the protocol, you are free to take your 10,000 DAI and use it as you wish. Why would you do this? Well, selling your ETH is a taxable event, and you may not be ready to pay taxes at this time. Alternatively, you may anticipate the value of ETH will increase radically, and thus, you do not want to sell your position, but you need to buy a new car or have other expenses in your everyday life.[12] Or you may want to loan out your DAI and collect interest on it while your ETH (hopefully) increases in value. Of interest to us is that if the value of ETH goes down and you no longer are collateralised at a rate of 170%, there are no bankers that sit and review your position and ask you to recollateralise your loan. The process is automatic. If your collateralisation drops below 170%, the smart contract automatically

---

[12] You would then convert those DAI into US dollars on a one-to-one basis on an exchange like Coinbase.

liquidates enough of your ETH to repay enough of your loan to bring you back under 170%.

For example, suppose that you deposited $17,000 of ETH in order to borrow 10,000 DAI. You are already at the minimum level of collateralisation. Now, suppose that ETH loses 10% of its value, leaving you undercollateralised. The smart contract automatically liquidates enough of your ETH to retire enough of your debt to bring you back within limits. Therefore, the protocol must liquidate enough of your ETH to recover $1,700 worth of DAI. Of course, in liquidating the ETH, your collateral is reduced as well, so to bring things into balance, additional ETH will be liquidated. But what is the role of the DAO in all this? One role is setting the minimum collateralisation rates, but as we are about to see, the DAO can also be called upon for more urgent matters.

In 2022, as a series of centralised banks and cryptocurrency exchanges collapsed – most notoriously Sam Bankman-Fried's FTX exchange – it was not lost on observers that DEXs like Uniswap and lenders like MakerDAO did just fine. With centralised platforms like FTX, the problem was that the operations of the company were not transparent. For example, no one really knew what FTX was doing with the deposited money (in hindsight, taking it and placing highly leveraged bets, it seems). The contagion across the crypto sector in 2022 revolved around the fact that one could not be sure that the backing collateral was actually where it was supposed to be. And in point of fact, it often was not.

In the case of MakerDAO, there were plenty of liquidations, but the collateral was where it was supposed to be – transparently so. The smart contract did its job and MakerDAO kept humming along just fine. Of course, this is not to say that things have never gone wrong for MakerDAO, a point we will return to shortly.

A third DAO example would be the DAO providing governance for Yearn Finance. Yearn is a 'yield aggregator', which means it accepts asset deposits in the form of specific cryptocurrencies and then deploys strategies to earn yield off those deposits. The contract executing the strategy is known as a 'vault'. Strategies vary but may range from loaning out deposits to placing them into liquidity pools like those in Uniswap with the goal of harvesting staking rewards. Yearn Finance also has a governance token, YFI, and holders of YFI can vote on the kinds of vaults that will be offered and the strategies those vaults will deploy. YFI token holders also vote on more general properties of the protocol, ranging from the number of YFI in circulation to the way in which developers are to be financially incentivised.

The examples we have given thus far are finance driven, and this is in part because many of the early applications of DAOs have involved decentralised finance. However, the applications for DAOs are certainly not limited to financial cases. There are as many possible applications as there are needs for coordinated human actions.

For example, VitaDAO is a DAO that is designed to support longevity research.[13] DAOs could be organised around any sort of scientific research project, and they need not involve the fundraising component. They could involve nothing more than votes on which research projects to pursue next without concern for raising funds. The rewards might include nothing more than the equivalent of gold stars for researchers who are aligned with community goals. The advantage of such DAOs would be that research could be driven by suitably aligned community members rather than by a centralised funding organisation like the National Endowment for the Humanities or the National Endowment for the Sciences. The idea would be that we could avoid the cronyism and patronage that takes place behind the closed doors of centralised funding agencies.

The examples we have discussed have involved some degree of human intervention, but as we noted, in theory, a DAO could be completely automated with no possibility for future human interference. For as long as there is a functioning blockchain, such a DAO would continue to perform its function consistently. Perhaps that function would be to serve as a decentralised financial exchange like Uniswap or maybe it would be to issue overcollateralised loans like MakerDAO. However, as we noted, *most* DAOs (for now, at least) involve human intervention, and understandably so. Conditions change. Even the underlying blockchains undergo changes. Therefore, there is typically a group of signatories with access to the smart contracts constituting the DAO. If enough signatories permit it, the developers can access and 'change' the smart contracts.[14]

---

[13] <https://www.vitadao.com/> [accessed 30 October 2024].

One case where this kind of human intervention was necessary involved the MakerDAO protocol, which we just discussed. In 2020, during the onset of COVID-19, there was a precipitous collapse in the value of ETH – a collapse so severe that the drop in ETH prices outran the ability of liquidations to balance the collateral ratios necessary to keep DAI at its $1 peg. Or as they say in crypto, DAI 'lost its peg'.

This was a case where the human element of the DAO became necessary. To prevent DAI from remaining below $1, the human DAO members took action. They voted to back DAI with USDC[15] (in addition to ETH) and found backers to inject this liquidity into the protocol.

It probably has not gone unnoticed that this voted-on fix was a step away from decentralisation – relying on an asset that is the creation of a centralised exchange and which relies on a centralised mechanism to maintain its stability. And in point of fact, the move remains controversial. Indeed, the decision took on some urgency in the wake of USDC itself losing its peg to the US dollar in March 2023.[16]

---

[14] We use scare quotes around 'change' because this is not strictly accurate. Once a smart contract is deployed, it cannot be changed. However, one can change a data field that the smart contract draws upon (for example, resetting collateralisation rates), and one can replace old contracts with a new version of the contract – i.e. redeploy the smart contract.
[15] USDC is a stablecoin maintained by a consortium that includes the payment company Circle and the centralised crypto exchange Coinbase.

USDC slipped from its $1 peg in 2023 because the assets backing the stablecoin (actual US dollars or short-term treasuries) must be stored somewhere. In this case, many of them were being held with Silicon Valley Bank, which succumbed to rising interest rates, undermining the value of its long-term US treasury bills and other long-term assets. When the bank collapsed, USDC (temporarily) lost access to its backing collateral. The members of MakerDAO subsequently discussed a return to using only native crypto assets as collateral but voted to retain USDC as part of DAI's collateral. The moral is that no store of value is completely safe and that centralised financial institutions are particularly unsafe – being as they are centralised and far from transparent.

Of interest to us at the moment is the role that DAO members play in the process. In this case, the governance occurred through a series of time-based polls. These votes took place online, and voting power was a function of the number of the protocol's governance tokens DAO members held. In this case, the governance token was MKR.

It should now be clear what responsibilities holders of a DAO governance token might have, but what is in it for the token holder? Why bother? Well, holding governance tokens usually conveys rewards beyond the ability to discuss and vote on policy changes in the protocol. Benefits might include a financial reward in the form of token airdrops or other forms of remuneration. In some platforms, like DEXs (although not Uniswap), governance token holders are rewarded with a percentage of the fees the protocol collects with each exchange transaction. Of course, holding voting rights also opens up the possibility of voting for other forms of rewards. The token-holding community could elect to issue more tokens (or burn tokens to make those remaining in circulation more valuable) or sell assets from the protocol treasury and distribute the proceeds to token holders. In protocols like the DEX Curve Finance – a decentralised platform for trading stablecoins – governance token holders can vote to determine which staked assets will yield the largest returns. Thus, the right to vote to increase the yield on staked assets provides the incentive to hold the governance token.

---

[16] Ashley Capoot, 'Stablecoin USDC Breaks Dollar Peg After Firm Reveals It Has $3.3 Billion in SVB Exposure', *CNBC*, 3 November 2023 <https://www.cnbc.com/2023/03/11/stablecoin-usdc-breaks-dollar-peg-after-firm-reveals-it-has-3point3-billion-in-svb-exposure.html> [accessed 29 October 2024].

There are a lot of details to be worked out regarding DAOs. The first issue relates to voting itself. DAOs offer the opportunity for us to explore promising alternative voting systems – many dating back to those proposed by Condorcet in the French Revolution. The voting could be ranked preference or head-to-head or any other traditional voting method. However, the question of whether there is to be one vote per person or whether the number of votes a member is eligible to cast is weighted by the number of DAO governance tokens they hold remains a major issue with DAO voting. The beauty of blockchain governance is that we can try all approaches and keep a record of their successes and failures. This, in turn, can help us to learn how to design better voting mechanisms. Blockchain communities can become laboratories for investigating democratic voting (and other) governance mechanisms.

If we wish to have a system in which each person is eligible to vote only once, then we introduce significant challenges: How can we ensure that there is only one person associated with each token? How can we be sure that some wealthy whale has not opened thousands of wallets, each containing one governance token? As DAO participants would put it, how do we make the protocol 'Sybil resistant'?[17]

Let us set aside the issue of Sybil resistance for now because there is a more pressing issue: What ensures that the developers and keyholders will actually carry out the policies voted on by the governance token holders? Here, it may seem that we have come back to where we started with traditional governance: How can we be sure that the people with their hands on the switch will do what we voted for? In legacy voting systems, there is little recourse. If representatives do not implement the policy you voted for, you are stuck. You have to wait until the next election in two or four or six years and hope for the best next time.

---

[17] The name is based on the 1973 book *Sybil* by Flora Rheta Schreiber, which fictionalised the story of a person with dissociative identity disorder.

With blockchain technology, the situation is different. Setting aside the question of whether there is a legal obligation for keyholders to execute voted-on policies (we currently have no idea – this has not been litigated), there is also the option of cloning the unfaithful protocol and starting a very similar protocol that does what the original was supposed to do. And protocols have been replicated in the past – or 'forked', in the language of the crypto community – for lesser reasons. Yearn Finance, Maker and Uniswap have all been forked many times already, sometimes in response to an unhappy vote outcome and sometimes just because someone saw an opportunity to make some money by cloning an established platform. The point is that there are exit strategies that allow communities to start over with the same basic platform if the developers and keyholders refuse to execute the wishes of the DAO. Keep in mind that smart contracts constituting the DAO are publicly visible on the blockchain. It is a relatively trivial matter to clone the protocol and start over with new developers and new keyholders.

However, at the end of the day, humans are typically part of the DAO equation, and there is no point in denying that these humans are possible points of failure. They inject all sorts of drama and conflict into the protocols, and it is difficult to engineer around drama. However, we can try, and we will discuss ways of doing that later. First, we have bigger fish to fry – or at least, more interesting fish.

Assuming that these concerns about developer compliance within DAOs can be answered or at least ameliorated, we can address the much more interesting question of what other kinds of things DAOs can do.

The question is interesting because there are many answers – indeed, there are many answers for every level of human governance.

Let us start with a very low-level example of governance – something like a condo board. A condo building could have its own DAO, with an online record of meetings, decisions and funding authorisations. It could also incentivise contributions to the community by making contributions from its treasury. Would this not be possible with a centralised platform? To some extent, yes, but this raises the question of who maintains the relevant server. In the future that we envision, the records are not merely distributed among the condo members but distributed globally – the records are outsourced to a distributed network that is scattered across the world and thus to individuals that have no interest in the affairs of a single condo and plenty of interest in preserving the integrity of the network as a whole. To put it another way, the idea is that a single blockchain can support many such condo associations and that the incentive to preserve the security and reliability of that blockchain swamps the pressures for corruption from individual condo associations.

This is not to say that the condo association could not create its own blockchain and issue its own coin. However, we would be more likely to trust an association in which the settlement layer for its token – the layer on which it is anchored – would be global and, by its very nature, immune to attempts at local-level corruption. Likewise, we would be more apt to trust an association that kept its records on a globally distributed file storage platform like IPFS (the InterPlanetary File System), a peer-to-peer file-sharing network for storing and sharing data in a distributed file system, or Codex, a decentralised file storage protocol developed by the Institute of Free Technology.[18] The idea is that much of the business of the DAO would be anchored in broader, more robust and more widely distributed platforms.

---

[18] We discuss Codex in greater detail in Chapter 14.

Here is another example of micro-level blockchain governance, which was executed in Brooklyn, New York. A neighbourhood established its own electricity microgrid, linking together a number of solar-powered homes in the area.[19] Because it was built on a blockchain, it provided users a transparent, immutable record of the energy contributed to the mini-grid from each home, the amount of energy consumed by each home and the amount of energy sold to neighbours on the grid (think of it as local peer-to-peer energy sharing). Additionally, it is presumably possible for the microgrid to sell energy to the local centralised power grid (in this case, Consolidated Edison). The endeavour also allowed the transparent and fair distribution of dividends to cooperative members. Beyond all this, it opened the door for the cooperative association to provide incentives to people who contributed to the network (for example, by onboarding new members or by contributing new technologies).

There are many other examples of DAOs that exist today, and we can hardly move on without discussing the DAOs created in the service of NFTs and their online communities. Although not all NFT communities operate through DAOs, many do. Those DAOs make decisions about the disposition of the treasury for the NFT project, policies for intellectual property rights for the NFT images, and plans for extensions and revisions to the NFT project. Typically, important decisions are debated and then voted on.

In subsequent chapters, we will envision a scenario in which many different levels of governance will reside on the blockchain and be under the control of DAOs. This means that perhaps not only your local electricity microgrid and favourite NFT project are under the control of a DAO, but possibly your local condo board is as well or perhaps your homeowner association (HOA) is or maybe formal layers of government will be, too – perhaps your county administration or your city or even your state. The typical citizen may end up being a member of multiple DAOs. Is that even feasible? We think so.

---

[19] <https://www.brooklyn.energy/about> [accessed 30 October 2024].

People who invest in multiple crypto protocols and NFTs are already familiar with this phenomenon. How does one keep track of it all? Well, one common way people keep track of events, proposals and policy debates today is to join a server on a platform like Discord for each of the DAOs in which they participate. For example, someone might belong to the DAO for Uniswap and MakerDAO and Yearn, as well as DAOs for some of their NFTs. The Discord server for each protocol has multiple threads open at any given time, providing information about official announcements, threads for making proposals, threads for social events, chatting and other relevant topics to the community. As currently designed, one can actively monitor all of the DAOs in which one participates, even if there are dozens of them.

Ideally, we would not have to access these DAOs through Discord. That is a centralised point of failure, after all. Fortunately, there are already tools that allow us to monitor multiple DAOs in real time, participate in those DAOs and vote on proposals. For example, the Institute of Free Technology, with which we are affiliated, addresses this need inside of its Status app – a mobile app that incorporates messaging, a wallet, a decentralised application browser and a community management tool (in effect, a decentralised version of Discord).

Potential applications for DAOs are only limited by our imaginations. Hopefully, we have provided enough examples of how DAOs work to anchor the next part of our discussion. We now want to turn to what are sometimes called 'impact DAOs' – DAOs that yield public goods and positive externalities and, in some cases, provide those goods and externalities in a regenerative way.

## 6.5   Impact DAOs and regenerative public goods

Blockchain technology, properly organised, has the ability to incentivise behaviours that help the community that organises around a particular blockchain. Or more accurately, behaviours that help the community to help the community. This incentivised behaviour is regenerative in that goodwill and good works are not expended in a one-and-done fashion but in a way that yields yet more goodwill and good works.[20]

Does this sound too good to be true? Does it sound like cultural ponzinomics? Such concerns are natural, but we think we can assuage them. Cooperation is possible, it can yield public goods and it can also yield more cooperation. However, it must be properly incentivised.

We can also incentivise positive externalities. This is to say that regenerative public goods can be extended beyond the local community. We can incentivise outcomes that help other communities and their members, and we can do so, again, in a way that is not one-and-done but in which positive outcomes continue to yield yet more positive externalities. Now, in saying this, we have to go deeper than slogans. We need to dive deeper into DAOs and how they have the unique capacity to bring about regenerative public goods.

Most, if not all, readers will be familiar with the tragedy of the commons. Given a public space, say a public pasture in the village, there is a tendency for people to use the shared space to its exhaustion as quickly as possible, the reasoning being that 'if I don't use these resources first, someone else will.' The end result is that the pasture is quickly over-grazed and left barren. The issue with the tragedy of the commons is that the game theory driving the outcome is zero sum, and essentially 'use it or lose it'. However, DAOs can provide ways of engineering the game theory to incentivise other behaviours.

To stick with our village pasture example, the village can incentivise villagers to use less of the public space or even to contribute more land to it. Imagine that using the pasture first required clearing more land to create public pasture so that the public space grew with its use. How feasible that is in the case of a real-world village is not entirely clear, but it is clear that there are many options for a DAO to provide viable incentives.

---

[20] For an excellent introduction to impact DAOs producing positive externalities see Kevin Owocki, *GreenPilled: How Crypto Can Regenerate The World* (2023).

The natural way for traditional governments to fund public resources is via taxation, but there are alternative options for a DAO. One strategy that is popular among DAOs is the generation of revenue by a small transaction fee. As noted during our discussion of DEXs, there can be a percentage captured in revenue with each transaction. This is, of course, a kind of taxation, but perhaps more palatable because it comes in the form of a fee for a specific service and is cognitively painless (you do not have to 'do taxes' – the money is collected automatically).

Alternatively, DAOs can issue governance tokens as rewards for behaviour, although this policy is inflationary and is really only feasible if the network is growing at a rate sufficient to absorb the newly minted tokens or, alternatively, has a token burn mechanism. We are apt to dismiss inflationary monetary policy, but Bitcoin, for now, still has inflationary emissions as a reward for Bitcoin miners. Meanwhile, Ethereum will continue to mint more ETH even as it burns ETH in transaction fees. Thus, Ethereum has found a way to incentivise people who stake their ETH while maintaining a non-inflationary money supply.[21]

Under certain circumstances, users can also contribute to their DAOs as a form of philanthropy, which has an advantage over giving to traditional charities in that the money trail is transparent, corruption resistant and capital efficient. There are no charity administrators to pay; the use of the gift is determined by an algorithm in a smart contract. To draw on our earlier metaphor, if you give money to build the community pasture, you can verify that the money will contribute to that goal.

Of course, where we are going with this discussion is towards the idea of regenerative public goods. The fundamental function of DAOs is to serve as a decentralised mechanism for coordinating people's behaviour via the protocol. If we want those behaviours to contribute to regenerative goods, this will be accomplished by offering the correct incentives to the behaviours that assist the protocol in its regenerative growth. How would this work?

---

[21] Non-inflationary in the sense that the supply of ETH should shrink over time, as more ETH is burned than created.

DAOs have been known to issue rewards to members who contribute to the onboarding of new members (helping newcomers, for example), teaching others about features of the protocol (in tweet threads, for example), assisting community members with technical problems or devising investment strategies (protocols like Yearn Finance provide rewards to users that come up with strategies for their vaults). Notice that each of these incentive structures is designed to help the protocol grow. However, the secret to success here is that we do not merely incentivise growth but a form of growth that leads to more growth.

The golden ticket to regenerative public goods is to incentivise DAO members to develop strategies that lead to more people contributing time and resources to the DAO in a way that brings about more DAO members sharing the same mission. There are a number of ways this might play out, but possible incentivisation activities include classes and tutorials that help users to not merely become contributors but to value regenerative public goods. Thus, educating DAO members about these values can supercharge the growth of the regenerative goods that flow from the DAO.

Education is just one example of a meta strategy (a strategy for creating new strategies) that DAOs can deploy. Another meta strategy would be to incorporate meta rewards, which would be rewards given to people that generate rewards for behaviours that contribute to regenerative goods. What would this come to? If there are programmes that successfully reward strategies that create positive outcomes, we can incentivise those programmes to continue doing so. We need not worry too much about vetting these programmes for future success as it has been established that retroactive rewards (rewards for past success) can be a successful way to incentivise future productive behaviours.

Blockchain-based DAOs are uniquely situated to carry out these various forms of incentive programmes. First and foremost, by making the incentive payments and the rewarded behaviour transparent and onchain, we allay the concern people have that they are not being fairly rewarded or will not be fairly rewarded or that the wrong people are being rewarded. Transparency is one of the keys to the success of blockchain governance, and it is certainly a key to the success of regenerative public goods.

As we noted earlier, there is also a class of DAOs that aims to bring about positive external outcomes. For example, outcomes that benefit the planet are clearly going to be desirable to many blockchain communities. Reflecting briefly on how such DAOs might be organised might be helpful.

Just as one can create onchain incentives to contribute to one's virtual community, contributions of external goods (external to the blockchain community) can also be incentivised. The interesting difference is that for positive externalities – and for positive contributions – one needs different strategies to confirm those external outcomes have been achieved and that the external contributions have been performed. This is a bit more complicated than confirming contributions and results within the DAO, where both are visible to DAO members. In the case of offchain contributions, some oracle or other form of decentralised confirmation tool would be required. We can even confirm contributions made on other chains, as there are cross-chain strategies for verifying states and actions.

There are currently several protocols designed to confirm states and activities in external blockchains. We will touch on how such cross-chain protocols work when we consider proof of storage. For now, we can say that the basic idea is to use 'zero-knowledge proofs'[22] to confirm certain activities or states on another chain – for example, the presence of financial reserves. Given such tools, one has transparent and reliable

---

[22] 'Zero-knowledge proofs' are strategies for mathematically proving the possession of information by an information processing system without revealing the information itself. See Chapter 14 for a more in-depth explanation of 'zero-knowledge proofs'.

confirmation that contributions to other blockchain communities have, in fact, occurred.

The possibilities for positive externalities are vast. For example, one might contribute financial resources to struggling online communities or provide programming assistance or instructional help and so on. The point is that insofar as cross-chain confirmations are possible, incentives can be structured in the same way that they are for internal incentives.

This means that those positive externalities can also be regenerative. The idea is the same as before, except with an eye to incentivising regenerative goods for those who are external to the home blockchain community. The objective is to reward projects that continue to generate positive outcomes or outcomes that are regenerative, and thus lead to additional positive outcomes down the road, whether through education or the rewarding of successful meta strategies.

Keep in mind that all of these DAO examples are also examples of human governance in action. In particular, they are examples of how blockchain technology can be leveraged to assist in human coordination. Just as the Byzantine Generals Problem was a problem of coordination, so too, problems of governance are problems of coordination. Things fall apart when people are not on the same page. For example, if people are waiting for their neighbour to contribute before they contribute, we have a classic coordination problem. The problem cannot be engineered entirely away; humans are complex social animals, after all, and coordinating them can be a bit like herding cats. Still, the right technologies can make the problem more manageable.

What makes DAOs special? What enables them to coordinate human affairs better than traditional governance mechanisms? Well, let us reflect. Coordination requires that people be on the same page at the same time. It also requires that everyone *knows* that everyone else is on the same page. These are goals that the blockchain is uniquely designed to accomplish. Beyond this, coordination requires trust. It requires that we trust that everyone has the same information and that the information has not been corrupted. It also requires that we can trust that what is supposed to happen under specified conditions *will* happen. These are also properties that blockchain technologies are uniquely designed to achieve.

Finally, however, coordination requires incentivisation. Like it or not, people want to know what is in it for them, and some coordination games do not properly incentivise our actions. In a simple prisoner's dilemma game, there is a strong incentive to defect – to not be coordinated.[23] DAOs can engineer coordination games in which people are rewarded simply for cooperating – for being in accord with the group. They can make cooperation the rational outcome and the attractive outcome.

## 6.6   Applying these technologies to human governance

In this and the previous chapter, we talked about the basics of blockchain technologies, smart contracts and DAOs. However, we have only glossed over how we might apply these technologies to the general problem of human governance and our vision of the post-Westphalian order. It is still too early to give the complete picture in this chapter because we have lots of ground to cover regarding the limitations of current forms of governance and the governance problems that we hope to solve. That said, perhaps it is not too early to provide a preliminary picture of how these technologies might be incorporated into methods of human governance.

To start with a simple example, let us consider a smaller unit of government – an HOA. As we saw in Chapter 3, HOAs are important players in the landscape of human governance. They are also a mess.

Understanding that HOAs are governance structures for residential communities, how would blockchain technology be applied to their governance? Well, one of the principal features would involve the use of blockchain-grounded immutable archives, which would be accessible to everyone in the community. If you are a member of an HOA, you can, at any time of the day or night, check online to see a record of the activities of the HOA governance. You can also be sure that the archives were not tampered with.

---

[23] Steven Kuhn, 'Prisoner's Dilemma', *Stanford Encyclopedia of Philosophy* (Stanford, CA, 1997) <https://plato.stanford.edu/archives/win2019/entries/prisoner-dilemma/> [accessed 29 October 2024]

As for the governance mechanism itself, it might involve some combination of smart contracts and human governance – smart contracts that would execute tasks like fee calculations and requests for payment, as well as incentives for certain kinds of positive behaviours. After all, our HOA may be intentionally structured to promote positive outcomes by incentivising members to support the wellbeing of the community. If human administrators are part of the HOA, they may be elected by the community, and the voting could take place onchain via a smart contract. Everyone could see that the election was fair and that no one had tampered with the votes.

The communications between HOA leaders would be onchain and visible to all community members. All such communications would be recorded on the blockchain and immutable for all practical purposes. However, beyond a record of votes and communications among leaders, there would also be a record of communications between community leaders and other citizens. This could involve complaints about service or the proper interpretation of community policies.

To be sure, there are laws today (like Florida's 'Sunshine' Law) that require governing officials to only communicate in public forums. Blockchain technologies would allow additional communication channels that could be widely accessible and, at the same time, impossible to tamper with. In other words, communications would not have to be limited to public meetings, but open communications could also be possible in online channels.

In short, blockchain technologies would serve as the central nervous system of the community, but a central nervous system to which everyone in the community has access. Does this mean that there will be no drama? No, but it means that actions, policies and disputes should be more transparent, and the record of those actions, policies and disputes should be more secure. Smart contracts will automate some functions, which as you will recall are themselves transparent and onchain. The execution of the smart contract will be visible to all.

Finally, blockchain technologies will afford ways to implement gametheoretic strategies in which people can be confident their contributions to the community will be rewarded in a fair and appropriately generous manner. No one will be taking advantage of their contributions.

This is just one illustration of how blockchain technologies might be applied to human governance. Our idea is that this basic concept can be extended to all forms of human governance – from nation-state equivalents (for example, cyberstates) to global organisations like the European Union and down to the tiniest governance structures (like the administration of a single condo building).

For example, consider a large governmental unit the size of the United States that is organised around blockchain technologies. The picture would unfold as in the case of HOAs, only on a much grander scale. Voting, for example, would be via smart contracts, the operations of which would be transparent to all. Some of the voted-for policies would also be instituted by smart contracts and thus guaranteed to take place (a tax cut or tax increase, for example). Incentive payouts for community contributions would also be transparent and, in some cases, guaranteed by smart contracts. As for the role of the governmental leaders (in effect, DAO leaders), their communications would be onchain, as would be the record of the interpretation of enacted policies.

You may have your doubts, which is understandable because we have yet to fully explore the abilities and limits of the technologies that we are discussing. Furthermore, none of this is to say that all problems of governance are solved, but it is to say that the situation can be improved significantly from where it is today.

Given the press that crypto has been receiving for the past decade, the claims made in this chapter may sound remarkable. Is crypto not supposed to be the favourite tool of drug dealers and terrorists and human traffickers? And above all, is it not supposed to be an ideal vehicle for money laundering? How can we say that blockchain technologies are the future of governance rather than the future of crime? In the next chapter, we take up the phenomenon of money laundering – or illicit financial flows – and build the case that, contrary to what the popular press exclaims, centralised governance is the source of the problem, and ultimately, blockchain technology will be the solution to the problem.

# WHY CENTRALISATION IS THE PROBLEM, AND CRYPTO IS THE SOLUTION

## 7.1 Preliminaries

On 6 December 2023, JP Morgan Chase's chief executive officer, Jamie Dimon, testified before the United States Congress, saying that crypto was a tool for 'criminals, drug traffickers [...] money laundering, tax avoidance', adding, 'if I were the government, I'd close it down.'[1] The irony is that just a few weeks after Dimon's congressional testimony, JP Morgan Chase was fined $348 million for 'inadequate trade reporting'.[2]

This was just the most recent fine in a string of enforcement actions taken against the world's largest bank by market capitalisation, including a fine of $920 million in 2020 for participating in fraudulent schemes involving precious metals and US Treasury bills.[3] That fine, in

---

[1] *Jamie Dimon: Government Should Close Down Crypto*, 2023 <https://www.youtube.com/watch?v=ujWR6t69UP8> [accessed 6 November 2024].

[2] Pete Schroeder, 'JPMorgan Fined Nearly $350 Million for Inadequate Trade Reporting', *Reuters*, 14 March 2024 <https://www.reuters.com/business/finance/jpmorgan-pay-nearly-350-million-penalties-inadequate-trade-reporting-2024-03-14/> [accessed 29 October 2024].

[3] U.S. Department of Justice, 'JPMorgan Chase & Co. Agrees To Pay $920 Million in Connection with Schemes to Defraud Precious Metals and U.S. Treasuries Markets', *Office of Public*

turn, followed over eighty regulatory fines against JP Morgan Chase for banking violations and other crimes dating back to 2003, with a total amount paid of over $39 billion.[4]

Of course, these were just the cases in which JP Morgan was caught and punished. In other examples, the multinational finance company avoided penalty thanks to important records 'accidentally' disappearing. In June 2023, the SEC was forced to file a cease-and-desist order against JP Morgan after it had deleted 47 million electronic communications. As the SEC complained in its filing, 'In at least twelve civil securities-related regulatory investigations, eight of which were conducted by the Commission staff, JP Morgan received subpoenas and document requests for communications which could not be retrieved or produced because they had been deleted permanently.'[5]

The problem is that fraudulent activity by JP Morgan Chase is merely the tip of the iceberg in the global financial system; it has not been an outlier. As we will see, banks all over the world are implicated in the same activities. It is not by accident, after all, that global banks have large offices in locations like Medellin, Colombia, and every other drug capital in Latin America.

However, banks are not the only bad actors when it comes to shady economic dealings. The problem actors include any centralised authority with control over money. The corrupt agents are not just banks and businesses but governments themselves. Sometimes, local governments are the bad actors, and sometimes, nation states are. If money passes through a centralised and non-transparent organisation, then that organisation is most likely a target for corruption and a particularly apt agent for what are known as 'illicit financial flows', or IFFs.

For example, many people know that the drug cartels impose a tax on businesses in many parts of Mexico – extortion, by another name. What people do not know is that these taxes are often paid directly to local municipalities as fees that are passed on directly to the cartels.[6] In other words, local governments have been captured by the cartels. Control of local governments is an ideal strategy for cartels because they can carry on the business of laundering money behind closed doors. In this and other ways, centralised authorities are ripe targets for those that traffic in IFFs – centralised governance is the best friend of criminals.

As we recounted in this book's Introduction the twentieth-century American bank robber Willie Sutton was once famously asked why he robbed banks, and he replied, 'Because that's where the money is.' However, that statement is no longer true. Some money can be found in banks, to be sure, and even more money is to be found in deals that banks are involved in and the transactions that they engage in, but as we will see, an even greater pile of money can be found flowing through governmental and nongovernmental organisations around the world. These are the places where twenty-first-century Willie Suttons operate because that is now 'where the money is'.

As we will see, fraudulent activity today takes numerous forms worldwide. There are many kinds of IFFs and many ways of hiding those IFFs behind the curtains of centralised governance. There is a staggering amount of dark money in the world, and once that dark money finds its way behind the closed doors of banks, governments and other centralised authorities, corruption is inevitable.

Our thesis in this chapter goes against what Jamie Dimon has said (and what mainstream media has repeated unreflectively): cryptocurrencies are not the cause of the problem. On the contrary, we believe that crypto is the only available solution to the problem. It is the only answer to what has become a worldwide blight of governmental and corporate corruption.

---

[6] One of this book's authors learned this while owning a restaurant in Mexico.

Crypto can be a problem solver here by making business and government transactions transparent and immutable on the blockchain (no more deletions of 47 million records to hide them from a dozen securities fraud investigations). Crypto takes control out of the hands of corruption-vulnerable central authorities and distributes it among all stakeholders. Furthermore, crypto provides tools to help automate government actions in smart contracts, making them transparent and reliable and eliminating counterparty risk. Finally, crypto accomplishes all this by moving on from centralised authority and putting critical governance functions on the blockchain.

We will discuss the positive solutions offered by crypto in some detail below, but before we get to these solutions, we need to understand the gravity of our current problem.

## 7.2   Quantifying corruption

Many people suspect that there is corruption afoot in big business and in our many layers of governing institutions, but it is not a trivial matter to locate and quantify that corruption. As noted above, one way of identifying and quantifying corruption is the metric of IFFs. While there is no single, agreed-on definition of IFFs, they generally include tax evasion, multinational tax avoidance, the theft of state assets and the laundering of the proceeds of crime, and they also cover a broad range of market and regulatory abuses, including payment for favours, drug smuggling and human trafficking.

IFFs thus constitute a basket of financial crimes, and by considering that basket of crimes, we can begin to put a dollar value on their cost. For example, the UN estimates that between 2% and 5% of global GDP ($1.6 trillion to $4 trillion) annually is currently connected with money laundering and illicit activity.[7] Notice that we are talking about trillions of dollars in IFFs, all occurring without the help of crypto and, we believe, being made possible because crypto is not widely used in our global financial system.

It is important to note that IFFs of all types are associated with either ineffective state functioning or illegitimate use of state power and are, without a doubt, an international problem – no country or region of the world escapes the blight of IFFs. Furthermore, capital outflows via IFFs can be considered lost GDP; IFFs reduce the revenue available to states and, ultimately, weaken the quality of governance. This means they weaken the ability of governments to crack down on crime and IFFs, creating a flywheel effect in which corruption begets more corruption.

In the *Research Handbook on Money Laundering*, Donato Masciandaro describes the societal costs of money laundering, outlining how every dollar of criminal proceeds reinvested can lead to more crime. In her introduction to the handbook, Brigitte Unger summarises the report thusly: 'Since money laundering can lead to an explosion of crime rates, it is a ticking time bomb.'[8]

In order to understand the issue, it is important we trace its origins, identify the criminals, and determine the scope of the problem and the costs associated with it.

---

[7] United Nations Office on Drugs and Crime, 'Money Laundering', *United Nations Office on Drugs and Crime*, 2024 <https://www.unodc.org/unodc/en/money-laundering/overview.html> [accessed 29 October 2024].
[8] Brigitte Unger and Daan van der Linde, *Research Handbook on Money Laundering* (Northampton, MA, 2013).

## 7.3   Who are the culprits?

If IFFs are a problem, it is natural to ask who is behind them. Are the bad actors crypto enthusiasts sending stablecoins on the blockchain? Apparently not, since IFFs were a problem before crypto even existed. As we will see, the bad actors with respect to illicit finance are big players in business (particularly finance) but also in traditional governance structures. For now though, let us focus on banks.

Although we opened this chapter with a discussion of actions taken against JP Morgan Chase, it is just one of the key offenders. Documents leaked from the US Treasury, known as the FinCEN Files, detail an astounding pattern of persistent abuse by five leading global banks, even after they were repeatedly caught. The International Consortium of Investigative Journalists (ICIJ) summarised the documents as follows:

> Secret U.S. government documents reveal that JPMorgan Chase, HSBC and other big banks have defied money laundering crackdowns by moving staggering sums of illicit cash for shadowy characters and criminal networks that have spread chaos and undermined democracy around the world.

According to the ICIJ's summary of the leaked documents, five global banks – JP Morgan Chase, HSBC, Standard Chartered Bank, Deutsche Bank and Bank of New York Mellon – 'kept profiting from powerful and dangerous players even after U.S. authorities fined these financial institutions for earlier failures to stem flows of dirty money.'[9] And we must couple this with the fact that the government rarely takes action anyway.

US agencies responsible for enforcing money laundering laws seldom prosecute megabanks that break the law, and the actions authorities do take barely ripple the flood of plundered money that washes through the international financial system. In those rare cases where the government has shown an interest in these banks' corruption, the banks largely ignored the government hand slaps. In some cases, they kept moving illicit funds even after US officials warned them that they would face criminal prosecutions if they did not stop doing business with mobsters, fraudsters or corrupt regimes.

---

[9] Alicia Tatone, 'Global Banks Defy U.S. Crackdowns by Serving Oligarchs, Criminals and Terrorists', *International Consortium of Investigative Journalists*, 20 September 2020 <https://www.icij.org/investigations/fincen-files/global-banks-defy-u-s-crackdowns-by-serving-oligarchs-criminals-and-terrorists/> [accessed 29 October 2024].

The consequences of all this corruption have been to prop up and finance some of the worst criminal actors on the global scene. Former Treasury sanctions official Elizabeth Rosenberg observed that banks like Jamie Dimon's facilitate this corruption by providing a mechanism for dirty money to 'slosh' around our financial system:

> The FinCEN files illustrate the alarming truth that an enormous amount of illicit money is sloshing around our financial system, and that U.S. banks play host and facilitator to rogues and criminals that represent some of America's most insidious national security threats.[10]

Too much of the business of centralised financial institutions takes place behind closed doors, so the temptation to alter the books or keep separate books or 'accidentally' lose the books is just too great.

## 7.4   Alleged attempts to deal with the problem

One might think that the gravity of all this corruption, the propping up of criminal actors and the global loss of wealth would lead to some attempt to solve the problem. Indeed, if one pays attention to congressional testimony about the dangers of crypto, one would think that government leaders are sincerely interested in this problem. However, the fact of the matter is that one can focus on where the problem is not (crypto) as a way of ignoring where the problem actually is (centralised governance and traditional finance's banking systems).

---

[10] Ian Talley and Dylan Tokar, 'Leaked Treasury Documents Prompt Fresh Calls for Updated Anti-Money-Laundering Regulations', *Wall Street Journal*, 21 September 2024 <https://www.wsj.com/articles/treasury-plugs-gap-in-anti-money-laundering-regulations-11600680611> [accessed 30 October 2024].

In a paper published in the journal *Policy Design and Practice*, Ronald F. Pol found that 'anti-money laundering policy intervention has less than 0.1 per cent impact on criminal finances'; more precisely, 99.95% of criminal proceeds are unaffected by anti-money laundering efforts.[11] Similarly, United Nations data for 2009 reported the figure for unaffected IFFs to be around 99.8%.[12]

One would think that with all the corruption in the world – and all the IFFs sloshing around in our financial system – finding and cracking down on corruption would be a trivial matter. Whether by design or incompetence, this is not the case. Indeed, Pol notes that 'compliance costs exceed recovered criminal funds more than a hundred times over, and banks, taxpayers and ordinary citizens are penalized more than criminal enterprises.'[13] In other words, whatever actions governments and banks are taking to crack down on corruption, they are not going about it in a cost-efficient way; for every dollar they spend on financial crimes, they recover one penny from the bad actors. And who is paying for these failing efforts? Ordinary citizens.

After seeing Jamie Dimon's unrelenting media campaign against crypto with the backdrop of JP Morgan's crimes, one begins to wonder if the campaign is not designed to deflect attention away from his own and similar organisations' actions. After all, if governments are concerned with cracking down on crypto, they are not focused on the real source of the problem – centralised organisations like large banks acting behind closed doors. It is also possible that Jamie Dimon believes what he is saying about crypto, but if this is the case, he is not really campaigning against IFFs as much as he is concerned that crypto will steal his grift.

---

[11] Ronald F. Pol, 'Anti-Money Laundering: The World's Least Effective Policy Experiment? Together, We Can Fix It', *Policy Design and Practice*, 3/1 (2020), 73–94.

[12] United Nations Office on Drugs and Crime, *Estimating Illicit Financial Flows Resulting from Drug Trafficking and Other Transnational Organized Crimes: Research Report* (Vienna, 2011) <https://www.unodc.org/documents/data-andanalysis/Studies/Illicit_financial_flows_2011_web.pdf> [accessed 29 October 2024].

[13] He adds the important caveat that 'The data are poorly validated and methodological inconsistencies rife, so findings cannot be definitive, but there is a huge gap between policy intent and results.'

Similarly, many in US Congress have, for all practical purposes, been bought by traditional finance. Large banks and other financial institutions are now the biggest donors to political campaigns in the United States. Thus, the campaigns against crypto are perhaps better understood as campaigns to protect donors from the world of traditional finance and, by extension, their money-making machine – some of it legitimate and, as we have seen, some of it very much in the service of criminal activity.

If this perspective is true, and it very likely is, then it is just one additional example of a worldwide approach to the application of anti-money laundering (AML) policies. These AML laws and policies should be designed to stop IFFs, but unfortunately, they are being deployed as tools to harass political and economic competitors. There are plenty of examples to pull from, but to get a glimpse at the international scope of the problem, we can begin with the use of AML policies by the Indian government to crack down on political enemies.

Amnesty International has argued that Indian authorities are exploiting AML laws to target civil society groups and activists and deliberately hinder their work. In a report titled 'Weaponizing counterterrorism: India's exploitation of terrorism financing assessments to target civil society', Amnesty revealed how the recommendations of the Financial Action Task Force – a global body responsible for tackling terrorism financing and money laundering – have been abused by the Indian authorities to bring in draconian laws in a coordinated campaign to stifle the non-profit public interest sector. These laws are, in turn, used to bring terrorism-related charges and, among other things, to prevent organisations and activists from effective fundraising.[14]

---

[14] Amnesty International, *India: Weaponizing Counterterrorism: India's Exploitation of Terrorism Financing Assessments to Target the Civil Society* (26 September 2023) <https://www.amnesty.org/en/documents/asa20/7222/2023/en/> [accessed 29 October 2024].

In particular, Aakar Patel, chair of the board at Amnesty International India, observed that 'Under the guise of combatting terrorism, the Indian government has leveraged the Financial Action Task Force's recommendations to tighten its arsenal of financial and counter-terrorism laws which are routinely misused to target and silence critics.'[15]

As you may have guessed, India is hardly the only place where AML laws and policies have been weaponised to assist despotic governments. The Open Dialogue Foundation published an article asking 'Can the EU's anti-money laundering reform help dictators?'. It argues that it not only can but that it does – that AML compliance can hurt civil society and that it does 'harm the rights of law-abiding customers, including those fleeing from or fighting authoritarianism.' The article goes on to provide a number of case studies to support its conclusion that 'politically-exposed organisations or individuals can become victims of the so-called false positives in AML compliance, which disproportionately affects low-profit customers.' For example, one class of victims of these policies are people trying to escape from despotic rule – AML policies prevent them from fleeing tyranny with their own money.[16]

As so often happens, AML policies that are alleged to crack down on crime – in this case, IFFs – do not really target the real bad actors (who are too powerful to bring to heel) but rather target individuals who are in no position to lobby against the policies. Perhaps you have noticed the difficulty in sending money to friends in other countries or had your bank account or PayPal account frozen temporarily for some arbitrary reason or other. This shows that the frictions are very real for little guys, even though they may be absent for larger players like JP Morgan Chase.

---

[15] Amnesty International, 'India: Government Weaponizing Terrorism Financing Watchdog Recommendations Against Civil Society', *Amnesty International* <https://www.amnesty.org/en/latest/news/2023/09/india-government-weaponizing-terrorism-financing-watchdog-recommendations-against-civil-society/> [accessed 29 October 2024].

[16] Lyudmyla Kozlovska, 'Can the EU's Anti-Money Laundering Reform Help Dictators?', *Open Dialogue Foundation*, 3 July 2023 <https://en.odfoundation.eu/a/578069,can-the-eus-anti-money-laundering-reform-help-dictators/> [accessed 29 October 2024].

In the paper 'Money Laundering', Michael Levi and Peter Reuter offer a diagnosis of the problem: current enforcement mechanisms fail to do the most obvious thing – follow the money – with predictably poor results. They reason, 'The regime does facilitate the investigation and prosecution of some criminal participants who would otherwise evade justice, but fewer than expected and hoped for by advocates of "follow the money" methods.'[17]

Rather than follow the money, the current system for controlling financial crimes utilises 'know your client' (KYC) methods that do not seem to be very effective against actual bad actors (drug cartels, corrupt governments and international banks) and create hassles for regular folks that just want to send a small amount money to a relative that needs help. This generates a problem of its own: such policies can create enough friction to drive otherwise law-abiding citizens underground; they push people to utilise the dark money economy. This is the conclusion drawn by Pierre-Laurent Chatain, Andrew Zerzan, Wameek Noor, Najah Dannaoui and Louis de Koker in a work titled *Protecting Mobile Money against Financial Crimes: Global Policy Challenges and Solutions*. In their view, 'Overly restrictive identification and verification processes in know-your-customer policies may push users back to the informal financial system.'[18]

Of course, low-income individuals are not the only victims of AML policies. We will ignore examples like Venezuela and Cuba, which are politically charged, but a good example was the policy of greylisting the Cayman Islands. Greylisting, in theory, merely involves closer scrutiny, but it can have a number of economic consequences for a country, including a reduction in foreign direct investment and increased scrutiny from financial institutions and regulators can bog the economy down in red tape. According to Andrew Perkins, writing in an article for the *Journal of Money Laundering Control*, greylisting the Cayman Islands was something that never should have happened – the nation was held to standards that were not asked of onshore jurisdictions, with unfair and economy-wrecking consequences.[19]

---

[17] Michael Levi and Peter Reuter, 'Money Laundering', *Crime and Justice*, 34 (2006), 289–375.
[18] Pierre-Laurent Chatain and others, *Protecting Mobile Money Against Financial Crimes: Global Policy Challenges and Solutions* (Washington, D.C., 2011).

Finally, as usual, these policies have been enacted with little concern for low-income individuals. Costs of money transfers have skyrocketed because of all the frictions introduced. Consequently, the people who most need money transfer services – for example, migrant workers who need to send their meagre income to families back home – are the ones who suffer. AML laws 'exclude low-income people from financial services through onerous regulations.'[20]

AML laws worldwide have not been effective at combating the IFF problem, recovering less than one cent for every dollar invested, forcing regular citizens to pick up the bill and placing an enormous burden on poor citizens who depend on money transfers to aid their families. At the same time, governments have not shown much interest in applying AML laws against large banks and other financial institutions. On the contrary, there have been instances of using the legal tools of AML laws to crack down on democratic movements and to support totalitarian regimes.

In summary, AML laws are powerful tools that governments wield for the purpose of fighting IFFs, but given that over 99% of criminal proceeds are unaffected by AML efforts, they clearly have not been effectively applied against the actual offenders, such as the world's largest financial institutions. However, they have been deployed to harass political and economic enemies and migrant populations. Needless to say, we believe that Elizabeth Warren's so-called 'war on crypto' is just the latest example of this ruse – directing AML policy against the crypto industry because it is perceived as a competitor to big banks.[21] It is, in fact, directing AML policy against the tools that can liberate citizens from tyrannical, despotic regimes around the world and, in the process, harming refugees and impoverished migrant populations.

---

[19] Andrew J. Perkins, 'Does Holding Offshore Jurisdictions to Higher AML Standards Really Assist in Preventing Money Laundering?', *Journal of Money Laundering Control*, 25/4 (2022), 742–56.

[20] Jennifer Isern and Louis de Koker, *AML/CFT: Strengthening Financial Inclusion and Integrity* (2009) <https://api.semanticscholar.org/CorpusID:167658451> [accessed 29 October 2024].

The case of Elizabeth Warren is a matter that deserves some paren­thetical attention, in part because of its tragic nature and in part because it provides a fine example of how regulatory capture works. Warren, for those who do not know, is a United States senator from Massachu­setts who gained attention early in her career for her campaigns against abuses by banks like HSBC. Subsequently, she has shockingly aligned with big banks' interests and with CEOs such as Jamie Dimon, as their joint act in his congressional testimony revealed.[22] Her 'anti-crypto army' is, in point of fact, fighting a war on behalf of traditional finance, with migrant populations being collateral damage. And we wish we could say the problem ended there.

## 7.5 Corruption all the way down

Earlier, we mentioned how cartels in Mexico rely on local governments to launder money and make their extortion collection policies more frictionless. No doubt, this takes place not just at the local level but at the national level as well. Certainly, on the global stage, there are nation states that are very much active in facilitating IFFs. Some nation states

---

may even rely on IFFs for their very survival.[23] However, it is worth focusing on lower levels of governance for a while.

Corruption is by no means limited to large governments. Local-level corruption, when considered in the aggregate, can be as significant as state-level corruption and IFFs. The key to understanding corruption is that it has nothing to do with whether the government is large or small, and it has nothing to do with the private sector versus the public sector. Large states, small states, large corporations and small corporations can all be corrupt. The secret ingredient in every case is centralisation. Centralisation gives someone sole control over the books, and this, in turn, invites abuse. Even if there are independent authorities to 'audit' the books, only the centralised authority can know if there are multiple sets of books and whether the auditor has the true set. In previous chapters, we went into the details about smaller-level governance structures like homeowner associations and the astounding level of corruption that takes place in those organisations – again, because they are centralised governance structures. Financial corruption is ubiquitous in centralised organisations.

The point we want to drive home here is that the problem is not just with bankers like Jamie Dimon nor with cartels nor corrupt municipalities in Mexico. The problem exists at every level of government in every part of the world, and the one common denominator that makes IFFs and other forms of corruption possible in every case is centralised control over records and a lack of transparency. Nothing good comes out of smoke-filled rooms. Somehow, we need to inject transparency into the system, from top to bottom. The question is, how do we do this?

---

[23] Eric Dante Gutierrez, 'The Paradox of Illicit Economies: Survival, Resilience, and the Limits of Development and Drug Policy Orthodoxy', *Globalizations*, 17/6 (2020), 1008–26 <https://www.tandfonline.com/doi/full/10.1080/14747731.2020.1718825> [accessed 29 October 2024].

## 7.6   Crypto to the rescue

We believe that transparency is the only answer to the catastrophic state of affairs in which dark money moves freely around the world. Michael Levi, in his 2015 paper 'Money for Crime and Money from Crime', echoed this sentiment:

> No-one could rationally think that AML controls in general or financial investigation in particular will 'solve' organised crime completely or eliminate high-level offending: for there even to be a chance to achieve that, there would need to be a step change in transparency and effective action against high-level corruption along all possible supply chains.[24]

But is crypto really the answer? How can this be? What of all the stories about money laundering that come from the likes of Jamie Dimon and Elizabeth Warren? These are fair questions, given what one hears in the media.

Perhaps we should look closer at the facts before we get into the details of how and why crypto will ultimately solve this problem. If we return to the topic of IFFs, then the question is naturally how crypto and blockchain technology can help with those. The answer, of course, is that IFFs are flows of money that are not visible to us. They take place behind closed doors. Once such transactions are placed on the blockchain, they are visible to all.

You do not need to take our word for this. You can go to any blockchain explorer, open it and follow the flow of money from any crypto wallet that might interest you. You can follow the money from that wallet to the next wallet and on and on until it leaves the blockchain and enters the shadowy world of traditional finance, which let us be honest, is the only point in the whole process where deception becomes possible.

---

[24] Michael Levi, 'Money for Crime and Money from Crime: Financing Crime and Laundering Crime Proceeds', *European Journal on Criminal Policy and Research*, 21/2 (2015), 275–97 <http://link.springer.com/10.1007/s10610-015-9269-7> [accessed 29 October 2024].

More importantly, blockchain technology offers us the possibility of programmable money. To illustrate the idea of programmable money, imagine that we designed a currency that was programmed so that if it was in the wallet of a minor, it could not be used to buy alcohol or tobacco products. The transaction would simply not go through; the money would not be functional for that purpose. This would not involve presenting IDs or complying with KYC checks but simply the idea that the money coming from the wallet of a minor would not work for certain purposes.[25] Applying this idea to IFFs, you do not actually need to rely on 'following the money'. You could program the money in such a way that it could not 'go dark' or be passed through the wallets of known criminals.

Sam Bankman-Fried, who is currently in prison for fraud committed when he was head of the crypto exchange FTX, is often associated with corruption in crypto, but none of his corrupt actions took place on the blockchain, nor could they. He did his dirty dealing behind closed doors at his centralised exchange. It was only there, in that centralised exchange, that he could take money from clients and repurpose it elsewhere. He created an environment where crypto could go dark. But programmable money could be engineered so that if it was placed in an exchange's reserves it could not be used for other purposes. One would not need to rely on the honesty of SBF. One could rely on the integrity of programmed money.

So, where do people like Jamie Dimon and Elizabeth Warren get the idea that crypto is a tool for money laundering? Presumably, from the idea that one might not know to whom a particular wallet address belongs. However, if someone is moving dirty money onchain, it typically comes from someplace offchain that made it dirty in the first place, and it typically goes to someplace new offchain. If it goes off-chain into the account of a terrorist organisation, then that is a clue that it is dirty. If it goes onchain, coming from a narcotrafficker, then that is again a clue that the money is dirty; with programmable money, that designation could not be laundered away. The good news is that once the money is onchain, one can follow it to its destination, or alterna-tively, one can program the money so that it shows itself as dirty given its origin, thus making it non-transferable to legitimate businesses (or to politicians). This is in marked contrast to the current system in which wealth is transferred in piles of cash or diamonds or artwork or gold or transfers of property or any other method of hidden wealth transfer that you can imagine.

---

[25] If one wanted, one could program the money so that funds transferred from the minor's wallet to a second wallet would not work for such transactions either.

Now, you might imagine that if there are centralised onramps to the blockchain like FTX, then these are tools by which dirty money can enter and leave the blockchain. However, this is easier said than done, for there are precious few onramps for crypto that can handle any serious monetary liquidity and those ramps that can (like the publicly listed Coinbase) are regulated. To be sure, an individual on the street can send you some crypto in exchange for cash, but this is not a serious problem in a world where trillions of dollars in dark money move about the planet with the help of nation states and global banking. Crypto and the blockchain are the only part of the entire process that is cur-rently pristine.

This is why, in this book, we make the case that the most important application for blockchain technology will not be for financial matters alone but rather for the business of human governance. It too must be made transparent, and it too must be placed on the blockchain, because until it is, governments will simply remain centres of power doing things behind closed doors, including the transmission of dark money around the global financial system.

## 7.7   The value of corruption-free governance

We have argued that crypto is not the problem with respect to IFFs but rather it is the solution. And we have argued that the problem is indeed massive – trillions (not billions, but trillions) of dollars of dark money slosh through the international financial system thanks to bad-acting banks and governments and other centralised authorities. However, even this fact does not do justice to the harm caused by IFFs and the opportunity costs of not having effective policies for dealing with them. In other words, we could have policies designed to actually fight IFFs as opposed to policies that protect the interests of large banks or policies that harm the interests of migrant workers who are simply trying to send money home to their families.

The World Bank's 2006 book *Where Is the Wealth of Nations?* highlights the profound impact institutions have on national prosperity. It found that the rule of law and human capital are the largest factors in the creation of wealth, dwarfing natural resource extraction and physical capital.[26] Furthermore, a study on institutional development and transaction costs published in the *Journal of Institutional Economics* found that a mere 0.1% reduction in transaction costs could quadruple a country's wealth. To put this into perspective, this is the difference between the financial health of Argentina and the financial health of Switzerland.[27] Optimising our institutional processes and eliminating the corruption that so naturally flows from centralised governance could not only halt the losses of trillions of dollars that are currently robbed from global GDP but also unlock vast economic potential worth additional trillions in value.

---

[26] World Bank, *Where Is the Wealth of Nations?: Measuring Capital for the 21st Century* (Washington, D.C., 2005) <http://elibrary.worldbank.org/doi/book/10.1596/978-0-8213-6354-6> [accessed 23 November 2023].

[27] Mitja Kovač and Rok Spruk, 'Institutional Development, Transaction Costs and Economic Growth: Evidence from a Cross-Country Investigation', *Journal of Institutional Economics*, 12/1 (2016), 129–59 <https://www.cambridge.org/core/product/identifier/S1744137415000077/type/journal_article> [accessed 29 October 2024]

## 7.8   Public confidence in governance

In the previous section of this chapter, our thesis was that eliminating IFFs could have a flywheel effect in that minor improvements in financial efficiency can have massive positive consequences. Additional efficiencies could flow from having a financial system that is more capital efficient – a system in which there were fewer frictions to the legitimate movement of capital. However, there is another flywheel effect that we could benefit from.

As matters currently stand in our world, governments do not inspire much in the way of confidence from their citizens. In countries like Mexico (which at the time of writing has the planet's twelfth largest economy), citizens' cynicism over criminal elements in governments cannot be advantageous for their national economies.

Understandably, seeing businesses that are clearly money laundering operations and skyscrapers bearing the names of banks that build their empires off the profits of the drug trade generates a lot of suspicion. Even in the United States, there is widespread public perception that governance at every level is corrupt and that financial institutions are equally corrupt (no doubt due to the ineptitude of governments in fighting corruption). Nor is this perception somehow mistaken. The people are right; their government systems are corrupt, non-transparent and not at all working in their interests.

We can even quantify this perception. Public confidence in governmental institutions has been in decline for decades, and it is happening everywhere. For example, in a report published by IPSOS (Institut Public de Sondage d'Opinion Secteur), it was found that France has reached a new historic low, where 82% of citizens believe the country is heading in the wrong direction. Great Britain experienced the biggest fall in optimism in the same month of the report, dropping fourteen percentage points.[28]

---

[28] IPSOS, *What Worries the World – March 2024* (March 2024) <https://www.ipsos.com/sites/default/files/ct/news/documents/2024-05/Global-Report-What-Worries-the-World-March-2024.pdf> [accessed 29 October 2024].

In the private sector, there is an equivalent to public trust in govern-
ment – the customer satisfaction (CSAT) score. Across a wide variety
of industries – be it finance, energy, technology, shipping or airlines
– industry average CSAT benchmarks are often found to exceed 70%,
essentially the inverse of the poor scores that people assign to their
national governments. The question, of course, is why governments do
so much worse in comparison to other institutions – even airlines, of
all things. One possibility could be the quite justified perception that
governments around the world are dens of corruption.

Transparency International's Corruption Perceptions Index report
monitors 180 countries and territories around the globe by their per-
ceived levels of public sector corruption. Its 2023 report stated that:

> Over two-thirds of countries score below 50 out of 100, which
> strongly indicates that they have serious corruption problems.
> The global average is stuck at only 43, while the vast majority of
> countries have made no progress or declined in the last decade.
> What is more, 23 countries fell to their lowest scores to date
> this year. [. . .] Both authoritarian and democratic leaders are
> undermining justice. The global trend of weakening justice sys-
> tems is reducing accountability for public officials, which allows
> corruption to thrive.[29]

Let us step back and remind ourselves why this is so important. We
began by pointing out that most of the wealth in the world is tied up
in the effectiveness (or lack of effectiveness) of traditional political gov-
ernance systems. We also saw that tiny changes in efficiency can have
enormous consequences on whether a government can be effective in
helping its people. Just a slight change in efficiency can affect whether
the economy is going to be equivalent to that of Switzerland or to that of Argentina in recent decades. And given the importance of good governance, we are brought back to the question of when we are going to do something to bring it about.

---

[29] Transparency International, *Corruption Perceptions Index* (2023) <https://www.transpar-
ency.org/en/cpi/2023> [accessed 29 October 2024].

The good news is that we already know what needs to be done. The first and most important thing to do is continue to develop blockchain technologies and apply them to all aspects of human governance, from the financial system to elections. Doing so will make government actions transparent, and applying the technologies to our financial systems will make them equally transparent. Crypto shines a bright light on activities that today take place behind curtains and in smoke-filled rooms with little to no accountability. The scourge of centralised governance has been a magnet for corruption for too long. It is time to tear down the curtains, kick down the doors of the smoke-filled rooms and shine the light of transparency on all aspects of human governance, including our financial system. Crypto is not the problem; it is our best and only solution to the problem.

Of course, there are many forms that future decentralised governance structures might take, including the idea of cyberstates, discussed in *Crypto Anarchy, Cyberstates, and Pirate Utopias*,[30] or what Balaji Srinivasan has called 'network states'.[31] Network states or cyberstates, or whatever we wish to call them, have received a lot of attention lately. In the next chapter, we will consider whether they are a promising alternative for the post-Westphalian era.

---

# C H A P T E R  8

# ARE CYBERSTATES THE ANSWER?

## 8.1  Preliminaries

In 1996, in the wake of the passage of the *Communications Decency Act* (CDA) by the United States Congress, the Internet found an unlikely hero in John Perry Barlow. Barlow was a Republican Wyoming cattle rancher, a former lyricist for the Grateful Dead and, not least, a cofounder of the Electronic Frontier Foundation. Fed up with the CDA's ham-fisted attempt to censor the Internet, Barlow wrote and uploaded his proclamation, 'A Declaration of the Independence of Cyberspace'. From its opening paragraph, the essay did not pull punches:

> Governments of the Industrial World, you weary giants of flesh and steel, I come from Cyberspace, the new home of Mind. On behalf of the future, I ask you of the past to leave us alone. You are not welcome among us. You have no sovereignty where we gather.

Barlow then doubled down on his thesis that terrestrial governments have no claim over the sovereignty of cyberspace:

Governments derive their just powers from the consent of the governed. You have neither solicited nor received ours. We did not invite you. You do not know us, nor do you know our world. Cyberspace does not lie within your borders. Do not think that you can build it, as though it were a public construction project. You cannot. It is an act of nature and it grows itself through our collective actions.

It concluded with a pledge to create a 'civilization of the Mind' in cyberspace. In order to accomplish this, the thought was that 'We must declare our virtual selves immune to your sovereignty' and further that 'We will spread ourselves across the Planet so that no one can arrest our thoughts.' And finally, it articulated the aspiration: 'May [our Civilization of the Mind] be more humane and fair than the world your governments have made before.'[1]

Inspired by this work and similar discussions in the Internet slipstream at the time, Peter Ludlow collected a group of essays into a collection entitled *Crypto Anarchy, Cyberstates, and Pirate Utopias*.[2] The title itself was a play on Robert Nozick's famous defence of libertarianism, *Anarchy, State, and Utopia*,[3] and it asked the question: What if virtual states could be formed that existed online – cyberstates? Would it be possible? Would it be a good thing?

Ludlow concluded that such states were indeed a possibility and that they were already in the formation process, but he expressed some pessimism about their ability to persist. He reasoned that these 'islands in the net' (using a phrase borrowed from the science fiction author Bruce Sterling[4]) might have their day, but that day would be fleeting. Terrestrial governments would surely use their power to eliminate them eventually. Thus, he closed his anthology with the classic underground manifesto by Hakim Bey, *T.A.Z.: The Temporary Autonomous Zone*.[5] Bey thought that these islands in the net, like the Caribbean pirate enclaves of the eighteenth century, would be fleeting postmodern utopias. They would form, dissolve and then reorganise elsewhere.

---

[1] John Perry Barlow, 'A Declaration of the Independence of Cyberspace', *Electronic Frontier Foundation*, 2016 <https://www.eff.org/cyberspace-independence> [accessed 6 May 2023].
[2] Ludlow, *Crypto Anarchy, Cyberstates, and Pirate Utopias*.
[3] Robert Nozick and Thomas Nagel, *Anarchy, State, and Utopia* (New York, NY, 2013).
[4] Bruce Sterling, *Islands in the Net* (New York, NY, 1989).

Ludlow's volume appeared in 2001 and, as David Hume once said of his most famous publication, *An Enquiry Concerning Human Understanding*, it initially 'fell stillborn from the press.'[6] But more than two decades later, with the rise of social media and web3 and blockchain technologies, people have begun to revisit its central thesis. Are cyberstates a possibility after all? And if so, is the outlook for their persistence more optimistic than it was two decades ago? We do not think they are the final stage in the progress of human governance, but we suspect they just might outlast traditional terrestrial nation states.

In this book's Introduction, we looked at the extensive role of legacy governments in our lives and, beyond that, the ubiquity of governance of some form or other in almost all aspects of our lives. In the previous chapter, we saw that the role of governments in the creation and preservation (and sometimes destruction) of wealth is massive. Of course, all of that was to set up a case that we want to make for an alternative to traditional nation states and to the traditional tools of governance that are used. Our promise was to make the case for blockchain governance. In Chapters 5 and 6, we discussed how blockchain governance will work (we will go into more detail later). Now, we want to introduce one particular form of blockchain governance: the cyberstate.

Cyberstates are governance structures organised around blockchain technology, with the idea that they will take on the role currently occupied by nation states, although with several advantages over contemporary nation states. Balaji Srinivasan has called them 'network states',[7] and while they are indeed organised around networks, we prefer to call them 'cyberstates', with a nod to Norbert Weiner's term 'cybernetics' and an even bigger nod to the Greek term *kybernetes*, or 'steersman' (metaphorically, 'guide' or 'governor'). In other words, cyberstates are not merely aimless networks of people and groups of shared interest, but they can provide guidance (direction) for those groups.

---

[5] Bey, *T.A.Z.: The Temporary Autonomous Zone*.
[6] Howard Darmstadter, 'David Hume at 300', *Philosophy Now*, 83, March 2011 <https://philosophynow.org/issues/83/David_Hume_at_300> [accessed 29 October 2024].
[7] Srinivasan, *The Network State*.

Cyberstates are superior to nation states because they can be organised around shared interests and values rather than arbitrary political boundaries, which are typically established by rivers and oceans and, quite often, by histories of armed conflict over natural resources. The cyberstate, by contrast, is geographically unencumbered. Its 'territory' is determined by its footprint in cyberspace, which is unlimited in scope and scale.

Cyberstates, like blockchain communities generally, also allow communities to become self-determining – a principle enshrined in the *Charter of the United Nations*, although not particularly respected by the existing political order.[8] Political self-determination means that people get to choose how they are ruled. The government they are subjected to is not determined by where they are born or which despot happens to take power over them. They have a say in what their government looks like and how it operates, and they also have an opportunity to move freely to a political system that they find more favourable.

Because cyberstates would be organised around information-processing networks, one could engineer tools to enhance commerce, voting and other aspects of governance. Indeed, one can build tools that allow cyberstates to help their citizens flourish economically and culturally. Cyberstates are relatively easy to join and relatively easy to exit. If the rulers of the cyberstate are not following your values, you exit. In reality, it is not quite this simple; as we will see in Chapter 9, exit is not frictionless, but we can engineer ways to minimise frictions.

---

[8] United Nations, 'Charter of the United Nations' (1945) <https://www.un.org/en/about-us/un-charter> [accessed 29 October 2024]

Certainly, there are a lot of questions that need to be addressed. Perhaps you have already thought of some of them. For example, what if there is no government that suits you? Or what if you choose a government that does not deliver on its promises? And forget about your cyberstate; who governs where you actually live in physical space? Who is in charge of security there? And who is in charge of fixing potholes and disposing of waste? Or for that matter, what if another cyberstate full of hackers comes after your cyberstate and drains it of its wealth? And what about money? And laws? And food? And everything else? We will get to these questions eventually, at least as they pertain to the more general idea of blockchain communities.

First, stepping back and taking the view from altitude, we can say that cyberstates are online communities, organised around blockchain technologies, that carry out functions usually associated with traditional nation states. This might include providing security; assisting community welfare (things like Medicare or Medicaid); playing a role in supporting national culture (think of the French government's support of the arts); and supporting business in the form of negotiating trade agreements, encouraging business development and so on.

Obviously, in the case of cyberstates, these familiar functions will be grounded in the digital realm and will thus look somewhat different than they do today. In the realm of security, for example, one principal function of the cyberstate would be to support the security of its core information infrastructure and avoid hacking assaults and 'zero-day exploits'.[9] Cyberstates would certainly be able to provide physical security as well, whether in the form of purchased security forces or simply leasing bases from which kinetic forces can keep lines of physical communication open (shipping routes, for example). At the end of the day, they are not virtual states but rather, as we stated, states organised around blockchain technologies. This means that they may have a very large footprint in the physical world. Blockchain technology is simply the central nervous system of the cyberstate, whether its activities are online or in the physical world.

_______________

[9] 'Zero-day exploits' are computer exploits that are thus far unknown – that is to say, known about for zero days. Therefore, they contrast with exploits that have been known for some time, such as thirty-day exploits. The problem with zero-day exploits is that, thus far, there has been no time to armour against them.

What then is the big difference between cyberstates and traditional nation states? The principal difference will be that in cyberstates, the governance activities of the state take place onchain. This means that important records, communications (including decision-making processes), voting, and government promises and policies are all on the blockchain and all very much visible to community members. Policies will take the form of onchain smart contracts. Votes may well be direct votes for those smart contracts. It will thus be an infrastructure grounded in blockchain technology, meaning its functions will be distributed yet cooperative.

More specifically, this means that, for a cyberstate, the economic rails will use a blockchain-based currency like bitcoin and an onchain decentralised financial system, its records will be immutable yet accessible to all, and its policies and intentions will take the form of smart contracts that are visible to all onchain. Contrary to John Perry Barlow's vision, it need not be some sort of disembodied state but may be a state very much grounded in the physical world, albeit organised with the use of technologies that afford decentralised cooperation.

As we noted, the general idea of a cyberstate has become more popular in the last two decades (certainly taken more seriously), and one recent version of the idea has been put forward by Balaji Srinivasan in his book *The Network State*. Since that project is well known in the online community, it is worth spending some time contrasting our position with his.

Srinivasan begins his book with a one-sentence definition of a 'network state'.

A network state is a highly aligned online community with a capacity for collective action that crowdfunds territory around the world and eventually gains diplomatic recognition from pre-existing states.

As we understand this statement, he is saying that there are three components to a network state:

1. A highly aligned online community with a capacity for collective action
2. Crowdfunding territory around the world
3. Gaining diplomatic recognition from pre-existing nation states

He follows with a paragraph-level definition, which goes into more detail on these key ideas.

> A network state is a social network with a moral innovation, a sense of national consciousness, a recognized founder, a capacity for collective action, an in-person level of civility, an integrated cryptocurrency, a consensual government limited by a social smart contract, an archipelago of crowdfunded physical territories, a virtual capital, and an on-chain census that proves a large enough population, income, and real-estate footprint to attain a measure of diplomatic recognition.

Our vision of blockchain communities intersects with Srinivasan's vision of network states in some respects, but we part company in key areas.

Let us start with the shared vision part. A blockchain community will definitely involve a highly involved online community with a capacity – indeed, a robust capacity – for collective action. We also agree with some of the development of this idea in the more extended definition. There should be an integrated cryptocurrency and a consensual government that utilises social smart contracts of the type we discussed in Chapters 5 and 6.

However, we do not see the centrality of 'a sense of national consciousness', nor do we even see the desirability of a sense of national consciousness. As Srinivasan points out, the etymology of the term 'nation' suggests the importance of place or group of birth ('nation' being from the Latin root *nasci* – to be born). Our view is that the nation is part of the core problem with nation states. Cyberstates (or network states) may choose to leave the idea of the nation behind. There are any number of ways a cyberstate might come together – through shared interests or beliefs or goals (including a shared interest in pluralism) – and this need not have anything to do with where we are born or whom we are related to by birth. People have no reason to organise themselves around their idea of national identity. However, this is not to say that they may not or will not.

We also take some exception to the idea that there must be a 'recognized founder'. There is nothing wrong with founders per se (one of the authors of this book is a founder), but in cases like Bitcoin, the founder may be anonymous and then disappear. Indeed, one might even make the case that it is better that way. On this line of reasoning, obsession with founders can be a way of introducing mythology into the foundations of our governments. For sure, it is hard to avoid the existence of founders. Someone has to get things up and running, after all. Still, there is no obvious reason to build the mythology or even importance of founders into the very definition of 'cyberstates'.

We do agree, as we said, with the promise of blockchain technologies for human governance. As we noted earlier, good governance, among other things, requires immutable, transparent records and economically sound money. While blockchain technologies do not guarantee these things, they certainly make them easier to achieve. As noted, this is a point on which we are in substantial accord with Srinivasan.

This brings us to clauses two and three of Srinivasan's one-sentence definition of a 'network state'. We have doubts about both of these components, and here, we are perhaps taking issue with the need to reproduce the image of the traditional state on the blockchain. In other words, we take issue with both the nation part and the state part of the nation state. First, we do not see the necessity of crowdfunding territory for a blockchain community. Such communities do not need to have, or even be interested in having, sovereign control over physical territory.

Second, we disagree that blockchain communities must somehow earn respect from or be recognised by existing nation states. In any case, it is not necessary for existing nation states to recognise blockchain communities as equal partners. It might be enough that they do business with blockchain communities, just as they do business with many nongovernmental organisations (NGOs), corporations and individuals throughout the world. Ultimately, what matters in blockchain governance is that community members are flourishing.

You are probably thinking that we glossed over that rather quickly, and we could not agree more. Thus, the remainder of this chapter is devoted to diving deeper into these questions. In Section 8.2, we will make the case that there is no need for blockchain communities to trouble themselves with the acquisition of physical territory. Doing so always exists as an option, of course, but ultimately, we will argue that while blockchain communities will be able to protect your individual property rights (should you so desire), it does not follow that they must acquire control of physical territory in order to do so.

In Section 8.3, we will visit Srinivasan's idea that cyberstates should seek out recognition from traditional nation states, and we will argue that while this is an option, there is not much point in it. Ultimately, we want to supersede the old boys' club of nations, not join it. In 8.4, we will take up the issue of national identity and argue that this can be a dangerous idea, which also should be optional. By the time we get to Section 8.5, Srinivasan's idea of a network state will be pretty well deconstructed, which is perhaps just as well because, as we noted in Chapter 2, the very idea of a state (like the idea of a nation) may have reached its expiration date.

## 8.2   Must blockchain communities have physical territories?

As we noted earlier, Balaji Srinivasan's definition of a 'network state' includes the idea that such a state would 'crowdfund territory around the world.' Srinivasan envisions the result as an archipelago of physical territories under the control of the network state.

Now, the first observation that we want to make is that this is not at all a crazy idea. It is entirely possible for blockchain communities to purchase chunks of physical land or perhaps other physical locations – offshore seasteading locations, for example. Our point is not that this would not work or even that blockchain communities should abstain from crowdfunded purchases of physical territory. Our point is merely that it is not necessary.

Why do blockchain communities not need physical territory? What kind of state-like governance does not have an inch of physical territory? People have to live somewhere, after all. However, where they live is quite a different question from where their community is located in physical space or even whether it is located in physical space at all.

As we write, approximately 1.6 million United States citizens live in Mexico.[10] Some of those citizens get into trouble from time to time, and if the trouble is serious enough, they will contact a United States consulate for assistance. If they wish, they can also receive regular updates from the US government about dangerous hot spots in Mexico and other countries where they might travel. They are, for all practical purposes, represented by a significant player should trouble arise. Chances are, if they are US citizens, they are much less likely to be harassed by the local police, the federal police, the local narco leaders or whoever happens to be in charge in that area.

This leads us to the following thought experiment. What if the United States did not have physical territory but still had the wealth and ability to project the power that it does? For example, suppose its wealth was all onchain and its military forces were stationed on leased property as they are today in military bases throughout the world. Indeed, as of this writing, the US has 174 military bases in Germany, 113 in Japan and 83 in South Korea. There are hundreds more scattered around the world in places ranging from Aruba and Australia and Bahrain and Bulgaria to Colombia, Kenya and Qatar, to name just a few. Worldwide, the United States has bases in more than seventy countries. Despite popular belief, none of those bases constitute US territory. Our point here is that if you have the financial resources, you can project a lot of military might without having complete sovereignty over or even owning physical territory.

---

[10] David Nadelle, '7 Reasons So Many Americans Are Moving to Mexico City, Starting With Cheap Rent', *Yahoo Finance*, 29 May 2024 <https://finance.yahoo.com/news/7-reasons-many-americans-moving-180139202.html> [accessed 29 October 2024].

What would be the point of having such a vast military presence if you had no territory to defend? Well, the myth about the US military is that it is there to protect US territory. In fact, the US military is vastly larger than would be necessary to protect its territory. Consider, for example, a conventional attack from the country that is arguably the second greatest threat to the United States: Russia. As we write this, the Russian Federation is struggling to maintain its supply lines as it struggles to seize territory in eastern Ukraine, even though Ukraine borders Russian territory.[11] Rivers like the Dnieper pose major tactical obstacles.

Meanwhile, should Russia choose to attack the forty-eight contiguous United States with conventional forces, it would not be as simple as crossing its border into Ukraine using already existing roads and rail systems. To execute a physical attack, Russia would first have to transport its forces thousands of miles. Assuming it had enough cargo planes and ships to do this successfully, it would still need to maintain logistical support across great distances. If Russia cannot handle the logistical challenges of attacking Ukraine, there is no way it can handle the logistical challenges of a conventional (non-nuclear) attack against the United States. Such an attack would be challenging for Canada or Mexico even if they had the offensive armies to attempt it. So, what is all that US military force for?

---

[11] Devin McCarthy, D. Sean Barnett and Bradley Martin, *Russian Logistics and Sustainment Failures in the Ukraine Conflict* (2023) <https://www.rand.org/pubs/research_reports/RRA2033-1.html> [accessed 29 October 2024].

The United States Armed Forces do not primarily exist to protect US borders but rather to ensure that the global financial system, global resources and global supply chains remain unmolested. This means that oil not only needs to be pumped freely in the Middle East but that it must be transported freely around the world. Similarly, it means that Apple's and Walmart's supply chains across the globe remain uninterrupted. As the 2020 COVID-19 outbreaks showed, supply chain interruption can have serious consequences for the global economy. And in the end, this is what US military forces are defending: the global economy. This is not a new phenomenon. Imperial navies like the British Navy long understood that one of their key roles was to keep international shipping lanes open – in their terminology, 'to protect the lines of communication'.

Let us return to our thought experiment, which showed that you could imagine being a United States citizen, enjoying the protections of its economic and military might, without the United States holding any physical territory at all. The thought experiment raises the question: Would it make any difference at all to the 1.6 million US citizens living in Mexico? Not likely.

In point of fact, we can extend our thought experiment. This would never happen, but surely it is logically possible that the United States could relinquish all its territory and its mission to protect its borders and focus all its attention on protecting the economic interests of US citizens worldwide. If you hold US citizenship, you would be protected (and of course, also required to pay taxes to the US). As with most thought experiments, we do not consider this a likely scenario, but it should nevertheless help us see what governments are really about – protecting the interests of their citizens and their business relationships around the world. The fetish for territorial sovereignty is a sideshow.

Now, we recognise that territory – land – is sacred for many people. Various political groups tie their identity to geographical locations. The Land of Israel is a case in point, as it is land that was believed to be promised to the descendants of Abraham. On very much the flip side, Nazi Germany advanced the concept of *Blut und Boden* (blood and soil), which emphasised the connection between German ethnicity (blood) and the land (soil).

A very different articulation of this idea is found in the book *Ohitika Woman* by Mary Brave Bird:

> *Maka ke wakan* – the land is sacred. These words are at the core of our being. The land is our mother, the rivers our blood. Take our land away and we die. That is, the Indian in us dies. We'd become just suntanned white men, the jetsam and flotsam of your great melting pot. The land is where even those Native Americans who live in the wasichu cities, far away from home, can come to renew themselves, where they can renew their Indianness. We have an umbilical cord binding us to the land and therefore to our ceremonies – the sun dance, the vision quest, the yuwipi. Here, the city Indians can relearn their language, talk to the elders, live for a short while on 'Indian time,' hear the howl of brother coyote greeting the moon, feel the prairie wind on their face, bringing with it the scent of sage and sweet grass. The white man can live disembodied within an artificial universe without ties to a particular tract of land. We cannot. We are bound to our Indian country.[12]

There is a discussion to be had about whether the attachment of national identity (or any identity) to physical territory is healthy or whether it is deeply problematic. Perhaps some of us can live within an artificial universe without ties to a 'particular tract of land', and others cannot. We are not here to take sides in such a debate. Our point, at the moment, is that physical territory need not be important for every community, conceding that it may be very important for some.

---

[12] Mary Brave Bird and Richard Erdoes, *Ohitika Woman* (New York, NY, 2014)

All of this may seem like a long-winded way to take issue with Srinivasan's idea of network states and the need to crowdsource physical territory, and for all we know, he would agree that states should not have to do so. However, our goal here is not to cavil with Srinivasan so much as it is to drive home the point that communities can be divorced from physical territory.

What does life look like for citizens of a landless community (whether you call it a state or not)? Well, it looks very similar to how it does now for the millions of expats worldwide. Some of them travel around the world as digital nomads, and some of them stay in the same place for the bulk of their expat lives. Some of them work online as programmers or graphic designers or writers – any type of work that can take place online. However, many others work in the service industry. There is absolutely no reason that they could not work in factories or be agricultural workers, although this typically involves expats from less wealthy to more affluent countries.[13]

There are plenty of options concerning the ways in which a landless blockchain community might help its citizens. Such a community might help with security and legal troubles, assuming citizens pay taxes to the blockchain community. Following our earlier thought experiment, it could provide health care and retirement benefits to its global citizens. In that case, it could use its clout to negotiate better prices for medicines and treatments, for example. It could negotiate deals with real estate developers and service providers on behalf of its citizens. It could negotiate tax treaties with terrestrial governments to ease the tax burdens on its citizens. There is really no limit to what such a community might be able to offer its citizens if it had sufficient clout. All services could be provided just as easily, even if there is no land or territory 'back home'.

---

[13] It is interesting that in this case, they are typically referred to as 'immigrant workers' while immigrants from wealthier to less wealthy countries are known as 'expats'. What is the actual difference? None.

A blockchain community could also negotiate tax-free special economic zones (SEZs) for businesses that its citizens create. Do you want to open a factory to make electronic vehicles or drones or breakfast cereal? There are plenty of existing SEZs around the world with rules negotiated with whichever sovereign authority controls the territory. A blockchain community, even if it had no territory of its own, could negotiate SEZs on behalf of its manufacturing sector, giving it an archipelago of manufacturing locations. Similarly, it could negotiate the leasing of farmland for its agricultural sector.

We are saying that a blockchain community need not depend solely upon a virtual economy simply because it is technically landless. If it can arrange the frictionless leasing of manufacturing and agricultural territory for its citizens, there is no reason that it could not have the manufacturing and agricultural base of the largest nation states. Blockchain communities could have sovereign control over physical territory. However, there is really no reason for them to need it, apart from certain religious or cultural demands that they should.

## 8.3  Should blockchain communities aspire to be diplomatically recognised by nation states?

We mentioned earlier that another point where we take exception to Srinivasan's network state is that, according to his definition, a network state should aspire to be recognised as an equal by existing nation states. Why do we take exception to this idea?

The first thing to understand is that traditional nation states do business with a wide array of entities beyond other nation states. It is not as though individuals and organisations are invisible to nation states. Indeed, nation states routinely enter into commercial arrangements with all kinds of private enterprises. For a very obvious case, think of defence contractors, which are organisations that work closely with the governments of the world. However, governments are constantly requesting bids for goods and services from corporations, individuals and NGOs.

Our point here is that if nonstate actors are already visible to existing nation states, it does not matter if some blockchain communities are recognised as being on a par with traditional nation states or not. Whether a traditional state wishes to hand the honorific title 'state' to a blockchain community or not hardly matters. The important question is whether traditional governments are willing to do business with blockchain communities or not, and the answer to this is that they absolutely will if the communities have something of value to offer them – for example, educated workers or investment opportunities or security arrangements or jobs for its people. Blockchain communities have a lot to offer, not only to their citizens but to the existing terrestrial powers that host their citizens.

So, our first thought is, who cares if a traditional nation state recognises your blockchain community as a fellow state or not? Who cares if you get a seat at the United Nations or not? On the other hand, we understand Srinivasan's sentiment. One puts a lot of effort into building a blockchain community; is it not reasonable to wish to be recognised as an equal by existing states?

At first glance, it may be natural to think this way, but on reflection, it is a somewhat backward way of thinking. Blockchain communities are not merely new kids on the global block. They are a big part of the future of human governance. If you are the future of governance, you do not worry about invitations to the old boys' club. You are going to form a new club. Our view is that nation states served their purposes or perhaps misserved their purposes but, in any case, their time has come and gone.

What then becomes of global governance and the United Nations? Well, there is no point in a blockchain community worrying about being invited into 'the family of nations'. Blockchain communities will be actors on the global stage, and besides traditional nation states, they will be engaged with many more entities. They will be engaged with international corporations, with countless NGOs, with all kinds of terrestrial authorities ranging from cities to SEZs and, in some locations, local cartel leaders and military strongmen.

Could we envision that cyberstates would be admitted into the United Nations someday? Sure, but as we asked, is that even such a big deal? Perhaps a bigger deal would be if the rise of cyberstates led us to appreciate other forms of governance around the world. Why not have an organisation to represent the interests of tribes or economic classes of people or religious groups or residents of SEZs or, for that matter, blockchain communities?

Maybe what we really need is an organisation that offers seats at the table to all modes of human organisation, including tribes and special interest groups – a League of Communities, perhaps. We are not making a proposal here; we are saying that the reflex to be seen and recognised as equals by nation states is a misguided one. Nation states are organisations built around obsolete technology. They are themselves becoming obsolete. We need not care what they think.

## 8.4   Should blockchain communities strive to have *national* identity?

This leads us to the final point where we take issue with Srinivasan's idea of a network state – the idea that there be 'a sense of national consciousness'. As before, we stress that there likely will be blockchain communities with a sense of national consciousness; there may even be communities with national identity tied to place of birth or ethnicity. However, there may well also be communities in which the sense of identity is nothing more than pride in being a member of the most efficient or most ethical or most pluralist community.

This is not just a quibble with Srinivasan, however, for we think that one of the principal benefits of blockchain communities is that they allow us to leave national identity in the proverbial dustbin of history. For traditional terrestrial states, located as they are in distinctive geographical territories, it is not surprising that there is a sense of ethnic or national identity in that area. It is the area where most of the citizens were born, after all. However, on reflection, this is something of an artifactual property of nation states. There are so many ways to organise a community; why should national identity be one that we prioritise?

There is also no question that nationalism can be a dangerous principle around which to organise a state or a community. It is not accidental that the term 'Nazi' is from *Nazionalist*, which is to say 'nationalist'. The Nazi party was simply the nationalist party.[14] And Nazi Germany is far from the only case where nationalism led to genocide and 'ethnic cleansing'. It is not even difficult to see why this should be so.

If one had a magic wand, perhaps it would be wise to use it to eliminate nationalist sentiment altogether, given its history. However, there is no such magic wand, and it is difficult to see how one can eliminate the sentiment that human beings have for ethically grounded national identities. For all we know, that sentiment could be baked into the selfish genes of human DNA.

It is also worth pointing out that certain forms of nationalism not only seem benign but perhaps even worth fostering. Should we take objection to some Native Americans identifying as members of the Lakota nation, for example? We also forget that national sentiment, not so long ago, was very much a progressive idea. For example, during the revolutions that swept through Europe in 1848, a lot of revolutionary spirit was driven by nationalist sentiment, which served as a foil against the Habsburg Empire.[15] Nationalist sentiment in Warsaw Pact countries also served as a counterweight against the former Soviet Union during its collapse in 1990.[16] Given that nationalism can be a positive force under some circumstances, and given that it is probably impossible to eliminate it in any case, we think that the best solution is to accept that some blockchain communities will organise around a sense of national identity.

---

[14] Strictly speaking, officially, the Nationalsozialistische Deutsche Arbeiterpartei.

[15] Pieter M. Judson, *The Habsburg Empire: A New History* (Cambridge, MA, 2016).

[16] Mark R. Beissinger, 'Nationalism and the Collapse of Soviet Communism', *Contemporary European History*, 18/3 (2009), 331–47.

In saying this, we are buying into the fiction that national identity is something real above and beyond a very amorphous sense of ethnic familiarity. The collapse of the Soviet empire also coincided with the collapse of the Socialist Federal Republic of Yugoslavia, with resulting atrocities and acts of genocide against Bosnians and Croatians. The Serbo–Croatian language became understood as separate languages – Serbian, Croatian and Bosnian – despite the fact that from a linguistic perspective, they are more similar than American English is to British English.[17] The differences between Catholicism and Orthodox Christianity, which played a role in the ex-Yugoslavian conflicts, seem just as artificial from the long-term perspective. Such differences can be a big deal if you want them to be (they were certainly a big deal in the Thirty Years' War) or if some demagogue convinces you that they are. Our point is that national identity is constructed from a collection of differences that may seem important but which, at the end of the day, are fleeting, random and unimportant.

It is also plausible to think that the intensity of ethnic identity in ex-Yugoslavia directly resulted from General Tito's attempt to wipe out cultural differences and impose a new national identity for Yugoslavia. In other words, it was his attempt to impose cultural conformity within terrestrial borders. This is, of course, a variation on the theme we discussed in Chapter 2 – a consequence of having diverse groups kettled together within a geographic territory. Only in this case, the idea was not to impose culture A over culture B but to impose some third universal culture C. Repressing cultural differences does not lead to happy outcomes.

---

[17] As linguists are fond of saying, a language is a dialect with an army and a navy. The point is that deciding when something is categorised as a distinct language is really just a political decision based on any number of factors.

Our point is that any difference in religion or language or culture or genetic makeup can be imagined to be a big deal and can be construed as an integral part of national identity. And maybe it is a big deal to some people. It is not possible to make it go away. Arguably, it can even be a positive thing for the preservation of some cultures. The optimal system of political organisation is thus not to eliminate cultural diversity nor repress the desire of people to organise around such principles but to ensure that groups, so organised, do not trample on the rights of others. We have already seen how blockchain communities enable people to do this. They allow people to organise themselves unconstrained by physical location, and they enable diverse groups to organise in different ways and around diverse principles even when they happen to be in the same physical territory.

None of this is to say that blockchain technologies can make atrocities and acts of genocide go away. As we saw, there is no stopping people from organising around perceived national identities, and there is likewise no stopping separate national identities from coming into conflict over physical territory. We can only show a path in which humans can organise themselves with indifference to physical territory but also in a way that allows people to exit from unfavourable governing situations.

## 8.5   Network states versus blockchain communities

So far in this chapter, we have taken issue with three key elements of Srinivasan's definition of 'network states', or as we would prefer to call them, 'cyberstates'. However, we find additional issues elsewhere in his definition. To be explicit about this, let us revisit his definition once more:

> A network state is a social network with a moral innovation, a sense of national consciousness, a recognized founder, a capacity for collective action, an in-person level of civility, an integrated cryptocurrency, a consensual government limited by a social smart contract, an archipelago of crowdfunded physical territories, a virtual capital, and an on-chain census that proves a large enough population, income, and real-estate footprint to attain a measure of diplomatic recognition.

In the previous three sections, we took issue with the need for a 'sense of national consciousness', the need for a 'recognized founder', the need for 'physical territories', and the need to 'attain a measure of diplomatic recognition'. Is there anything left here worth keeping? Well, it is not at all clear that governance requires 'an in-person level of civility', nor is it clear to us that 'moral innovation' should be part of the bargain – a blockchain community might be morally conservative or even reactionary. Even the need for an 'integrated crypto-currency' is not entirely clear, as a blockchain community could rely on entirely external cryptocurrencies like BTC or ETH.[18] Yet, there remain elements of Srinivasan's project that are worth salvaging.

Our principal disagreement with Srinivasan is with his focus on national identity and the notion of a state as an important part of the formula. As we have already indicated, national identity is a tricky business, and there is no reason it needs to be folded into blockchain governance. As for statehood, as we argued in Chapters 2 and 3, the very idea of a state is unravelling across the globe. While we do not see the need for things like civility and moral innovation in successful governance, we can agree with Srinivasan about one very important point: human governance should be implemented with blockchain technologies at its very core. As we said several times earlier in this book, those technologies will serve as the central nervous system of future human governance.

---

[18] Or more likely, restaked versions of these. The idea would be that sovereign communities could use restaked versions of BTC and ETH as their layer of economic security. Bad actors in the community would thus be putting these very valuable staked assets at risk. For an introduction to restaking, see <https://www.ledger.com/academy/what-is-ethereum-restaking>.

However, once we move from talking about states – or cyberstates and network states, for that matter – to talking about blockchain communities, new questions arise. How do we join these communities, and just as importantly, how do we leave them? And finally, what if a community insists that we leave? As we will see in the next chapter, these are not simple questions with easy answers.

# CHAPTER 9

# EXIT, EXILE AND ACCESS

## 9.1  Preliminaries

Many people can relate to the following experience: you do not feel at home somewhere, and you want to leave, but you cannot. The obstacle does not have to be a powerful king or some other form of tyrannical government, like the Soviet Union with its Iron Curtain and Berlin Wall. You might be a slave on a plantation facing a death sentence if you try to escape. Or it might be an economic obstacle. Perhaps you do not have the resources to exit your situation. Or maybe there is a cultural barrier to exit – for example, linguistic or religious barriers to movement. Such barriers or frictions come in different degrees. The issue is always whether it is worth the effort to escape. Often, when it is a matter of preservation of a culture or tradition or individual freedom, people will sacrifice everything to overcome these barriers. This being said, one wants to minimise such barriers where possible, as they are not just barriers to movement but to self-determination – they prevent us from moving to a jurisdiction with a political system that aligns with our values.

Exit is not the only situation to consider. There is also the issue of exile. Sometimes, people do not wish to leave, but they are forced to. There are also many examples of this phenomenon, and it spans across cultures. Human history is strewn with examples of forcibly removed people: Jews, Palestinians, Native Americans, Kurds, Armenians, not to mention Slavs and then Africans pressed into slavery; the list goes on and on.

One lesser-known example of exile would be the so-called Acadians, who were French-speaking people who lived in Acadia – what is now Eastern Canada's Maritime provinces, as well as parts of Quebec and parts of what is now the state of Maine in the United States. During the French and Indian War (known to Canadians as the Seven Years' War), British colonial officers suspected that Acadians were aligned with France and indeed found some Acadians fighting alongside French troops. The British, together with the help of New England legislators and militias, carried out Le Grand Dérangement (the Great Expulsion) of the Acadians between 1755 and the mid-1760s.[1] It was not pretty.

Most Acadians were deported to the British American colonies, where some were put into forced labour or servitude. Some Acadians were deported to England, some to the Caribbean and some to France. Some of the Acadians who were removed to France were then recruited by the Spanish government to migrate to Luisiana (today, Louisiana). These Acadians settled into the existing Louisiana Creole settlements, sometimes intermarrying with Creoles, and gradually developed what became known as Cajun culture.[2]

It may be that the British had some reason for moving the Acadians. They certainly were not loyal to the British Crown and some of them did fight alongside the French. However, the issue is that if you absolutely have to force someone out, there must be more humane ways of doing it. It is estimated that one-third of the Acadians perished during their resettlements. The event was traumatic.

---

[1] John Mack Faragher, *A Great and Noble Scheme: The Tragic Story of the Expulsion of the French Acadians from Their American Homeland*, 1st ed. (New York, NY, 2005).
[2] This is reflected in the phonological similarity between 'Acadian' and 'Cajun'.

That traumatic event was, of course, nothing compared to what the Native Americans endured. One famous example from the 1830s is the 'Trail of Tears', which was the forced relocation of Native Americans from the Southeast region of the United States. Among others, the displaced tribes included the Cherokee, Creek, Chickasaw, Choctaw and Seminole. They were moved west of the Mississippi River, and estimates based on both tribal and military records suggest that approximately 100,000 people were forced from their homes during the period and between 15,000 and 30,000 people died during the involuntary journey west.[3] The cost in lives is not the only reason it was called the Trail of Tears, of course. It was also traumatic in that people were forced out of their ancestral homes, and their communities and cultures were destroyed.

We have mentioned examples of exit and exile so far, but to set up our discussion in this chapter, we must introduce one more class of situations: those involving access. Sometimes, you find yourself on the outside looking in, and you want to petition to enter into some new community. Typically, the motivation might be economic, or it might be that you are fleeing political oppression where you now live, or it might simply be that you are already culturally aligned with the place you are petitioning. As we write, many countries are facing difficult decisions about accepting migrant populations that want to enter and live within their borders. Of course, this is not restricted to people petitioning to enter the European Union and the United States. Many Venezuelans are currently petitioning for entry into Colombia, Brazil and Ecuador. Other cases include the Rohingya refugees petitioning to enter Bangladesh; Sub-Saharan Africans to North Africa; Haitians to South America; Central Asians from Uzbekistan, Tajikistan and Kyrgyzstan to Russia; and Eritreans to Sudan and Ethiopia. The question, of course, is whether these people have a right to petition and enter, what the limits of those rights are, and at what point those requests become unreasonable. More specifically, we want to know how those rights extend to blockchain communities.

---

[3] Encyclopedia Britannica, 'Trail of Tears', *Britannica*, 2024 <https://www.britan nica.com/event/Trail-of-Tears> [accessed 29 October 2024].

Our goal in this chapter is to consider exit, exile and access from the perspective of blockchain communities. What rights and responsibilities do blockchain communities have when it comes to permitting exit, carrying out exile and granting access?

It seems logical that people should have the right to exit their blockchain communities, but how do people exit them? Under what conditions can they do so? For example, do people get to take their assets with them when they exit? What are your obligations when you belong to a blockchain community? Can you just up and leave, or should that be discouraged? Or for that matter, what if a blockchain community wants to kick you out? Should that be allowed? And if so, what are the rules for doing so? And what if there is a blockchain community you want to join, but it creates barriers to doing so? Is that allowed? Again, what are the rules?

These questions must be addressed upfront because they pose critical constraints on how we engineer blockchain communities but also on how we establish relations between them – ultimately, on how we build the ecosystem in which they will reside.

## 9.2  Exit

As we noted in this chapter's introduction, the bold promise of blockchain communities is that there can be an exit strategy for us if we find that the values of our blockchain community do not align with our own. However, exit is not entirely frictionless, nor is it clear that it should be. This will lead us to take a closer look at the barriers to exit – social, economic and so on – as well as at the conditions that facilitate exit.

Exit also implies that there is an alternative place for us to exit to, and this leads us to the question of onboarding: How easy should it be for people to be onboarded to new blockchain communities, or at least to create new blockchain communities that align with their goals? The journalist and philosopher Hannah Arendt was stateless for nineteen years after World War II. That sounds like an excessively long time. What is a reasonable obstacle to movement when we talk about blockchain communities? We will explore this question in depth in the following two sections.

A guiding philosophy that we endorse is the principle that there should be a plurality of governance mechanisms and that changes in governance should not require violent revolutions and wars and extended periods of political strife but should be as frictionless as possible. However, this idea requires some critical reflection.

To understand the subtleties of the exit problem, consider the position of someone who may have been a citizen of a single blockchain community for all or most of their life. Perhaps most of their friends and family are fellow citizens. Let us assume also that most of their business relations are with fellow citizens. They have paid taxes their entire working lives, and there is an understanding that the blockchain community will assist them when they retire. Clearly, there is a sense in which such a person *can* exit, but can they exit without being economically ruined?

The problem is particularly easy to see with regard to the benefits that may be owed the person after a lifetime of paying into what they supposed was their community's retirement programme. A community *can* allow a person to cash out on exit, but nothing we have said thus far guarantees this. Can blockchain communities withhold from retirees who exit the community? For that matter, can they prohibit persons who exit from taking their current assets with them? Suppose, for example, a blockchain community had its own currency, let us call it Community Coin (CMTY), and suppose a citizen who wished to exit had all of their assets tied up in CMTY. Even if the state allowed you to take your CMTY with you, would CMTY even have value outside the community?

Of course, golden handcuffs come into play independently of the question of asset transfer. Consider a person who is not close to retirement but who runs an online business providing computer programming services to members of the blockchain community. Certainly, the shared community relationships will have included business relations with fellow citizens. What happens when a person exits from a community because of a disagreement over policy or a shift in the moral compass of the community? Can the community block future collaborations between such a person and its members? Ideally, many of these rules and terms of agreement are present and are independently verifiable onchain prior to one's joining the community. However, while a blockchain community's legitimacy depends on the strength of its commitment to its stated principles and rules, communities do sometimes fail in their principles, and rules are often ignored and rewritten. What is the recourse for the community member who cannot abide such failures?

One might argue that the existence of golden handcuffs should have been obvious to someone who joins a blockchain community based on the transparent smart contracts grounding the community's governance. However, some people are born into their communities and do not make this choice. Other people may make choices under economic or political duress. Others still may choose to join a community and find that the governance shifts beneath them, negating their earlier decision.

Meanwhile, a blockchain community might argue that it is well within its rights to make exit difficult. It may insist that good governance requires a long-term commitment from its citizens and thus that people cannot be allowed to walk away freely. Perhaps they make the aforementioned golden handcuffs for their citizens explicit; you are told you can leave but that you cannot take your assets with you.

The first issue we need to address in response to these concerns is whether it is actually possible for a blockchain community to block the movement of assets. There can certainly be blockchain communities in which there are no golden handcuffs and from which assets can be extracted freely. Still, there is no guarantee that everyone will belong to such a community. It may be that for certain kinds of blockchain communities, such freedom to exit with assets is problematic. The question is, what kind of obstacles do these golden handcuffs represent?

If the citizen holds their wealth in bitcoin or some other globally used cryptocurrency, it is difficult, if not impossible, for a cyberstate to block movement, assuming the assets are self-custodied. If the assets are locked into a governance contract, then matters become more complicated. On entry to a community or while enjoying citizenship in a blockchain community, one might have to commit assets to a smart contract. In effect, the asset would be placed under the control of the community for some specified period of time. It might be that the community wants leverage against the citizen in case of bad behaviour, or it might be that the community needs the assets as collateral in seeking loans for financing community investment. One can imagine many possibilities here, each involving conditions in which a citizen has assets that are under the control of the community and which, hypothetically, could be withheld on exit.

As we noted, one does not have to enter blockchain communities that have such a policy, but some people may find themselves in such a position through no fault of their own. Do we need general guidelines against such contracts? It is definitely a topic for conversation. On the one hand, one expects people to avoid communities with such policies, but on the other hand, we all make mistakes and picking the wrong community with the wrong contract should not be a fatal decision. The penalty for selecting a bad community should not be a life sentence in that community.

Many blockchain communities will have their own tokens, some of which may serve as within-community currencies. They may also have governance tokens, which will work similarly to how DAO tokens work today; their primary function would be to provide a mechanism through which blockchain community members can vote on governance matters. Both kinds of tokens can be used to incentivise and reward behaviours in the community. Someone might earn those tokens in any number of ways during their time as a citizen of the blockchain community.

To keep this discussion as focused as possible, let us again suppose that a blockchain community has a governance token called Govern Coin (GC). Let us say that GC has no utility outside our hypothetical community. The token has plenty of value within the community but not much outside of it. Are there any options for the person to exit with those GC assets or something of equivalent value? It depends.

Under some circumstances, there might be a market within the community itself to exchange GC for BTC or ETH or some other asset. Alternatively, if people are hoping to enter the community and wish to have assets when they land, there might be an external market for GC.

Problems will arise, however, if GC utility is tied to a single individual using Sybil-resistant strategies of some sort (recall from Chapter 6 that 'Sybil resistant' in the context of token-based voting means that only one human person has access to and use of the token). For example, suppose that access to GC and its utility was tied to biometric data, or perhaps a private key was assigned to individuals so that only the key holder could ever access the tokens. Would this mean that the tokens will not be portable? Presumably, but there is a caveat.

If someone exits their community and another community, let us call it NewCommunity, wanted to recruit them, NewCommunity might consider the holdings of Community Coin to be evidence of a loyal community member and incentivise them with an equal value of New-Community Coin. However, in this case, it is not so much that the asset itself moved but rather that NewCommunity is paying incentives to individuals who showed evidence of excellent community involvement. There is, however, no guarantee that such payments will be offered. So, there are limits. Not all assets can be extracted from a blockchain community that is being tyrannical. There will always be scenarios in which the value one has created within a community cannot be fully extracted from that community.

If you find yourself in a particularly tyrannical blockchain community, either by choice or birth, it is entirely possible that the community will have economic levers to restrict your freedom of exit. However, this needs to be understood in contrast to the situation in traditional nation states when *they* become tyrannical. The former Soviet Union is a good case in point. Exiting the Iron Curtain (for example, by trying to cross the Berlin Wall) was difficult and often deadly business. In short, instituting real freedom of exit is not trivial since leaving a successful blockchain community requires leaving behind the benefits that its citizenship confers.

So far, we have been concerned with the movement of assets, and we were speaking of assets with monetary value, but of course, there are also cultural assets. Let us take up the issue of the cultural importance that a particular cyberstate might have. Earlier, we spoke in the abstract of the Lakota nation, but now, to illustrate our point, let us imagine that there was a blockchain community engineered around a shared Lakota culture. Let us imagine that our blockchain community was a platform designed to facilitate the sharing of Lakota history and language and which provided entry to real-world cultural events and ceremonies. Now, imagine a small group of Lakota youth, say in their twenties, who are very much interested in learning and nurturing Lakota culture, who viewed our hypothetical blockchain community as the ideal platform for doing so, and thus built their life around business and social contacts in the community. To what extent is exit feasible for such people?

Let us suppose that our Lakota friends very much enjoy their life in the Lakota blockchain community but cannot ethically abide by its current direction. They consider its policies unethical but are consistently outvoted on political matters. Let us suppose they become outcasts in their community and punitive actions are taken against them (socially punitive or economically punitive, it does not matter).

They begin to feel like second-class citizens. They feel oppressed. But can they really exit?

To be sure, there is a hypothetical sense in which our Lakota friends are free to leave and start a new Lakota blockchain community. But how easy is that really? Do they have the technical ability to do so? Can they get enough people to follow them so that the new community can achieve critical mass? These questions are important because one of the key selling points of blockchain communities has been that they provide ways of avoiding a minority group being kettled in a space with another group that does not share their attitudes.

Presumably (and hopefully), we are not looking at a situation like that faced by Immaculée Ilibagiza during the Rwandan genocide. It is not going to come to that. Hopefully, the adversaries of our Lakota friends would not be waiting outside with machetes. They might even be somewhere on the other side of the world. Still, it is clearly a case where diverse interests are kettled together. They are not being kettled together because they inhabit the same geographical location but because they share certain important cultural interests. Yet, at the very same time, important cultural differences exist.

Cases like this do not even require cyberstates organised around an identifiable culture. Our friends might have found themselves in a pluralistic cyberstate that, for some reason, tolerates sexism (or some other form of discrimination). Let us say that they lobby, unsuccessfully, for a prohibition on such behaviour. However, their values are otherwise aligned with those of the community, and their friends and business relations are all contained in their cyberstate. Is an exit for them frictionless? Or rather, is it frictionless *enough*?

The short answer is that the deployment of blockchain communities will be an imperfect technology if the goal is to eliminate frustration with being on the losing end of a political fight. However, eliminating all pain and frustration was never the goal. The goal was to prevent conflicting values from being kettled together, with passions rising to the point that wars and acts of genocide are the order of the day. The goal is to provide people an exit and a potential new home before extreme situations manifest.

Of course, for some individuals, it is not a question of exit. They get kicked out. We could call it involuntary exit, or we could call it exile. And the question becomes, what can we say about cases where individuals are exiled from blockchain communities?

## 9.3  Exile

Many people know of Hannah Arendt for her reporting on Adolf Eichmann's trial for war crimes and her observations about 'the banality of evil'.[4] Fewer people know about her personal history. In 1933, she was arrested in Berlin by the Gestapo and subsequently escaped to Paris. When France began to fall in World War II, she was placed in an internment camp named Gurs, located near the town of Pau in the French Pyrenees, from which she fled when she heard rumours that the camp residents would be turned over to the Gestapo. Following her escape, she biked, walked and hitchhiked to a friend's home near Toulouse. From there, she made her way to Marseilles, then Spain and then to Lisbon, where she found passage to New York City. From 1933 to 1951, when she finally became a US citizen, she was officially stateless.

It was thus not by accident that Hannah Arendt became concerned with the rights of stateless people – persons in exile. In 1946, she first published an essay, which was later reproduced in her book, *The Origins of Totalitarianism*. In it, she spoke of 'the right to have rights', by which she meant that if you do not belong to a community, you have no functional rights at all. If you are exiled and are stateless, it is as though you have fallen through the cracks and no longer live in a world of rights and norms. In her view, 'We become aware of the existence of a right to have rights (and that means to live in a framework where one is judged

---

[4] Hannah Arendt and Amos Elon, *Eichmann in Jerusalem: A Report on the Banality of Evil* (London, 2006).

by one's actions and opinions) and a right to belong to some kind of organized community'.

What is it like to be stateless and no longer belong to an organised community? Arendt describes the phenomenon this way: 'Once they had left their homeland they remained homeless, once they had left their state they became stateless; once they had been deprived of their human rights they were rightless, the scum of the earth.'[5]

Here, we get to the deep point that Arendt is making. The idea is that rights really only make sense in the context of a community, for only fellow community members can negotiate *de facto* rights with you. De facto rights are not something you simply have; they are something that we have to forge together. Here is how Arendt puts it:

> We are not born equal; we become equal as members of a group on the strength of our decision to guarantee ourselves mutually equal rights. Our political life rests on the assumption that we can produce equality through organization, because man can act and change and build a common world, together with his equals and only with his equals.[6]

If you are not part of a community, you no longer enjoy the conditions under which your equal rights can be forged. You still have a natural right, but that is the right to be part of a community where these rights can be forged.

We can dig even deeper here. Part of Arendt's point is that while you might have the right to free speech, if you are not part of an organised community, then you are basically just howling at the moon. You have speech, but you do not really have the 'right to opinion', by which she means that you do not have the kind of speech that matters in getting things done. 'Opinion', as Arendt uses the word, is speech that takes place in a context in which what you say can (potentially) have uptake in policy decisions. Similarly, in her view, you have the right to freedom when you are alone, but you do not have the 'right to action', which means that if you do not belong to some form of community, what you do does not have real consequences. So, in the big picture, many of our rights do not materialise, or put better, are not causally effective unless they are realised in the context of a community. Otherwise, they are inert.

In the world of nation states, the problem of the stateless and of exiles has become massive. In her book *Exile, Statelessness, and Migration*, Seyla Benhabib observed that at the end of 2016, there were estimated to be 65.6 million refugees worldwide.[7] Or to put it in other terms, one in every 113 people in the world is some form of refugee – they are stateless, outside the realm of rights. Now, to be sure, not all of these people were exiled by the state in which they lived. Some were displaced by war or other factors (and many were born into these conditions), but the core issue is that these are all individuals who have become untethered to the institutions in which meaningful rights are forged. The terminology for these people varies. They may be referred to as 'refugees', 'asylees', 'internally displaced persons' or 'stateless persons'. Whatever name we give to the displaced, their situation is dire.

Part of the promise of blockchain communities is that they can provide safe landing spots for displaced people. Specifically, they can provide communities in which people can express opinions and act in ways that have consequences. Refugees should not have to wait for new nation states to be forged for them or for old nation states to accept them; they should be able to form blockchain communities even when they are without territorial homes. They should be able to form their own blockchain communities as soon as they are needed.

---

[7] Seyla Benhabib, *Exile, Statelessness, and Migration: Playing Chess with History from Hannah Arendt to Isaiah Berlin* (Princeton, NJ, 2018).

Ideally, one can find asylum without bouncing from country to country and then living in legal limbo for decades. However, that is the ideal situation, and we need to be clear that if states are at liberty to exile members, we must ensure that everyone has a plausible landing spot and that the landing spot comes without delay. Among other things, this means that the exile should be fair.

So, what should happen when blockchain communities wish to exile citizens who are bad actors or are perceived as not contributing? Clearly, being banned from a blockchain community would not be as onerous as what happened to Arendt and millions of other refugees from Nazi Germany, or as what is still happening to Palestinians today, or as onerous as what happened to the Native Americans or the Acadians. However, there are still very real costs to being exiled from a blockchain community.

At first glance, it seemed reasonable that a blockchain community should be free to force out whomever it wishes, subject to its own principles. However, the matter is fundamentally just as problematic as a blocked exit. It is one thing to ask an exiled citizen to surrender their governance tokens, but what if the request is to no longer do business with their partners in the community? And what if those business partners are critical of the exiled person's livelihood? It may be that the global community will expect that banned citizens receive a suitable compensation package on forced exit, even if the banning was in accord with the laws and principles of the blockchain community. A banning, after all, has consequences outside the virtual and physical borders of the blockchain community.

It seems we must find a path that allows blockchain communities to exile individuals, but that is fair to the exiled individuals – we need to protect their 'right to have rights'. We have said that blockchain communities should be free to conduct their own affairs, but that freedom only makes sense if exiled people have the ability to land on their feet. There are certain universal principles that blockchain communities must respect if their very idea is to be workable. And having individuals and minority communities cast out without functional resources is not a workable solution. It is the sort of failure that might be tolerated in the age of nation states but which is not an option in the age of block-chain communities. So, what can be done? What kinds of safety nets must be made available to the exiled?

The fundamental issue is that if we are to allow exile, we must allow frictionless access to a new home. It took Hannah Arendt decades to acquire her US citizenship. Refugees from Palestine and elsewhere have had to wait even longer to find homes. Indeed, as Ben-habib points out, there are 'temporary' Palestinian internment camps in Lebanon that have been in operation since 1948. The Dadaab refu-gee camp in Kenya, with 420,000 refugees, has been in operation for twenty years.[8] In view of the loss of rights of the stateless persons living in these places, this is an embarrassment to all of humanity. It is now time to do better. But how?

## 9.4  Access

We concluded the previous section by saying that exiled people should have the right to gain access to a blockchain community (and thereby the 'right to have rights'). Should they, therefore, have the right to access *any* blockchain community? Do you get to choose your landing spot? Or do you merely have the right to *petition* for entrance? But what kind of a hearing would this entail? Or do you merely have the right to form a new blockchain community with others?

There are several issues to address here. First, does the right to access entail the right to start-up resources? And if so, who is responsi-ble for providing those start-up resources? The community from which the refugee is exiled? The community in which the refugee lands? Other blockchain communities? Next, the question involves whether you have the right to enter *any* community or merely some communities. If you can only feely enter *some* communities, *which* communities? Finally, if there is a right to petition, what does that entail? What counts as a serious consideration of your petition?

---

[8] Ibid

Should people be free to join any blockchain communities they wish? Presumably not; there are lots of good reasons to restrict access, not least of which is the problem that blockchain communities built around minority cultures and values could get swamped by bad actors joining from a majority culture with the express goal of undermining the community. Beyond that, it seems reasonable for communities to control conditions for entrance.

There are, of course, concerns here familiar from our world of exclusive clubs and neighbourhoods that block entry from undesired groups. There is a widely shared attitude that exclusion from minority organisations is permissible, but exclusion from majority organisations – typically, from access to power – is problematic. The reflex of this problem in the case of blockchain communities is that we are inclined to say that our hypothetical Lakota blockchain community is within its rights to restrict entry to tribal members but that a powerful majority blockchain community should not be so restrictive. Is this attitude well placed?

It certainly makes sense to allow blockchain communities to be organised around ethnic groups, and it certainly makes sense that blockchain communities should not be in the business of freezing people out of power, but clearly, these two ideas come into conflict, just as they do in traditional nation states. A blockchain community organised around European culture could hardly avoid being a blockchain community organised around wealth and power in today's world.

We can offer some preliminary reflections on how this tension might be resolved. For example, we can distinguish between blockchain communities that are organised around culture and blockchain communities that are organised around economic opportunities. Individuals could belong to both, just as we have dual citizenships today. For example, our Lakota friend could belong to a blockchain community organised by Lakota and a separate blockchain community organised around economic concerns. The key idea would be that economic communities must be inclusive, and culture-based communities may be exclusive.

It is not even obvious that this would have to be enforced as an intercommunity principle. Assuming a plurality of economic communities, it is difficult to imagine that any of them, much less all of them, could flourish by being restrictive. On the other hand, it may be that some cultural communities need to address the economic welfare of their members. So, one may want to allow that cultural communities can provide economic assistance up to a certain economic threshold.

These are all questions that will remain open, but for now, we can close off our discussion by noting that blockchain community exit, exile and access, while certainly not free of frictions and certainly not free of hard decisions, will at least be less painful and violent than what we are used to in the age of nation states.

However, once we have reduced the frictions of exit and access, and once we have minimised the hardship of exile, we find ourselves confronted with new questions. If humans are organised around communities, and communities may overlap, and if the movement of citizens between communities is fluid, what will governmental sovereignty look like in this picture? Will there even be a unified notion of sovereignty? Or will this lead to many new kinds of overlapping sovereignties? These are the questions we turn to in the next chapter.

# RETHINKING SOVEREIGNTY

## 10.1 Preliminaries

In Chapter 8, we laid out why we think that the very notions of statehood and nationhood must be abandoned, along with that of the nation state itself. They are relics of the Westphalian era. As we explained, there is much that we agree with in Balaji Srinivasan's idea of the network state. However, we concluded that the best parts of his proposal can be captured by the idea of a blockchain-organised community.

To be sure, there are some elements to the idea of the network state that seem to simplify things for us. For example, if network states hold physical territory and are recognised as equals of traditional states, then they should inherit the territorial sovereignty of traditional nation states. However, this simplification is an illusion. There is nothing gained by trying to cling to traditional notions of nation-state sovereignty. If we rethink governance for the post-Westphalian order, we have to rethink sovereignty as well, and while there will be a notion of sovereignty moving forward, it will look nothing like the kind of sovereignty that nation states exercise today. In this chapter, we will try to understand why.

## 10.2   Post-state sovereignty

There are many aspects of sovereignty, but the aspect that is most salient in the public imagination is that of territorial control. However, even today, when we talk about territorial control, there is no uniform notion of control. A nation state might have well-defined borders, but it typically cedes much of the control over its territory to regional governments (for example, provincial governments), local governments, individuals and organisations. The nation state might have the last word, but only within well-prescribed limits.

So, sovereignty is something of a shared project. There is no single sovereign that controls everything – contrary to Hobbes' perspective in *Leviathan*.[1] The philosopher Gottfried Wilhelm Leibniz (whom we discussed in Chapter 5, in our discussion of archives, and again in Chapter 6, in our discussion of smart contracts) had ideas about this subject, too. Living in the age shortly after the Peace of Westphalia, he envisioned a picture of sovereignty quite different from that of Hobbes.[2] His idea was that there is not one kind of sovereignty but that there are many overlapping varieties tasked with control over different things.

In an article titled 'Social Sovereignty', Robert Latham points out that nation states are not natural homes for sovereign control and that it was a long, slow process for them to acquire the sovereignty that they did:

> The strengthening of claims over peoples and places by kings
> and the states they built – a process I label *inclosure* – took centuries and unfolded quite unevenly and in different forms. European states, for example, did not easily constitute national economies and most internal economic relations and practices in Europe were controlled by local authorities before the 19th century, despite early efforts to regulate foreign trade and to establish kingdom-wide coinage.[3]

---

[1] Thomas Hobbes, *Leviathan or The Matter, Forme and Power of a Common Wealth Ecclesiasticall and Civil* (Toronto, ON, 2016).
[2] For discussion, see Pinheiro, 'Leibniz on the Concepts of Archive, Memory, and Sovereignty', 309–21; William F. Drischler, *Leibniz Contra Westphalia: Conceptual Underpinnings of Globalized Lax Sovereignty* (2015); and Janneke Nijman, 'Leibniz's Theory of Relative Sovereignty and International Legal Personality: Justice and Stability Or the Last Great Defence of the Holy Roman Empire' (New York University School of Law, 2004) <https://iilj.org/wp-content/uploads/2016/08/Nijman-Leibniz%E2%80%99s-Theory-of-Relative-Sovereignty-and-International-Legal-Personality-2004-2.pdf> [accessed 11 November 2024].

If we keep pressing on this idea of different sovereignties, then we may ask what makes something sovereign control as opposed to just plain control, and the answer has to be that there is no real interesting difference. Sovereign control is simply control that lies in the hands of a governing body or system of relations and which cannot easily be revoked. States that exist within a federation of states (for example, states within the United States of America) will have sovereign control over some things, and due to customs or constitutional orders, it is not easy to revoke that control. Sovereignty thus extends down to smaller and smaller levels of government and all the way down to individuals, who enjoy (or at least should enjoy) some forms of self-sovereignty.

We often assume that physical territory must be under the control of some sovereign nation state, but there is really no reason for this to be so. Any level of human governance, down to the level of self-governance, has elements of sovereign control, and there is absolutely no reason why this dynamic conception of sovereign control would not apply to the control of physical territory. Specific entities might have sovereign control over military action or taxation or religion or culture or architecture or commerce or any combination of the above. Sovereignty is divided. It is not tied to nation states.

Latham argues that the complexity of sovereignty goes far deeper than questions of how sovereignty is divided between governmental units. In his view, it is a mistake to think that sovereignty is restricted

---

to agents of government at any specific level. Sovereignty can also (and often does) reside in social relations. In this vision, sovereignty is not controlled by some identifiable decision maker or governmental body but rather by the social relations of the community. In fact, Latham observes that this form of social sovereignty must precede territorial sovereignty, as territorial sovereignty is knit from a tapestry of preexisting social relations:

> In grasping the implications of that rise for changing understandings of sovereignty, we should first question the close association of sovereignty with territory. State-based territoriality emerged only after states, for centuries, deployed forces in the organisation of judicial, administrative, constabulary and military realms. Political territory formed out of social and political spaces that became increasingly bounded by royal and feudal claims and rights, language, and systems of economic production and distribution. For example, by the time the 13th century came to England, an emerging English state was deploying courts, officials, taxes, codes, records and symbols.[4]

All of this means that when we talk about cyberstates or any blockchain community having sovereign control over either physical or non-physical territory, we are being necessarily vague about the form of sovereignty we are talking about, and we are being future-directed. Sovereign control can be maintained at every level of human and individual governance and in all manner of social relations. Sovereign control of physical territory need not be the natural prerogative of nation states nor any other kind of state nor even of governing institutions at all. On the contrary, states may well choose to be indifferent to whomever has sovereign control over physical territory, whatever that ends up meaning at the end of the day.

---

[4] Ibid.

In contemplating what lies ahead for our future and that of the nation state, numerous scenarios have been proposed. A National Intelligence Council (NIC) report offers one potential scenario in which 'The nation-state does not disappear, but countries increasingly organize and orchestrate "hybrid" coalitions of state and non-state actors which shift depending on the issue.' It foresees an 'increasing designation of special economic and political zones within countries'.[5] In this scenario, while nation states do not vanish into obsolescence, their role and function evolve significantly. Greater emphasis would be placed on creating special economic and political zones within countries – areas distinguished by their unique regulatory frameworks designed to foster economic growth and innovation.[6]

The NIC-report scenario has parallels to 'neomedievalism' in political theory, first introduced by Hedley Bull in his 1977 work *The Anarchical Society: A Study of Order in World Politics* as a means to describe the diminishing authority and control of nation states in the context of an increasingly globalised society. To Bull, this system would 'avoid the classic dangers of the system of sovereign states by a structure of overlapping structures and cross-cutting loyalties that hold all peoples together in a universal society while at the same time avoiding the concentration inherent in a world government.' He elaborates on the consequences of such a view:

> It is also conceivable that sovereign states might disappear and be replaced not by a world government but by a modern and

---

[5] National Intelligence Council, *Global Trends 2030: Alternative Worlds* (Washington, D.C., 2012).

[6] Thinkers such as Tom W. Bell (*Your Next Government?*), Patri Friedman ('Dynamic Geography: A Blueprint for Efficient Government' <https://patrifriedman.com/old_writing/dynamic_geography.html> and <https://www.seasteading.org/>, Hans-Adam II (*The State in the Third Millennium* [London, 2009]) and Balaji Srinivasan (*The Network State*) have investigated these ideas further, viewing future governments akin to service providers. They propose that more efficient institutions can be discovered through competitive governance in small territorial enclaves – using special economic zones to effectively establish a 'startup sector' for governance.

secular equivalent of the kind of universal political organisa-
tion that existed in Western Christendom in the Middle Ages.
In that system, no ruler or state was sovereign in the sense of
being supreme over a given territory and a given segment of the
Christian population; each had to share authority with vassals
beneath, and with the Pope and (in Germany and Italy) the
Holy Roman Emperor above. The universal political order of
Western Christendom represents an alternative to the system
of states which does not yet embody universal government.[7]

In his 1999 book *Legal Rules and International Society*, Anthony
Clark Arend claims that Bull's identified trends had become even more
pronounced by the end of the twentieth century. Arend argues that the
emergence of a 'neo-medieval' system would have profound implica-
tions for the creation and operation of international law.[8]

Aside from the forces of globalism, mass migration and multicul-
turalism increasingly undermine nation state sovereignty and interstate
forms of governance. Migration and cultural pluralism lead to dynamic
cultural values within a national territory, with the result that such val-
ues no longer neatly correspond to nation state boundaries. Stephen J.
Kobrin, in his 1998 paper 'Back to the Future: Neomedievalism and
the Postmodern Digital World Economy', argued that the sovereign
state as we know it – defined within certain territorial borders – was
about to change profoundly, if not to wither away, due in part to the
digital world economy created by the Internet. Kobrin saw cyberspace
as a trans-territorial domain operating outside of the jurisdiction of
national law.[9]

---

[7] Hedley Bull, *The Anarchical Society: A Study of Order in World Politics*, 3rd ed. (Basingstoke, 2002).

[8] Anthony C. Arend, *Legal Rules and International Society* (New York, NY, 1999).

[9] Stephen J. Kobrin, 'Back to the Future: Neomedievalism and the Postmodern Digital World Economy', *Journal of International Affairs*, 51/2 (1998), 361–86 <http://www.jstor.org/stable/24357500>.

## 10.3   Formalising a new vision of sovereignty

We have argued that the idea of sovereignty is coming under pressure and that this is due to a number of factors, including the fact that the very distinction between state and non-state actors will be porous in the future we envision. In fact, we believe it is already porous. What, fundamentally, is the difference between an official state and some other organisation? It certainly is not the population or the size of its economy. We ultimately need a new way of formalising this new idea of sovereignty.

Monaco, which has been an independent state since 25 February 1489, has a 2.02-square-kilometre footprint, with around 39,000 total residents.[10] The Republic of Nauru covers an area of twenty-one square miles and has a population of approximately 10,000 inhabitants.[11] Tuvalu covers an area of twenty-six square kilometres and has a population of just over 10,500 people.[12] These states dwarf Vatican City, which is officially a nation state, although it does not have a seat in the UN. It has a territorial footprint of 0.44 square kilometres[13] and is home to between 700 and 800 people of which only 618 are recognised as Vatican citizens.[14]

You might argue that while these states are tiny, they have a lot of wealth. However, this is not true for them all (for example, Tuvalu), and it is worth emphasising that there are even very large nation states with small economies. Indeed, there are nation states that have economies that are smaller than those of online video games.

---

10 World Health Organization, 'Monaco - Statistical Data', *European Health Information Gateway* <https://gateway.euro.who.int/en/country-profiles/monaco/> [accessed 5 December 2024].
11 Central Intelligence Agency, 'Nauru', *The World Factbook*, 2024 <https://www.cia.gov/the-world-factbook/countries/nauru/> [accessed 5 December 2024].
12 Department of Economic and Social Affairs, *2023 Demographic Yearbook* (2024) <https://unstats.un.org/unsd/demographicsocial/products/dyb/dybsets/2023.pdf> [accessed 5 December 2024].
13 Encyclopedia Britannica, 'Vatican City', *Britannica*, 2024 <https://www.britannica.com/place/Vatican-City> [accessed 5 December 2024].
14 Stato della Città del Vaticano, 'Popolazione', *Vatican State*, 2018 <http://www.vaticanstate.va/it/stato-governo/note-generali/popolazione.html> [accessed 5 December 2024].

As early as 2002, Ted Castronova argued that the then-popular video game *Ultima Online* had a larger economy than some nation states, and if we looked at the per capita gross national product of *Ultima*, we would find that it held up quite well against established nation states. As Castronova noted, 'The nominal hourly wage is about $3.42 per hour, and the labors of the people produce a GNP per capita somewhere between that of Russia and Bulgaria.'[15] More interestingly, and setting aside per capita GDP, there are nation states which have a lower overall GDP than some online virtual worlds. Back in 2002, Castronova pointed out that Liberia's entire GDP was smaller than that of *Ultima Online*. Who knows how it would compare with subsequent, much larger virtual worlds like *Fortnite*?

All of this prompts us to ask: Might there not be a successor virtual world with perfectly legitimate claims to being a peer to a traditional nation state? The discriminating criterion cannot be the size of an economy or population or a significant land footprint.

There are already special economic zones (SEZs) larger than some nation states. Some of them have more people, and some have greater wealth. King Abdullah Economic City, an SEZ in Saudi Arabia, is supposed to eventually house two million people in an area the size of Washington, D.C. If that happens, both its population and landmass will dwarf many nation states. All of this prompts the question: What is the difference between an SEZ and a state? Could the answer turn on who has sovereign control over the territory? It seems not. The notion of sovereign control over territory is being deconstructed before our eyes. Governments already cede some sovereignty to SEZs, and it should not go unremarked that they often cede control of territory to United States military bases, which among other things, bring their own policing and judicial mechanisms. Our point is that just as the traditional notion of a nation state cannot be grounded in its land footprint or population or wealth, it cannot be grounded in sovereign control, because sovereignty itself is a very complex and dynamic concept – one that deserves some reflection.

---

[15] Edward Castronova, *Synthetic Worlds: The Business and Culture of Online Games* (Chicago, 2006).

Leibniz, as we noted in the opening section of this chapter, argued that there is no single unified notion of sovereignty. He argued that matters are not as simple as Hobbes imagined in *Leviathan*, with a single sovereign ruler holding control over everything. Rather, different kinds of sovereignties overlap each other. His thought was that the Church might hold some sovereignty, the Habsburg Empire a different kind of sovereignty and principalities yet another. Let us dive deeper into the idea of mixed sovereignties.

We first need to see that many variables are in play here. It is not enough to say that a state or community has sovereignty over some territory. It may have sovereignty over one kind of activity but not another. And that sovereignty could be limited by time (perhaps it does not have taxation authority during regularly occurring tax holidays, for example). And for that matter, it could be limited by what we could call a context – here, taking context to be a certain set of circumstances. For example, one might have the authority to raise a militia only during times of conflict or to issue taxes only in times of economic growth. Contexts, so understood, would differ from times in that the sovereignty is not on a regular schedule but is contingent on certain conditions holding. It might also be limited to a certain domain, which could be either physical territory or network territory. And even when this is narrowly circumscribed, the sovereignty might only extend to a particular group of people.

Our goal here is not to provide the most general formalisation of sovereignty but rather to develop one that is relevant for communities organised with the help of blockchain technologies. As we will see, even with this circumscribed concept of sovereignty, there are many variables at play. For example, a community might have sovereignty over an activity like taxation for a group of people that are citizens, but only involving certain activities (say retail sales), in a domain that is defined by the network and only in certain contexts (say when the activities take place during times of economic growth). A physically overlapping community might have sovereignty over different persons or activities (visitors and gambling, let us say, and during wartime). You can be as fine-grained as you wish in carving up the space of sovereignty. One community might have sovereignty over public health issues, another over territory and another over public welfare, but all are indexed to specific domains and contexts, persons, and activities.

As a step to formalising this, let us say that a community *c* is sovereign over activity *a*, for a group of persons *g*, in the domain of activity *d*, in context *x*. So, sovereignty is a five-place relation like this:

$$\textbf{Sovereign } (\textbf{\textit{c, a, g, d, x}})$$

This formalism, in and of itself, fails to solve problems of conflicting sovereignties. Still, it is a tool that can help us identify points of conflict and paths to avoiding conflict by dividing sovereignties appropriately. It is a tool to help us visualise the ways in which we can set the relevant parameters of sovereignty.

As a next step in understanding sovereignty, let us distinguish *de facto* sovereignty from *de jure* sovereignty.[16]

A community c has de facto sovereignty over activity a, for a group of people g, in domain d and context c, just in case the community has a largely unchallenged ability to exercise control of activity *a*, for a group of persons *g* and so on.

_____________

[16] 'De jure sovereignty' refers to an authority to govern that is morally well grounded, but may not exist in reality, while 'de facto sovereignty' refers to an authority to govern that exists in reality but may or may not be morally well grounded. We return to the concepts of de jure and de facto in Chapter 11 in the context of the rights of responsibilities of blockchain communities. For further elaboration, see John Tasioulas and Guglielmo Verdirame, 'Philosophy of International Law', *The Stanford Encyclopedia of Philosophy*, 2024 <https://plato.stanford.edu/archives/sum2024/entries/international-law/> [accessed 29 October 2024].

A community *c* has de jure sovereignty over activity *a*, for group of persons *g*, in domain *d* and context *c*, just in case the community has either moral or legal justification for uniquely being in a position to exercise unchallenged control of activity *a*, for group of persons *g* and so on.

Given these definitions of sovereignty, it should be clear that there are an unlimited number of ways to divide up sovereignty and unlimited ways in which sovereignties might intermesh with each other. In the simplest case, there is just one Hobbesian authority that has sovereign control over everything. However, as sovereignty can be divided, and different kinds of domains and activities and groups of individuals can overlap in unlimited ways, there are likewise unlimited possible interwoven sovereignties.

This leads us back to the question of nation states. Nation states do not all exercise the same kinds of sovereignties; as we have seen, they can and do cede control over some activities and some territories and some individuals. While we might think that physical territory is a very special domain, it is certainly not the only way to fill the domain variable. One might, for example, identify the domain with a network or a membership in a blockchain community.

What all of this means, from a theoretical perspective and we believe from a practical perspective, is that no single form of sovereignty is special. There are just different ways to set the parameters. If this is right, then perhaps the idea of a nation state is not incoherent, as it is just one of many ways to set the input variables for sovereignty. However, the bigger question then becomes this: Since different communities will have different sovereignties, is there any point in setting a few hundred of those communities apart as being somehow special? A better outcome, we think, is to set aside the idea of states altogether and think in terms of communities, each with different kinds of interests and different forms of sovereign control. Whether they organise with each other will depend on their interests.

## 10.4   Territorial sovereignty revisited

At many points in this book, we have discussed territorial sovereignty, which on the one hand, could indicate a sovereign that has control over everything within a physical territory but, on the other hand, could indicate very limited sovereignty. We also saw that there may be many communities within a physical territory, each with different kinds of sovereign control. All of this raises the question of what happens when we get to the nitty gritty of people on the ground within the same physical territory.

This brings us back to cases like the Rwandan genocide and, as we write this, the conflict between Israel and Hamas within Gaza. What happens when people of conflicting values occupy the same physical footprint? Is conflict not inevitable? And are nation states not required to ameliorate these conflicts?

However, these questions have things backwards. Perhaps it is our fetish for states and state control and state territorial sovereignty that gives rise to these conflicts and to the human rights abuses that extrude from them. That certainly appears to be the case in the Rwandan genocide, which stemmed from having diverse tribes kettled together within terrestrial borders established by nation states attempting to install the Westphalian order where it had no business being installed.

But how can this possibly work? If people belong to different networks or blockchain communities, they still must come into contact with each other in places like Palestine, so do we not have the same old problems? But upon closer inspection, we really do not have the same problems.

It is true enough that people of diverse values and religious and economic principles will, at times, inhabit the same physical space, but this only becomes a problem if we imagine that control of physical territory gives one warrant to control the values and principles of the individuals who live and do business there. To illustrate this idea, consider the case of an international hotel in a large city. As you pass through the lobby, you will doubtless see many people with diverse values, religious beliefs, economic ideologies and so on. However, it is exceedingly rare to see conflicts between such groups break out in these spaces because it is not the business of international hotels to impose values or religious beliefs or economic ideologies on their guests. Guests are welcome to stay regardless of their beliefs as long as they pay their bills and do not create trouble for other guests. Each guest must still abide by the principles of their governments and pay their taxes and so forth, but that is no business of the hotel.[17]

Hotels are not the only places where the terrestrial authority is indifferent to the values and beliefs of the people passing through. Airports, for example, ordinarily do not care about your values or the income tax wherever you come from.[18] They just expect you to pass through without causing trouble for others. Such authorities, whether they be hotels or airports, are not ordinarily considered to be tyrants for the simple reason that their portfolio of demands is very limited: whatever your values, take care of your business and do not create problems for us or other guests.

But now, we want to ask why it is unthinkable to imagine that terrestrial authorities – that is, authorities in charge of maintaining order within a physical location – could have limited portfolios. What if they were not in charge of enforcing religious beliefs or moral codes or raising taxes to provide welfare for others or raising armies? What if they were just in the business of maintaining only enough physical order so that overlapping online community members within the same space can conduct their business?

---

[17] We are aware, of course, that hotels have a long history of discrimination and bad acting. Here, we are speaking more of an idealisation of a fair-dealing hotel.

[18] Again, we note that this is an idealisation – in this case, of how an airport could be.

This vision requires that sovereignty over physical territory – the role of keeping order there – is not the province of any single state but is rather state-indifferent. The parties in charge of maintaining order within a physical space could be recorded on a decentralised global registry, with no one power having authority over the registry. Once there is a decentralised record of territorial control, we can entertain new ideas about the nature of the control itself. A global community might well accept the presence of local policing for reasons of safety but reject attempts to control cultural and ethical values within the territory.

And here, finally, we get to the deep point. Blockchain technologies point the way to strategies for completely decentralised-yet-cooperative ways of organising ourselves, and this can apply to more than just economic cooperation through protocols like Bitcoin and Ethereum. It can apply to the very idea of property control. This leads us to the topic of decentralised property registries.

## 10.5  Decentralised property registries

Throughout this book, we have been talking about how governmental sovereignty is fragmenting. One lesson is that fragmentation by itself does not solve the problems we face. However, it might also be that, for certain purposes, fragmentation works against us. Sometimes, solutions must be global in scale. An obvious case in point would be cryptocurrencies like BTC and ETH, which are, without a doubt, global currencies. They are not designed to be used only in certain countries or certain communities. The Bitcoin and Ethereum protocols are both global in their reach.

Often, when people talk about 'globalisation', they use the term to mean something negative since many forms of globalisation also come with centralisation. However, if something can be global yet decentralised, like BTC and ETH, then that is fundamentally a win because it avoids introducing a central global authority. Indeed, for certain purposes, like sending money to friends around the world, we definitely want the protocol to be global. It would be senseless for all cryptocurrencies to be tied to small individual communities or specific terrestrial locations. Even if there were markets to exchange our individual local currencies, those markets would have to be global in their scope. Fundamentally, we need some applications – like payment applications – to be global yet still decentralised. Sharding, as described in Chapter 3, is counterproductive for such applications.

What other applications are possible for such global, non-sharded technology? One application would be in the role of negotiating trade agreements and, for that matter, any sort of international agreement. The whole idea of a negotiation is that there are parties with competing, if not conflicting, interests. To balance these interests and ensure affected parties have a voice in the proceedings, they must form a kind of umbrella community, inclusive of the parties in conflict, in order to negotiate the terms of the agreement. The idea is that negotiations are easier to accomplish when records of agreements and the goals thereof are transparent to all interested parties and, indeed, are not under the exclusive control of one party to the negotiations. Centralisation is an obstacle to agreements between parties with distinct interests and goals.

In this section, we want to look at another candidate for something that could benefit from being global yet decentralised: property registries. That might initially sound like a crazy idea. After all, we are used to property records being kept locally – with a title company, for example, or as in Mexico several years ago, with a notaría. We have already covered the difficulties with those sorts of property registries. They are points of failure and corruption, and in many locations around the world, they do not provide security for property owners. We have touched on this topic several times, but it is an important one, and it is time for us to put together a positive proposal for how property might be registered on a global blockchain. First, however, let us revisit the horror shows that constitute the systems of property registration existing around the world today.

Transparency International, a nongovernmental organisation (NGO) that studies corruption in land administration services, reports that globally, one in five people has had to pay bribes to land administration officials to maintain their property rights. In Africa, the reported number is one in two,[19] and some reports suggest that the number of property owners that must pay bribes in Africa is closer to 100%.[20]

In 2019, the African Development Bank met to discuss this problem. In that meeting, Ivory Coast Minister of Justice Sansan Kambile observed, 'Without land tenure security, and the various implications, no development can be sustainable.' He was not wrong about this. In many countries, the principal economic opportunities are agricultural; if land ownership and other elements of land administration are not stable, then there is no stable foundation for economic development. In this vein, Josefa Leonel Correia Sacko, Commissioner for Agriculture and Rural Economy at the African Union Commission, argued that corruption in the land sector was fueling military conflicts across the African continent.[21]

We can see that land administration services are critically important even if not always performing successfully, but what are they exactly? Well, they involve a lot of things, but we can mention some key activities that fall under such services.

Obviously, land administration involves land registration – i.e. creating an official record of land ownership and rights, like a title company would do in the United States. Then there are land tenure regularisation programmes, which aim to clarify and formalise land rights, particularly in areas with customary or informal land tenure systems and where use planning involves the systematic allocation and regulation of land for different purposes, such as residential, commercial, agricultural or conservation. For example, when we discussed Disney's Reedy Creek District, we noted that the district had certain rights, which were subsequently taken away under Governor DeSantis (for example, the right to build a nuclear power plant). Of course, another important administration service is land valuation and taxation. Land valuation is the process of determining the market or assessing the value of land for various purposes, including taxation, compensation or land acquisition.

---

[19] Transparency International, 'The Impact of Land Corruption on Women: Insights from Africa', *Transparency.org*, 2018 <https://www.transparency.org/en/news/impact-of-land-corruption-on-women-insights-from-africa> [accessed 7 September 2023].
[20] African Development Bank Group, '2019 Conference on Land Policy in Africa: Technology and Innovation Will Help Speed up Removal of Land Sector Corruption in Africa' (2019) <https://www.afdb.org/en/news-and-events/press-releases/2019-conference-land-policy-africa-technology-and-innovation-will-help-speed-removal-land-sector-corruption-africa-african-development-bank-32901> [accessed 7 September 2023].
[21] Ibid.

Land administration services also often include mechanisms for resolving land disputes and conflicts. This can involve formal judicial processes, customary dispute resolution systems or alternative dispute resolution methods to address conflicts over land rights, boundaries or access. There are obvious elements to this, such as disputes over property boundaries, but also less obvious elements, such as water rights. Such services were nowhere to be found when the haciendas seized land from Mexican peasants, nor when Zapata's men retaliated by seizing the haciendas during the Mexican Revolution, highlighting that these disputes, if not equitably resolved, can lead to armed conflict. Finally, land information systems encompass the collection, management and dissemination of land-related data. These systems store and analyse information on land parcels, ownership, transactions and other relevant information to support decision making and provide public access to land information.

Blockchain-based archives can be useful to all of these elements of land administration. Land registration should be obvious. Land tenure regularisation also can be put onchain, where it is visible to all community members or it might be part of a database with an AI frontend that citizens could query regarding land policies. Then there is the matter of land use planning, in which it is critical that everyone be on the same page as to what the plans are. There might also be secure records of proceedings in the planning stage, which itself could take place in the context of a DAO. Land valuation and taxation could actually be partly automated through the use of smart contracts. Once there is a sale of property in an area, the property values of nearby land could be recomputed and published. Again, the process should be transparent. Land dispute resolution, in some parts of the world, could definitely benefit from a public record of how previous disputes were resolved, thus establishing precedent. However, it might also be possible to automate some of these decisions with AI. As for the land information systems, in our vision, they would take the form of a blockchain archive drawing information from decentralised oracles.

There are many things that blockchain technologies could contribute here, but there is just one problem. How on Earth do we get people to actually apply these technologies and trust them? In other words, we are back to the bootstrapping problem. Here is a simple way to put the point: it is one thing to say that we have a record of property boundaries and a ledger that keeps a record of who owns what, but what good is that record and ledger if terrestrial governments refuse to recognise their veracity? What if legacy governance structures decide to stick with their old ways? Once the technology is adopted, everything is great, but how do we get it to be adopted?

We can add a layer of complexity to the problem by pointing out that it would be ideal if such registries of information were not merely local and not merely national, but global. Just like BTC and ETH have become global currencies, there is a strong case to be made that we want land administration services – or at least the record-keeping part – to be global as well. Note again that when we say global, we do not mean to say that they should be under some global central authority; we rather mean to say that they should be globally decentralised like Bitcoin. There would be no single repository of information, but rather, it would be shared, verifiable information that is accessible to all via the blockchain.

There are really two problems at this point. The first problem is how do we get governments around the world to make use of this recordkeeping technology? This can only be accomplished by explaining the technology, what it does and how it can solve a lot of headaches for everyone involved in land administration services, from HOAs, SEZs, and state and local governments to special taxing districts. Everyone can benefit from a secure archive, particularly if that archive is global in scope and has the security profile of the Ethereum or Bitcoin protocols. The registry might be on a layer-two protocol or reside on a data availability platform like Celestia or even a smaller chain. Nevertheless, it would still have a settlement layer on an established blockchain like Ethereum.

To be sure, even existing governments look for ways to improve their record keeping. It was not long ago that they converted to computer databases, and we earlier gave the example of Mexico moving on from its shady notaría system. Therefore, it is not unthinkable that these technologies will be adopted, particularly if NGOs and others are willing to step in and help in places like Africa where the technology is desperately needed.

There remain additional problems, however. You probably have already thought of some of them. For example, a blockchain archive cannot prevent a local warlord or cartel leader from holding a gun to your head and forcing you to sign over your property. The fact that your property is registered on the blockchain does not help if corrupt agents target the process of adding and erasing information under the guise of 'transfers of property'. On the other hand, if we have enough information onchain, we can begin to see patterns that point to someone trying to manipulate or extort people into surrendering property. For sure, it is something that has happened throughout human history, and many a Western movie has told the story of corrupt land barons on the prairie stealing from others. But while warlords and narcos and land barons cannot be eliminated, having the information onchain can help us spot points of corruption early on, and it provides a pattern and a record should prosecution take place at a later time.

As with most of the claims we make in this book, we are not saying that blockchain technology solves every problem; what we are saying is that it solves some problems – indeed, some very serious problems – and it mitigates the effect of other problems. Given the likelihood of statistical and AI analysis of blockchain databases, it should be possible to identify today's exploitative land barons early on. It should also be easier to identify legal precedents and query whether actions are consistent with land tenure regularisation. There is a lot that can be accomplished if we put our minds to it.

Nor do we think it is crazy to suppose that we will put our minds to this problem. People do agree to global standards for communications and airline flights, for international shipping, for banking services, and the Internet itself. These things can be done, and they have been done in recent memory. We are optimistic. In the meantime, there is more conceptual groundwork to lay, and this begins in the next chapter with our initial discussion of the rights and responsibilities of blockchain communities.

# THE RIGHTS AND RESPONSIBILITIES OF BLOCKCHAIN COMMUNITIES

## 11.1 Preliminaries

In 1789, during the French Revolution, the National Assembly of France approved a document known as the *Declaration of the Rights of Man and of the Citizen*. Initially drafted by George Washington's protégé, the Marquis de Lafayette, with assistance from US Declaration of Independence author Thomas Jefferson (then serving as US Ambassador to France), the final version was written by the Abbé Sieyès.[1] It incorporated a great deal of Enlightenment philosophy, including ideas from the philosophers John Locke and Jean-Jacques Rousseau, as well as inspiration from the US Declaration of Independence. To this day, it is a model for our understanding of human rights and is considered a valid document to which to appeal in constitutional and international law.

The document has been of seminal influence, but its very existence raises questions. Where do these rights come from? Who makes sure

---

[1] As was customary of the period, the titles Marquis de Lafayette and Abbé Sieyès are not personal names but reflect the social roles of Gilbert du Motier as a nobleman and Emmanuel Joseph Sieyès as a cleric, respectively

we are free to exercise them? The rights also seem to apply to individuals; could we envision a similar document that applied to communities rather than individuals? What would it look like? And would we have the technology necessary to protect these rights?

## 11.2   De jure and de facto rights

Let us begin by making a distinction between the de jure rights that someone might have – for example, rights that God might grant to them – and the de facto rights that they are actually able to enjoy. For example, if God were the foundation for our rights, God might grant us the right to liberty, but state authorities might take that right away in practice. In this case, we might say that one has a God-given, de jure right to liberty but no corresponding de facto right, as the state has stripped it from us. Sometimes, we might say that the de jure right is still there, although we are not able to exercise that right.

There are probably as many theories of the origins of de jure rights as there are thinkers on the topic. We have mentioned the idea that they might be God given, and this idea is widely held (as evidenced by the common use of the phrase 'God-given rights'). Its most famous advocate would be Saint Aquinas. In a nutshell, the idea is that all rights that we have are conferred by God.

This is not the only view of the origins of rights, of course. We earlier mentioned John Locke and Jean-Jacques Rousseau, who had an influence on the drafting of the *Declaration of the Rights of Man*. Locke's view was that individuals have natural rights to life, liberty and property, which are inherent and inalienable.[2] Rousseau, on the other hand, grounded de jure rights in his concept of the social contract and the 'general will'.[3] Utilitarians, like Jeremy Bentham, thought that rights would have to extrude from the idea of maximising happiness for the greatest number of people. Others have advocated any number of other approaches to the grounding of rights.

---

[2] Alex Tuckness, 'Locke's Political Philosophy', *The Stanford Encyclopedia of Philosophy*, 2024 <https://plato.stanford.edu/archives/sum2024/entries/locke-political/> [accessed 29 October 2024].

We aim to remain neutral on the source of the de jure rights of individuals. For current purposes, it does not matter where they come from; we just need to agree on what some of those rights are – for example, the right to liberty or the right to political self-determination. Once we agree on what those individual rights might be, we immediately get to the question of whether we will be allowed to exercise them. Or to put it another way, do we have the corresponding de facto rights?

This leads us to the question of what sorts of systems are actually capable of protecting our individual rights. Can monarchies do it? Can oligarchies do it? Can democracies do it? Can anything involving traditional nation states do it?

It will probably not surprise you to learn that we think nation states are, by their very nature, not capable of protecting the de jure rights of individuals. They can hardly be expected to do so when individuals with conflicting values are kettled within their borders. For that matter, we doubt that any centralised authority can do it either – the dangers of corruption and lack of transparency are just too great in centralised systems. The rights of individuals invariably take a back seat to the interests of centralised authorities.

It will also not surprise you to learn that we think that any system of governance that is actually capable of defending our rights – making sure that we are free to exercise those rights – must allow us to exit untenable situations, and this, in turn, requires that they must allow us to be informed of the actions of the government and what kinds of ways of living are possible under alternative governance structures. There is a series of responsibilities that governing institutions have if

---

they are to preserve our ability to exercise our de jure rights – to make those rights de facto rights.

Here is where things get interesting. If governing institutions are responsible for protecting our de facto rights, they must be given the resources to do so. 'Ought' implies 'can'. Therefore, if they are to fulfill that responsibility, they must also possess the means – and thus the rights – necessary to do so. However, these are not yet de facto rights for communities. If we are to build an alternative governance structure around the concept of blockchain communities, we must ensure that the underlying technology stack can ensure these de facto rights.

On this view, we are not so much worried about how the rights of individuals are grounded, but we are worried about how the rights of communities are grounded. At least some of those rights are grounded in a community's right to carry out its responsibilities, including its responsibilities to individuals. If individuals have a right to exit, for example, then communities have the responsibility to make that a de facto right, but then communities, in turn, must have rights of their own. And if those community rights are also to be de facto rights, then we need to ensure that the underlying technology enables those de facto rights as well.

In other words, blockchain communities will have significant rights of their own, and the question is how we can ensure that those rights become de facto rights, not merely aspirational ideas. Our aim here is to offer a template for discussing these de jure community rights, with an eye to later reflecting on how we can organise the technology stack for blockchain communities to afford them these abilities and rights – in other words, to make them de facto rights.

## 11.3   The Declaration of the Rights of Man

To begin this discussion, it might be useful to review the *Declaration of the Rights of Man* and use it as a template for how we could address the issue of the rights of communities. However, it will also be a guide;

if we endorse the individual rights proposed, then we need to think about what must be afforded to blockchain communities in order to preserve those rights. For ease of reference, we reproduce the declaration's articles here in full.

> ARTICLE I. Men are born, and always continue, free and equal in respect of their rights. Civil distinctions, therefore, can be founded only on public utility.

> ARTICLE II. The end of all political associations, is the preservation of the natural and imprescriptible rights of man; and these rights are liberty, property, security, and resistance of oppression.

> ARTICLE III. The nation is essentially the source of all sovereignty; nor can any individual, or any body of men, be entitled to any authority which is not expressly derived from it.

> ARTICLE IV. Political liberty consists in the power of doing whatever does not injure another. The exercise of the natural rights of every man, has no other limits than those which are necessary to secure to every *other* man the free exercise of the same rights; and these limits are determinable only by the law.

> ARTICLE V. The law ought to prohibit only actions hurtful to society. What is not prohibited by the law, should not be hindered; nor should anyone be compelled to that which the law does not require.

> ARTICLE VI. The law is an expression of the will of the community. All citizens have a right to concur, either personally, or by their representatives, in its formation. It should be the same to all, whether it protects or punishes; and all being equal in its sight, are equally eligible to all honours, places, and employments, according to their different abilities, without any other distinction than that created by their virtues and talents.

ARTICLE VII. No man should be accused, arrested, or held in confinement, except in cases determined by the law, and according to the forms which it has prescribed. All who promote, solicit, execute, or cause to be executed, arbitrary orders, ought to be punished, and every citizen called upon, or apprehended by virtue of the law, ought immediately to obey, and renders himself culpable by resistance.

ARTICLE VIII. The law ought to impose no other penalties but such as are absolutely and evidently necessary; and no one ought to be punished, but in virtue of a law promulgated before the offence, and legally applied.

ARTICLE IX. Every man being presumed innocent till he has been convicted, whenever his detention becomes indispensable, all rigour to him, more than is necessary to secure his person, ought to be provided against by the law.

ARTICLE X. No man ought to be molested on account of his opinions, not even on account of his *religious* opinions, provided his avowal of them does not disturb the public order established by law.

ARTICLE XI. The unrestrained communication of thoughts and opinions being one of the most precious rights of man, every citizen may speak, write, and publish freely, provided he is responsible for the abuse of this liberty, in cases determined by law.

ARTICLE XII. A public force being necessary to give security to the rights of men and of citizens, that force is instituted for the benefit of the community and not for the particular benefit of the persons to whom it is entrusted.

ARTICLE XIII. A common contribution being necessary for the support of the public force, and for defraying the other expenses of government, it ought to be divided equally among the members of the community, according to their abilities.

ARTICLE XIV. Every citizen has a right, either by himself or his representative, to a free voice in determining the necessity of public contributions, the appropriation of them, and their account, mode of assessment, and duration.

ARTICLE XV. Every community has had a right to demand of all its agents an account of their conduct.

ARTICLE XVI. Every community in which a separation of powers and a security of rights is not provided for, wants a constitution.

ARTICLE XVII. The right to property being inviolable and sacred, no one ought to be deprived of it, except in cases of evident public necessity, legally ascertained, and on condition of a previous just indemnity.[4]

The first thing to notice about this document is that, in fact, not all of the articles apply to individuals. Article XV, for example, speaks of the right that a society has to the accountability of its administrators (it appears to be an example of a group right). Article XVI is not really a right but rather a responsibility, and it is a responsibility of the society rather than individuals. It suggests that if a society does not guarantee the rights of individual citizens, it is a false society – or in any case, it is without a true constitution. Article XIII also seems to be more of a right for the society than for individuals. Or perhaps it is a hybrid in that the society has a right to the service of individuals, but those individuals have the right to an equal call to service.

Article III is puzzling in that it appeals to the nation for the grounding of sovereignty and suggests that all rights stem from the

---

nation. This needs to be placed in context, as it is properly understood as an alternative to rights stemming from the divine right of the king. Still, today, it strikes us as a peculiar place to ground individual rights.

Since our focus here is more on communities than individuals, we will want to develop a different kind of constitution. We ultimately propose a *Declaration of the Rights of Communities*, understanding communities broadly to include blockchain communities and, hypothetically, any other similarly organised group of individuals (decentralised yet cooperative). However, to get at the rights of communities, we first need to discuss what their responsibilities are, then reflect on their de jure rights and, from there, think about whether the underlying technology stack can underwrite the corresponding de facto rights.

## 11.4   On the responsibilities of blockchain communities

The responsibilities of blockchain communities are of two kinds. There are external responsibilities and internal responsibilities. As an example of an external responsibility, we might argue that communities do not have the right to infringe upon the rights of other communities and, thus, the citizens therein. With regard to internal responsibilities, we have argued that it is critical that citizens have the right to exit.

We are not in a position to offer an exhaustive survey of responsibilities, as they could potentially involve any aspect of human governance. However, it might be useful to discuss a handful of examples to give some sense of the nature and scope of such responsibilities. Let us set aside, for the moment, what should happen when responsibilities are violated and who or what should enforce the commitment to these responsibilities. For now, we just want to focus on the responsibilities themselves, ignoring how, or even if, they are to be enforced.

Let us take up the issue of internal responsibilities first, and a good place to begin is with a case we have already discussed – the right to exit and the community responsibilities that extrude from that right. The right to exit is surely core to an understanding of how blockchain communities can be viable governance structures. However, if the right to exit is to be more than an empty platitude, we need to give it some teeth. In Chapter 9, we saw that if we are serious about exit, then citizens should be free to exit with the resources that they have acquired. Furthermore, we saw that the right to exit is also hollow if people are in the dark about what their community leaders are up to. Thus, viable exit requires transparency on top of everything else.

If we accept these additional requirements, it seems reasonable to cast them as internal responsibilities for blockchain communities. If the right to exit is fundamental, and if exit must be viable, then communities have the responsibility to ensure the conditions of a viable exit. In Chapter 9, we noted that this transparency would apply to any information that might be relevant to a person's desire to exit. To put it another way, if knowing some piece of information would be relevant to a person's exit decision, then the person has a right to know that information. This too opens the door to a class of responsibilities for blockchain communities. Transparency is not merely a good; it is a responsibility.

What does this responsibility entail? There are many possibilities here. For example, it is reasonable to think that persons have a right to know how their tax contributions are being spent. They presumably should have a right to know if their government is engaged in criminal acts. They have a right to know if the rights of fellow citizens are being trampled. All of these questions might be relevant to an exit decision.

This is just one example, or family of examples, related to exit, but it is enough to help us formalise the structure of rights and responsibilities for blockchain communities. Again, this formalisation is not cast in bronze. It is designed to initiate conversation. However, we believe it is a reasonable place to start.

*I. Responsibilities to individuals*

1. Communities have the primary responsibility to ensure individual exit. This responsibility entails additional responsibilities:

2. Communities have the responsibility to be as transparent as possible about their actions, plans and goals. Reliable records of actions, plans and goals should be easily accessible to all community members.

3. Communities have the responsibility to enable members the right to access information outside of the community.

4. Communities have the responsibility to ensure that community members do not forfeit their ability to exit even if they wish to do so.

5. Communities have the responsibility to ensure that their members retain access to the resources necessary to exit, should they choose to do so in the moment.

*II. Responsibilities to other communities.*

1. Communities have the responsibility to not interfere with the conduct of any community that is operating in a way that respects the rights of other communities and the ability of their members to exit.

2. Communities have the responsibility to be accurate and transparent with communities with which they wish to collaborate.

As stated, this is not a definitive list but the initiation of a conversation, and one can see that interesting questions already arise.

Blockchain communities have responsibilities to their individual members, and many of those responsibilities are tied to the right to exit. However, they also have responsibilities to other communities and the members of those communities. Blockchain communities have the right to attend to their own affairs as long as they are not interfering with others, but they also have the responsibility to honour the rights of other communities. If someone has a right, there is naturally a responsibility to respect that right.

This means that while there is a great deal of leeway in what kind of culture a blockchain community might inculcate, there can be no culture that infringes upon other cultures – even those of competitors. As usual, this leads to certain complicated cases. For example, the forced attack on another community is not permitted but there are more subtle issues surrounding, for example, proselytisation. You cannot force a community to adopt your values by the sword, but can you blanket them with propaganda?

As we have set things out, a community has a right to control the information that enters it, but not over the right for members to venture out and explore the information space. This suggests limits to proselytisation, whether it be religious or cultural. Insofar as a community allows you into their information space, you play by its rules, but if someone ventures out, there really are no rules – proselytise away. The responsibility is then to respect the boundaries set by the individual community. A community can understandably set limits on the practice of proselytising within its network.

All of this raises the question of what it means to restrict the flow of information internally but permit someone to step outside the information sphere. And the answer, of course, is that from a technical point of view, there really is no interesting difference. The information is flowing by; the question is whether you are going to sample it or not. One way to think of it is like parental restrictions on online content. You can turn it on or you can leave it off, but even when you turn the filter on, the information is there; you are simply choosing not to view it.

This, of course, leads to objections. How does one preserve a culture when there is so much information working against it that is only invisible because of some easily toggled community web nanny? The answer is that cultures can only be preserved by their desire to survive; they cannot survive by virtue of the involuntary ignorance of their community members. Any culture that requires involuntary censorship and forced ignorance to ensure its survival is perhaps not a culture worth preserving.

In Chapter 9, when we discussed exit and exile, we considered the right to petition to enter another community. This suggests that communities have a corresponding responsibility to consider such petitions in a fair manner. It may also suggest a responsibility to help make exit and access as frictionless as possible. This may include a responsibility to financially support individuals in exile and those seeking access.

So far, we have been talking about the responsibility that communities have towards their individual members, but as we noted, communities also have responsibilities towards other communities. One area where this arises is in the issue of respecting the culture of alternative communities. This is the inverse of the responsibilities that communities have to make knowledge of other cultures accessible. However, those other cultures, in turn, have the responsibility to be forthright in the information they provide. Just as a failure to exit out of ignorance is problematic, an exit based on misinformation is likewise undesirable.

Again, there are no easy answers here. Who decides if the information provided is misinformation or a ticket to escape oppression? Ultimately, there can be no centralised authority to judge when a community is being helpful and when it is being misleading. Litigation of these questions will have to take place on a case-by-case basis by the ecosystem of communities. The framework we are offering here does not pretend to circumvent hard questions. The goal is rather to remove the resolution of those questions from centralised authorities. In some cases, these disputes can be resolved by negotiating on a peer-to-peer or community-wide basis. In other cases, the dispute may well escalate. Communities may even come into open conflict.

In the next chapter, we will get to the question of how blockchain communities are optimal for resolving disputes, and then we will get into the business of what happens when their methods of dispute resolution fail. However, before we get into those matters, we need to stress one key point. As we have emphasised, blockchain communities have important responsibilities – both to their members and to other communities. Thus, there is the inevitable question of what happens when a blockchain community does not live up to its responsibilities. Our knee-jerk reaction may be to think that we must have some sort of oversight authority, but exactly who or what has oversight over these failures of responsibility? If we reject appeals to centralised authority, then the answer to this question is going to look very different from what we are used to.

## 11.5   Decentralised oversight of blockchain communities

In the previous section, we saw that blockchain-based communities have responsibilities to individuals and to other communities. But what happens if they shirk those responsibilities or act in a way that is contrary to them? For example, what if they interfere in the conduct of another community, or what if they create barriers to exit for their citizens? Or to put it another way, what if they violate the rights of other communities and individuals?

We are now leaving behind questions of what is right and what is wrong and beginning to ask questions about what is to be done if norms are violated. As we saw in the previous two sections of this chapter, there are arguably norms applying both to the rights of individuals and the rights and responsibilities of blockchain communities. Regarding individual rights, many of them extrude from the right to exit and the right to access. Regarding communities, many rights extrude from the idea that they should be free to exist unmolested as long as they do not harm other communities. Two questions immediately arise. Firstly, who or what determines when an agent or community has violated the rights of other agents or communities? Secondly, what is to be done when an agent or community violates the rights of other agents or communities?

For obvious reasons, we do not advocate a centralised authority to judge when communities are acting in violation of norms. Just as clearly, we do not advocate a standing (or even temporary) centralised authority to act against the violation of such norms. The guiding philosophy of the project in this book is that centralised authority is an inadequate solution to problems of governance in today's world. We certainly do not want to rely on centralised authorities regarding this critical lynchpin to human governance. Fortunately, we do not have to.

The first observation that must be made is that in a world of decentralised governance, the preferred way to resolve disputes is not to 'run to the parents' – to a centralised authority. Disputes and misunderstandings often can be resolved on a peer-to-peer basis without appeal to a higher authority. As we will see in the next chapter, blockchain technologies will play an important role in making such dispute resolution more successful. The idea will be that conflict negotiations involve forming a kind of community – even if it is a community in conflict – and if such a community in conflict is organised using blockchain technologies, there will be many tools that can help resolve matters. For example, disputes about violations of norms can be assisted by shared records of negotiations over what each party to the dispute thinks the relevant norms are and how they want those norms to be respected.

For now, however, let us suppose that there has been a breakdown in the peer-to-peer attempt at dispute resolution. For example, imagine that there is a blockchain community that persists in violating fundamental norms regarding the right to exit or the freedom for a community to remain unmolested. Depending upon the severity of a violation of norms and assuming the exhaustion of negotiated solutions to norms, what is to be done? Communities may organise themselves into temporary alliances to take action against communities that are violating norms. Following the general principle that governance should be decentralised, this suggests that punitive alliances should be fleeting – they should be in place just long enough to respond to imminent threats to norms and no longer.

But what should, and for that matter, what *can* such an alliance do? For most blockchain communities, even those that supersede the role of states, it is unlikely that state identity will be tied to the control of territory. Success for a blockchain community (including cyberstates) is a direct function of its ability to network successfully with other communities. Being a pariah endangers this successful networking and thus endangers the ability of such a community to prosper.

This point deserves emphasis as it is perhaps a hidden feature of blockchain communities. They prosper thanks to network effects and having partners with whom they can collaborate. In the contemporary context of terrestrial nation states, sanctions are a crude instrument for success because traditional nation states have direct control over physical territory and the resources therein. They also have access to kinetic forces, which may be brought to bear against civilian populations. However, blockchain communities will not project force in the same way. Indeed, their most powerful mode of action is to disconnect and isolate bad actors – a strategy that is more successful when engaging blockchain communities than when engaging terrestrial nation states. It is easier to disconnect someone from a communication network than to remove them from a physical territory.

There are, of course, lots of questions that arise here. How do we sanction other blockchain communities? What does disconnection and isolation look like? What happens when such a community seeks to retaliate with either cyberwarfare or kinetic attacks? Or for that matter, what happens when a blockchain community comes into conflict with a traditional nation state? We will dive deeper into these questions in Chapter 13. However, for now, we want to leave this conversation with one takeaway: there are clear norms that govern the conduct of all blockchain communities – surrounding the right to exit, for example – but judging that something is a violation of these norms and taking action in response to a such a violation is not the province of centralised authority. Violations of norms can only be recognised by individuals and individual communities, and their principal mode of action will be to collaborate with other communities to isolate the offending community.

## 11.6   On the rights of blockchain communities

We have touched upon the responsibilities of blockchain communities, and we have also given our first gloss over the kinds of political and governmental tools that blockchain communities can be expected to have in carrying out those responsibilities. The flip side of these responsibilities is that blockchain communities will have rights as well. The political and governmental tools that are available are the very tools that can help de jure rights become de facto rights. They are the tools that ensure we not only have rights but that we can exercise them.

Of course, we are not only interested in preserving individual rights but also in preserving community rights. And we want those community rights to be not merely de jure rights but de facto rights. So far, however, we have not explained what those community rights are. In what follows, we offer our own *Declaration of the Rights of Communities*. It is informed by the responsibilities that communities have, both to individuals and to each other, and it is also informed by the tools that communities need to be afforded in order to carry out their responsibilities. Insofar as we are interested in the de facto rights of communities, we observe that those de facto rights are necessarily circumscribed by what the technology stack can actually provide.

While this declaration focuses on the rights of communities, this is not to suggest that individuals have no rights or even lesser rights than communities. On the contrary, community rights extrude from the responsibility to protect individual rights.

Libertarians are apt to focus on the rights of individuals, and communitarians are apt to focus on the rights of communities. However, we think it is important to keep in mind that you cannot realistically have one without the other. Communities can preserve and nurture the rights of individuals, and communities really cannot function well without freely acting members. Together, they create a system in which individual rights and responsibilities serve community interests, and community rights and responsibilities serve individual interests.

Without further ado, and again, to open the discussion, we present the *Declaration of the Rights of Communities*:

ARTICLE I. All communities should be of equal status in the eyes of the technology stack. The same rules should apply to them all.

ARTICLE II. One of the goals of political association between communities is to contribute to the flourishing of those communities and the conservation of their rights, which include their liberty, property and safety, and their ability to defend themselves.

ARTICLE III. The principle of sovereignty of a community resides in its internally shared goals and attitudes. No individual or body may exercise authority over the community in a way that is in conflict with these goals and attitudes.

ARTICLE IV. Community liberty consists of a community being free to do anything that does not infringe upon the rights of its members nor harm its members nor other communities and their members.

ARTICLE V. The constraining principles of blockchain communities only have claim over community actions that are harmful to or infringe upon the rights of others. Communities may not be otherwise constrained in their policies and actions.

ARTICLE VI. The constraining principles of blockchain communities are the expression of the wellbeing of the ecosystem of communities. They must be the same for all communities and be applied equally to all communities.

ARTICLE VII. No punitive action may be taken against a community except to prevent harm to others, as determined by the constraining principles of blockchain communities. Actions otherwise taken against communities are themselves harmful and must be responded to as such.

ARTICLE VIII. Punishments and sanctions against communities must be strictly and evidently necessary for the wellbeing of the ecosystem of communities. No community shall be punished by sanction or other methods except under a constraining principle established before the alleged offence and applied by appropriate decentralised authorisation.

ARTICLE IX. Every community is presumed faultless until it is declared culpable by appropriate decentralised judgment. If a community is found at fault, it must not be punished more severely than is necessary to correct the behaviour, and any punishment in excess of that threshold is itself in violation of the constraining principles of blockchain communities.

ARTICLE X. No community may be punished for its opinions, doctrines, policies, principles and attitudes except insofar as they cause harm to its members or to other communities.

ARTICLE XI. Free communication between communities and between communities and community members may not be prohibited nor in any way obstructed.

ARTICLE XII. If communities have a need to use terrestrial force or virtual force, the authority to use that force stems from that of the community it serves and the ecosystem of communities it inhabits. The use of force may not serve the interests of individuals merely because they are in a position to control said force.

ARTICLE XIII. Communities have the right to raise funds from community members in order to provide community services and administration, provided the fundraising is approved by the community and the services are needed, desired and possible to deliver.

ARTICLE XIV. Communities may discern the need to pay fees for inter-community administrative functions (adjudicating disputes between states or communities, for example). In that case, they have the right to determine the proportion, basis, collection and duration of these fees.

ARTICLE XV. The community has the right to transparency of the actions of every administrative agent of the community (including virtual agents) and from any agent in a higher-level governance structure in which the community participates.

ARTICLE XVI. Any community that does not respect the rights of other communities and citizens therein forfeits its standing and recognition as a legitimate community agent in human affairs.

ARTICLE XVII. The property rights of communities and their members must be recognised, provided they are appropriately documented and consistent with the constraining principles of blockchain communities. Even communities that do not recognise property rights internally must recognise the property rights of communities (and members of those communities) that do.

As stated, this is not the final word about the rights of communities but an initial proposal for discussion – it is the Lafayette draft, if you will. Whichever direction the discussion unfolds, certain ideas should be salient. Communities, whether organised by DAOs using blockchain technologies or some other strategy, are very much actors on both local and global stages, and they are reifications of our group attitudes, goals, beliefs and cultures. They are the vehicles by which human cultures survive and flourish. They have rights, and those rights ought to be respected.

A couple of comments on this draft are also in order. First, when we say that 'all communities are of equal status', we are, of course, not saying that they are of equal wealth and power. We are saying that the governing principles for communities are neutral with respect to the values and principles of those communities. If, for example, several communities were to reside on the Ethereum protocol, the thought is that Ethereum is neutral towards them – the same rules apply to them all.

We have scrupulously avoided the term 'laws' (as deployed in the *Declaration of the Rights of Man*) in favour of the expression 'constraining principles of blockchain communities'. We do not envision these principles as being codified as official law somewhere but rather as involving the nature of the underlying blockchain technology stack plus our working understanding of the values that are necessary for the ecosystem of blockchain communities to thrive – for example, values that help ensure the right to exit and the right to access and whatever considerations flow from those basic values in conjunction with the technology stack.

None of this should be taken to be a promise of a mode of governance that is free of oppression or conflict. Decentralised groups are, of course, capable of harm. The key is to minimise potential harm insofar as possible, and this minimisation strategy leans heavily into decentralisation and the understanding that the most effective action to take against a rogue networked cyberstate or other blockchain community is a consensus-based collaborative effort to isolate it until it respects the norms required for the ecosystem of blockchain communities to flourish. The following two chapters will detail how all of this might work.

# HOW BLOCKCHAIN COMMUNITIES WILL COLLABORATE

## 12.1  Preliminaries

We concluded the last chapter by reflecting on the responsibilities of blockchain communities, raising the question of how such communities might collaborate in responding to some other community or nation state that acts against these responsibilities (for example, a community that prohibits exit). The short and obvious answer is that, first, adversely impacted communities should attempt to engage the bad actor community in dialogue. This would, of course, be true of any dispute. When two or more parties are in dispute, they first need to engage in discussions to attempt to resolve their problems.

Of course, dispute resolution is not the only reason that different blockchain communities might need to engage with one another to achieve some common goal. There are as many reasons for nation states and communities to communicate and collaborate as there are forms of coordinated human endeavours; the goal could be to resolve conflict or head off impending hostilities, but it might also be to negotiate a mutually desired trade agreement, establish new global communications standards, or articulate shared approaches to AI or the environment or religious freedom or, frankly, anything.

In the following two sections of this chapter, we present a more general picture of how blockchain communities can engage in collaborative efforts. Then, in Section 12.4, we return to the specific issue of how to proceed when blockchain communities are in conflict. As will become clear, we do not see a big difference between these two kinds of cases. Whether one begins from a point of agreement or disagreement, the basic tools required are the same. However, some may find an element of our position paradoxical. Our central thesis in this chapter is that when two communities are engaged in negotiations or some other form of shared communicative project, they have, in effect, created a new umbrella community to carry out the negotiations. In the case of two well-aligned communities working out an agreement of mutual interest, this might seem natural enough. However, the paradoxical element is the idea that this is also true when two communities are in conflict. Even in the act of disagreeing with each other – even if parties are engaged in the exchange of epithets – they have already formed a community of sorts. Our goal is to provide the tools that will guide the disagreeing communities in resolving their dispute more successfully.

Before we get into the advantages of blockchain technology for the purpose of negotiation and other collaborative projects, we want to dive into what is known as 'relational contract theory'. This is an approach to thinking of contracts not as one-off business deals but as opportunities to construct productive and long-lasting relationships with the other party of the agreement. To put it another way, you do not just negotiate to close a single business deal. The best way for either party to accomplish their long-term goals is to use the negotiation process to build a kind of intentional community with the other participants. Why do we begin with contracts? Because, if you think about it, treaties and other transnational agreements are kinds of contracts (even if grander in scale). Our hunch is that by beginning with a discussion of relational contracts, we can gain insights into other forms of agreements and negotiation processes.

In Section 12.2, we will briefly tour relational contract theory and how blockchain technology and DAOs are well suited to its execution. In Section 12.3, we will extend this idea from talk of contracts to talk of grander agreements like treaties between communities and nation states. Finally, in 12.4, we will turn to the issue of what happens when we are not merely trying to negotiate a simple business contract or an agreement between two friendly parties but instead are attempting to resolve a serious conflict – perhaps even a conflict that has already become kinetic.

## 12.2   Relational contract theory

In 2019, *Harvard Business Review* published a fascinating article by David Frydlinger, Oliver Hart and Kate Vitasek titled 'A New Approach to Contracts', with the sub-heading 'How to build better long-term strategic partnerships'. The article began by discussing how a long-term business relationship between the IT supplier Dell and logistics firm FedEx had recently reached breaking point, with both parties attempting to abide by a 100-page document that contained lots of 'supplier shall' statements. Neither party was happy with the agreement, and each side felt they were somehow being taken advantage of by the fine-print details of the contract. The two parties subsequently decided to abandon the idea of a traditional contract, realising that a well-functioning business relationship between them was more important than nailing down all the details in advance within the context of an adversarial negotiation. The solution was to opt for 'relational contract theory' – a relatively new idea,[1] although its seeds were first articulated in a 1969 paper by Ian Macneil titled 'Whither Contracts?'[2]

---

[1] David Frydlinger, Oliver Hart and Kate Vitasek, 'A New Approach to Contracts', *Harvard Business Review*, 1 September 2019 <https://hbr.org/2019/09/a-new-approach-to-contracts> [accessed 10 July 2023].
[2] Ian R. Macneil, 'Whither Contracts?', *Journal of Legal Education*, 21/4 (1969), 403–18 <https://www.jstor.org/stable/42891974> [accessed 16 July 2023].

The basic idea of a 'formal relational contract', as articulated in the *Harvard Business Review* article, is that it 'specifies mutual goals and establishes governance structures to keep the parties' expectations and interests aligned over the long term.' More importantly, given those interests, it is 'designed from the outset to foster trust and collaboration', and finally (also of interest to us), 'this legally enforceable contract is especially useful for highly complex relationships in which it is impossible to predict every what-if scenario.'[3]

Why are these matters of interest to us? Firstly, because they are precisely the sorts of things that DAOs and blockchain technologies are designed to facilitate. Those platforms provide tools that enable us to create a shared record of goals – particularly long-term ones – and a permanent record of those goals and the history of their changes. The goals can be updated, but the original goals remain visible to all. Similarly, blockchain technology is designed to facilitate trust, in part by using 'trustless' technologies like smart contracts.[4] Finally, blockchain technologies can handle the representation of systems and organisations of arbitrary complexity. You simply cannot build an organisation that is too complex for a language that is Turing complete.

We mentioned the issue of impossible-to-predict scenarios, and it is worth reflecting on where they can arise – essentially, everywhere. Frydlinger, Hart and Vitasek note that they naturally occur in 'complicated outsourcing and purchasing arrangements, strategic alliances, joint ventures, franchises, public-private partnerships, major construction projects, and collective bargaining agreements.'[5] The reason this speaks to the wisdom of seeking contractual relations instead of mere formal classical contracts is that we want the contractual relationship to evolve over time in response to conditions on the ground but also due to our deeper understanding of the goals and aspirations of the parties to the contractual relation.

---

[3] Frydlinger, Hart and Vitasek, 'A New Approach to Contracts'.
[4] The scare quotes are because nothing is truly trustless. As we will see in Chapter 15, the most we can hope for is distributed trust.
[5] Frydlinger, Hart and Vitasek, 'A New Approach to Contracts'.

One of the problems that relational contract theory is designed to mitigate is what the authors of the paper call 'shading' – a phenomenon in which one party begins to feel it is being taken advantage of and decides to get even by taking advantage of minor loopholes in the contract to get an edge over the contractual partner, which rarely goes unnoticed and leads to an escalating war of shading episodes until both parties withdraw from the contract.[6] The theory behind relational contracts is that both parties want to make it a priority for the contractual relationship to succeed and for both parties to benefit from the relationship. In this case, one of the things that the parties to the contract will do is to articulate their goals from the relationship to each other so that both parties have a sense of what is important to the other and what will contribute to a positive business relationship.

You can probably already guess how blockchain technology and DAOs can help with an approach to contracts such as this – the description of the record keeping, the dynamic nature of the relationship and the call for building a kind of community sound a lot like the features of a DAO. However, before we get into the details, we want to say more about the process of negotiating contracts and other agreements and what kinds of tools and conditions facilitate it.

One critical element discussed in the literature on contract negotiation is maintaining a record of proposals. The idea here is that instead of scribbling notes in no particular order, the best practice is to keep clean records of all parties' proposals and the times at which they were made. Firstly, this is advantageous because it avoids rehashing things that have already been agreed upon or rejected. Furthermore, observing what has met with agreeable responses can help gauge which future proposals might be accepted or rejected. As usual, and for many reasons, good record keeping is key, and of course, in this case (as in many other cases), one wants the record to be immutable and recognised as valid by all parties to the negotiations so that one party cannot unwind from agreed-to positions. We might add that the record of proposals should also contain information about the interpretation that all parties have assigned to the letter of what has been agreed upon. The archive itself is not everything; as Jacques Derrida pointed out, we also have to consider its interpretation.[7]

---

[6] Ibid.

We agree with the general philosophical perspective of relational contract theory and the idea that contracts are not just one-off pieces of paper. Those signed pieces of paper are just a small formal part of dynamic, long-lasting relationships between interested parties. In some sense, the contract is never completely set in stone, just as the business relationship is never set in stone. Even the bronze Tabula Alimentaria that we discussed in Chapter 5 had to undergo changes over the years it was displayed as conditions on the ground changed. Therefore, it is not just the negotiation of the contract that is going to involve a record of proposals, but the entire life of the contract is going to demand some way of recording proposed changes, proposed interpretations, disputes over interpretation and other details of interest to the concerned parties.

There are a number of ways in which blockchain technologies can assist us in negotiating and developing relational contracts. The leading idea is that if each contract – or each contractual relation – is assigned a DAO in which the parties to the contract and relevant mediators are participants, then we immediately have a venue in which proposals can be discussed and in which guiding philosophies and goals can be articulated. If those discussions are recorded on the blockchain (as they should be), then we also have a shared, immutable history of the proposals made (literally a record of proposals) as well as a forum in which those proposals can be discussed, approved and rejected.

---

[7] We are aware, of course, that this process does not bottom out. A record of interpretations is yet another piece of archival information that is subject to interpretation. While this process never bottoms out, a few iterations of it should resolve many concerns and result in a greater understanding about the agreement being forged.

Turning to the issue of trust, one key advantage of onchain collaboration is that it provides onchain verification of actions and intended actions. Take, for example, a negotiated economic policy or a trade agreement in which parties agree to purchase a good or service at a specific price. The history of trade agreements is littered with broken agreements and aspirational promises that were not fulfilled. However, if the agreement can be encoded in a smart contract, then there is no issue as to whether the promise will be kept. If the agreement is onchain and automated in a smart contract, it will be executed once it is digitally signed and the required conditions are met. In negotiating the details of a smart contract, one can eliminate a number of concerns about whether some policies will be executed and what measures need to be negotiated to ensure the proper execution of the agreement. The lines of code in the smart contract guarantee the fulfilment of many of the agreement's terms.

Let us consider a hypothetical example of this. Suppose that we agree to supply one million widgets at one price in exchange for ten thousand barrels of oil at a different price. Will we deliver on our promise? We can trustlessly guarantee that we will because we can write a smart contract in which, upon receipt of an oil contract at a certain price, we release our widgets from inventory. Indeed, the contract might even include logistics so that delivery methods and times are specified in the contract itself. If necessary, the oil could be released in stages depending on where our shipment rests in the logistical chain. Smart contracts take us from the age of 'trust but verify' to an age of 'trustless with continuous verification'.

We can give more concrete examples as well. Many treaties involve agreements for payments of reparations and the like. One could achieve a more trustworthy agreement if the funds were deposited in a smart contract and were dispersed over time as per the agreement. Or for example, agreements about territorial boundaries could have more bite if the property ownership was recorded onchain via a global property registry of the sort we discussed in Chapter 10. However, this talk of treaties may seem out of place; are treaties not different from contracts? Not in any sense that is relevant to this discussion.

## 12.3   Bringing DAOs to treaties and treaty organisations

If one thinks about treaties (let us say between two nation states), they are, in effect, contracts between the parties to the treaty. There are obviously some differences between treaties and typical contracts. Most contracts, for example, are written within an established legal framework. Treaties often involve agreements between nation states and are therefore outside the scope of an external or higher legal framework. However, they are very much the same in other relevant respects. Party A agrees to do X, and party B agrees to do Y, with various specified conditions, verification methods and enforcement mechanisms. More to our point, if relational contract theory makes sense in the realm of contracts, something very much like it makes even more sense in the realm of international or global treaties. Ideally, one does not want treaties to be one-off efforts but instead wants to be in a relational treaty agreement in which the parties to the treaty remain in contact and can articulate their goals and concerns. If you think about it, the historical practice of arranging marriages between royal family members as a part of a treaty was a version of this idea – by marrying into the treaty partner, one is saying that the goal is a long-term relationship with that partner.

We believe that the process of successful negotiation involves the creation of a temporary umbrella community. However, if we apply relational contract ideas, then the thought would be that we want the umbrella community to endure for some period of time – perhaps, indefinitely.

We may find that a temporary umbrella community becomes a permanent community because agreements involve the creation of collaborative efforts, and these collaborative efforts will evolve into communities in their own right. A snowball effect can follow: as these communities make progress in finding common ground and building shared institutions, they tend to grow closer together. They will not necessarily be permanent communities, although they might evolve into that, for better or for worse. They are fundamentally only in existence for the agreed-upon life of the collaborative agreement. Still, agreements can be extended, and if agreements are modelled on relational contracts, we would come to think of them as being part of an agreement relation, which could have an unlimited lifespan.

This point applies vividly to the act of negotiating a treaty between two communities – for example, between a cyberstate and a nation state or between two blockchain communities that are engaged in an exchange of goods and services or that wish to put an end to hostilities between themselves. In negotiating a treaty, one has already created a kind of meta community – typically, a community grounded in an understanding between the parties to the negotiations that there are shared values with respect to the goals of the treaty, the methods of verification, the methods of enforcement and so on. Treaties – or at least treaties of any consequence – are the product of people working together with shared values and goals.

Blockchain technology allows us a fresh perspective on treaties and other agreements. Of course, treaties, being agreements, require methods of verification, and sometimes verification is simple (was the purchased grain delivered or not?) and, sometimes, not so simple (were those uranium centrifuges dismantled or not?). Successful treaties and agreements are not negotiated across communities so much as they are negotiated *within* a community, which means that a kind of community must exist that includes both parties to the treaty. It might be just the immediate parties to the treaty or it might include other facilitators as well (arbitrators, inspectors and so on).

The problem is not that such communities are impossible to create but merely that they are very difficult to nurture. There are always many reasons for the parties to the treaty to be distrustful of each other and doubtful as to whether the terms of the agreement will be honoured. In some cases, the inability to carry out effective record keeping can be a hindrance to the successful negotiation of a treaty.

And yes, we are back on the topic of record keeping because it is just as important in negotiating and resolving conflict as in anything else. As we noted earlier, one standard practice in negotiations is to maintain a record of proposals, which tracks the history of proposals made and agreed to over time. It is critical that both sides agree to the record of proposals – that they be able to see the progress made and are on the same page about what has been agreed to. If this is not done, progress made can be lost by one or more parties. Furthermore, having a shared record of proposals and agreements can bring to the fore disagreements about the interpretation of those proposals. You cannot clarify something if you have lost track of what it is that needs clarification.

Our point is that a major part of resolving conflict involves having a clear and immutable record of what has been agreed to and how it has been interpreted, a history of what both parties have tended to agree to, and when possible, onchain mechanisms for the execution of the ultimate agreement. To put things into perspective, let us review how traditional economic treaties between nation states are enforced and then reflect on how blockchain-grounded treaties would carry out the same functions.

Economic treaties are verified and enforced through various mechanisms (subject to the treaty itself). Some of the most common mechanisms include the following:

1. Self-reporting: This involves regular reporting by the signatories on the implementation of the treaty provisions, including data on trade flows, investment patterns and other relevant economic indicators.

2. Monitoring: This typically involves some independent organisation to carry out the monitoring. For example, international organisations such as the World Trade Organization (WTO) and the International Monetary Fund (IMF) typically play a key role in collecting and analysing this kind of data.

3. Dispute resolution: Economic treaties typically include provisions for resolving disputes between signatories, for example, through a neutral third-party mechanism such as a panel or court. The dispute resolution process may involve the submission of evidence and arguments, hearings, and the rendering of a final decision. There is a veritable alphabet soup of such courts that are in the business of settling trade disputes. These include the WTO's Dispute Settlement Body, the International Centre for Settlement of Investment Disputes (ICSID), the International Chamber of Commerce (ICC), the International Court of Arbitration (ICA), the United Nations Commission on International Trade Law (UNCITRAL), Arbitration Rules (an organisation that provides a framework for resolving commercial disputes through arbitration), the European Court of Justice (ECJ), the NAFTA Dispute Settlement Body, the International Trade Commission (ITC) and the International Commercial Court.

4. Peer review: Some economic treaties may include provisions for peer review in which signatories conduct periodic reviews of each other's implementation of the treaty and provide feedback and recommendations.

5. Enforcement mechanisms: Economic treaties typically also include enforcement mechanisms, such as sanctions or fines, to deter signatories from violating treaty provisions. These may also take the form of trade restrictions such as tariffs or quotas and suspension or termination of the treaty.

In short, treaties involve a combination of monitoring, reporting, dispute resolution, peer review and enforcement mechanisms. The effectiveness of these mechanisms can depend on many factors, ranging from the good faith and political will of the signatories, their ability to deliver on their promises, and quite often, the ability of international organisations to successfully monitor compliance. And we have yet to even get to the issue of enforcement (how do we make someone pay the fines if they lose a trade dispute case in an international court?).

All of this brings us back to treaties that are grounded on the block-chain and the question of how different they might be. Some elements might be similar, but others are not similar at all.

Let us begin with the strategy of self-reporting. The weakness of self-reporting is, of course, that it opens the possibility of deceptive reporting. However, there is the additional problem that the data that the reporting community believes to be accurate may not be accurate at all. This could come about for many reasons, ranging from bureaucrats attempting to turn in friendly numbers to keep their bosses happy to the simple inability of states to accurately gather the requested information.

With blockchain technologies, self-reporting should be a far more successful enterprise. Onchain information is difficult to falsify since it must be consistent with the information at every step of the supply chain and manufacturing process. It is not enough to falsify the number of widgets you have made because that number must also be consistent with the information at each stage in the manufacturing process, the supply chain for that manufacturing process, inventory control and, ultimately, the shipping of the product. All of these numbers are visible to everyone with access to the blockchain, so if it is your blockchain (i.e. the blockchain on which your community is based), misrepresenting the numbers is quite difficult.

This leads us to the issue of monitoring other communities. There are several solutions here. If the information to be monitored is onchain, then one can provide access to the information onchain. For example, in the case of widget production, one can provide access to onchain information regarding production, the supply chain, inventory control and so on. If that information is proprietary in certain circumstances, it is possible to provide zero-knowledge proofs that can establish the reliability of the output information by proving that the numbers were not falsified somewhere along the line.

This connects with more general advantages of utilising zero-knowledge proofs for other purposes. For example, a treaty might require that a certain percentage of state resources must be committed to some social cause. However, the community may not wish to provide information concerning its total resources. With a zero-knowledge proof, it can still prove that it has provided the requisite percentage of resources to social causes.

Returning to the issue of exit, here again, if communities have a treaty obligation to provide the ability to exit for their citizens, cross-chain proofs should be possible to determine whether citizens in other communities have resources available to them should they choose to exit.

Again, this would be provided by zero-knowledge proof procedures that would query whether the DAO contract of the monitored community indeed unlocks agreed-to resources to citizens on their exit.

This brings us to dispute resolution. There may still be a need for international courts to adjudicate disputes in some cases, but their task should be much easier given the onchain data available to the courts. The jobs of the alphabet soup of international organisations should be easier, but will such organisations be all that necessary? At least regarding treaties and agreements that take the form of smart contracts, it is hard to see how disputes could even get started, given that the code in the smart contract executes the agreement automatically. That said, there may be cases in which there are hacks to a contract or in which a contract is entered into under duress, so such international organisations may remain somewhat relevant.

Smart contracts are better designed for economic disputes, where it is easy to see how the relevant information can and will be available onchain. However, even in a dispute about territorial control, smart contracts can be deployed.

The key to extending smart contracts to all manner of treaties, agreements and subsequent dispute resolution involves the application of oracles that can monitor the relevant actions. For example, if an agreement says that adversarial forces must withdraw from a given territory or withdraw a particular class of weapons from the territory, oracles can take the form of bots that monitor the area for relevant information. To assure all parties to the treaty that only the relevant information will be sampled, the bot's software would be under the observation of members of the DAO for the treaty. In other words, one could engineer 'observers' that pay attention only to the relevant parameters and are blind to everything else and that report only to the shared DAO. This represents an advantage over human observers who may be biased or who may steal classified information not relevant to the treaty. Bots, under the control of the DAO, observe only what is relevant to the treaty.

Are there limits to this kind of monitoring? Presumably, yes, particularly in areas that require human judgement. However, treaties and agreements can be written in such a way that the parameters of the agreement are within the capabilities of the bots. Suppose, for example, that we wanted to enter into a treaty regarding the treatment of prisoners. Using current AI technology, it should be possible to train a bot to recognise acceptable and unacceptable behaviours, and a treaty might be programmed to abide by the judgement of the bot, as trained for the specific purposes of the treaty. The bot reports to the DAO smart contract, which then affirms whether the treaty is being observed or not. If the treaty is not being observed, the enforcement mechanisms kick in.

Let us stay with our example of the treatment of prisoners (although the same strategy could be deployed for other human rights abuses). If a suitably trained monitor bot reports abuses to the DAO, the smart contract automatically acts on that information. Perhaps the programmed response is to issue a warning (at least initially), or perhaps it is to issue a fine. This might mean that the treaty involved the staking of resources in the contract as a guarantee that its terms would be observed. The smart contract could deduct this fine automatically. Or the smart contract might issue other punitive actions. It could also issue a complaint to traditional dispute resolution mechanisms. There are many ways to do this. The participants will engineer the necessary specifics into each treaty and agreement as the relational treaty evolves.

All of this discussion of treaties and other agreements was designed to establish a framework for how blockchain communities might collaborate with each other, not just on economic matters but on matters of fundamental community rights and individual rights. In Chapter 11, we talked about the rights and responsibilities of blockchain communities, and the issue now becomes how those rights are to be enforced – both the rights of individuals and those of communities.

You may not have noticed it, but all of the rights and responsibilities outlined in the previous chapter are amenable to being represented in the kind of smart contract agreements we have discussed here. Take, for example, the right to exit, which we just mentioned. Once established, the parameters for exit can be monitored via zero-knowledge proofs, and if there is a violation of the principle, that violation can be reported to the DAO around which the signatories are organised.

If it is not clear, we envision the various communities, organised around different principles and values, having some shared values – possibly those values encoded in our list of rights and responsibilities for communities. That opens the opportunity for the kind of meta DAO that we mentioned earlier. Such a community of communities will also be organised around smart contracts.

Just as a first-order community will have strategies for incentivising positive collaboration among its members, so too will the meta DAO, which will have the usual DAO strategies for incentivising collaboration among *its* members. These incentives might involve carrots, and they might involve sticks, but engineering collaboration at this meta level is not different from engineering it at the lower level in any interesting way.

None of this is to say that we will not encounter conceptual limits regarding this form of meta-DAO governance. As we will see in Chapter 15, there are significant issues surrounding the limits of oracles to reliably report real-world events and record those observations onchain. Before we get to those conceptual limits, however, there is a more pressing question that needs to be addressed.

What if there are communities that want no part of the international order we envision here? For example, imagine a community that refused to allow exit, or that interfered, unjustifiably, in the affairs of another community, or that carried out human rights violations against its citizens or those of other communities. What happens if a community refuses to play nice or refuses to cooperate? This leads us to the topic of strategies for dealing with noncooperative communities, and ultimately, this leads to the topic of what happens when communities enter into conflict and war.

## 12.4   Bringing DAOs to conflict resolution

There is a famous quote from the philosopher Ludwig Wittgenstein, in which he says, 'If a lion could talk, we wouldn't be able to understand it.'[8] What Wittgenstein means by this is that communication requires more than the ability to articulate words or to even have a syntax for those words. Communication requires that we have a shared community with shared practices. We would not understand the lion because we do not share the community and practices of lions, whether they be in zoos or hunting gazelles on the savanna. However, there is another way to look at this. If we are communicating with someone, even if we disagree about everything, we can only do that because we share a kind of community and set of practices. It does not matter how hopeless the disagreement appears or how deep it goes. It does not matter if it has degraded into an endless exchange of insults. You are still communicating – something you cannot do with Wittgenstein's lion – and for that to be possible, there is already a community in place.

---

[8] Existential Comics, 'Wittgenstein's Lion', *Existential Comics*, 2018 <https://existential-comics.com/comic/245> [accessed 30 October 2024].

It can even be argued that disagreements are critical components to coming to a common understanding with someone. You may have noticed that if you get into a disagreement with someone, you sometimes come out of it with a deeper understanding of your own position. However, you may also have noticed that a certain level of understanding must come before disagreement is possible. College teachers have long recognised this about their students. If the students passively nod their heads to the lecture, there is a good chance they are not understanding. However, if they are raising objections, they actually do understand what you are saying – or at least, they are understanding well enough to engage. So perhaps disagreement is a path to understanding and vice versa. First, there must be understanding, from which disagreement emerges, and this, in turn, leads to better understanding and more disagreement, and the cycle continues.

Our point here is not to get philosophical about the nature of disagreement and understanding but simply to point out that disagreement itself assumes some form of understanding in the background, and it similarly can lead to (and we would argue is necessary to) future understanding. We bring this up to highlight that when we establish platforms upon which to conduct negotiations between communities in conflict, there is nothing at all bizarre about saying that in those negotiations, despite all of their differences, the parties to the disagreement constitute a working community. We might even say that they constitute a permanent community because as long as people communicate well enough to insult one another, they are part of a shared community.

More generally, when two separate communities find themselves in conflict, they still also constitute a kind of joint community. That is to say that even if two parties are diametrically opposed to one another, and even if they are engaged in some form of kinetic conflict, if they are communicating with each other for any purpose – even if it is to degrade each other – they ipso facto also form an umbrella community encompassing both of the communities in conflict. This may seem paradoxical. How can two communities in dire conflict – even to the point of warfare – jointly constitute any sort of community? As contradictory as it may seem, the two parties in dispute, even if they are yelling at each other, are still engaged in a kind of communicative practice, and they are appealing to ideals and principles that they imagine the other party either shares or ought to share with them. This suggests more than a little common ground.

What may seem odd about our perspective is that we often think of communities in conflict as being very separate communities. We fail to see that, at the end of the day, there is still a broader community encompassing both of the communities in dispute. Maybe that broader community is global or regional in scope, or maybe it only encompasses the two parties in conflict. Whatever the scope, those two parties, once engaged with each other and communicating (even if it is to trade insults), have formed a community, even if it is initially just a community of conversational participants. Why is this important? Because if it is a community with the shared goal of resolving the conflict and not much more than that, we want to be able to provide the tools to facilitate that process, and one of the best tools that can be provided in this and similar cases is a DAO.

In earlier sections of this chapter, we observed that when two separate communities, each organised around their own DAO, need to negotiate or resolve some conflict, the best way to do this is to build an umbrella blockchain community for the purposes of the negotiation. This umbrella community might consist solely of the communities in dispute or representatives of those communities or the parties in conflict along with other interested parties and stakeholders.

Just as we saw that there is not really an interesting distinction between contracts and treaties, so too, there really is no interesting difference between conversations between parties in conflict and those in accord. In each case, there is some sort of goal. In the case of parties in conflict, the goal might be the end of hostilities. Or if there is no desire to end hostilities but merely a mutually shared desire to trade insults, that is good enough. Blockchain technologies and DAOs are as useful to communities in conflict as they are to communities in accord.

For example, it can only be helpful to ask parties in conflict to clearly articulate their principles and values and what their complaints against the other party are. If one of the parties finds the other to be evil, that suggests an assumption about some shared values or principles. Even if fundamental principles are, at the end of the day, incommensurable, it is still valuable to have those principles clearly articulated, if for no other reason than to know what sorts of proposals are unlikely to be successful. More specifically, if two parties are in conflict to the point that they do not trust each other, there is all the more reason to take the dispute to the blockchain so that both parties to the dispute will know that the records are immutable and that trustless agreements are possible in the form of smart contracts.

While we would like to think that the application of blockchain technologies described in this chapter can resolve many disputes – or at least, take the edge off them – we certainly would never say that they can help us prevent disputes from becoming kinetic. Disputing parties will go to war. Maybe blockchain communities will have to go to war as well. The question is, what will that look like? We will explore this more thoroughly in the next chapter.

## CHAPTER 13

# WHEN BLOCKCHAIN COMMUNITIES ARE IN CONFLICT

## 13.1  Preliminaries

Nation states, quite obviously, repeatedly find themselves in conflict and, all too often, those conflicts escalate. They can lead to trade wars and sanctions and, ultimately, to kinetic warfare – conflicts in which bullets fly. Of course, such conflicts are not limited to nation states. Empires go to war, as do communities and factions and clans. Theoretically, any organised group of humans is capable of engaging in warfare. Sometimes, the conflict is kinetic, but in some cases, there are forms of warfare like psychological operations (PSYOP) in which one party attempts to impose its will on another through non-violent means. Such actions may involve less bloodshed but are still an important part of warfare.

The dawn of blockchain communities will not be the end of warfare – neither the end of PSYOP nor kinetic conflicts – but as we saw in the previous chapter, there is reason to believe that such conflicts will be less frequent, less deadly and easier to resolve in a peaceful manner. We have already seen several reasons why this might be so.

Blockchain communities make it easy for individuals to exit peacefully; they are not defined with respect to physical territory, and their interests are less about controlling territory and more about engaging in fruitful networked relations. Communities in conflict also have access to neutral, reliable, corruption-resistant platforms upon which to negotiate solutions to their differences. But of course, given that blockchain communities are ultimately populated by fellow human beings, bad actors are inevitable. Violent conflicts are inevitable. Trouble will inevitably come.

This chapter is about what happens when military conflict materialises. As we will see, there are several forms of conflict that might be categorised as military conflict. In the first place, there are conflicts that involve PSYOP. Are these really military conflicts even if no physical weapons are used? It has certainly been recognised since Sun Tzu wrote *The Art of War* in the fifth century BCE that there are non-violent aspects to warfare. Sometimes, it is tricky to determine if a particular action or strategy counts as being an act of warfare or not, but that is fine. There will be borderline cases.

When we think of military conflicts, we are apt to think of situations that involve what we would recognise as the deployment of physical weapons – guns, knives, bombs, chemical attacks, biological attacks and so on. However, there are also cases where the conflicts are digital in nature. These might be independent of PSYOP and might include attacks against a military target or its command-and-control system using various hacking strategies, including computer viruses and attempts to armour against such attacks in response. Borrowing a term from the title of a book by Major Jason P. Lowery of the US Space Force, we can call this *Softwar*.

In this chapter, we devote Section 13.2 to PSYOP, 13.3 to kinetic warfare and 13.4 to softwar. On the one hand, we want to know how blockchain communities can armour themselves against such attacks. On the other hand, we want to know how blockchain communities can impose these forms of force when necessary.

## 13.2   Blockchain communities and PSYOP

If a blockchain community has a small or distributed terrestrial footprint, enemies are apt to attack using modes of asymmetric warfare, using economic attacks and otherwise working to undermine the beliefs and values of the citizens of the target community. In point of fact, blockchain communities may be more vulnerable to this sort of attack because their identity is not tied to a shared physical location but to a shared (or perceived to be shared) set of beliefs and values.

In 2008, the United States Army authored a then-classified field manual on the conduct of asymmetric warfare. Entitled *Army Special Operations Forces: Unconventional Warfare*, it described strategies for attack and defence involving non-kinetic asymmetric warfare. One of its key ideas illustrated how both conventional (kinetic) and irregular approaches to conflict have the same ultimate goal: to control the decision making of the target community.[1]

The point is that both conventional warfare and irregular warfare have precisely the same goal – to get the opposing government to do as you wish. In conventional warfare, that is accomplished by using the military to control the outcome through violence. In irregular warfare, the goal is to manipulate the population of the target state to achieve the same outcome through (ideally) less violent means.

We will get to the issue of conventional kinetic warfare in the next section (for there will still be such a thing in the future we envision). However, cyberstates and other blockchain-grounded communities will also have to confront irregular warfare – external forces attempting to manipulate a community into compliance with their external goals and, sometimes, goals in conflict with those of the target community.

Now, in our view, there is nothing wrong with an external community attempting to persuade your community to change or institute some policy. However, there is a very murky line between attempts at persuasion and attempts at manipulation – one can go too far. The army field manual does not distinguish between modes of persuasion that are within acceptable bounds and modes that are outside those norms, but it does give us some examples of methods of population control and manipulation that we would surely classify as hostile.

---

[1] United States Department of the Army, *Army Special Operations Forces Unconventional Warfare* (Washington, D.C., 2008).

For the moment, let us set aside the hard problem of distinguishing between acceptable external interventions and hostile interventions. Instead, let us consider a more pressing issue: What is a community to do when it encounters what is (let us assume) an external and hostile PSYOP attack? How is it supposed to arm itself against such attacks?

Before we go deeper into how a community can protect itself, it will be useful to go into more detail about the nature of these attacks. Here, we can use the army field manual again as our starting point. First, we need to emphasise a previously made point: irregular warfare (IW) is not about control of territory; it is about control of people. From the manual:

> IW focuses on the control or influence of populations, not on the control of an adversary's forces or territory. Ultimately, IW is a political struggle with violent and nonviolent components. The struggle is for control or influence over and the support of a relevant population.

The question of *how* a population is to be controlled immediately arises. Again, from the manual:

> IW operations also employ subversion, coercion, attrition, and exhaustion to undermine and erode an adversary's power, influence, and will to exercise political authority over a relevant population. What makes IW 'irregular' is the focus of its operations (a relevant population), its strategic purpose (to gain or maintain control or influence over), and the support of that relevant population through political, psychological, and economic methods.

This finally takes us to the issue of psychological operations, or what we have been calling 'PSYOP':

> [We define] PSYOP as 'planned operations to convey selected information and indicators to foreign audiences to influence their emotions, motives, objective reasoning, and ultimately the behavior of foreign governments, organizations, groups, and individuals.'[2]

Again, the key point is that the target is the people, not the territory; the objective is to control their actions through various forms of manipulation. While simple persuasion might work, the point of PSYOP is that more underhand strategies are also in play.

One thing to keep in mind is that blockchain-based communities are difficult targets for any sort of hostile attack. They are, as the intelligence community calls them, complex adaptive systems. They are decentralised. There is no single central node that you can take out in order to defeat the system. You presumably have to take out multiple points of control. However, the system is adaptive; it can heal itself even while your attack is ongoing. Blockchain communities, being complex adaptive systems, make extremely difficult targets.

Complex adaptive systems are difficult to successfully attack, but they are not invulnerable, and they are not a type of adversary that is unknown. There are many nonstate actors that constitute such complex adaptive systems, from drug cartels to terrorist networks. You typically cannot kill them by cutting off the head, although there are famous exceptions.[3] Nor are standard kinetic attacks likely to be effective. Their existence, in some cases, has the blessings of sovereigns and nation states that sometimes use them as pawns on a global geopolitical chessboard.

---

[2] Ibid.

Likewise, blockchain communities are not invulnerable, and there are known attack strategies targeting complex adaptive systems like blockchain communities (or any decentralised group). For example, there is a game plan set out by Ed Waltz in his paper 'Means and Ways: Practical Approaches to Impact Adversary Decision-Making Processes'.

Firstly, we need to ask what it means to say that a blockchain community is a 'complex adaptive system'. Among other things, it means you cannot easily predict the outcomes of the actions against nodes in that community. More precisely, as Waltz puts it, 'the organization's behavior cannot be predicted by models of the properties of the actors nor by a simple linear combination of them.'[4]

This does not mean you cannot attack such a system, but it means you must attack it by stressing the system itself, and you do this by targeting individuals in a way that has systemic consequences. That is, if you want to degrade a complex adaptive system, you can apply pressure in the following ways (again, following Waltz):

> *Defection*: Recruit, train, and establish individuals within the organization and the supporting population capable of conducting resistance and dissent activities leading up to, if required, a coup d'état.

---

[3] One notable exception would be the elimination of Pablo Escobar, which resulted in the end of Colombia's Medellín Cartel. See Santiago Neira, 'El fin del Cartel de Medellín: La muerte de Pablo Escobar y el surgimiento de Los Pepes', *Infobae*, 6 December 2023 <https://www.infobae.com/colombia/2023/06/12/el-fin-del-cartel-de-medellin-la-muerte-de-pablo-escobar-y-el-surgimiento-de-los-pepes/> [accessed 30 October 2024].
[4] Ed Waltz, 'Means and Ways: Practical Approaches to Impact Adversary Decision-Making Processes', in *Information Warfare and Organizational Decision-Making* (Boston, MA, 2007), 89–114.

*Division*: Conduct psychological operations to disrupt the unity of purpose, discipline, and agreement between government bodies and between the government and population groups.

*Deception*: Protect the true intentions of the counterorganization operations by operational security while revealing selected opposition activities and simulating false activities that cause the target to believe a different adversary plan and approach is being implemented.

*Diversion*: Divert or misdirect attention from the true sources of the opposition; secure third-party support to divert attention from the primary source of the attack.[5]

More conceptually, any force attacking a decentralised community would want to generate a feedback loop by attacking nodes to weaken the network in a way that further debilitates the targeted nodes.

Suppose we think of a node as a prominent individual or organisation within a blockchain community. We can select a handful of important individuals, apply stress to them and observe the effects on the network. Similarly, we can manipulate the network itself (in this case, the community itself) so that it acts against the targeted individuals. In this way, the feedback loop is initiated, and the health of the community is degraded.

This strategy for attack is all the more poignant if we consider the literal nodes of a computer network like Ethereum. There is a finite number of nodes assembling blocks for the proof-of-stake ledger, and if one targeted enough of these nodes, one would put a lot of stress on the network. Given sufficient resources, such an attack is possible. However, the idea here is that you want to do more than simply target the nodes; you also want to observe how the network responds to these attacks and organise your subsequent attacks with this information in hand. For example, you might find that as the originally targeted nodes go offline, certain nodes become more important. Therefore, they should be the targets of your subsequent attacks.

---

[5] Ibid.

Now obviously, there are lots of ways to initiate attacks on blockchain communities. There are as many ways to attack a community as there are ways to attack individuals (and combinations of individuals) within the community. The real question is this: Can a blockchain community be constructed in such a way that it is fault tolerant with respect to these attacks? Note that we are not just talking about run-of-the-mill failures within the network. Now, we are worried about attack strategies from outside the community. Does Byzantine fault tolerance still work for these attacks? Here again, the answer has to be that 'it depends'.

On the bright side, most of the attack vectors against such a network involve misinformation and deception, and a large portion of that is focused on encouraging people to doubt the integrity and honesty of their government and governmental actors. However, this strategy is only effective if there is some reason to think that important nodes within the community are being dishonest or corrupt, but in a well-constructed blockchain community, it is inherently difficult to manufacture this sort of distrust. If all-important community actions are onchain, and if all financial activity is onchain, and indeed, if those actions are the provable consequence of community voting and the execution of auditable smart contracts, then how is the deception supposed to work? It would have to rise to the level of getting people to doubt the reliability of blockchain technology itself. Of course, you can always get people to doubt proven technologies – there are flat-Earthers and people who challenge the Moon landing, after all – but our point here is that this is the degree of suspicion and doubt that has to be injected into a community. Furthermore, it may well be that the rise of conspiracy theories about the Earth being flat and the Moon landing being a hoax are fueled by very legitimate distrust of institutions of knowledge more generally. Bad acting on the part of governments and academic communities may well have opened the door to all forms of suspicion and doubt. Thus, it may be that blockchain communities (where deceptive governments are harder to maintain) will not yield the same propensity for conspiracy-theory-level suspicion.

Still, there are also things that can be done to protect communities against external attack. A community that wishes to be resistant to attack must ensure that community members have sufficient technological literacy to understand its foundational technologies. They need to not just *have* technologies that are corruption resistant, but they must understand *why* they are corruption resistant. If a community wishes to survive external attacks, it will need community members to understand its foundational technologies in addition to its foundational values.

This would apply not only to information-theoretic attacks but to more aggressive hacking efforts. Presumably, in a world of blockchain communities and cyberstates, known vulnerabilities are shared between communities, but certain powerful communities might have the resources to acquire zero-day exploits and use them against selected vulnerable communities.

It is difficult to see a way around this, but it has to be observed that the situation cannot be any worse than it is now, and there is at least some hope that if blockchain communities share similar technologies, then solutions to exploits would also be shared. By this, we mean that if everyone depends on the integrity of the network, then they have an interest in sharing knowledge of zero-day exploits (and their patches) with everyone on the network.

Similarly, no blockchain community under unjust attack is alone. There will always be similar communities with similar values, and if the attack can be identified, then other communities may choose to come to its victims' defence, just as defenceless nation states are often protected by a broader community of nations. Presumably, treaties between communities will be part of the manner in which blockchain communities protect themselves. Bad actors can be isolated, and any isolated community in a networked world is going to quickly starve itself of resources in terms of economic partnerships, cultural partnerships and other opportunities.

This is significantly different from the situation today, wherein just a handful of world powers have the capacity to lock an adversary out of the world's credit markets, freeze their assets and so on. As things stand, one power (for example, the United States) or a handful of powers, can lock a community out of the world's family of communities – that is to say, deny the target community's ability to maintain necessary economic and cultural relations. In our scenario, a community can be frozen out of some relations, but that is determined by the consensus of the communities in the network, not the decision of a centralised authority.

## 13.3   Kinetic attacks on blockchain communities

So far, we have been thinking in terms of attacks on blockchain communities that involve irregular warfare and, in particular, information-theoretic attacks. These might involve misinformation or propaganda or injecting false narratives among blockchain community members.

For many cyberstates, these might be the only viable vectors of attack since many cyberstates and virtual communities will not have physical territory, or alternatively, their physical territory may be widely distributed and not easily accessible or, for various reasons, not worth the cost of attacking with kinetic resources. But what of blockchain communities that have physical territory of some value? Suppose, for example, that a community held territory rich in natural resources like oil or rare earth minerals, and another actor used force to take the territory. Or imagine state-sponsored pirates that targeted the trade routes of the blockchain community. What then?

Here, we need to be clear on all the possible considerations in play. It is not as simple as in the standard nation-state model, where a state has well-defined territory that it must defend. Indeed, as we saw in Chapter 10, there is a serious question about whether states would continue to have sovereign control over physical territory in the age of blockchain communities. There may be an entirely different level of governance for that.

What do we mean by this? One possible outcome is a situation in which states are in charge of the wellbeing of their citizens but in which territory is under the sovereign control of no single state (just as today Bitcoin and Ethereum are under the control of no single state) and there is a transnational decentralised title registry that tracks ownership of land independently of questions of sovereignty. With this way of thinking, no singular state would have sovereign control over any piece of physical territory, and sovereignty would be distributed among a global group of communities.

Of course, even if there was a regime like this for territorial ownership, it does not forestall the possibility of rogue actors wishing to seize property by force, ignoring the claims of ownership recorded by the global community. For example, let us say a community purchases territory near a valuable oil field, and some group then invades the valuable property, seizes it and lays claim to the natural resources, ignoring global consensus on ownership and mineral rights. The first thing that must be observed is that such a property is certainly no less defensible in this scenario than it is now. Traditional claims of sovereignty would not have made this property and its resources more easily defended. Presumably, it will be possible to purchase military assets (just as it is now), to build bases (just as it is now) and recruit professional soldiers (just as it is now).

However, there is more to the story. If sovereignty is distributed and recognised globally, then preserving secure title to physical territory is also in everyone's interest. In any case, it is certainly in the interest of every other individual and community that owns title to territory. This is because the failure to preserve the title of one community undermines the holdings of every other community. This, in turn, suggests that there are always a host of natural allies to any community that is a victim of this sort of land seizure by force. Indeed, just as it is in every Bitcoin user's interest to recognise the authenticity of the ledger recording their ownership of bitcoin, it is likewise in every property owner's interest to recognise the authenticity of the ledger that records their real estate ownership. No one can afford to say that a specific wallet address is invalid, for doing so invalidates every other wallet address.

The situation we are describing is different from that of today in the following way. Presently, when a state is invaded, it is invaded by another state with a competing claim to sovereignty. What complicates this is that the sovereignty claim may involve any sort of justification, ranging from ethnic identity to some historical connection (real or imagined) or simply a vision for a new world order. All of these options and the resulting conflicts are possible because state claims of territorial sovereignty are inherently bogus. Such claims are justified by nothing of substance. They are based on invented histories, fables and imperial aspirations. We believe it is time to retire the traditional notion of states as territorial sovereigns. Instead, we think it is time to embrace the notion of context-dependent, overlapping sovereignties of the form articulated in Chapter 10. The effect of this reconceptualisation of territorial sovereignty is that many of the justifications for warfare are undermined – they are exposed as being empty. And one hopes that being thus exposed, they will no longer serve as triggers and fuel for kinetic conflicts.

## 13.4  Softwar

In the previous two sections of this chapter, we looked at how blockchain communities will engage in two forms of warfare: PSYOP and kinetic warfare. However, there is a third kind of warfare that does not cleanly fall within these two categories – let us call it softwar or, if you prefer, cyber warfare, and let us take it to involve hacking attacks, including worms, viruses, denial-of-service attacks and so on. Such attacks might accompany a kinetic attack (for example, with the goal of disrupting command and control operations), and they might be part of a PSYOP effort (for example, preventing a community from responding to a disinformation campaign). However, they are nevertheless a different animal. For that matter, digital warfare may be more than a different animal; it may, in the fullness of time, become the most important form of warfare. For blockchain communities, cyber warfare attacks are certainly a significant concern.

In his fascinating book *Softwar*, Major Lowery develops a theory of why softwar is not merely important but urgently important as it may ultimately become the future of warfare. To understand why, we need to dive deeper into his analysis.

Lowery has an interesting take on warfare, which is that it is primarily a tool of decentralisation. If you find yourself in a situation under the control of a system of governance that you do not consider ethically valid, and if there is no way to change that system or to exit, then you either have to accept the centralised authority or find some alternative way to resist it in order to achieve decentralisation. In some instances, the only alternative is to raise the physical costs of centralisation to the point where it is no longer worth the effort to fight to maintain centralisation.[6]

Parenthetically, we should not let it pass unremarked that kinetic warfare is also a predominant tool for governmental centralisers. History is full of empires that used force to establish centralised control over vast – sometimes global – territories. From the kings of Mesopotamia through the Roman Empire and the Spanish Conquistadors to the present day, kinetic warfare has been a popular tool of centralisers.

For the moment, let us entertain Major Lowery's 'war-as-decentralisation' thesis. It is certainly reasonable to argue that many colonial revolutions were successful because the revolutionaries were able to use physical means to raise the cost of maintaining a centralised government so much that the central authority decided it was no longer worth the investment. For example, the British Empire could have no doubt maintained its war against the American colonies much longer than it did but decided it simply was not worth the cost of doing so. A similar story played out in South America when Simón Bolivar resisted the Spanish Crown. Could the Spanish Crown have maintained its control of South America? Well, would it have been worth the expense?

---

[6] Jason Paul Lowery, *Softwar: A Novel Theory on Power Projection and the National Strategic Significance of Bitcoin* (2023).

This then is the key idea: to break centralised authority's control over oneself, one must raise the costs of maintaining control. Warfare – kinetic warfare – is the method used for imposing such costs. Now, the question becomes, what is the analogous way of raising costs in the case of digital warfare? Here, your initial thought might be that we can impose a cost by using various firewalls and software security measures to protect a community against hackers. However, Major Lowery claims that this is actually only effective in the short term – until an innovative hacker comes up with a new zero-day exploit to attack your digital infrastructure. The thing is, there are always exploits to be found. Simple security shields are not enough. One must impose an actual cost.[7]

For example, there are techniques for fighting spam in which a transaction on the network requires a micropayment. You can spam a million addresses if you want, but those micropayments add up, imposing a cost significant enough to make spamming not worth the trouble.

As Major Lowery notes, Bitcoin's proof-of-work protocol does essentially the same thing.[8] One can try to seize control of the Bitcoin network, but the barrier is not the need to break through some firewall. There is no firewall to break through. Everyone already has access to the network. The cost arises from the attempt to acquire enough hashing power to take control of the network. As we noted earlier, quite apart from the cost of buying enough miners with enough hashing power, this means paying for enormous amounts of energy – something in the order of the cost of Finland's energy usage.

---

[7] Ibid.
[8] Ibid.

Another way to look at this is that the energy demands of proof-of-work protocols can become the new version of kinetic warfare, at least as far as digital warfare goes. As Lowery observes, this is not some small piece of the warfare pie, but in the fullness of time, it will be the biggest piece of the pie and maintaining proof-of-work protocols is the new city walls or the new Patriot missile systems.

So far, it sounds like proof of work is a defensive strategy. Can it also be deployed as an offensive strategy? – for example, if a block-chain community or collection of blockchain communities chose to act against a rogue community that engaged in slavery or committed egregious environmental pollution or prohibited exit for its citizens? The answer to this question is not entirely clear.

On the one hand, if you had majority control of a proof-of-work network, it would be relatively easy to impose constraints on the network and censor network actors. For example, if there are bad actors on the global stage, and if we had the majority of hashing power in some critical blockchain application to which a bad actor needed access, then we could indeed punish them. We believe this is part of why Major Lowery sees an urgency in the United States having a robust presence in the crypto scene – if the US could control such critical networks by controlling a majority of the hashing power, it could punish its enemies very effectively.

Regarding Major Lowery's proposal, the first question has to be whether proof of work is really the deciding factor. Why not proof of stake? It is understandable why someone like Major Lowery, who comes from a military background, would see a similarity between proof of work and kinetic military activities. Both involve the expenditure of energy. In the case of traditional warfare, the energy is expended in the form of chemical energy released in the firing of shells and bullets or in the detonation of grenades and bombs. Thus, the potential force is maintained in chemical energy until it is released on the battlefield in the form of kinetic energy. Similarly, tremendous energy is expended in the transportation of forces to the battlefield and in the logistical supply chain. Again, we can think of fuel as being chemical energy that is translated into kinetic energy when troops and supplies are delivered to the battlefield. The energy expended in proof-of-work protocols is supposed to play an analogous role to the use of energy in the military context. But does it?

We are sceptical. Is the expenditure of energy in proof of work really what is doing the labour here, or is it merely an artefact of some deeper principle? We are inclined to think the latter. To see why we are sceptical, consider the question of whether a proof-of-stake protocol could not accomplish precisely the same thing as proof of work in this context. For example, in the case of proof of work, we noted that it simply would not be worth the trouble to acquire enough hashing power and use enough energy to take control of a proof-of-work network. Ultimately, this is simply to say it is not worth the financial cost of doing so. The energy is there if you want to pay for it, and the hashing power is hypothetically there if you have unlimited resources to pay for it. But is it worth the cost?

Of course, we can ask precisely the same question about a proof-of-stake protocol. Is it worth the expense to acquire enough of a cryptocurrency – let us say ETH – and stake it with the goal of gaining control of the Ethereum network? As of this writing, there is the equivalent of $90 billion staked in the Ethereum protocol.[9] Based on the assumption that you would have to purchase enough ETH and stake it to seize control of the network, you would be looking at a monumental expense. Keep in mind that any effort to acquire 50% of all staked ETH would send the asset's price skyrocketing like never before – even attempting to corner 10% of the staked ETH market would have an enormous impact on ETH price. It is very difficult to grasp just how expensive it would be to execute such a project. You also face the prospect of being

penalised – having your staked assets frozen or even seized if you are a bad actor in the network.[10]

We think the moral here is that the fundamental axiom of conflict is not the application of physical energy but rather the expenditure of value. When warfare is expensive, it is often because one side of the conflict determines that it is worth the expenditure of resources to try and impose their will. In the age of empires, it could be worth the expenditure because there was treasure to be gained from winning the war. In the current era, which we could call the neoliberal era, the prizes are free markets and all the financial benefits that flow from controlling those markets and the natural resources that can be unlocked. If the potential market is small, it may not be worth the trouble to subdue it.

What we are suggesting then is that the fundamental axiom of warfare has little to do with the amount of kinetic forces released or the amount of energy consumed and everything to do with questions of value and the expense (by whatever means) of acquiring something of value (be that mineral rights or open markets or treasure). Kinetic warfare with expenditure of energy is one form that conflict can take but certainly not the only form, and at the end of the day, it is not a reliable measure of power.

We can see this by reflecting on rich nations throughout history. They certainly did not always maintain standing armies and would purchase mercenary forces when necessary. They could also bring about their goals through what neoliberals today call 'soft power', which is just another way of saying that they can exert their financial clout in numerous ways – for example, by buying influence or gifting money or offering favourable trade relationships or, on the flip side, by threatening economic penalties and so on. Traditional warfare is another version of this – a threat to financially penalise the adversary if they do not conform to one's will. Proof of work is just one way of imposing costs. But then, so too is proof of stake.

---

[10] In this case, the assets would not be frozen by a centralised authority but by consensus of the nodes on the network.

All of this leads us back to the question of what recourse there will be to act against bad-actor blockchain communities. As noted earlier, the principal recourse is similar to soft power solutions widely used today. However, such soft power will be decentralised rather than centralised. There will be no global cop, like the United States or a centralised authority like the UN or the World Bank, but rather the collective action of a global community of blockchain communities.

As we noted, any nodes run by a criminal actor would be subject to penalty. In the case of proof-of-stake nodes, like those supporting Ethereum, staked assets could be forfeited. In the case of proof of work, blocking the offending party from the network is a possibility. Such forms of digital excommunication for a rogue community would be extreme but certainly possible in cases that required an extraordinary response – for example, in acting against a neo-Nazi empire.

Certainly, there is a danger that blockchain groupthink could take things too far and be too liberal in its use of excommunication, but as we said, such decisions would no longer be in the hands of a single power nor in the hands of a body like the UN, in which power resides in the votes of arbitrarily constructed and obsolete nation states. The power would lie in a decentralised network of global blockchain protocols like Bitcoin and Ethereum and whatever other protocols may prove critical to the conduct of government business.

To be sure, there will be cases in which global blockchain communities do not act in accord with our desires. It will certainly be frustrating to communities like the United States that believe in their 'exceptionalism', and it will be equally frustrating to actors that somehow manage to be frozen out of the global community because of their actions. We cannot promise that everyone will be happy with the results (an impossibility), nor can we even promise that the global community will always make the correct decisions. All we can really promise is that the decision will be decentralised, the actions will be decentralised and that it will be driven by the core values of the stakeholders in a global community.

But what if physical action is also necessary? What if removing an actor from the global community is not enough? What if it is a neo-Bronze-Age empire with no digital footprint causing physical harm to others? If the bad-actor community has a physical presence in the world, then we can bring the same kinetic forces to bear on it as we do today. As a last resort, physical harm may be met with physical force. However, if the bad-actor community has a distributed or nonexistent physical foot-print, there is little that one can do in terms of traditional military and policing actions. You cannot monitor or censor them online, and you cannot locate them in physical space either. So, what is the solution?

The first question that needs to be asked is how much harm such a community can actually do. If they have a minimal physical foot-print, there is likewise a limit to what they can do in terms of physi-cally oppressing individuals or harming the environment. This is not to say harm is impossible, but it is to say that environmental and human rights abuses are harder to execute by a community that is scattered around the world and without any territorial control. It is not, of course, unimaginable. Terrorist organisations, after all, are structured in this way, as indeed are many drug cartels. For that matter, there might be a distributed online community that maintains household slaves. The question is, what could be done in such cases?

Let us start with the example of the community keeping slaves. Assuming that our slaver community existed, there would be real-world victims in need of help. What then is to be done? There is no obstacle to initiating police action on a case-by-case basis, nor even to taking advantage of assassination markets if it came to that. If there are real-world victims of a community and its failed morals, there are real-world policing solutions, although this would be a new form of policing – decentralised in organisation and backed by globally shared values. The same form of decentralised police action would apply to the cases of terrorist organisations and drug cartels as well.

Of course, in the case of terrorist organisations and drug cartels, we are talking about complex adaptive systems which require a distinctive vector of attack. The optimal strategy is to attack such bad-actor communities in much the way Waltz envisioned. Once such a community is infiltrated and can be thus monitored, it will be possible to stress its network by targeting individuals (or individual nodes), assessing the effects on the network, developing new targets and proceeding in this manner. The effectiveness of such efforts would be a function of the strength of consensus among the global community that the harm is great and that kinetic action or PSYOP or softwar (or any combination of all three) is required. Again, all of this assumes that efforts to negotiate solutions have failed despite the application of blockchain technologies to assist those efforts.

Of course, a sufficiently wealthy and powerful terrestrial community might have the resources to forestall any attempt to police its actions and the actions of its individual members. In such a circumstance, kinetic warfare responses may not be a viable solution. However, it is very difficult to imagine that a community could acquire that much wealth while being cut off from the global blockchain community, here assuming that blockchain technology will be (with the help of AI and other incorporated technologies) the global engine of wealth. How much value in resources can an isolated bad-actor community actually sacrifice to outlast punitive action in these cases? As with all conflicts, including kinetic warfare, there are no guarantees. In each case, the targeted community has to decide if its continued behaviour is worth the price.

We have ended this chapter on a nuanced note, which is appropriate given the conceptual limits we encounter when we begin to think about bad actors and the strategies for dealing with such actors in a decentralised, post-nation-state world. However, this is not the only case where we encounter the conceptual limits to what can be achieved. In Chapter 15, we will explore some of those limits. Before we pursue that conversation, we need to dive deeper into the nature of the technology itself to better understand both its promise and its limits. We resume this exploration in the next chapter.

## CHAPTER 14

# A DEEPER DIVE INTO THE TECHNOLOGY

## 14.1  Preliminaries

Back in Chapters 5 and 6, we laid out the basic technical tools that underlie blockchain technologies. We discussed how proof of work and proof of stake worked, how smart contracts worked, and so on. That was enough for us to discuss some of the general conceptual issues we have addressed thus far, but along the way, we frequently promised that we would eventually go into more detail about those technologies. In this chapter, we make good on that promise.

In saying this, we do not mean we are going to provide the only way of fleshing out the technical details. In point of fact, there is more than one technological solution; there are many promising technologies that already exist or that are being developed or that someone will think of soon enough. Our goal here is to describe the strategies that we like best, which include the tools that have been developed by the Institute of Free Technology, with some words along the way about the technologies that inspired them as well as our current understanding of the technological landscape.

These strategies are the ones that we have selected to talk about, given the needs, interests and abilities of existing blockchain communities and their members. You may have identified other needs, interests and abilities and may prefer a different set of strategies. That is to be expected. Our goal here is simply to lay out one approach to the available technologies and explain how it works and why people were motivated to develop this package of technologies. You are, of course, free to disagree with us and suggest that alternative technologies would do a better job addressing the limitations we discussed. We would welcome the mere existence of such debates, as it would be yet more evidence of the vitality and promise of blockchain technologies and their role in securing better governance and better lives for all of us. There is no single correct path to human flourishing. There are many paths, although we believe they all run through some form of decentralised blockchain technology.

## 14.2   Durable, corruption-resistant, transparent archives

In Chapter 5, we discussed at length the importance of secure archives for human governance. As we saw, archives preserve the history of government decisions, of property ownership, of culture and of relations with other communities. Indeed, as we saw, it is arguable that the archive comes before the community and the state, or in any case, that the community and state cannot exist without the archive. Furthermore, if a state or community cannot reliably maintain records of its actions, then it undermines the ability of its members to know if they should exit because the community administration no longer conforms to their values. Similarly, if the records are not accessible, community members lose the ability to understand the relations their community has entered into, the actions it has taken and, again, whether it is conforming to community values.

Just as we saw that archives are very important, they are also very vulnerable. As we have seen in this book, in the past, they have been destroyed by revolutionaries, drug lords, conquistadors, counterrevolutionaries, invading armies, earthquakes and fires. They can also be lost, corrupted with bad data, hidden and otherwise made inaccessible (or incriminating parts can be made inaccessible). So, the issue is that for any sort of community, and certainly for blockchain communities and cyberstates, we desperately need some way to make such archives secure, incorruptible and accessible.

The solution that we alluded to in Chapter 5 was to deploy decentralised archives that would be Byzantine fault tolerant. The idea is that there will not be a single point of failure; one would have to take down a vast portion of the network to destroy the archive. Nodes in the network could be targeted, but the network could survive such attacks. Of course, that is a very general claim. It is time to go into more detail.

There are certainly many available options for distributed archives, including decentralised storage networks such as the InterPlanetary File System or IPFS (which is connected to the cryptocurrency Filecoin), Storj, Arweave and Sia. Some of these efforts have very good ideas for which we will advocate. To date, however, none of them offer a complete package of desirables.

Let us begin with some of the limits that a decentralised file service might run into. One problem, of course, is that nodes within a network are of varying quality. We are talking about hardware here, and hardware failures happen often, without warning, and it sometimes seems that they happen at the worst possible times. One solution to the problem is to replicate the data at multiple locations in the network as often as possible. On one extreme, this would mean reproducing everything at every node. However, this strategy is wildly inefficient, and it has a centralising effect in that only very large-capacity nodes can hold all the information relevant to the conduct of, for example, a cyberstate. The network would consist of a handful of nodes that had the ability to supply Amazon or Google levels of data storage.

You might think that we could get by with less redundancy if we can quickly repair a failure in the network. For example, a network in which two copies of every piece of data exist might be viable if you could repair the information loss the second a node went down. You would have to rebuild the redundancy as soon as you lost it. The problem is that two nodes that happen to have the same information might collapse at the same time. There is also the problem that constantly pinging nodes to see if they are online and available has a computational cost of its own.

Therefore, the solution seems to be that we want more redundancy than two copies of the information. However, whether the solution is to create three or ten or one thousand copies depends on two things: how quickly you need to identify the data loss, and how quickly you need to repair it.

There is also an additional problem here, which is that just because a node is supposed to have a certain piece of information does not mean that it does. A node in the network might claim to be keeping a piece of information safe in order to receive incentives of some form (i.e. payment), but it might be acting dishonestly. Thus, we need to assure ourselves that the nodes actually have the information they are supposed to. This will generate difficulties when the information that the node is supposed to hold is confidential – medical records, for example. Therefore, we want to know that they have the relevant information without having to see the data itself.

Another concern is the issue of how incentives are supposed to work. Obviously, a distributed archive only works if people support the network and maintain the nodes that they are supposed to, but if the incentive structure is suboptimal, we might find that the incentives are only appealing (or mostly appealing) to very large data centres. This would have the effect of centralising the network all over again.

Therefore, there are many potential issues, and we are not the first to worry about these matters. Let us discuss some of these concerns and connect them with previous attempts to allay them.

Earlier, we stated that we do not want (actually, cannot have) a distributed network with a huge number of redundancies. There has to be some redundancy, of course, but critically, we also have to know when there is a data retention failure. To put it another way, you have to repair data loss in the network, but before you can do that, you have to know that the data has been lost.

One solution, adopted by the Codex decentralised file storage protocol developed by the Institute of Free Technology, relies on 'erasure coding'. 'Erasure coding' builds upon an old idea known as 'Reed-Solomon codes', which were developed by Irving S. Reed and Gustave Solomon, staff members of MIT Lincoln Laboratory, in 1960. Their seminal article was titled 'Polynomial Codes Over Certain Finite Fields', and their idea ended up having many applications over the years, most notably in compact disks.[1]

The basic idea is this: to find out if a given piece of data has gone missing, you do not want to have to keep looking for the data itself or its absence. It is far more efficient to sprinkle some tracking information into the data – markers for the relevant data, if you will. For example, a given block of medical data might come with a code. You do not search for the medical data itself. You just search for the marker. If you ping a node for the marker and get no response, you have reason to believe that the data has gone missing, at least temporarily. Maybe the node has gone down. Or maybe bit rot has corrupted the medium of memory. It does not matter what happened; you just need to know that the information might have gone missing.

Of course, this only actually works if the node being pinged is a good citizen and is not trying to deceive you, for it is possible to dump the data that is supposed to be preserved and keep the relevant tracking marker. Some sort of auditing mechanism is required.

While erasure coding supports data loss detection, this alone may not be enough in Byzantine decentralised systems. Malicious nodes might try to implement a wide range of strategies to pretend they are storing information. Why would they do this? Maybe they are trying to cut costs and want to reduce expenditure on storage and bandwidth, but at the same time, they want to receive payment for data storage. Or maybe they are politically malicious and are part of a plot to disappear important cultural or legal information across the network. We need to occasionally probe or audit the information that was allegedly held.

---

[1] Irving S. Reed and Gustave Solomon, 'Polynomial Codes Over Certain Finite Fields', *Journal of the Society for Industrial and Applied Mathematics*, 8/2 (1960), 300–304 <https://www.jstor.org/stable/2098968> [accessed 16 January 2024].

This is a well-studied problem in the academic literature, and it goes by names such as 'proof of custody' and 'proof of space-time', among others. Most of the existing solutions rely on a frequent random sampling of data blocks across the whole dataset. During this process, storage nodes have to provide clear evidence that they are in possession of the data they say they hold.

These mechanisms are widely understood today, and for the most part, they do the job they are supposed to do, but with a couple of caveats. First, there is an issue of efficiency. There is a huge computational cost to conducting data audits – going through the data line by line. Then, for decentralised archives, there is the problem of what you are going to compare that data against to see if it is accurate. You cannot use a centralised register of information because then you have a *centralised* solution to the problem – the 'official' record has become centralised and thus becomes an immediate point of vulnerability. Beyond this, there is the issue discussed earlier in this section, which is that sometimes the information you are auditing is private. Take the example of medical records (or state secrets or whatever you like). You would like to pass the audit without releasing the information being stored to the auditor.

Codex and other distributed databases leverage zero-knowledge proofs to prove the possession of information. Given that we have mentioned zero-knowledge proofs multiple times without much in the way of elaboration, let us now provide more detail about these groundbreaking cryptographic protocols.

At its most abstract level, a zero-knowledge proof is simply a way of proving you have certain information or a certain ability without giving that information away or without exercising the ability. Note that this is stronger than simply supplying a marker or code as evidence that the data is held. This is much closer to a mathematical proof that you hold the relevant data.

Suppose that we have a computer program that we want to sell to you – let us say an image analysis program that can identify buried treasure (yes, this is a completely fictional example, but stay with us). You want us to show you that the program works, so you ask us to run it to demonstrate its capabilities. Sneakily, during our demonstration, we run the very computation that you want to have performed; the demo itself will reveal the location of the buried treasure. One solution is to carry out demonstrations in other domains, but you might argue that no test is really reliable unless you can see that it functions as promised in the domain in which you intend to use it. What are we to do about your demands for proof of ability?

This is where zero-knowledge proofs come in. Let us say we want to prove to you that our program can do what we say it can without giving away the thing of value (let us say we are selling a program to perform some task). We can do this in the following way. We provide a proof that anyone who can do the task you have in mind, let us call it Task 0 – a computation of some kind – can do this only if they can perform a certain logically related task (possibly including a collection of subtasks). Let us call this Task 1. To reiterate, Task 1 and Task 0 are related mathematically; in terms of computer science, Task 1 is reducible to Task 0. The proof relies on a challenge to see that the individual can perform Task 1. Since performing Task 1 implies the ability to perform Task 0, then we can (with reasonable probability) assume that the individual has done (or can do) Task 0. Keep in mind that Task 0 is the task you are interested in. It is the one for which you are ready to pay. The other task is not worth anything to you per se beyond the fact that it establishes that we can carry out Task 0. You can ask us to carry out enough of these computationally related tasks so that you are satisfied that our program can do what we claim it can. In effect, you issue a series of challenges to us in order to prove that our program can perform the tasks we claim it does.

Zero-knowledge proofs come in many forms. We are inclined to favour succinct non-interactive arguments of knowledge – better known as 'zk-SNARKs'. The concept of SNARK-type proofs has been in the literature for several years, and they have also been implemented by protocols such as Zcash and the zero-knowledge rollups (ZK-rollups) used by the Ethereum protocol in so-called 'layer-two applications' (for example, protocols like Polygon zkEVM). Here, we can think of a layer-two protocol as a separate blockchain that performs operations rapidly but uses a more decentralised layer-one protocol like Ethereum as its secure settlement layer.

The key difference between a zk-SNARK and a standard zero-knowledge proof procedure (like we used to verify our fictional treasure-hunting software) is that the zk-SNARK is, as the name suggests, non-interactive. This means that you do not need to issue a series of challenges to inductively establish the proof of knowledge. The idea is that a common reference string that both the prover and verifier share is sufficient to achieve computational zero-knowledge without requiring interactions.[2] The non-interactive nature of the proof also removes the necessity of direct communication between the proving party and the validating party. This means that anyone can take a zk-SNARK proof and validate it with the same level of confidence that any other party has. This would be particularly helpful to independent third parties that might suspect the proving party and the validator are colluding. Finally, non-interactive proof procedures are particularly useful in the case of blockchain protocols, as constant interactive queries would be computationally expensive.

A moment ago, we mentioned the ZK-rollups that are used by Ethereum layer-two protocols, and these are useful as well, for they offer proofs of the validity of batches of transactions instead of transaction-by-transaction proofs. This can be beneficial if we want to query multiple databases for proof of possession of certain data without requiring proof of every single query. Batched queries are more efficient.

---

[2] Manuel Blum, Paul Feldman and Silvio Micali, 'Non-Interactive Zero-Knowledge and Its Applications', in *Proceedings of the Twentieth Annual ACM Symposium on Theory of Computing* (Chicago, IL, 1988), 103–12 <http://portal.acm.org/citation.cfm?doid=62212.62222> [accessed 30 October 2024].

So far, we have been talking about failure detection, and we have argued that data loss can be detected in several ways. We can detect benign data loss through erasure coding, and we can detect malicious data loss through remote auditing using zk-SNARKs. However, it is not enough to simply detect the missing or corrupted data. There is also the question of what you are going to do about it. Data loss needs to be repaired, but it needs to be repaired efficiently.

One thought would be that as soon as you detect a loss of data, you should immediately repair it, but this turns out to be an inefficient solution. Usually, when a node goes down, it is a temporary problem, such as a power loss or required maintenance. Rather than engage in a computationally costly and potentially unnecessary repair process immediately, it would be better to wait to see if the node comes online again with the data intact.

That said, nodes do go offline permanently and data degrades for any number of reasons. We believe a strategy known as 'lazy repair' is best positioned to balance these concerns. The basic idea is that you can tolerate data loss for a while because you have enough redundancy in your system, but when that data loss crosses a certain threshold, you must initiate repairs and restore the system's necessary redundancies.

Codex, for example, implements a strategy that starts with an idea from the previously mentioned Reed and Solomon work – in particular, a strong Reed-Solomon algorithm in which multiple data blocks from a dataset can be lost before the dataset becomes irretrievable. This means that the network can tolerate multiple missing blocks and still be able to reconstruct the whole dataset quickly. This allows us to implement a bandwidth-efficient, 'lazy' recovery technique like that just described. The basic idea is that when your redundancy falls below a certain threshold, you execute the repair strategy. For example, where seven copies are routinely maintained, one might tolerate a loss down to four copies, after which the creation of additional blocks is triggered. This strategy saves the network from a constant desperate churn to repair every single piece of missing data.

Where does this leave us? The thought is that we can use a basket of existing technologies to not merely identify and repair data loss in the network but also to identify those losses and repair them in the most efficient way possible. It will not matter if those losses of data are accidental or caused by a malicious actor, we will be able to audit the data and repair it as necessary.

All of this leads to an important question: Who conducts and pays for all this work, and how are they incentivised? The answer, of course, is that the network itself has to carry out these activities, which need to be hard coded into the design of the network. The incentives are those we have discussed throughout this book in terms of the value of having secure archives. The costs are borne by whomever happens to participate in the network. It is not a cost that is felt directly by network members, although perhaps indirectly in terms of greater transaction fees.

However, in addition to the general role of the network in securing the archives, individual nodes have to do their part, too – nodes that need to be incentivised (in addition to time commitments, there are hardware and electricity expenses associated with running a node). Few people are going to run a node on a distributed network without some sort of compensation. Hypothetically, a network could demand that its members carry their weight by maintaining archival nodes, but another possibility is to incentivise people to run nodes by paying them through a system of rewards.

This leads to the last big issue regarding archives: What does the incentive structure look like? This is not a trivial issue at all because we want the incentives for nodes to be designed so that they help the network grow in desirable ways. As we noted earlier, a poorly designed incentive structure might simply pay the most money to those node operators storing the most data. However, this leads to giant data repositories and something that looks a lot like the centralised Internet we see today, with powerhouses like Amazon maintaining giant server farms. This, in turn, gives them a kill switch to shut down many protocols should they choose to do so or be asked to do so by nation states and other powers. What is the solution?

One idea is to structure the incentives so that they are very high for smaller stores of data and that they diminish as the set of stored data increases in size. There can be further incentives for data that is not widely replicated within the system or for data that is prioritised for some reason (medical records come to mind).

Now clearly, this strategy sacrifices some efficiency for more decentralisation. However, such a sacrifice is well worth it for all the reasons addressed in this book. Decentralisation leads to many goods, including securing communities against corruption – a phenomenon with astronomical costs. Another way to put the point is that forsaking decentralisation to save a small amount of financial resources is penny wise and pound foolish.

## 14.3   Decentralised, secure communications

Effective states and communities need not only secure and transparent (when appropriate) archives but also private rails upon which their community members can communicate. Community members may wish to communicate with each other about business strategies or inventions or ideas for new technologies. They accordingly need to know that their communications on these matters are secure.

However, communities also need to have a secure means of communication to discuss political matters. The fact that blockchain communities and cyberstates find people to be largely politically aligned does not mean that no important political disputes arise. People need to be able to discuss these in private until they feel ready to go public with their ideas.

There is, obviously, an interesting asymmetry here in that the state itself should have open and transparent communications while individuals need private and secure communications. Persons in state positions will have access to private communications protocols, and these can be abused for purposes of state business. This is a problem we will return to but in the meantime, for us to even begin this discussion, we need to talk about secure communications.

You might think that we have secure communications now, and to some extent, we do. However, there are important limitations in the current system of communication networks and communication technologies. Much of the infrastructure for encrypted communications is highly centralised, with all of the dangers that this entails. For example, while we can communicate using Pretty Good Privacy (PGP), and while this affords us the protection of military-grade encryption, there are points of failure. The key server, for example, is housed in a centralised location. Our existing communications network can refuse to carry encrypted communications. Or they can refuse to carry encrypted communications from a particular source. Furthermore, although we can use encrypted communication protocols, it is still possible for others to know that we are the ones communicating.

It is worth reflecting on why this is important. Sometimes, the meta information of who is talking to whom is even more important than what they are talking about. A lot can be extracted from this metadata. There is the network of people communicating, the time they are communicating, a reasonably good idea of the amount of information they are communicating and one can determine who is at the centre of the group of people communicating – the hub, as it were. This is an issue for any sort of communication of political strategy, but it is also an issue for business dealings. Business leaders might wish to communicate about a potential merger without signalling that they are communicating with each other.

Therefore, there are many issues to contend with. Perhaps the biggest of them involves the possibility of nodes being restrictive about the information that they let pass. Should nodes have the ability to censor the flow of information through the network? One would hope not, although censorship in decentralised nodes is not at all unheard of.[3]

Of course, we have technologies that allow us to encrypt our communications, so the problem is not that someone might read our communications and then censor them. The problem is that someone might censor our communications based on their origin or destination. Someone might choose to censor nodes that originate in right-wing or left-wing or politically agnostic communities. Or they might censor communications that are addressed to such communities. Or they might choose to censor communications of a certain size because they indicate a certain degree of interest in contemporaneous political events. Can nodes do that? Sure, they can do it now because communications must propagate through the network, and this means that each node in the network is responsible for passing packets of communicated data around the network.

Fortunately, protocols like Waku – developed by the Institute of Free Technology – address these problems. The desideratum is to find a strategy in which network nodes pass communications while being blind not just to the content but also to the source, the destination and the amount of information. How can we do this?

The central idea behind Waku is that packets of information passed through the communications network are encapsulated in a way that does not reveal content or a known author or recipient. The nodes must simply pass the encapsulated information on. But how do we obscure the metadata?

The metadata can be blurred in several ways. First, to disguise the amount of information being sent, short messages can be filled with additional junk information. Large messages can be broken into pack-ets. All of these messages will not be routed directly from point A to point B but rather will be routed randomly through the network in a way that is blind to their ultimate destination. All a node needs to know is if a particular capsule is for that particular node. It can ping it using zero-knowledge proofs. If it is for them, the capsule will verify this; if it is not, the capsule will not verify ownership, and it will be passed on to other nodes.

---

[3] For example, Bitcoin developer Luke Dashjr (Luke-jr) has advocated that Bitcoin transactions involving 'ordinals' be censored. See Frederick Munawa, 'Among Bitcoin Developers, Debate Is Raging Over Whether to Censor Ordinals BRC-20s', *CoinDesk*, 5 December 2023 <https://www.coindesk.com/tech/2023/05/12/among-bitcoin-developers-debate-is-raging-over-whether-to-censor-ordinals-brc-20s/> [accessed 30 October 2024].

Now clearly, we want the transfer of information to be efficient despite all this, and one of the key ways to achieve this is to organise the network along the lines of a scale-free network. Such networks are designed with several densely integrated hubs connected to many local nodes but also connected to other centralised hubs. Scale-free networks like this are familiar in nature – the human brain and airline flight networks being cases in point. For example, most airlines have a handful of centralised hubs, each of which is densely connected to regional airports. This sort of network (explained by Duncan J. Watts in his book *Small Worlds*) gives rise to the 'six degrees of separation' phenomenon. Such networks not only seem to emerge from natural phenomena and human activity but are wildly efficient from a math-ematical point of view.[4]

Therefore, we can think of a communications network for a block-chain community as having this sort of organising principle, with com-munications encapsulated and passed throughout the network quite rapidly thanks to heavily interconnected hubs and with the metadata obscured thanks to the sender being known only to the addressee and the addressee only known to the sender. Even the amount of infor-mation within the capsule will also be unknown, as small batches of information can be sent with junk information, and large batches of information will be split up randomly. The idea would be to have each capsule contain precisely the same number of bits of information.

---

[4] Duncan J. Watts, *Small Worlds: The Dynamics of Networks Between Order and Randomness* (Princeton, NJ, 2003).

At the end of the day, this strategy, as embodied in Waku, provides a private and secure means for network citizens to communicate with each other while obscuring who is communicating with whom, how much is being communicated and, if we wish, obscuring the time of communication as well. Most importantly, there will be no opportunity for nodes in the network to censor the information they are passing along, as they will have no idea what its content is, who its sender is or who its receiver is.

## 14.4   Cryptocurrencies and sound monetary policy

We can hardly conclude this chapter without saying something about cryptocurrencies. We have avoided talking about cryptocurrencies throughout this book (except to use them as illustrations) because we want to highlight newer applications for blockchain technologies – in particular, those applications that support decentralised governance. However, cryptocurrencies are, of course, not peripheral to this project. They are essential to any attempt to provide decentralised-yet-cooperative human governance.

This should not be surprising. It would be foolish to offer a platform for decentralised governance and not apply it to currencies. Traditional fiat currencies and traditional finance have all the drawbacks that we have discussed throughout this book – they offer up centralised points of failure, are vectors for attack and are magnets for corruption. And, let us be real: any account of human governance that avoided the fiscal sector of human governance would hardly be taken seriously. Managing monetary policy and regulating the financial sector is one of the principal elements of governance today and has been a central element of governance for millennia.

The fundamental question here is not whether decentralised-yet-cooperative governance should avail itself of cryptocurrencies, but rather what form of cryptocurrencies should be utilised. We can sharpen that question in the following way: Should blockchain communities utilise off-the-shelf, globally recognised cryptocurrencies like BTC and ETH, or should individual communities mint their own tokens with which to carry out their business? The answer is: Why not both?

In the first place, it would be foolish for any blockchain community not to allow the usage of core cryptocurrencies like BTC and ETH. They, at least as of this writing, are monetarily sound; BTC will eventually reach its hard-capped supply limit of 21 million coins, and as we write this, ETH is sometimes already deflationary – more ETH is burned in a transaction than minted as staking rewards.[5] Furthermore, the more widely the currency is circulated and the more widely decentralised a cryptocurrency network, the safer it is. We have already discussed the monumental amount of resources it would take to initiate a successful attack against either of these networks. One would either have to acquire massive amounts of mining resources (in the case of BTC) or stake a massive amount of ETH. Plus, the nodes for both protocols are already widely scattered around the globe. Given these advantages, why would one opt for a community-specific cryptocurrency?

There is utility to having individual blockchain communities issue their own cryptocurrencies. In doing so, the community is maintaining its own ledger and recording value that is inherent in the operation of the community. Such cryptocurrencies might be used to measure stake in a DAO or they might be used to reward people making contributions to the community or they might be used to carry out business that is of interest to the community in a cost-efficient way. Thousands of web3 protocols have issued their own tokens for these and other reasons. Doing so is fairly simple, and as the protocol grows, such issuance can give rise to the accumulation of wealth by holders within the community.

---

[5] <https://beaconcha.in/burn> [accessed 30 October 2024].

The problem with a community-issued cryptocurrency is that if the community is smaller, it means the value of the network will be less and a 51% attack of the type discussed in the previous chapter is always a possibility. Is there a way to enjoy the community-specific features of a community-based cryptocurrency while simultaneously preserving the security and safety provided by coins like BTC and ETH? Certainly. The trick is to anchor the community-based cryptocurrency in the more decentralised global coins, using the latter as a kind of settlement layer.

There are many ways to ground a community-based cryptocurrency in a protocol like Ethereum, for example, by issuing the community-based cryptocurrency as a layer-two coin – something analogous to coins like Polygon's POL and Arbitrum's ARB, albeit presumably smaller in scope. In any case, one idea would be to issue the coin as a kind of ZK-rollup of the form we discussed earlier in this chapter.

Here is the idea. Let us say that our blockchain community issues a layer-two coin called Community, or CMTY for short. That, in effect, means that we have a layer-two ledger that carries out computations, records information, and keeps track of transactions and who owns what for our community. However, if it is a ZK-rollup, we can also do two other things. Firstly, we can use zero-knowledge proofs to prove that the transactions are being carried out according to some specified set of rules. Secondly, we can assemble those layer-two transactions into batches and, periodically, permanently record the history of those transactions on the layer-one network – in this case, on the Ethereum blockchain. Thus, we get the best of both worlds. We get a ledger, smart contracts and so on that are dedicated to our community and its interests, and we can offer proof that those transactions and computations are valid. However, we can anchor the results of those transactions on the Ethereum blockchain and benefit from that blockchain's robust security, which is, in part, a product of its global node distribution.

A related strategy is to borrow from the Eigenlayer protocol and 'restake' ETH into the new token. Staying with our example of CMTY, this means that one first stakes one's ETH in a protocol like the liquid staking service Lido, yielding stETH, which represents staked ETH. In doing so, you have staked your ETH to help secure the network and will earn financial rewards for doing so, but now you have a token – stETH – that is tremendously valuable. One could then take that stETH and restake it as the security layer for CMTY. In this way, it would be extremely costly to try to seize control of the CMTY network. There would be real losses if the staked assets had to be surrendered as a result of penalties levied against such bad actors.

If that sounds too good to be true, we must confess that there are limitations to this strategy. In the next chapter, we will see that there are conceptual limits to decentralisation, and layer-two protocols certainly reveal points of centralisation. The first problem begins with bridging information and cryptocurrencies from layer one to layer two. As we write this, there are no established decentralised cross-chain bridging solutions. Beyond this, layer-two protocols often operate on a single or a small number of dedicated servers. This means that even though we can monitor the operations and query for proof of their validity, if those transactions take place on a single server or a small number of servers, there are points of physical attack. Now, perhaps these difficulties can be overcome with layer-two redundancies, but this would lead to sacrificing the efficiency that we come to expect from layer-two protocols (if you have carried out transactions on Polygon or Arbitrum, you know how much faster and cheaper they were than similar transactions on Ethereum). Our point here is not that there are no solutions but simply that there are tradeoffs to any solution – security for efficiency, for example.

We could continue delving into technical issues, but you probably already grasp our central point. There are technologies that are already well understood and some newer ones that can be combined in different ways to pursue the goals that we outlined in the first thirteen chapters of this book. Because there are conceptual limits to what we can accomplish, and because there will be tradeoffs in any strategy we pursue, it would be foolhardy to say that there is one single best strategy that we should follow exclusively. Indeed, different blockchain communities will doubtless settle on different technology stacks.

Acknowledging that there is no single best approach to blockchain technologies, we can still offer tools that we feel are optimal for building such communities. In this chapter, we have already mentioned Codex and Waku and, previously, Status. However, there is one piece of the technology stack that we have not mentioned yet – Nomos. We have waited this long to introduce Nomos because, at its core, Nomos is what this book is about. Nomos is a layer-one blockchain that is optimised for human governance. It is a platform designed to help us build blockchain communities that meet the desiderata that we have discussed in this book.

We encourage readers to explore and apply these technologies, adjusting them as necessary and adapting them to their needs. You may have entirely different ideas about how to proceed, but that is also completely fine. We are not merely advocates of decentralisation for finance and governance but for the enterprise of building out new technologies as well. As we stated earlier, there is no best way to proceed here, and if there is, we certainly do not know which way that would be – no one does.

You can learn more about the elements of the technology stack we prefer at the following websites: Codex (https://codex.storage/), Waku (https://waku.org/), Nomos (https://nomos.tech/) and Status (https://status.app/). Those interested in these technologies and the principles behind them may also find value in the rest of the Institute of Free

Technology's portfolio (https://free.technology/). Feel free to use these technologies or ignore them as you see fit.

There are some additional issues that we need to address. These recommended technologies, like any technologies, do not operate in isolation. They are technologies designed to be used by humans but by humans that hold certain values. Nothing works if we are not successfully aligned with our technologies, or perhaps, it would be better to say if they are not aligned with us. However, there are other conceptual issues that we also need to address – for example, whether anything can truly be trustless and whether anything can be truly decentralised. We turn to these fascinating conceptual issues in the next chapter.

# CHAPTER 15

# CONCEPTUAL LIMITS OF BLOCKCHAIN GOVERNANCE

## 15.1 Preliminaries

So far, we have focused on the promising aspects of blockchain governance. We discussed how it might work and what it might accomplish. However, no technology is without limitations. There are, of course, technical limitations to getting it up and running – fixing bugs and whatnot. However, there are also conceptual limitations to what it can accomplish, and it is important that we address some of them.

Blockchain governance is supposed to be 'trustless', but as we will see, there are limits to how trustless a crypto protocol, or really anything, can be. At some point, everything bottoms out with human beings. Second, there is the issue of centralisation and whether anything can be fully decentralised. Nothing really can, nor is it clear that it would be desirable, even if possible. Third, there is the issue of transparency. We have spoken about how information on the blockchain is visible to all, but there are important caveats that need to be added. The information is certainly there, but how many people actually have the ability to interpret it? We take these and other issues up in the following sections.

## 15.2   Nothing is 100% trustless

On 20 July 2016, about a year after Ethereum launched, Vitalik Buterin announced a hard fork of the protocol. By making that announcement, Buterin shattered certain tightly held assumptions about the future of trust. He also incensed many people.

To understand how, we first need to discuss trust and its place in the fabric of our lives. Trust might be in short supply these days, but we have no choice but to rely on it. We trust schools and babysitters to look after our children. Some still trust banks to hold our money and to transfer it safely for us. We trust insurance companies to pay us should we encounter some disaster. When we make a large purchase – such as a house – we trust our solicitors or an escrow company to hold the funds until the transaction is complete. We trust regulators and governments to make sure these institutions are doing what they are supposed to be doing.

Sometimes, however, our system of trust fails us. There are runs on banks. People lose faith in currencies issued by nation states. People stop trusting their political institutions because of the chicanery, short-sightedness and general incompetence of the self-interested people running the show.

Blockchain technology is often characterised as being 'trustless', meaning that we no longer need to trust fellow humans – we can trust the algorithm. However, when viewed on a conceptual level, this is not accurate. An example alluded to in this section's opening can help us understand why.

In April 2016, the first DAO was created. It was called simply 'The DAO', and about 11,000 people contributed a total of $150 million to take part. Contributors to The DAO believed they had bought a share in a virtual hedge fund that would invest in other companies and ventures. Anyone wanting to receive funding from The DAO had to submit a proposal online in the form of a self-executing contract that

DAO shareholders would then vote upon. If approved, The DAO was programmed to automatically transfer the agreed allocation of ETH.

In theory, shareholders did not have to worry about the good intentions of The DAO's employees, for this was the type of DAO that had no employees; nor need they have to worry about the competence of supervisors or executives because there were none; nor did they have to ask lawyers to go over the fine print, for there was no fine print to go over. They would have no need to trust courts and police and attorneys to enforce the contracts because the contracts did it themselves. All they had to do was look at the software code in the smart contracts, see what the program (that is, the organisation) would do and choose whether or not to buy in.

On paper, the scheme appeared to be flawless. However, it quickly transpired that it was not. On 17 June 2016, someone – we still do not know who – successfully hacked The DAO. The hacker syphoned off the equivalent of $50 million into a second DAO contract that they had deployed, which subsequently became known as the 'Dark DAO'. When this flaw in the code was detected, other stakeholders used the same exploit to move the remaining ETH into a third DAO, known as the 'White Hat DAO'. Then, all the existing accounts in the three DAOs were frozen.

But what to do with the money in the Dark DAO and the White Hat DAO? Some argued that, as the hacker was only doing what the software allowed, the ETH in the Dark DAO rightfully belonged to the hacker. And why was one DAO called 'Dark' and the other 'White Hat' – were both hacks not undertaken using the same code? And was the code not the law?

This brings us, finally, to what infuriated people – the fork (actually, the forks, for there were two forking options: a soft fork and a hard fork). A soft fork would be backwards compatible, meaning that the nodes that did not upgrade their software would still be able to operate. However, the hard fork option was another matter entirely. Among other things, it would undo previous transactions (reverse them, in effect). In this case, it would take the money back from the Dark DAO and the White Hat DAO and put it back in the hands of the hood-winked investors.

But in a trustless universe, who decides if the fork happens? This is where the miners enter the story. At the time, Ethereum was still a proof-of-work protocol, and miners did the grunt work of sealing transaction data into blocks. While Buterin and the Ethereum Founda-tion could *propose* a fork, ultimately, the decision was in the hands of the Ethereum miners. They were the ones who had to mine the revamped Ethereum code and keep the whole system running.

On 20 July 2016, Buterin announced that the miners had accepted the hard fork and were now mining with the updated code. The reality was that *most* of them had. A number of holdout miners and Ethe-reum users were outraged by the decision to hard fork the protocol. In their view, the hard fork undermined the core principle of Ethereum, which was, after all, to bypass all the meddling humans – the corrupt bureaucrats and politicians and board directors and CEOs and law-yers. Code was supposed to be law. If you did not see the weakness in the software, that was your problem since the software and all its code were publicly available.

Thus, some Ethereum miners refused to run the updated software and instead stayed with the original Ethereum protocol, which they redubbed Ethereum Classic. You would think that would be the end of it, but no. Shortly after the hard fork of Ethereum and the network split that created Ethereum Classic and what we now call simply Ethereum, a further round of technical problems were identified with the Classic protocol. Soon, there was a counter-proposal to hard fork Ethereum Classic, which led to the inevitable threat by the true believers that they would respond with an Ethereum Classic Classic.

Such are the perils of supposedly trust-free technology. It might make for good marketing copy, but as we have seen throughout this book, blockchain technology incorporates many components that involve trust. First, you need to trust the protocol of the cryptocurrency and DAO. This is not as simple as saying, 'I trust the maths', for some actual human (or humans) wrote the code and hopefully debugged it, and are we not at least trusting them to get it right? Well, in the timeframe leading up to The DAO, maybe they did not get it right.

Second, one has to trust the stakeholders (whether miners or validator nodes) not to destabilise the protocol with a hard fork. One of the objections to the hard fork was that it would create a precedent that the code would be changeable. However, this objection exposes an unmentioned universal truth – it was always changeable: the immutability of the blockchain is entirely a matter of trusting other humans not to fork it. Ethereum Classic Classic would be no more immutable than Ethereum Classic, which was no more immutable than Ethereum. At best, the stakeholders – all humans – were showing that they were more trustworthy about not forking the blockchain. At the same time, they obviously *could* change their minds about forking at any time. In other words, if Ethereum Classic is more trustworthy, it is only because the humans behind it are more trustworthy.

Third, if you are a non-technical individual buying into Ethereum or The DAO or any other DAO, you are being asked to trust the people who review the algorithm and tell you what it does and whether it is secure. However, those people – computer scientists, say – are hardly incorruptible. Just as you can bribe an accountant to say that the books are clean, so too can you bribe a computer scientist to say that the code is clean. Moreover, you are putting your trust in whatever filters you apply to select that computer scientist. Was it university or professional qualifications? A network of friends? The testimonials of satisfied customers? – which is to say, the same method by which people selected Bernie Madoff as their financial advisor.

Finally, even if you had it on divine authority that the code of a DAO was bug free and immutable, there are necessary gateways of trust at the boundaries of the system. For example, suppose you wrote a smart contract to place bets on sporting events. You still have to trust the news feed that tells you who won the match to determine the winner of the bet. Or suppose you wrote a smart contract, the terms of which stated you were to be delivered a truck full of orange juice concentrate. The smart contract cannot control whether or not the product is polluted by lemons or some other substance. You have to trust the humans in the logistics chain and the humans at the manufacturing end to ensure your juice arrives unadulterated.

Can these gateways to the system not be trustless as well? Can smart contracts not someday be written to contain code to call for robotic orange pickers and robotic juice concentrate makers who would summon their robotically driven trucks to deliver the orange juice concentrate straight to our door? Yes – in theory. However, imagine the task of reviewing the code to ensure that every step in the process had not been corrupted by a bug that uses security failures to hijack trucks or that gives false approvals to adulterated orange juice. Perhaps we could write second-order programs to automate the testing of the first-order programs – but why do we trust those? Do we ultimately need automated program-tester testers? Where does it end?

By now, the answer should be obvious: it ends with other humans. Blockchains do not offer us a trustless system but rather a *reassignment* of trust. Instead of trusting our laws and institutions, we are being asked to trust stakeholders and miners and programmers and those with sufficient programming skills to be able to verify the code. We are not actually trusting the blockchain technology; at the foundational level, we are trusting the people who *support* the blockchain. In the end, we have to trust people. Therefore, blockchain technology is not trustless; it is rather a kind of distributed trust. We do not trust a single centralised organisation, but we trust a large network of individuals to continue to do the right thing.

## 15.3   Nothing is 100% decentralised

In the previous section, we talked about trust and the Ethereum protocol. Let us stay with the example of Ethereum while we pursue another question: Can any blockchain network be truly decentralised?

Ethereum is now a proof-of-stake protocol. The greater the value of assets staked on a validating node, the greater the chances of that node being permitted to select the next block. But who exactly is assembling those blocks? And how does that process actually work? And is Ethereum actually decentralised? As we will see, decentralisation comes in degrees.

Let us start by getting into the nitty gritty of the Ethereum protocol. What happens when you engage in a transaction on the Ethereum network? Let us say you are sending ETH to a friend or you are buying ETH with DAI or you are depositing ETH and DAI into a smart contract – maybe you want to contribute it to the liquidity pool on a decentralised exchange like Uniswap. The first step is that you confirm your transaction in your wallet. Maybe at that point, you visit a blockchain explorer like Etherscan to get an idea of how long your transaction will take. Maybe you are waiting to see the green message 'Success!' What is going on while you are waiting for that to happen?

As it turns out, a *lot* is going on. If you have made transactions on Ethereum, you have probably noticed that sometimes your transaction resolves quickly and sometimes it seems like an eternity. Maybe, in your frustration, you increased the gas fee (the transaction fee) you were willing to pay to make it happen. While you were sitting there, waiting for your transaction to process, a lot was going on. For starters, your proposed transaction was sitting in the 'mempool' (short for memory pool).

The mempool is where proposed transactions wait, hoping to be selected to be put into blocks. Picture it this way. Think of proposed transactions as people hoping to get on the blockchain train. Let us say that there are several train station employees tasked with selecting bundles of people to load into a train car, but only if they have been chosen to load passengers onto the next train. For each train, only one of those bundlers is awarded the job, and different bundlers may have different strategies for selecting groups of people to load onto the train.[1]

Obviously, the bundlers want to make as much money as they can, and one obvious way to go about it is to load the people who are waving the most money at them. If you are not offering money, you are probably going to have to wait until a loader selects you to board the next train or the next one or the one after that. However, it is not only about who waves the most money. Sometimes, you can gain a different kind of edge.

The train loaders know more than just who wants to get on the train; they also know their intended destinations. This opens up new opportunities to profit. If you notice that a lot of people are trying to buy passage to Albany, for example, then you now have a piece of valuable information. Maybe a lucrative business deal will occur there. With that information, you can buy up some of the tickets in advance and sell them at a scalper's markup – that is, you can 'front run' the transactions people are trying to make.

In the parlance of Ethereum, this is called 'maximum extractable value', or MEV. Now, we can argue about whether there is something wrong with trying to capture MEV. On the one hand, we want people to be incentivised to do the work of forming blocks. On the other hand, it would be preferable if people did not front run our intended transactions. However, in the case of Ethereum, two things must be noted. First, as matters stand, there are not that many people constructing blocks for Ethereum (just a few thousand). Second, those who do it have a tremendous amount of power. This leads one to question just how decentralised things actually are.

---

[1] For a crude visualisation of this metaphor, see <https://txstreet.com/v/eth-btc> [accessed 30 October 2024].

To sharpen our discussion, let us move from our metaphor to a description of what exactly is occurring on Ethereum. The train loaders in our metaphor are nodes on the Ethereum network. We have mentioned them extensively in this book, but what exactly are nodes? As of this writing, approximately 6,500 Ethereum nodes are scattered around the world.[2] They play the role that our Byzantine generals and our scattered rulers of Paxos did. However, to understand them more deeply, let us start with Ethereum client software. There are several varieties of this software, but let us say you download a client software called 'go-ethereum' – or more commonly, 'Geth' – install that client on your computer and use it to connect your computer to the Ethereum network. Now, you have a node up and running. Therefore, the node can be thought of as an operational instance of the Ethereum client software.

Digging a little deeper, we should point out that there are different kinds of nodes: full nodes, light nodes and archive nodes. Full nodes download all blocks from the blockchain, storing them on their hard drive. This permits users to verify transactions directly. Light nodes only download those blocks pertaining to their own account balance and, as such, might be used as user wallets. Archive nodes store all data from every block ever created and build an archive of past blockchain states. These are used by blockchain explorers like Etherscan.

Each full node has approximately twenty peers on the network with which they are directly connected. When your proposed transaction arrives at a node, the node propagates it to its peer nodes, and those nodes propagate to their peers until your proposed transaction is scattered across the network. However, the transaction is not yet on the blockchain. It is still in the mempool. Or more accurately, it is in the mempools because, technically, each full node has its own mempool, and the composition of each mempool will vary from node to node.

The work of the nodes is to package transactions together into blocks. In the days of proof of work, these nodes would involve miners

<hr>

[2] <https://www.ethernodes.org/> [accessed 20 April 2024].

that would compete for the prize of getting to submit the official block. Today, because of proof of stake, you assemble the blocks in the hope that it will be your turn to propose the next official block.

The problem is that each of these nodes is a potential point of failure. Now, of course, that is to be expected because, theoretically, blockchain technology is designed to be Byzantine fault tolerant. However, what if some centralised authority were to identify each of those 6,500 nodes and apply pressure to their operators? For example, suppose that the centralised authority demanded that they censor certain transactions from their mempool and pressured nodes into censoring blocks that contained those unwanted transactions (for example, payments to WikiLeaks). That is, what if the nodes were forced to never assemble blocks that contained the undesirable transactions?

We do not need to speculate about such a possibility because it is happening right now as we write. The United States government recently sanctioned Tornado Cash, a protocol that inputs several crypto transactions from different sources and 'mixes' them so that the flow of money cannot be tied to a particular wallet. When the US government cracked down on Tornado Cash and arrested its principal developer (on the grounds that it was a money laundering tool), it was able to take individual Tornado Cash servers offline and, obviously, detain the people running those servers.[3] As for Ethereum nodes, 6,500 is not an impossibly large number of targets for a nation state like the United States. For example, the FBI, in partnership with local and state law enforcement, arrested approximately 6,000 alleged violent criminals and gang members between 1 May and 2 September 2022.[4] Of course, any such operation would be international in scope, but there is little to constrain the United States from organising such an enforcement operation on a global scale (the Tornado Cash action shows that the United States can and will act outside its borders in collaboration with the law enforcement agencies of other countries).

---

[3] Sanction Scanner, 'Tornado Cash: A Crypto-Mixing Service Now Blacklisted by the US Treasury', *Sanction Scanner*, 2024 <https://www.sanctionscanner.com/blog/tornado-cash-a-crypto-mixing-service-now-blacklisted-by-the-us-treasury-675> [accessed 30 October 2024].
[4] U.S. Department of Justice, 'FBI and Law Enforcement Partners Arrest Nearly 6,000 Violent Criminals This Summer', *Office of Public Affairs*, 2022 <https://www.justice.gov/opa/pr/fbi-and-law-enforcement-partners-arrest-nearly-6000-violent-criminals-summer> [accessed 30 October 2024].

To be clear, we are not saying that governmental actions against the Ethereum network are inevitable or even likely. Given our proximity to the crypto sector, we hope they are highly unlikely. However, our goal here is to explore the conceptual limits of decentralisation, and it cannot be denied that conceptual limits exist. In the ideal world, perhaps everyone would be running a node on their smartphone so that there would be billions of nodes scattered around the world. That would certainly be close to being safely decentralised. However, it needs to be asked if a hyper-decentralised network like that would even be optimal.

As we noted in the previous chapter, recent work on the science of networks suggests that the most efficient networks are so-called 'scale-free' networks, which are networks in which not everything is equally interconnected but in which there are certain important hubs that have many network connections. As we highlighted, this is the design of air traffic routes, for example, in which there is a group of heavily interconnected airline hubs. Your brain is organised this way, too. And it is because of this network architecture that we get phenomena like six degrees of separation.[5]

If the Ethereum network (or any network, for that matter) was completely flat and every node connected to the same number of local nodes with no node being more densely connected, we would be looking at a very inefficient network – so inefficient, one might wonder if it was even functional. To illustrate this problem, imagine that airlines eliminated their hubs so that there were no direct flights from New York to Chicago or to Atlanta, but that each flight had to be to the nearest commercial airport. A flight from New York to Los Angeles would be a nightmarish experience with dozens of stops. It is much better to have some degree of centralisation in the form of heavily connected hubs.

---

[5] For an accessible introduction to network theory, see Watts, *Small Worlds*

Another way to decentralise would be to connect every airport with every other airport. However, this would generate inefficiencies of its own. Again, it would be an absolute nightmare to have every single airport in the country having direct flights to Los Angeles and, for that matter, to Sioux Falls, South Dakota. Therefore, the most efficient network needs to allow some degree of centralisation. This example illustrates the tradeoff between decentralisation and efficiency. On the one hand, complete decentralisation eliminates points of vulnerability but, on the other hand, too much decentralisation can make a network functionally useless.

We might decide that some degree of network centralisation is a good thing or perhaps even a necessary thing, but there is another kind of centralisation that we need to consider – the power and influence of founders and leading figures. We could call this 'social centralisation', and the concept deserves further discussion.

## 15.4  Social centralisation

We have been talking about blockchain technologies as though they offered a level playing field in which everyone, in the abstract, had the same power and influence. Of course, this cannot be the case in practice. Leaders do emerge, and by their very nature, leaders have more power than others. Do we, therefore, end up with a kind of oligarchy of technological elites?

An example will help illustrate this concern. When Status released its token back in 2017, the event generated so much traffic on Ethereum that it ground the network to a halt. In the midst of the network congestion, Vitalik Buterin directly contacted the founders of Status (one of whom coauthored this book) to find out what they were doing and to see if the problem could be mitigated.

Today, Buterin does not have much in the way of direct control over the Ethereum protocol, but he has plenty of indirect control simply by virtue of the credibility he has established by being the network's founder and by virtue of the proposals and ideas he continues to put forward. He plays a huge role in the direction of Ethereum. It is doubtful – impossible to believe, really – that Ethereum could have converted to proof of stake had Buterin not wished it. Of course, the case of Buterin and Ethereum is not unique. Important and charismatic founders can be found throughout the crypto sector, and to some degree or other, every founder has a huge influence on their protocol. Examples include Charles Hoskinson (Cardano), Andre Cronje (Yearn Finance, Fantom), Hayden Adams (Uniswap), and the list could go on and on. It is doubtful that a DAO would work against the wishes of one of these founders if they had stated a clear preference for a decision.

One might think that Bitcoin is an exception to this phenomenon because its founder, Satoshi, appears to have vanished, but there are still important Bitcoin developers and stakeholders that carry enormous clout. Famous examples include Gavin Andresen, Wladimir J. van der Laan, Pieter Wuille, Cory Fields and, of course, the Bitcoin gadfly Luke Dashjr. Indeed, as we write this, Luke Dashjr is involved in a campaign against ordinal inscriptions, claiming they are a 'bug' and a 'fraud' and an 'exploit',[6] and some observers have complained that he is using his clout as a developer to lobby against a Bitcoin feature has become reasonably popular (particularly with Bitcoin miners).

Again, to be clear, we are not saying that Luke Dashjr or any of these other figures are steering Bitcoin in a bad direction. We are simply saying that they have a lot of social clout in the Bitcoin world and that, conceptually speaking, a cult of personality is possible, and one can see how such a cult of personality might form.

---

[6] <https://x.com/LukeDashjr/status/1732204937466032285> [accessed 30 October 2024].

According to some accounts, every attempt to organise humans has this element of hidden oligarchy. Back in Chapter 2, we mentioned Bertrand de Jouvenel's concerns about the power and violence of nation states. However, he also observed that there was typically an oligarchy – perhaps hidden – exercising control in such states.[7] The notorious German political theorist Carl Schmitt made a similar observation about groups that purported to be egalitarian. During periods of stability, it may appear that the group is free of hierarchy, but then, in times of crisis, we see power brokers emerge and take charge. In Schmitt's view, they were always there, and they always had the power.[8] Like Buterin contacting the founders of Status, the hidden leaders only show their hand when things have fallen into disarray.

Our point here is that there may be a practical, if not entirely conceptual, limit to how egalitarian a protocol can actually be. Even on online message boards, there tend to emerge board 'leaders' that carry a lot of clout.[9] The same could be said for crypto influencers on online platforms like X (formerly Twitter). The political landscape is not entirely flat in blockchain communities.

It is an interesting question as to whether such new oligarchies are a good thing or a bad thing. Human organisation seems to require leaders, and good leaders do lead to successful communities. Furthermore, some people are natural leaders and some people look for leaders. That will be true in the blockchain world, just as it is in the traditional world of centralised authorities. However, there are important differences between leadership in the blockchain world and leadership in the centralised world.

---

[7] Jouvenel, *On Power*.

[8] Carl Schmitt, 'The Tyranny of Values, 1959', *Counter-Currents*, 2014 <https://archive.ph/flqkJ> [accessed 16 April 2023].

[9] There is a great study of this phenomenon by the early Internet writer Carmen Hermosillo, aka Humdog, in her essay 'History of the Board Ho' (see Carmen Hermosillo, 'The History of the Board Ho', *The Alphaville Herald*, 2004 <http://alphavilleherald.com/2004/05/the_history_of_.html> [accessed 16 January 2024]).

Centralised governance serves to establish moats that protect and reward leaders who do not deserve their elevated position and who do not serve our interests. Part of the problem is that in a world of nation states, where exit is not a realistic option for the vast majority, many people are, by necessity, kettled into spaces with people who do not share their values and with leaders who are not aligned with those values.

However, centralised governance also tends to build moats protecting economic classes. Thanks to the corruptibility of centralised banking and the inevitable debasing of national currencies, many wage earners are frozen out from ever being financially secure. Meanwhile, people who hold physical assets can tolerate the debasing of their national currency just fine. Although it is not impossible for people to turn an hourly wage into vast wealth or simply economic security, that path is far from frictionless. Furthermore, if leaders are drawn from the class of asset holders, then they can hardly be expected to represent the interests of wage earners in the long run. Even if they provide entitlements to wage earners, they typically debase the currency in doing so. Thus, the real winner is the asset-holder class.

There are many more moats than this protecting the interests of our de facto oligarchy in the age of centralised governance. Important economic information can be siloed to the advantage of those in the know. For example, business secrets about things like impending mergers can be kept within a narrow circle. In the decentralised blockchain world that we envision, such moats would be less common. Assets like BTC and ETH provide ways to avoid the debasement of currencies. The transparency they bring to governance means that no one gains an edge by hoarding secrets. And because technological developments in crypto are typically onchain and visible to all, it is easier for people to catch up technologically.

In the end, we cannot say that there will be no leaders in the blockchain world, and if you wish, you can call these leaders the new oligarchy. However, the difference is that now the interests of the oligarchy are better aligned with our own – not just better aligned but demonstrably better aligned because of the transparency of their onchain actions and, in particular, the transparency of the smart contracts and online protocols they deploy.

Finally, we believe that the pathway to being part of the new oligarchy is more frictionless and more merit-based than before. Of course, we recognised that similar claims were made with the emergence of liberalism during the eighteenth century, and there is no doubt that this represented a great step forward. However, to keep advancing towards an egalitarian society, we need to dismantle the information siloes and fill in the moats of the centralised world.

As we said, there are going to be conceptual limits here. There will be a new class of leaders, and leaders will inevitably fail us, and deserving people will inevitably be left out of leadership positions. We cannot solve for this conceptual limit with technology alone, but we can make the situation much better than it is today and perhaps move ever closer to delivering on the egalitarian promises of the eighteenth-century revolutions. As we will see in the next chapters, we will also need a realignment of our values if we want our new technology stack to work with us instead of against us.

## 15.5   The epistemological limits of oracles

In our discussion of trust at the beginning of this chapter, we briefly glossed over an important issue that deserves more attention. We mentioned the difficulty of ensuring that the data structures in computer code are reliable representations of what is going on in the so-called real world. If you missed the point, we understand; it came up during our discussion of how one might come to trust a smart contract that promised the delivery of orange juice concentrate. The problem, of course, was that oranges grow on trees and are not data structures in a computer. Smart contracts can only act on computational data structures, and this means that at some point in the process, information about real-world physical oranges and real-world orange juice concentrate needs to be reliably recorded onchain.

We described this as an issue about trust, which of course, it is. However, it also reveals a deeper conceptual issue regarding the connection between events in the real world and representations in smart contracts and blockchain ledgers. Humans also have this problem and applied to humans, philosophers refer to it as a problem of our knowledge of the external world. How can we rely on our sense organs to reliably tell us about the external world? In Chapter 6, we introduced oracles, which are like sense organs connecting the blockchain to the external world. As we might expect, this raises all sorts of philosophical issues.

The problem of blockchain oracles is difficult because computations that take place onchain typically (although not always) involve information about the extra-blockchain world. You can have a stablecoin that tracks the value of the US dollar, but it requires an oracle to inform the smart contract as to the actual price of the US dollar relative to that of a specific cryptocurrency. Or you might want to engage with an online betting market to place bets on the outcome of football games or elections. Once again, one relies on there being a reliable source of information offchain and a reliable tool to move that information onchain. There are, of course, organisations that do this, most notably Chainlink, but these organisations raise questions of their own.

The first and most obvious question is really an extension of the conversation that we had in the second section of this chapter – oracles potentially present points of centralisation and, thus, vulnerabilities. To understand the problem, consider this. We could have a perfectly decentralised blockchain (yes, perfectly decentralised is impossible, but this is just a thought experiment); however, if all the offchain information being imported to the chain is coming from a single source, what did your decentralisation actually accomplish? Sure, a perfectly decentralised blockchain ledger would be difficult to tamper with, but if you can take control of the oracle and corrupt the feed of the value of the US dollar or of gold or of the current price of Amazon stock or of the price of orange juice, then the information imported onchain is corrupted.

In this chapter's opening section, we also saw all of the hoops that one would have to jump through in order to resolve this problem. In effect, one would have to continually make more and more real-world information into onchain information. And how does one do that? To answer that, let us remain with the orange juice example, as it provides a good illustration of the problem and of the obstacles you would have to overcome in order to solve it. Let us say that a smart contract involves the exchange of a certain amount of a US dollar stablecoin for a certain amount of orange juice delivered to your store. How does the smart contract know that the orange juice has been received? Well, someone has to record that information onchain.

Now, we must question whether the information is reliable and one way to make it more reliable is to expand the scope of the information imported to the blockchain. For example, we might include shipping information about orange juice as it leaves the factory. However, the problem does not end there. How do we know that it was orange juice that left the factory?

It is not like there is nothing to be said about the orange juice leaving the factory. We certainly would have quality control people entering information about the purity of the orange juice in the factory, but how do we trust those human points of failure? We can keep expanding the amount of onchain information. We can input the information onchain at each step in the manufacturing process, beginning with taking oranges into inventory and sending them to the juicers. To assure ourselves that those records are also reliable, we can track the orange shipment from the orange groves, and we can presumably also record information from the harvesting equipment – if desired, modern harvesting equipment can record the GPS location for every orange that it picks down to the very tree from which it picked it. However, that information could also be corrupted because any recording of information can be corrupted. We could utilise an army of bots and drones to verify the authenticity of the information being recorded at harvest and perform audits on the quality control equipment. However, then we might need drones to audit the drones and so on. It plays out just as it did in our discussion of trust above.

While we never arrive at a completely secure foundation of Cartesian certainty, at each iteration of the process, we make it progressively less likely that our information is corrupted. Maybe you can get the store delivery service to lie about receiving orange juice, but you also have to get the trucking service to lie about the delivery and the initial factory to lie about loading the trucks. And you need the factory to lie about producing the orange juice and the quality control people to lie about the measurements they took. And you have to get the shipping service to lie about the delivery of the oranges to the factory and the farm to lie about picking the trees and loading the trucks, and have someone to lie about the AI drones that audit various parts in that whole process. In other words, it would take an epic conspiracy to successfully execute such a scheme.

Now, of course, we recognise that there are points of weakness in this process. Maybe one dishonest trucker can disable his GPS, deliver quality orange juice to a different customer, and deliver inferior orange juice or convince those overseeing the shipping and receiving processes to say that the orange juice was received and record it in inventory. However, it is not trivial business because the audit trail is massive. At the end of the day, it is highly probable that the oranges were harvested, delivered to the factory, shipped out of the factory and delivered to the store, with information recorded onchain at every step of the audit trail. As noted above, we also have to trust fellow humans at many points of that audit trail. However, the longer the audit trail, the more reliable the information that ends up being fed into the smart contract. This is because a smart contract oracle does not need to merely input the record of receipt of orange juice; a good oracle will peer as deeply into the audit trail as necessary to mitigate financial risks.

Earlier, we mentioned Chainlink, which is an oracle network. It is worth noting that it uses a different strategy from the one outlined above. Chainlink oracles are actually more like small networks of individuals who are tasked with providing reliable information. People participating in these oracle networks need to stake assets, which can be forfeited if they provide incorrect information. However, we can take this general idea and expand on it.

For example, if you wanted an especially secure node for football game scores for a high-stakes gambling service, you might want more than a news feed from a service like the BBC because you might worry that a single feed could be hacked. You might also want more than one pair of eyes reporting to your oracle. Therefore, someone in your oracle network might go to the games, someone might monitor the news service and someone else might query the league office for scores. Each of these individuals would operate a node, and each of them would stake resources which would be forfeited if they reported an incorrect score. If each of our human observers operates a node, then the oracle is actually a network of these nodes. Engineered correctly, such an oracle will be decentralised and will survive individual points of failure. It may well be an organised blockchain community itself.

With oracles like those following the Chainlink model, we are back in the business of trusting other humans, but we have somewhat decentralised that trust, and we have made sure that there is a penalty for providing false information. Therefore, it is a peculiar kind of 'trust' in which a Sword of Damocles hangs over the person we trust (they could lose their stake if they fail to report accurate information), and we are also distributing that trust among others in the oracle network, each with a Sword of Damocles hanging over their heads. Furthermore, incentives encourage nodes to perform reliably, and nodes that report accurate information are rewarded (oracles are a service for pay, after all).

What is the upshot of all this? The conceptual limit is that there cannot be absolute certainty about the information that is being fed into the blockchain via oracles, but by putting more and more information from the audit trail onto the blockchain, we gradually increase the probability that the information is reliable. Similarly, as oracles add more nodes, they become more robust – ideally, they become Byzantine-fault-tolerant networks themselves. These protocols are not vapourware but are being brought online today as we write this. To be sure, there is a conceptual limit to the reliability of oracles, but it seems to be one with which we can live if networks and oracles are correctly designed. The blockchain cannot have perfect knowledge of the external world. However, robust decentralised networks can achieve a level of veridicality that centralised information aggregators cannot.

## 15.6   Freedom and conservatorship

On 1 February 2008, American pop singer Britney Spears was involuntarily placed under a conservatorship with her father, Jamie Spears, and attorney Andrew M. Wallet as the conservators. The conservatorship lasted until November 2021 and raised questions over conservatorship laws in the United States and more than a few cries of 'Free Britney' from her fans. At its core, the idea behind a conservatorship is that the law can determine if an individual, for reasons of age or deteriorated mental health, cannot take care of their person or finances. In Britney's case, the concern was both for her person and her finances. The results of the conservatorship were perhaps predictably troubling, as Britney brought a lawsuit with charges of mistreatment, coercion and conflict of interest. On 7 September 2021, Jamie Spears filed to terminate his conservatorship, although Britney's fans will tell you that this was not out of respect for Britney so much as to avoid discovery, depositions and, potentially, a trial involving Mr. Spears.[10]

Our concern here is not with Britney per se (although we are happy she has been subsequently freed) so much as it is with all the other people put in this position and also with the very unclear line between people who need conservatorship and those who do not. Finally and most importantly, we are concerned with the conceptual issue of when blockchain communities are entitled to permit conservatorship over their citizens. Here, we are returning to an issue raised much earlier – the rights individuals within blockchain communities have to know about external communities, the rights they have to exit their community and so on. Do blockchain communities have the right to deny the rights of seventeen-year-old citizens to explore other communities? Do they have the right to deny exit? And if the age of liberation is not seventeen, when is it? Is age the only reason to insist on this sort of conservatorship, or should other considerations like emotional health be factored in, too? Can a blockchain community restrict the right of exit and access to individuals like Britney Spears?

---

[10] Elizabeth Wagmeister, 'Britney Spears' Father Felt Pressure to Terminate Her Conservatorship. But What's Next for the Pop Star?', *Variety*, 9 October 2021 <https://variety.com/2021/music/news/britney-spears-conservatorship-jamie-spears-next-steps-1235059326/> [accessed 30 October 2024].

Now, obviously, this is not really an issue about blockchain technology, but it certainly is an issue about the rights of community members, and whatever the technology grounding a community, there is an issue to be addressed here. The issue cannot really be escaped by blockchain communities. The question is, do blockchain communities have something to contribute to discussions about conservatorship? Do they add a new factor to be considered? Is there hope for future Britneys in the form of blockchain communities? Maybe.

In the first place, as noted throughout this book, decentralised blockchain technology offers us more transparency, and presumably, this sort of transparency could only help people like Britney by ensuring that the details of their conservatorship are kept secure and accessible to the relevant auditors. It should not be necessary to go through a lengthy and expensive discovery process to investigate whether a conservatorship is being handled correctly. Recall that Britney's father released her from her conservatorship when faced with the fact that he might have to be deposed in a court of law and have his records subpoenaed – in other words, he was concerned that his actions would become transparent, at least to the court.

Once again, we can see the importance of transparent, immutable records. One feature of such records is that they allow the entire community to understand the consequences of decisions being made with respect to conservatorship. The community's decisions about conservatorship can be better informed, whether those decisions are being made on behalf of a pop star or an entire age group of children. Although decisions can be better informed, there is no guarantee that bad decisions cannot be made. It will be up to the communities to make these decisions for better or worse, and here we see the conceptual limits for blockchain communities; bad decisions can always be made by fellow humans. The best that blockchain technology can do is help these decisions be better informed and their consequences better understood.

## 15.7   The conceptual limits of transparency

In previous chapters, we carved out a broad class of things that citizens have a right to know about the conduct of their blockchain communities, but there are limits to what can and should be accomplished with respect to transparency. Total transparency may not be viable, for it is reasonable to think that community governments have to keep certain information secret – either for reasons of citizen privacy or security or economic strategy. For example, consider strategies for negotiating trade agreements – you do not want your maximum offer price to be public knowledge. Here, we get into interesting and difficult territory. If there are to be secrets, who will classify them as secret? And whoever it is, we seem to be back with a central authority and the consequent abuses that extrude from a central authority keeping something secret.

The dangers of allowing centralised secrets should be obvious. There have been countless examples of politically embarrassing facts, moral failures and politically damaging facts being classified as state secrets. In one of the most famous examples, the *Pentagon Papers*, US military analyst Daniel Ellsberg leaked confidential internal documents that acknowledged that the war in Vietnam was not winnable by the US.[11]

However, perhaps a blockchain community can maintain secrets without centralising them. We naturally suppose that state secrets must be determined by a central authority because otherwise, how are they kept secret? If everyone knows it, it is hardly a secret. Therefore, it seems fair to assume that there must be a single official secret keeper. However, maybe this is not the case. Perhaps decentralised secret keeping is a possibility. One way to achieve this is via a protocol known as 'Shamir's Secret Sharing', named after a 1979 paper titled 'How to Share a Secret' by Adi Shamir.[12]

Shamir's Secret Sharing involves breaking encryption keys into pieces and distributing them to different members of a community. It then takes a specified number of community members to reconstruct the key. The theory is that the key will only be reconstructed for a specific purpose and only with the approval of a large enough group of community members.

One issue that Shamir's Secret Sharing does not resolve is the problem of *classifying* secrets. Can that process be distributed as well? One idea is that we can decentralise the classification process if we, as a community, publicly identify the kind of information that we think should be classified and then leverage smart contracts and discreet oracles to gather that information and keep it secret until the community agrees to release it. This, in turn, requires that the information being gathered is offchain since any onchain information is, by its very nature, very public. Thus, we would need smart contracts that could access black box oracles and store the relevant information in their own black box, only to be accessed upon some community-agreed-upon actionable purpose.

There are all kinds of challenges that flow from such a strategy. Firstly, we have already talked about the fact that oracles tend to be points of centralisation. However, the other issue is that the strategy requires that certain crucial state secrets be entirely in the hands of black box oracles in which no human knows their contents. We can use zero knowledge proof strategies to ensure that the oracles are behaving correctly in their classification of secrets, but even if we know enough about the oracles to be satisfied that they are good actors, they may still be vulnerable to attack.

Finally, there is the problem that tensions between the need for governmental transparency and citizen privacy will arise. For example, we often hold that leaders, like presidents, should make their medical conditions and financial holdings public, but in a blockchain world in which potentially anyone can become a leader of importance, the line between leader and citizen is very thin. Perhaps bots can be enlisted to review medical records and financial holdings, but as before, this raises questions about the trustworthiness and safety of the bots.

## 15.8   The limits of voluntary forfeit

A good case can be made that certain moral rights are voluntarily forfeitable. A Puritan community might forfeit its right to carnal pleasures or even the right to music and other forms of entertainment. After all, during the Reformation, the city of Geneva banned musical instruments and singing in harmony. You can imagine a community banning rock and roll or reggaeton music. However, even here, we get into interesting questions. If music has been effectively banned, how do you know enough about music to fully grasp what it is that you have agreed to forfeit? This seems to suggest that one must have the ability to temporarily step outside of voluntary self-censorship and witness how the rest of the world carries on.

The issues here are perplexing. A community may think it is in its best interest to protect members from outside temptations, and we can concede that such outside threats may, in fact, be threats to the community. On the other hand, intentional communities do survive even while quite obviously surrounded by communities with very different morals. The groups that accomplish this are too many to list exhaustively, but obvious examples include Hassidic Jews, the Mormons, the Quakers and the Mennonites, all of which survive quite successfully even while surrounded by peoples with quite different moral practices.

Thus, while access to the external world (and the visibility of alternative lifestyles) presents a certain level of threat to community survival, it is not a mortal threat, and we would suggest that any threat is trumped by the need for people to intelligently pick the community and system they want to live in. Citizens of North Korea may believe they live in the best country in the world, but this belief is forged with the help of their ignorance of the world around them, thanks to government censorship. They did not enter into those conditions voluntarily, but voluntary or not, no government should keep its population ignorant of the rest of the world. Access to knowledge of other communities is a right that cannot be forfeited because if the right to knowledge of alternatives is forfeit, the right to exit is undermined.

How far does this right to knowledge extend? Must a community provide a fair and balanced picture of the external world to its community members, or is it free to spin the external world as being defective or even depraved? The difficulty, of course, is that there is no fair and balanced picture of the external communities, and even if there was, there would be no neutral authority that could dictate what that picture is. Thus, it seems that while communities cannot restrict access to the outside world, they have the right to spin the merits of the external world as they see fit. Fair-minded communities will strive to be balanced, but individuals must find their own way to navigate through whatever propaganda their community generates. Acquiring knowledge from outside the community is not a frictionless enterprise.

Thus, individuals have the right to exit and the right to government transparency, and they also have the right to peer beyond their communities and sample external communities in a (mostly) unfiltered fashion. These rights cannot be forfeited. But might other rights be forfeitable? To illustrate with a case we discussed earlier, could one join a community in which one is a voluntary slave?

Here again, we bump up against the issue of demands that are generated by the right to exit. Even voluntary slaves cannot surrender their right to knowledge of the external world and their right to knowledge of internal facts about their government. It follows that they cannot forfeit their rights to basic education or to virtually visit other communities as a free person or do anything that might allow them to know what they are missing out on given the choice they have made.

There is also the issue of whether you can forfeit your right to the material resources to exit and start over somewhere else. Maybe volunteer slaves must post a deposit sufficient for them to move on should they choose to do so one day. To some extent, this means that pure voluntary slavery is not an option, given that the capacity for exit must always remain available and may not be surrendered. So too, the knowledge of exit possibilities and alternative life experiences must remain available and may not be surrendered.

To push a little further on this point, let us imagine someone who chooses to be ignorant of alternative life possibilities. They buy into the culture of their community and voluntarily take a pill or engage in mental exercises that steel themselves against acquiring knowledge of alternative cultures and possible life experiences. It is inevitable that we encounter difficult cases on the margins, and quite clearly, these count as difficult cases. Taken to its limit, freedom means freedom to surrender freedoms and freedom to be ignorant. However, we all recognise cases in which we wilfully did not consider options and came to regret those choices. What then of cases where we are never in a position to question those earlier decisions? Does this not make us prisoners of decisions made by our past selves?

This suggests that people should always have the tools and resources to unwind their past choices, and that communities have the responsibility to provide these tools and resources, even when doing so appears to conflict with the integrity of the community. The right to exit trumps community integrity.

This chapter has been about conceptual limits, and accordingly, we are not saying we can solve these conundrums. Hard decisions are coming. As we said, the goal of this chapter is to understand the limitations of what can be accomplished with the deployment of decentralised blockchain protocols.

By now, it should be clear that no technology stack by itself can solve for all of these conceptual problems. The bigger issue is that we can really only expect these technologies to work if community values are correctly aligned. There is no victory in distributed trust if everyone is untrustworthy. Similarly, there is no victory in a decentralised network if every node is left vulnerable. Of course, there is no victory in a fixed monetary supply of 21 million BTC if miners and developers do not defend that maximum supply. Human beings need to be value-aligned with the technology stack. We turn to this issue in the next chapters.

# ARE BLOCKCHAIN COMMUNITIES INEVITABLE?

## 16.1 Preliminaries

In this book, we have made the case that autonomous blockchain communities can do many great things for us. We argued that they could minimise human conflict by minimising the phenomenon of diverse groups being kettled together within the same nation state. We have argued that they can minimise corruption by introducing decentralised, immutable records and that they are resistant to internal and external attacks through the deployment of Byzantine-fault-tolerant strategies. We have argued that they can avoid economic failures by relying on decentralised blockchain currencies, and finally, we have argued that blockchain communities can be harnessed for regenerative public goods and positive externalities. By now, hopefully, they sound like a great idea. The question is, are they even possible?

Scepticism here is not surprising. The picture of governance we are painting is radically different from the picture of governance to which we are accustomed. We are, after all, accustomed to nation states that have established physical territorial boundaries and that are granted sovereignty over that territory. We are accustomed to those institutions and other centralised institutions (such as the United Nations and the Organization of American States and the International Monetary

Fund) calling the shots in our world. These are the institutions that create the laws that govern us, that control our currencies and economies, that go to war, that tax us, that control our movements on planet Earth, and so forth. Nation states are found on every piece of territory on Earth. They are ubiquitous. We were born into this system, as were our parents and grandparents. It is quite hard to imagine things being any other way. Is all this talk of cyberstates and sovereign blockchain communities not simply too pie-in-the-sky to be taken seriously?

It is certainly true that none of us alive have known another international order, but as we observed in the introduction to this book, this Westphalian order was not always here. More importantly, changes in human governance often arrived in the context of people not being able to imagine any other way. However, new ways of governing did emerge.

There was a time, not very long ago, in which monarchies gave way to democracies. These shifts in governance may have seemed wildly implausible at the time. Even the shift from an absolute monarchy to a constitutional monarchy with minimal constraints on the ruler was considered wildly implausible at the time. Of course, it seemed that way because that was simply not the order of things that people were used to. Kings had divine rights – until they didn't.

Perhaps the most interesting element to all of this is that when those great transitions in the form of human governance took place, the spark for change was often something that might have appeared insignificant and trivial. However, the other remarkable thing is that when change finally came, it seemed so obvious that it was almost as if the new order already existed. And perhaps, in a way, it already did. If that sounds paradoxical, stay with us; it should make sense by the end of this chapter.

## 16.2   Seeds of cybergovernance now

In 1847, in Paris, a number of banquets were held. Each of them was a social and cultural event, but mostly what we might call a 'vibe' today. Within a year, King Louis-Phillipe would fall from power.

These Parisian banquets were copied elsewhere in Europe and ultimately contributed to the many revolutions that swept across the continent in 1848. The banquets, although social gatherings, were considered subversive and often banned. But why? Why ban a little party? Why ban a vibe?

The banquets were considered subversive because they brought people together under an attitude – an attitude opposed to centralisation of authority, an attitude opposed to the top-down imposition of cultural norms. Thus, the very act of gathering socially was subversive as, of course, was the motive of the gathering.

Our point here is that the seeds of a new order of decentralised blockchain governance might not be what you expect. They might stem from a series of social events rather than an organised political movement. Let us consider a possible scenario to see why this might be so.

Balaji Srinivasan has argued that the NFT community Friends With Benefits (FWB) might be an example of an organisation that evolves into a more robust community and possibly even a cyberstate (what Srinivasan calls a network state). The membership requirement for the group consists of holding a certain number of the cryptocurrency FWB and answering questions about your occupation and interests.

The group members regularly maintain dialogue within chat platforms like Discord, and beyond this, there are regularly scheduled 'ask me anything' chats hosted by group leaders. There are informal meetings of group members in various cities, but the big events are the large social gatherings in different cities around the world. On the face of it, there is nothing more to it than that – just people chatting and organising parties.

But let us take a closer look. There is an actual governance structure underlying the decisions about where to hold the next social events. It is an example of participatory online democracy. More to the point, the community is not grounded by a shared interest in parties so much as shared views about the importance of decentralised technologies in all aspects of its members' lives.

Indeed, if you dive deeper into the various archives within the FWB platform, you will find plenty of writing about cyberstates, using blockchain technologies for regenerative public goods and so on. It is a group for holding social events, but even that task can be highly political in and of itself. As Hakim Bey wrote in his classic essay *T.A.Z.: The Temporary Autonomous Zone*:

> Let us admit that we have attended parties where for one brief night a republic of gratified desires was attained. Shall we not confess that the politics of that night have more reality and force for us than those of, say, the entire U.S. Government?[1]

Or let us take a similar example. There is no overt political message to the Bored Ape Yacht Club (BAYC), which appears to be an NFT collection that is driven by online gaming and social events like its annual ApeFest. However, there is a message behind BAYC culture – behind its vibe. Indeed, it is arguable that the criticism of BAYC stems not from it being a 'scam' but from its rejecting top-down culture. It is a nascent cultural movement that celebrates community-based culture.

Are we saying that these NFT-based communities will spawn the blockchain communities and cyberstates of the future? No, our point here is that no one knows what the exact drivers of the new forms of human governance will be. Those who attended the banquets of 1847 may have had no idea what these events were leading to. They had a diverse set of political views, but they shared a vibe. However, nothing is just a vibe. Nothing is just a party. Sometimes, they are doors to the unimagined future.

Keeping in mind that new forms of governance might emerge from unlikely places, let us consider some alternative scenarios that follow a different path. Let us imagine that current blockchain communities grounded in shared economic interests evolve into something that takes on the roles that states hold today.

---

[1] Bey, *T.A.Z.: The Temporary Autonomous Zone*.

Consider an example like Uniswap and its DAO, membership of which is contingent on holding its UNI token. To be sure, Uniswap is an important platform, and it may well become the largest and most important trading platform in the world – eventually eclipsing the NASDAQ and NYSE trading platforms. This could happen because Uniswap provides a decentralised platform that cuts out middlemen and is capable of hosting any sort of trade. If you can tokenise an asset, you can trade it on Uniswap, and you can trade it without any centralised authority or needless intermediaries. Just as significantly, because it is an automated market maker protocol, it algorithmically determines prices based on demand and available resources in its liquidity pools. If Uniswap does indeed become the largest trading platform in the world, then its DAO will surely become a politically significant player on the global stage.

Of course, there is a big difference between being an important trading platform (even the most important trading platform) and becoming as powerful as a state. Still, if you think about it, if Uniswap becomes that important, it will render many of the key functions of the nation state otiose. The code in the Uniswap smart contracts will take on many of the responsibilities of the state, including auditing transactions, and enforcement of trades will become automatic – agreed-to trades will happen, whether you want them to or not.

So far in this chapter, we have been talking as if we are gazing into the future – imagining scenarios that are grounded in the present but still very much speculative. However, if we step back and take a broader view of matters, we will find that this is not really speculation but rather an adjustment of our way of understanding the present. If we know what we are looking at, we will find that many blockchain communities are already here and already playing an important role in life today.

To see this, let us look closer at the Ethereum protocol. As we write this, there are around 6,500 Ethereum nodes running around the world.

All of those nodes have agreed to participate in the network and, thus, have agreed to its technical requirements. They have also agreed to be fair players in that they understand that bad actors will be penalised.

Now, one might say that this is not very impressive because Ethereum is nothing more than a specialised network of computers, but the reality is that it is much more than that. It is also a community (already established), and its decisions, arrived at collectively, already play an important role in the welfare of network participants as well as in the positive externalities that network members are trying to achieve. Or to put it another way, the Ethereum community seems to share a group consensus that it wants to build a better world, but it is also here, now, today, working on behalf of the interests of network participants and here, now, today, it is building out positive externalities consistent with the values and ethical principles held by community members.

Perhaps this point needs further elaboration. Another way to put it is that the Ethereum protocol is not merely *like* a blockchain community or a cyberstate, and it is not merely a platform that *will* give rise to such governance structures. It is already such a governance structure, and it is already working on behalf of community members, and it is already building out the new legal architecture for a post-nation-state world.

We sometimes think that laws and computer code are very different things, but as Lawrence Lessig observed in his book, appropriately titled *Code: And Other Laws of Cyberspace*, computer code can and should be thought of as a form of law. More precisely, we should think of the moderation of behaviour (whether by governments or individuals) as being circumscribed by a number of conditions, of which traditional law is just one. As Lessig points out, we do not pass laws against stealing skyscrapers because they are too big for someone to snatch and run away with. The laws of physics constrain the set of possible behaviours here. Similarly, certain cultural norms might constrain behaviours, as may market forces. Finally, architecture can constrain behaviours (walls, for example, can control where you cannot walk, and bridges can allow you to walk over gorges that you otherwise might not be able to pass over). However, there is also computer architecture and software code that play a very significant role in our world. While such code is not law in the traditional sense, it is still functionally equivalent to traditional laws. It directs the behaviours of individuals and organisations, in some cases, restricting what can be done and, in other cases, enabling actions that might not otherwise be possible.

Lessig illustrates the situation with what he calls the 'pathetic dot'. Where the dot can and cannot go (and presumably what it can and cannot do) is not determined by a single thing but by a confluence of factors, including the aforementioned laws (physical and legal), norms, the architecture of its world (physical and computational) and market forces.[2] For our purposes, the important factor is the computational architecture of the world.

On this last point, the Ethereum protocol is not so different from any major Internet platform. Google and Facebook are also shaping the movement of the pathetic dot. The difference is that when traditional Silicon Valley corporations do this, they do so in a top-down manner. They are our versions of the Westphalian-era kingdoms, imposing their will from a position of centralised authority. Recognising that code is going to shape our world for better or for worse, we much prefer that the reach of the code should be limited to the community for which it is written and that it should be written and understood and supported as a group effort within that community.

Two points deserve to be considered in isolation here. The first relates to the role that Silicon Valley corporations currently play in shaping the legal order of our world – in determining the topology of the spaces in which the pathetic dot can freely move. As we saw in Chapter 13, Major Jason Lowery articulated an extreme version of this idea in his book *Softwar*, arguing that those with control over our software technologies constitute a kind of tyrannical elite. As he puts it:

---

[2] Lawrence Lessig, *Code: And Other Laws Of Cyberspace*, 1st ed. (New York, NY, 1999).

Cyberspace is a globally-adopted belief system that is radically transforming the way society organizes itself, in much the same way that agrarian abstract power hierarchies did. Just as agrarian society led to the formation of empires, so too does cyberspace appear to be leading to the formation of cyber empires, complete with the threat of oppressive rulers rising to the top of the hierarchy.[3]

Does he have in mind people like Microsoft's Bill Gates and Meta's Mark Zuckerberg as being these oppressive leaders? Presumably, yes, although it needs to be noted that they are not acting as traditional tyrants did, with police and armies doing their bidding, but with software code being the shock troops for this new form of tyranny (here, we are not endorsing Lowery's conclusion, just attempting to articulate it). Lowery goes on to hypothesise that 'humanity is going to become so tired of being systematically exploited at unprecedented scales by computer networks by an elite, tyrannical, and technocratic ruling class, that they are going to invent a new form of digital warfare and use it to fight for zero-trust, permissionless, and egalitarian access to cyberspace and its egalitarian resources.'[4]

Major Lowery is on active duty in the military and tends to view the world in pugilistic terms – or at least more pugilistic than we do. Many revolutions throughout human history have occurred without the use of warfare or really anything metaphorically like warfare. The agricultural and industrial revolutions come to mind. Sometimes, people simply see a better way to live their lives, and they adopt the new technology. We hope that is the case here. In fact, we can do more than hope because we can see it happening around us today.

This returns us to the second point made above: DAO-based decentralised blockchain protocols are already shaping their communities,

---

[3] Lowery, *Softwar*.
[4] Ibid.

thereby shaping the futures of their community members and, ultimately, shaping our futures as well. What this means is not that platforms like Uniswap and the Ethereum protocol will become cyberstates or anything closely resembling states. What it means is that platforms like Uniswap and their smart contracts, and protocols like Ethereum and its infrastructure, will replace many of the functions of states. The result is probably not something like a state but rather something entirely new.

For example, in previous chapters, we discussed ways in which the infrastructure of a blockchain protocol can be designed so that you cannot identify the source or destination of any message moving through the network. Network nodes would, therefore, not be in a position to censor other nodes on the network or even censor the transactions that individuals were attempting on the network. Now, this technology, if implemented, would have far-reaching consequences for its network. It would effectively prevent censorship, to be sure, but it would also make it very difficult to economically isolate an opponent on the network. When every packet of information looks the same, your options for censorship and embargo are quite limited. Alternatively, the system could be engineered so that every transaction is tagged with a source and a destination, and this would certainly make censorship and embargo possible. If the network was value-aligned to be censorship-happy, then one could expect quite a bit of such activity.

Our point here is that, to some extent, the future of blockchain networks is very much open ended and being determined today by active members of those communities. Those communities are building out the architecture of their future. If we think of ourselves as being in a position akin to Lessig's pathetic dot, then blockchain communities are today building out the computational architecture that will determine the fate of those pathetic dots within their respective communities.

This sort of scenario is not just the case for DAOs on the Ethereum network, but if you think about it, the same can be said about the Bitcoin protocol. To be sure, there is a very robust community surrounding the Bitcoin protocol, and there are open debates about the future of the network that either go nowhere or result in some form of consensus or, alternatively, a fork of the network. We previously mentioned the ongoing dispute over whether the Bitcoin protocol should allow ordinals, but such a debate is not new to Bitcoin. Between 2015 and 2017, the Bitcoin community engaged in a debate that subsequently became known as the *Blocksize War*, and it was chronicled in a book by the same name.[5]

It is important to recognise that the community surrounding the Bitcoin protocol is very much like the blockchain communities we have been discussing. Despite the hype, Bitcoin did not fall from the sky (or even from Satoshi) in immutable form. There have been and continue to be robust debates about the future of Bitcoin. Sometimes, these debates lead to stalemates and, in turn, to forks of Bitcoin (for example, BSV, which stands for 'Bitcoin: Satoshi's Vision'). The key thing to note here is the very thing we have been talking about throughout this book. The Bitcoin community, like all good decentralised blockchain communities, has no single leader with decision-making authority. Future changes to the protocol are the result of recorded debates and, hopefully, consensus. When consensus cannot be achieved, members may freely exit and, if they so wish, create a new protocol by forking the original. And behind it all, there is a set of values (and vibes) that guide the contours of the debates. However, this is the situation today. What is coming tomorrow?

Eventually, enormous resources will fall into the lap of communities like the Uniswap DAO, and it will be up to the DAO to determine how those resources will be used. For sure, some will be used for future development of the platform, but is it implausible to think that DAO members might want to allocate resources to external concerns, such as assisting refugees or developing renewable energy or fighting human trafficking or supporting any other causes that might be of interest to DAO members? If a DAO is capable of taking up external causes, it is also certainly capable of taking up the personal concerns and interests of DAO members. Is there any reason the individual rights of DAO members cannot and would not be protected anywhere in the world?

---

[5] Jonathan Bier, *The Blocksize War: The Battle for Control Over Bitcoin's Protocol Rules* (2021).

You might think that the above scenario sounds plausible but object to the fact that there is nothing inevitable about it, and for sure, there is nothing inevitable about a specific scenario playing out in detail. However, if we are content with thinking in terms of broad trends, then the inevitability becomes apparent. New technologies do get adopted, although not always in the form that we expect. Thomas Edison thought that the principal application of the phonograph would be for business secretarial purposes, serving as a kind of Dictaphone. He did, in fact, mention entertainment and music as possible applications, but those were not the most significant potential applications in his view.[6] Similarly, Edison thought the future of electricity was direct current, but as we know, Nikola Tesla's invention of alternating current carried the day.[7]

The point is that no one is omniscient about details. However, when you have a revolutionary new technology, you can see that *something* is inevitable, even if you do not know the exact form or even the ultimate use of that technology. Edison was correct in thinking that electricity would be ubiquitous; what he did not know was the ultimate form of delivery. He was similarly correct in thinking that the phonograph would be an important invention; he simply did not know in what form. Likewise, when the Internet was initially developed by DARPA, few could see the form it would ultimately take.

---

[6] Library of Congress, 'History of the Cylinder Phonograph', *Inventing Entertainment: The Early Motion Pictures and Sound Recordings of the Edison Companies* <https://www.loc.gov/collections/edison-company-motion-pictures-and-sound-recordings/articles-and-essays/history-of-edison-sound-recordings/history-of-the-cylinder-phonograph/> [accessed 30 October 2024].

[7] Department of Energy, 'The War of the Currents: AC Vs. DC Power', *Energy.gov*, 2014 <https://www.energy.gov/articles/warcurrents-ac-vs-dc-power> [accessed 30 October 2024].

At its inception, the initial thought about blockchain technology was that its principal application would be as an economic tool. Indeed, in the very first sentence of the Bitcoin white paper, Satoshi describes Bitcoin as a 'payment system'. And for sure, economic concerns drove the development of Bitcoin. The economic troubles in 2008 were very much on Satoshi's mind and there is no doubt that the problems surrounding centralised finance were very much a driving force behind his efforts.

We hope we have made it clear that we think that blockchain applications will be much more extensive than Satoshi imagined – or at least, more extensive than articulated in their white paper. For sure, financial uses of blockchain technology will be important, but financial transactions are only one small piece of the puzzle that is human governance, and ultimately, human governance writ large is going to be the most important application of blockchain technologies.

## 16.3   How we can nurture blockchain governance

Let us suppose you agree that blockchain governance is a good idea, and we can already find nascent versions of these future governance structures today. Is there something we can do to help them evolve into the governance structures we are looking for? And if decentralised blockchain communities are indeed inevitable, is there anything we can do to make their adoption as frictionless as possible?

Clearly, any such effort is going to involve a heavy dose of community participation. Simply by participating in a decentralised blockchain community, one can put one's hand on the tiller at critical moments. Using the cases we discussed in the previous section, we can say that in each instance, the task involves expanding the mission, projects and strategies of the blockchain community and moving them in a direction not initially envisioned.

For example, in the case of the FWB community, we can imagine a scenario in which the documents about cyberstates, already archived and discussed by FWB members, are taken to be not just ideas to be discussed but aspirational goals for the FWB blockchain community. This would be an aspiration to evolve from a blockchain community that creates social events with positive vibes to a blockchain community that has its heart set on evolving into a cyberstate of some form – an organisation that does more than entertain its members but enables their flourishing by providing many of the services that nation states do today.

In a similar vein, a DAO designed for economic interests might expand its portfolio as well. The Yearn Finance community might decide that in addition to voting on creating vaults with investment strategies, they might take on the role of representing the interests of their DAO members, stepping in for them as advocates in some cases, becoming involved in the purchase of physical territory and the management of that territory for their community members, and so on. The community could expand its portfolio to international trade and manufacturing and, ultimately, the flourishing of their DAO members. One can even imagine mergers between DAOs here. For example, a social-based NFT community might merge with an economically based DAO. Alternatively, one might just build a community from scratch that had all of these features.

What then is the key to frictionless adoption? Participation seems to be the crucial element. The more one can participate and, when needed, 'touch the tiller' on these projects, the sooner they can mature into the robust decentralised blockchain communities that we envision.

## 16.4   Why the technology is doable

If you have reached this far in the book by reading the earlier chapters, then you already know that the technology is doable. We *have* the technology. Still, let us review those technologies now that we have some aspirational goals in place and some hints at how we might approach those goals.

Recall that the principal needs for blockchain communities are secure archives, decentralisation with Byzantine fault tolerance, ways for people to collaborate in these communities, transparent administration of these communities and corruption resistance. Meanwhile, members have the rails to communicate privately with each other and their business partners, and there are also economic rails, such as cryptocurrencies, in place for this to happen.

All of these technologies exist to some extent today, and in Chapter 14, we provided some very specific open-source versions of them. However, it is worth thinking about how these technologies might be (more rapidly) adopted. In other words, how do we facilitate getting from here to there?

Happily, for existing blockchain communities, the necessary technologies have already been adopted or are at least familiar to community members. For example, let us suppose that the members of FWB acquired the aspiration to be a full-on cyberstate or at least a player on a global scale. What they already have is a blockchain-based DAO. What they need to incorporate are robust voting mechanisms, a secure private communication system for their citizens and an official blockchain-based currency for their community. Clearly, these are already off-the-shelf technologies. Thus, rather than finding the necessary technologies, the real task is to direct those technologies towards the community's aspirational goals.

This direction does not require new technologies but instead new attitudes to go with existing technologies. Communities need to want to use those technologies to expand the footprint of their blockchain community. This is to say that they need to leverage the technologies that they have in order to contribute to the flourishing of their community members, and this will ultimately lead them to take on many of the functions that have historically been the province of nation states and other levels of human governance.

We can already see this movement in the form of cryptocurrencies like bitcoin, which take over the role that government-issued fiat currencies used to have. However, if communities want their members to flourish, they will also work to secure their economic interests, create conditions for shared culture to thrive, and provide social securities and services on a global scale.

Let us illustrate this with an example of what FWB members could do if they wished. They could secure land, or if they had several members located in cities around the world, they could represent the interests of their community members in those cities or with whichever terrestrial authorities controlled the land where they lived. They could formally arrange business relationships (this already happens informally) and agreements. They could help citizens establish businesses in special economic zones. They could facilitate members with security problems. There is a lot that they could do. Indeed, the bigger question is whether there is something they could not do.

We began this section by asking whether the technology was doable, but in the end, we have seen that technology is not really the issue. The issue is whether a blockchain community has the desire to leverage existing technologies to provide progressively more robust services to its community members – ultimately, taking on roles that resemble those of existing Westphalian states. In other words, the technology is here. The question is whether we have the will to leverage it.

## 16.5 Why people will try to develop blockchain communities

We concluded the previous section by asking whether blockchain communities of any stripe are going to have the will to take on more and more services for their community members. In this section, we are going to argue that they definitely will. This is not to say that decentralised blockchain communities are a certainty, but it is to say that people *will* try to develop them. The kinds of decentralised blockchain communities that we have talked about in this book are not merely inert academic ideas. We do not know exactly what forms they will take or how they will take the forms that they do, but people will keep leveraging available technologies to build them out.

This prediction does not flow from any special features of blockchain communities or our vision of cyberstates. It rather flows from the simple fact that ideas for human organisation, no matter how foreign sounding in the beginning, eventually sound less foreign and, in the fullness of time, are eventually implemented in some form – for better or for worse.

This is not to say that all of the ideas attempted have lasted or that they have been helpful; it is rather a point about humans wanting to improve their lot in life and their willingness to try new things to accomplish that. So strong is the human desire to attempt new orders of political organisation that they will attempt them even faced with threats from the powers that be in the form of potential imprisonment, torture and execution. It is an uncanny human trait to want to keep trying new social orders. Now, there is clearly a competing human trait to preserve the status quo, and this is where many conflicts are born, but in the end, new technologies for human organisation are always attempted.

The events surrounding the French Revolution illustrate this capacity vividly. When the French Revolution began, it was simply an attempt to get Louis XVI to accept a constitution and not much more. However, when that request met with violent resistance, the resulting cauldron of ideas generated many projects and theories of governance. Some of those projects did not get very far – the Paris Commune being a case in point. Snuffed out early on in 1871, it lasted less than two months. It went on to inspire a number of thinkers, however, and was inspirational to future governance structures well into the twentieth century.[8]

The same is true for individual thinkers. In 1755, Étienne-Gabriel Morelly published *The Code of Nature*, a pamphlet in which he proposed a utopia in which 'Nothing in society will belong to anyone, either as a personal possession or as capital goods, except the things for which the person has immediate use, for either his needs, his pleasures, or his daily work.'[9] This was well before the French Revolution and in an era when the dominant political debate was between absolute monarchists and constitutional monarchists. However, his ideas were noticed by Engels, Marx and Proudhon, and they were eventually put into effect, for better or for worse.

---

[8] Norman Hampson, *A Social History of The French Revolution* (Hoboken, NJ, 2013).

It is very difficult to think of political ideas that have not made their way to adoption eventually, and enough have been adopted so as to suggest that those that have not been attempted will be adopted eventually. This brings us to what we consider one of the great advantages of the framework we are advocating. It provides a substantially more friction-free way of incorporating new political ideas and studying their success. While some may celebrate great bloody revolutions, we have a strong preference for velvet revolutions, and cyberstates and blockchain communities provide a platform for these non-violent social upheavals. If people wish to implement Morelly's utopia, they are free to try, so long as people within that utopia have the right and ability to exit.

Bending this discussion back around to blockchain technologies, people have been willing to try anything to implement a new political or social order, even if that involves the murder of millions of innocents. One hopes that they would take a path of lesser resistance if they could. And this is yet another reason why blockchain technologies will be deployed. There are lots of revolutionary ideas out there. They can incubate in blockchain communities and take full form in cyberstates, and this can be accomplished without spilling blood. We have the technology to attempt to bring about the flourishing of different forms of governance. It seems inevitable that it will be used as such.

---

[9] Étienne-Gabriel Morelly, *Code de La Nature, Ou La Véritable Esprit de Ses Loix* (London, 2018).

Of course, just because people will attempt to build decentralised blockchain communities (indeed, just because they are attempting it now) does not mean that these attempts are guaranteed to be successful. Nothing is guaranteed in this world. As we saw in the previous chapter, the issue for any technology is that if it is to be successful, the technology must be aligned with our values. This is an issue of such gravity that we dedicate our next and final chapter to it.

# CHAPTER 17

# VALUES AND THE TECHNOLOGY STACK

## 17.1 Preliminaries

In Chapter 15, we touched on some of the conceptual limits of blockchain technology. For example, we saw that nothing is entirely trustless and that the blockchain and protocols living on it are not trustless so much as they involve a rethinking of trust. We no longer trust centralised banks and government institutions to manage our money, but we trust a broad range of community members. It is, if you will, distributed trust.

However, as we saw, distributed trust is still trust. We trust the community of Bitcoin developers and miners not to lose their collective minds and start inflating the Bitcoin supply beyond its cap. Aliens might arrive tomorrow and brainwash us all into forking Bitcoin and inflating it in the same way that central banks and governments inflate fiat currencies. It would not even take aliens. A generation from now, public pressure might prevail on developers and miners to abandon the 21 million BTC limit for 'the betterment of society'. Indeed, as we write this, Greenpeace is lobbying for Bitcoin to adopt a proof-of-stake protocol.[1]

Similarly, we saw that nothing is entirely decentralised and that decentralisation is more of an aspirational goal than something that is here with us today. However, we trust developers and miners and stakers to pursue the vision of a more decentralised blockchain rather than seek ways to make protocols more centralised.

This means that if blockchain technologies are to work, and thus, if blockchain governance is to work, the people responsible for maintaining the technology stack must uphold certain values. If they do not, the project collapses. This may sound like a disheartening note, but it is an important point that we introduced in the introduction to this book: technology does not exist in a vacuum. If it is technology, it is designed to be used by people in an expected way. There are behavioural norms that technology users are expected to follow. For example, our air traffic control system incorporates advanced technologies involving radar and computers, but it also assumes that air traffic controllers and maintenance personnel will follow certain norms in their use of those technologies. The air traffic controllers will use the technology to keep the aeroplanes at a safe distance from each other and prevent them from colliding.

The human element permeates every aspect of our technologies, not just their proper use. We trust coders not to hide malicious bugs in the code. We trust chip designers not to engineer in critical failures. We trust maintenance personnel to replace old parts with new parts and not the other way around. We do not think about this much because these sorts of occurrences are rare. However, it would not be this way if we were different creatures.

---

[1] Tyler Kruse, 'Change The Code: Not The Climate – Greenpeace USA, EWG, Others Launch Campaign to Push Bitcoin to Reduce Climate Pollution', *Greenpeace USA*, 2022 <https://www.greenpeace.org/usa/news/change-the-code-not-the-climate-greenpeace-usa-ewg-others-launch-campaign-to-push-bitcoin-to-reduce-climate-pollution/> [accessed 30 October 2024].

Creatures with different values might find it impossible to build, use and maintain the technology stacks that we use in our world. Their technologies might look entirely different, or it might be that their ability to cooperate is so degraded that it makes technology as we know it impossible. Aliens from other planets, having different values, might have technologies that we simply could not operate reliably.

The point is that it does not make sense to try and engineer away the human parts of our technology stack because the technology stack is larded through and through with human points of contact and thus relies on assumptions about norms of human behaviour, just as it relies on scientific norms governing the properties of silicone and rare earth minerals. Just as we rely on the proper curation of metals in our technology stack – the proper annealing of steel, for example – so too the human element of the technology stack must be curated to preserve certain fundamental values if the technology is to work.

Therefore, in this chapter, we are concerned about the specific values people should cultivate if they are to be part of the development, maintenance and use of the technologies we have discussed in this book. The issue goes much deeper. We do not merely want stakeholders to value decentralisation and commitment to sound monetary policy; we want them to value – better, to preserve the values of – government transparency, the immutability of records, the privacy of individual communications, the right and ability to exit, and so on. No technology can guarantee the preservation of specific values if the people involved in maintaining it do not share those values. People are part of the technology stack, and their values will thus infuse the technology itself. There is no point in trying to divorce the person from the technology.

Before we get into the kinds of values that are necessary to maintain our technology stack, we should perhaps reflect on values in general, for the whole point of blockchain technologies and blockchain communities is that we want platforms that allow people to express their values and have them respected. For example, certain blockchain communities might value freedom of expression more, others might value family more and others might value some form of patriotism more.

The design of blockchain technologies should allow these different values to flourish. However, the meta-level question that we need to address is: What values must we have in order to maintain the technology stack of systems that can allow these different values to flourish in diverse blockchain communities?

Again, we are considering values at two different levels. Level one: What values should blockchain communities nurture? Level two: What values must we uphold in order to maintain the technology stack that can accomplish the level-one goals for multiple communities with diverse sets of values?

## 17.2   Level-one values

For the most part, we have spoken in the abstract about blockchain communities nurturing values, offering examples like family, patriotism, freedom and inclusiveness. However, this still leaves open the question of what values are and what are the different values we are talking about.

There is certainly room for discussion here. The philosophical debate over what constitutes value is not exactly settled, but we can begin to understand the problem by looking at empirical work that offers a fairly tight definition and a menu of values that appear to be universal and which we would thus want to flourish within a blockchain community.

Now, it is important to point out that even if these values are universal, it does not follow that every community will give them equal weight. While it may be that all communities value family and inclusiveness, some communities will value one more than the other. Thus, when we talk about different blockchain communities having different values, it would be more accurate to say that they rank values differently. With this caveat in mind, let us consider some empirical research into this topic.

Perhaps the most well-known contemporary research into values has been conducted by Shalom Schwartz and his collaborators. Schwartz defined 'values' as 'conceptions of the desirable that influence the way people select action and evaluate events.' We can argue about whether this is the optimal definition, but it is at least tight enough to drive his research, which has led to a kind of taxonomy of values – in particular, what Schwartz and his collaborators considered to be 'basic individual values'.[2]

Schwartz argued that these universal values would correspond to three different kinds of human needs: biological needs, social coordination needs, and needs related to group welfare and survival. After surveying more than 25,000 people in forty-four countries across a range of cultures, Schwartz found that there are fifty-six specific universal values that fall into ten categories of universal value. The resulting taxonomy can be paraphrased as follows.

*Power*: authority, leadership, dominance, social power, wealth

*Achievement*: success, capability, ambition, influence, intelligence, self-respect

*Hedonism*: pleasure, enjoying life

*Stimulation*: daring activities, varied experiences, exciting life

*Self-direction*: creativity, freedom, independence, curiosity, choosing your own goals

*Universalism*: broadmindedness, wisdom, social justice, equality, a world at peace, a world of beauty, unity with nature, protecting the environment, inner harmony

---

[2] Shalom H. Schwartz, 'Universals in the Content and Structure of Values: Theoretical Advances and Empirical Tests in 20 Countries', in *Advances in Experimental Social Psychology* (Cambridge, MA, 1992), xxv, 1–65 <https://www.sciencedirect.com/science/article/pii/S0065260108602816> [accessed 24 January 2024].

> *Benevolence*: helpfulness, honesty, forgiveness, loyalty, responsibility, friendship
>
> *Tradition*: accepting one's portion in life, humility, devoutness, respect for tradition, moderation
>
> *Conformity*: self-discipline, obedience
>
> *Security*: cleanliness, family security, national security, stability of social order, reciprocation of favours, health, sense of belonging[3]

Now, these are the values that Schwartz considered universal, and there are surely other values that are important to some cultures that do not make this list. For example, Schwartz tested for spirituality and finding the meaning of life as possible universal values but did not find that they were recognised in all cultures. We have argued that blockchain communities should be in the business of helping different cultures protect and nurture their values, and we do not mean to suggest that only universal values should be recognised as important for blockchain communities. However, universal values are a good place to start; if you cannot accommodate universal values, you cannot accommodate marginalised values either.

The first thing to observe is that even though these values are claimed to be universal, they can clearly come into conflict. Benevolence can clash with security at times. All of the values listed above can come into conflict under the relevant circumstances. Thus, as we mentioned earlier, the issue is not that different cultures necessarily have different values so much as they disagree over the order of importance attached to these values. What ranks higher? – family or benevolence, tradition or universalism? Everyone considers family to be important, but is it more important than being benevolent to neighbours and people at risk across the world? That is a community-by-community decision. Within communities that are neutral on the matter, it can be an individual-by-individual decision. We can also ask questions about how much more important one value might be than another. Is the value in question somewhat more important or much, much more important?

---

[3] Ibid.

We should consider the question of how blockchain-based communities are able to facilitate the flourishing of these values. Consider, for example, a blockchain community that wishes to cultivate the value of family. There are lots of options here, some of them financial and some of them more informational in nature. Let us focus on the latter group first. If the idea is to allow families to stay in contact despite geographic separation, this is easily facilitated through trustless communication protocols like Waku, which we discussed in Chapter 14. If the goal is to preserve a history of family records in a secure way, then a record system like Codex (again, discussed in Chapter 14) is easily leveraged for this task.

Alternatively, if the demand is to allow parents more time for child raising or providing child care, then the blockchain community should be optimised for the allocation of resources to this sort of goal. Of course, traditional communities can do this as well, but blockchain communities can be optimised specifically for such ends, including all of the values in Schwartz's taxonomy. Once values are articulated, blockchains and DAOs can be organised to nurture those values and to prioritise some values over others.

## 17.3   Level-two values

This brings us to level-two values, which as you recall, speak to the values that are necessary to preserve the level-one values and the technology stack necessary for that. To put it another way, we are interested in what the global architecture of blockchain technology must be if it is to preserve the values discussed in the last section and what the values must be for that global technology stack to be successful.

Perhaps the following is a good way to illustrate the task. We saw that certain values are important to all communities and, hence, to blockchain communities. Different communities will prize different values more highly, but our technology stack needs to be agnostic about this ordering. This leads to the question of how we build a value agnostic system and what meta values are necessary to maintain that agnostic system.

Note that when we say the technology must be value agnostic, we are not saying it should be value free. To the contrary, we are saying that it should provide a platform upon which those values, whatever they are, can flourish. If you value family most of all, then the first-order technology stack should provide the resources for you to express that value and also to act in support of that value – it should allow you ways to do things that allow families to flourish. We have already looked at some of the things that might be done to facilitate the flourishing of families, but our interest now is not in this one case but in the idea of a platform that can do the same for any ranking of values.

What this means is that there are elements to which the first-order technology stack must be agnostic (for example, the ranking of values across communities) and other values around which it should be structured (for example, the idea that each community can exist unmolested insofar as it observes key norms, in particular, the right to exit and fair procedures for exile and access). All of these things, in turn, require a second-order technology stack with corresponding second-order values – those values being a commitment to decentralisation, self-sovereignty, security and privacy.

Here, we get to the fundamental concern: these second-order values cannot be assumed to exist; they do not emerge out of nothing. We can clearly see that they are necessary, but we also need to see that we have not only the right but also the responsibility to inculcate these values not only within our communities but across our communities.

Obviously, this can lead to tensions. There may well emerge communities that harbour a religious fervour in opposition to decentralisation or in opposition to a particular ranking of values. For example, communities might emerge in which it is felt that there is only one valid ranking of values – a ranking that places religious interests above all others, perhaps. The existence of a handful of such communities is not fatal to the project (after all, it is Byzantine fault tolerant), but were everyone to take on such values, the project would collapse, not from a failure of the technology per se but from a failure of the values of the global community to support the necessary technology stack.

Any future for a world of blockchain communities, each community being self-sovereign and each community respecting the rights of its citizens to access and exit, is going to be a world in which people are educated in the importance of these values. Such education needs not be mandatory, and it does not even need to be universal, but people interested in the success of global freedom and human flourishing would do well to consider making such education a cornerstone of their community values.

We mentioned it earlier, but it is also worth raising again that another part of this educational effort should be basic literacy in how blockchain technology works. This does not mean one has to get deep into the code (although it certainly would not hurt), but it does mean that people need to understand how decentralised cooperation is possible, how transparent immutable records are possible, and what immutable smart contracts are and how they too are possible. More to the point, it should be the right of every community member to learn about these technologies as deeply as they care to. If someone wants to understand the technology at the level of code, they should feel entitled to that knowledge, and communities should make that knowledge accessible via online tutorials or classes or whichever means is most appropriate. If people do not understand the technology, they cannot trust it, and if they cannot trust the technology, once again, the project collapses.

No technology operates in a vacuum, and if a technology is to be successful, it must be aligned with human values. However, just as technology is not set in stone and can and should be modified and improved, so too the human component of our technologies cannot be considered set in stone. If we want this or any other technology to be successful in the long run, we would be well served by incorporating the human element into the equation. This entails the will and means to provide education about core technologies and the values that serve as foundations for the technologies and governance structures that furnish our world.

## 17.4   Beyond Westphalia

The lesson of this chapter is that we must be mindful of the importance of values and how they are integrated into the technology stack. Every human technology has humans built into it at some point, and the technology thus relies on humans behaving in accordance with certain norms. For example, the technology behind automobiles requires that people use the steering wheel to stay in their proper lanes and not use it to deliberately drive into oncoming traffic. In this respect, decentralised blockchain technology is really no different than automobiles; it has to make certain assumptions about human beings and the kinds of values they hold as they use the technology.

However, if blockchain technology can be properly aligned with our values (and vice versa), then there is great promise in this project. For the first time in history, we can engineer political systems in which people are coordinated without the need for centralised authorities and centralised methods of control. We can be decentralised yet cooperative.

Similarly, we no longer have to be kettled together within physical boundaries established by rivers and oceans and previous human conflicts, but each of us can choose a governing system with which we are aligned – a system that, by its very design, will be transparent in its operations yet grant us privacy in our private affairs.

It is, to be sure, a bold vision of the future. However, given advancements in our understanding of decentralised systems, it is no longer a utopian vision that is out of our grasp. It is within our means, and it is not merely possible but inevitable that, very soon, these new forms of political governance will begin appearing on the political landscape.

When they arrive, they will provide us relief from the failures of the Westphalian order, and they will offer creative alternatives for humans to govern themselves in effective-yet-self-sovereign ways. If we are mindful of the conceptual limits of these new governing technologies and of our responsibilities within these new systems, they will usher in a new era of decentralised-yet-cooperative governance. We should be optimistic about the promise of these new forms of governance. After all, we have nothing to lose but the tyranny of centralised governance, its corruption and all of its barbed wire fences.

# ABOUT THE AUTHORS

Jarrad Hope came to Bitcoin in early 2011 through agorism, counter-economics, and crypto anarchy. Seeing that Bitcoin could operate a monetary policy in a hostile environment, he began to view public blockchains as a voluntary social order, one that did not depend on a monopoly of violence. From there, he participated in early attempts to generalise the Bitcoin script to advance institutional libertarianism, ultimately becoming an early contributor to Ethereum. While advancing privacy technologies through the development of the end-to-end encrypted and peer-to-peer private messaging client and super app Status, Jarrad realised that privacy technologies are not enough and now advocates for self-sovereign crypto networks and the realisation of a latent cypherpunk dream, the cryptostate.

Peter Ludlow entered the world of philosophy through a deep interest in linguistics, the philosophy of language and digital technologies. His early work in artificial intelligence and natural language processing showed him the cooperative part of language comprehension – an idea explored in his book, *Living Words*. This led him to make significant contributions to our understanding of how meaning is a shared, collaborative enterprise. As a leading voice in the philosophy of mind and language, Peter has authored and contributed to influential works on the intersection of technology and society, including the seminal anthology on how cyberspace is poised to impact human organisation, *Crypto Anarchy, Cyberstates, and Pirate Utopias*. His current focus is on the potential for digital platforms to foster self-sovereign communities and new, decentralised-yet-collaborative social orders.